\documentclass[aps,prb,twocolumn,superscriptaddress,floatfix]{revtex4-1}
\usepackage{amsmath,amsthm,amssymb}
\usepackage[dvipdfmx]{graphicx}
\usepackage{amsmath,bm,array}
\usepackage{color}
\usepackage{ifthen}
\usepackage{comment}

\newcount \commentout
\newcommand{\1}{\mbox{1}\hspace{-0.25em}\mbox{l}}

\def\ket#1{|#1\rangle }
\def\bra#1{\langle #1 |}

\newcommand{\calH}{\mathcal{H}}
\newcommand{\calO}{\mathcal{O}}

\newcommand{\calJ}{\mathcal{J}}

\newcommand{\eff}{\mathrm{eff}}

\usepackage{braket}

\renewcommand{\Im}{\mathrm{Im}\,}

\newcommand{\sgn}{\mathrm{sgn}\,}

\newcommand{\muB}{\mu_{\rm{B}}}

\newcommand{\Eq}[1]{Eq.~\eqref{#1}}

\newlength{\figwidth}
\newlength{\figlarge}
\begin{document}
\title{
Topological $d$-wave superconductivity in two dimensions
}

\author{Youichi Yanase}
\email{yanase@scphys.kyoto-u.ac.jp}
\affiliation{Department of Physics, Kyoto University, Kyoto 606-8502, Japan}
\affiliation{%
  Institute for Molecular Science, Okazaki 444-8585, Japan
}%

\author{Akito Daido}
\email{daido@scphys.kyoto-u.ac.jp}
\affiliation{Department of Physics, Kyoto University, Kyoto 606-8502, Japan}

\author{Kazuaki Takasan}
\email{takasan@berkeley.edu}
\affiliation{%
 Department of Physics, University of California, Berkeley, California 94720, USA
}%
\affiliation{
Materials Sciences Division, Lawrence Berkeley National Laboratory, Berkeley, California 94720, USA
}

\author{Tsuneya Yoshida}
\email{yoshida@rhodia.ph.tsukuba.ac.jp}
\affiliation{Department of Physics, University of Tsukuba, Ibaraki 305-8571, Japan}

\date{\today}
\begin{abstract}
Despite intensive searches for topological superconductors, the realization of topological superconductivity remains under debate. Previous proposals for the topological $s$-wave, $p$-wave, and chiral $d$-wave superconductivity have both advantages and disadvantages. 
In this review, we discuss two-dimensional topological superconductivity based on the non-chiral $d$-wave superconductors. It is shown that the noncentrosymmetric $d$-wave superconductors become topological superconductors under an infinitesimal Zeeman field without fine-tuning of parameters. Floquet engineering for introducing the Zeeman field in a controllable way is also proposed. When the two-dimensional noncentrosymmetric superconductors are stacked to recover the global inversion symmetry, the field-induced parity transition may occur, and the high-field odd-parity superconducting state realizes various topological phases depending on the stacking structures. Two-dimensional heterostructures of strongly correlated electron systems, which have been developed by recent experiments, are proposed as a platform of the high-temperature topological superconductivity and the interplay of topology and strong correlations in superconductors.
\end{abstract}
\pacs{
}
\maketitle

\section{Introduction}\label{sec:introduction}
Quantum many-body states, such as magnetic, density-wave, and superconducting states, have been traditionally classified by symmetry. Symmetry breaking allows emergent responses and specifies universal criticality of the phase transition. On the other hand, recent advances in the condensed matter physics shed light on a new aspect of quantum phases; they can be classified based on the topology. Indeed, the concept of topology is a useful tool to classify both gapped and gapless phases by combining it with symmetry. The so-called topological periodic table~\cite{Schnyder_classification_free_2008,Kitaev_classification_free_2009,Ryu_classification_free_2010} classifies topological insulator/superconductor~\cite{Kitaev2001,Kane-Mele2005,Qi-SCZ_review} with global symmetries, namely, time-reversal symmetry, particle-hole symmetry, and chiral symmetry. Later it has been recognized that the crystalline symmetry may enrich the topological properties, and the concepts of topological crystalline insulator/superconductor~\cite{Fu_TCI2011,Sato2016_review,Sato2017_review} as well as higher-order topological insulator/superconductor~\cite{BBH_science,Schindler2018} have been established. 

Although the realization of topological insulators has been demonstrated in materials, the presence of topological superconductors in nature remains under debate. This is because the condition for the topological superconductivity (TSC) is hard to be satisfied and the experimental method to detect signatures of TSC is limited. 
Unlike topological insulators, topological superconductors in some classes host Majorana quasiparticles at boundaries or defects~\cite{Read-Green,Ivanov,Kitaev2001}, and potential applicability to fault-tolerant topological quantum computation has been intensively studied~\cite{Kitaev2003,Nayak2008}. 
Thus, identifying topological superconductors is highly awaited from the viewpoints of both fundamental science and future technological innovation. 

Superconductors are traditionally classified by the relative angular momentum $l$ of Cooper pairs, as $s$-wave, $p$-wave, $d$-wave, $f$-wave {\it etc}~\cite{LeggettRMP}. They are characterized by the inversion parity and spin degree of freedom. In the usual setup, $s$-wave, $d$-wave, and other superconductors with even $l$ are even-parity spin-singlet superconductors, while  $p$-wave, $f$-wave, and other superconductors with odd $l$ are odd-parity spin-triplet superconductors. Previous searches of topological superconductors mainly focused on the $s$-wave superconductors~\cite{Sato-Fujimoto2009,Sato-Fujimoto2010,Sau2010,Lutchyn2010,Alicea2010} and odd-parity superconductors~\cite{Kitaev2001,Read-Green,Ivanov,Sato2010,Fu2010} such as $p$-wave and $f$-wave ones. Both of them have advantages and disadvantages. The $s$-wave superconductors are ubiquitous in the sense that they are found in most superconducting materials. However, to realize TSC we usually need sizable spin-orbit coupling and magnetic field as well as fine-tuning of parameters~\cite{Sato-Fujimoto2009,Sato-Fujimoto2010,Sau2010,Lutchyn2010,Alicea2010}. 
On the other hand, the odd-parity superconductors are strong candidates of intrinsic topological superconductors because fine-tuning of parameters is not needed under some conditions~\cite{Kitaev2001,Read-Green,Ivanov,Sato2010,Fu2010}. However, the odd-parity superconductors are very rare in nature. 
Although recent progress due to extensive studies is approaching the TSC of these classes~\cite{Mourik2012,Nadj-Perge2014,He2017,Yanase2017,Daido2019,Ran_UTe2_2019,Ishizuka_UTe22019,Sato2016_review,Sato2017_review,Fu2010,Sasaki2011,Levy2013}, it is still an on-going issue. 

In this review, we discuss another strategy for realizing TSC, focusing on $d$-wave superconductors. The $d$-wave superconductivity is the most typical class of superconductors in strongly correlated electron systems~\cite{Yanase2003}. Thus, the $d$-wave superconductors may be a promising platform of TSC if it is realizable without fine-tuning of parameters. So far, the $d$-wave superconductors have not been considered as a strong candidate for intrinsic topological superconductors because the gapless excitation makes the bulk topology ill-defined unless the chiral $d$-wave superconductivity spontaneously breaks time-reversal symmetry. Although there are previous proposals for chiral $d$-wave superconductivity, such as in graphene~\cite{Nandkishore2012} and SrPtAs~\cite{Fischer2014}, it is rare like the odd-parity superconductivity. Below we explore the TSC based on the ordinary non-chiral $d$-wave superconductors. 
This is a concise review of our works in this direction.

Our proposals are based on the crystal structure which globally or locally breaks space inversion symmetry. In the globally inversion asymmetric (noncentrosymmetric) $d$-wave superconductors, a Zeeman field opens the gap in the excitation spectrum, making the bulk topology well-defined, and the TSC is realized without fine-tuning of parameters (Sec.~\ref{sec:Daido}). As for a source of the Zeeman field, in addition to the ferromagnetic proximity effect and external magnetic field, Floquet engineering using the circularly polarized laser light is discussed (Sec.~\ref{sec:Takasan}). 
For locally noncentrosymmeric superconductors, the parity transition in the superconducting state and resulting odd-parity spin-singlet superconductivity are predicted (Sec.~\ref{sec:Yanase}). Analysis of multilayer superlattices reveals various topological superconducting phases protected by the mirror reflection symmetry. Interestingly, trilayer superconductors may host stable Majorana fermions, and quad-layer $d$-wave superconductors are a testbed for the reduction of topological classification by interactions (Sec.~\ref{sec:Yoshida}). 
For an experimental platform, heterostructures of high-$T_{\rm c}$ cuprate superconductors and heavy-fermion superconductors are mainly discussed. Moreover, a recently discovered bulk superconductor CeRh$_2$As$_2$ is also discussed as a candidate for parity transition and TSC.

\section{Topological d-wave superconductivity}\label{sec:Daido}
In this section, we discuss a general mechanism to realize TSC in noncentrosymmetric nodal superconductors.
The obtained results are used in the subsequent sections.
The idea is to focus on the gap nodes of {\it e.g.} the $d$-wave superconductors.
Such a linear dispersion of the quasiparticle energy is similar to the Dirac electrons in graphene, and thus called the Bogoliubov-Dirac quasiparticles.
It is shown that the Bogoliubov-Dirac quasiparticles acquire a mass gap and the degeneracy at the gap nodes is lifted by applying the Zeeman field, owing to the interplay with the inversion-symmetry breaking.
We can obtain a large Berry curvature near such massive Bogoliubov-Dirac points, leading to TSC with a nontrivial Chern number.

This section is organized as follows.
In Secs.~\ref{subsec:2A} and~\ref{subsec:2B}, we introduce the notations and obtain the expression of the mass gap acquired by the Bogoliubov-Dirac quasiparticles.
The parity mixing of the superconducting order parameter plays an essential role.
In Sec.~\ref{subsec:2C}, the obtained assembly of massive Bogoliubov-Dirac quasiparticles is shown to have a nontrivial Chern number for spin-singlet superconductors with a small admixing of the spin-triplet component.
Thereby, two-dimensional (2D) TSC and its associated Majorana edge states are obtained.
In Sec.~\ref{subsec:2D}, we discuss various Majorana edge states realized in the presence or absence of the Zeeman field.
Chiral and unidirectional Majorana edge states are realized under the Zeeman field, while Majorana flat bands are realized in the absence of the fields.
Their relation is discussed based on the evolution of the bulk Bogoliubov-Dirac quasiparticles under the Zeeman field.
Section~\ref{subsec:2E} is devoted to the extension of the obtained results to three dimensions.
We find that Weyl superconductivity is realized from line-nodal spin-singlet-dominant noncentrosymmetric superconductors by applying Zeeman fields.
We discuss experimental platforms in Sec.~\ref{subsec:2F},
and summarize the discussion of this section in Sec.~\ref{subsec:2G}.

\subsection{Model for noncentrosymmetric superconductors}
\label{subsec:2A}

Let us begin with a general theory for field-induced TSC in noncentrosymmetric systems. 
 The model is 
\begin{align}
\calH&=\sum_{\bm k \sigma} \xi (\bm k) c_{\bm k \sigma}^\dagger c_{\bm k \sigma} 
+\sum_{\bm k \sigma \sigma'} (\alpha \bm g (\bm k) \cdot \bm \sigma)_{\sigma \sigma'}
c_{\bm k \sigma}^\dagger c_{\bm k \sigma'} 
\nonumber \\
& - \sum_{\bm k \sigma \sigma'}(\mu_B\bm H \cdot \bm \sigma)_{\sigma \sigma'}
c_{\bm k \sigma}^\dagger c_{\bm k \sigma'}
\nonumber \\
& + \sum_{\bm{k} \sigma \sigma'} \Delta_{\sigma\sigma'}(\bm{k}) c^\dagger_{\bm{k}\sigma}c^\dagger_{-\bm{k}\sigma'} +\mathrm{H.c.},
\label{eq:model_sec1}
\end{align}
where $c_{\bm k \sigma}$ is the annihilation operator of electrons with momentum $\bm k$ and spin $\sigma (= \uparrow, \downarrow)$. The first term is kinetic energy measured from a chemical potential $\mu$. The second term is a spin-orbit coupling characteristic of the systems lacking inversion symmetry. This term is regarded as the momentum-dependent Zeeman coupling with the spin polarization axis at $\bm k$ specified by the $g$-vector, $\bm g (\bm k)$. When the time-reversal symmetry is preserved, the $g$-vector has to be antisymmetric in momentum, $\bm g (\bm k)=-\bm g (-\bm k)$. Thus, this term is called anti-symmetric spin-orbit coupling (ASOC). 
The third term represents the Zeeman field, whose potential origins are discussed in Sec.~\ref{subsec:2F}. 
The superconducting gap function is here  phenomenologically introduced as $\Delta(\bm{k})=(\psi(\bm{k})+\bm{d}(\bm{k})\cdot\bm{\sigma})i\sigma_y$, with $\psi(\bm{k})$ and $\bm{d}(\bm{k})$ being real. 
Note that the even-parity spin-singlet component $\psi(\bm{k})$ and odd-parity spin-triplet component ($d$-vector) $\bm{d}(\bm{k})$ are allowed to be admixed because of the broken inversion symmetry.
We adopt the notations $H\equiv|\bm{H}|$, $d(\bm{k})\equiv|\bm{d}(\bm{k})|$, $g(\bm{k})\equiv|\bm{g}(\bm{k})|$, and  $\hat{g}(\bm{k})\equiv\bm{g}(\bm{k})/g(\bm{k})$.
In the following part, we assume the conditions,
\begin{align}
\mu_{\rm B} H, \,\, |\psi(\bm{k})|, \,\, d(\bm{k}) \ll \alpha g(\bm{k}), 
\label{eq:weak_coupling}
\end{align}
for $\bm{k}$ near Fermi surfaces.
These conditions are satisfied in most of the existing noncentrosymmetric superconductors.

In the following, we discuss the energy spectrum and the topology of Bogoliubov quasiparticles.
For this purpose, it is convenient to rewrite \Eq{eq:model_sec1} with the Nambu spinor $\Psi^\dagger_{\bm{k}}=(c^\dagger_{\bm{k},\uparrow},c^\dagger_{\bm{k},\downarrow},c_{-\bm{k},\uparrow},c_{-\bm{k},\downarrow})$ and introduce the Bogoliubov-de Gennes (BdG) Hamiltonian:
\begin{gather}
  \mathcal{H}=\frac{1}{2}\sum_{\bm{k}}\Psi_{\bm{k}}^\dagger \mathcal{H}_{\mathrm{BdG}}(\bm{k})\Psi_{\bm{k}}+\text{const.},\\
  \mathcal{H}_{\mathrm{BdG}}(\bm{k})
=\begin{pmatrix}\mathcal{H}_N(\bm{k})&\Delta(\bm{k})\\
\Delta(\bm{k})^\dagger&-\mathcal{H}_N^T(-\bm{k})\end{pmatrix},
\label{eq:model_BdG}
\end{gather}
where $\mathcal{H}_N(\bm{k})=\xi(\bm{k})+(\alpha\bm{g}(\bm{k})-\mu_B\bm{H})\cdot\bm{\sigma}$ is the Bloch Hamiltonian of the normal state.
The spectrum of Bogoliubov quasiparticles is given by the eigen-energies of $\mathcal{H}_{\mathrm{BdG}}(\bm{k})$, while TSC can be discussed based on its eigenstates.

In this section, we proceed with discussions without specifying $\xi ({\bm k})$, $\bm g (\bm k)$, $\psi(\bm{k})$, and $\bm{d}(\bm{k})$ characterizing the model.
Later, we set the form for 2D $d$-wave superconductors and show the topological superconducting phase. 
Classification and basis functions based on the crystalline point group can be found in Ref.~\cite{SigristUeda_91} for superconducting gap functions and in Refs.~\cite{Frigeri_Dthesis,Smidman2017_review} for the ASOC. 

\subsection{Gap generation}
\label{subsec:2B}

We focus on gapless superconductors, which accompany nodal excitation in the Bogoliubov quasiparticle's spectrum. 
The key to realizing TSC is the gap generation due to the paramagnetic effect. The finite excitation gap allows a well-defined bulk topological invariant, and the geometric properties of nodal quasiparticles may give rise to the nontrivial bulk topology. 

\begin{figure}
    \centering
    \includegraphics[width=0.25\textwidth]{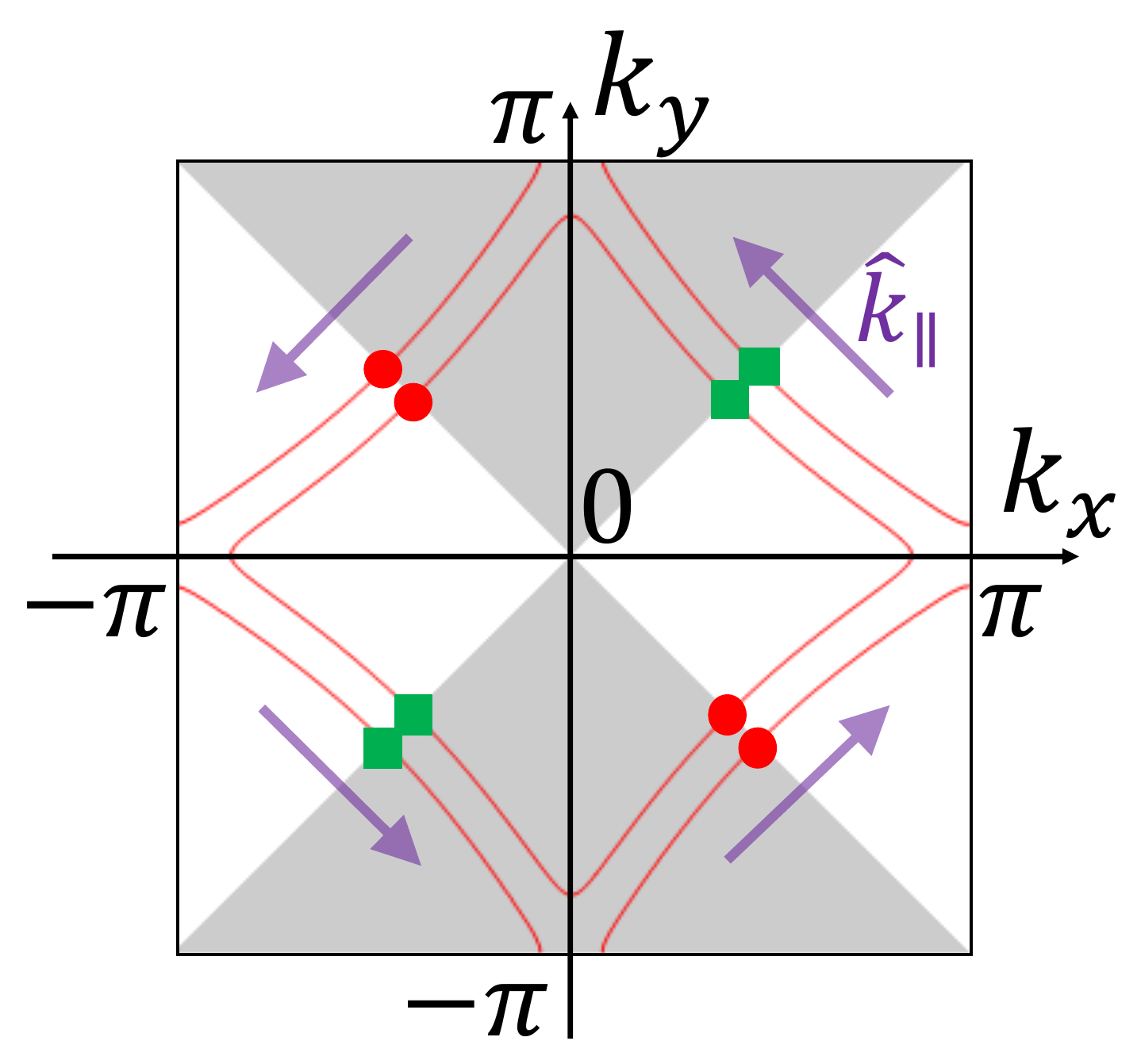}
    \caption{Fermi surfaces and point nodes of the noncentrosymmetric $d$-wave superconductivity for $H=0$, modeled by Eqs.~\eqref{eq:model_dwaveSC}.
    Spin-split Fermi surfaces are shown by red lines.
    Red closed circles and green closed squares represent nodes with the winding number $W(\bm{k}_0)=+1$ and $-1$, respectively.
    The purple arrows show the direction of $\hat{k}_\parallel$.
    The shaded region indicates $\psi(\bm{k})>0$.}
    \label{fig:FermiSurface_D+p}
\end{figure}

We begin by clarifying that nodal excitation is stable in the presence of the time-reversal symmetry, and therefore, time-reversal-symmetry breaking is indispensable for the gap generation.
In our model, the time-reversal symmetry is preserved for $H=0$. On the condition~\eqref{eq:weak_coupling}, the energy dispersion of the Bogoliubov quasiparticles is obtained as\cite{Smidman2017_review,NCSC_book,Daido2016} 
\begin{gather}
  \mathcal{E}_\gamma(\bm{k})=\pm\sqrt{E_\gamma(\bm{k})^2+|\psi(\bm{k})+\gamma\bm{d}(\bm{k})\cdot\hat{g}(\bm{k})|^2},
  \label{zeromagspectrum}
\end{gather}
with the normal band energy with the helicity $\gamma=\pm$,
\begin{gather}
  E_\pm(\bm{k})\equiv\xi(\bm{k})\pm\alpha|\bm{g}(\bm{k})|.
\label{eq:spin_splitting}
\end{gather}
Thus, $\psi(\bm{k})\pm\bm{d}(\bm{k})\cdot\hat{g}(\bm{k})$ is interpreted as the gap function of the helicity $\pm$ bands.
A gap node exists when the relation
\begin{gather}
  E_\pm(\bm{k}_0)=\psi(\bm{k}_0)\pm\bm{d}(\bm{k}_0)\cdot\hat{g}(\bm{k}_0)=0,
  \label{eq:node_cond}
\end{gather}
is satisfied for a point $\bm{k}_0$ on the Fermi surface with the helicity $\pm$.
For instance, a $d_{x^2-y^2}$-wave superconducting state with $\psi(\bm{k}) \propto \cos k_x - \cos k_y$ and ${d}(\bm{k}) = 0$ has nodal points along the $\bm k \parallel [110]$ line, if the line crosses a Fermi surface (See Fig.~\ref{fig:FermiSurface_D+p}). 
These nodes are stable even when a finite $\bm{d}(\bm{k})$ is taken into account, as can be readily examined for the model given later [Eqs.~\eqref{eq:model_dwaveSC}].

The stability of the nodal points is ensured by topological protection. 
The nodal points in 2D systems, which correspond to line nodes in three dimensions, are actually robust in time-reversal invariant spin-singlet superconductors. A protecting topological invariant is the one-dimensional winding number defined by \cite{Schnyder2011,Sato2011,Yada2011}:
\begin{align}
  W(\bm{k}_0)\equiv\Im\oint_{C(\bm{k}_0)}\frac{d\bm{k}}{2\pi}\cdot\nabla_{\bm{k}}\ln \det q(\bm{k}),
  \label{winddef}
\end{align}
where $C(\bm{k}_0)$ is a sufficiently small loop running anticlockwise around the nodal point $\bm{k}_0$, and $q(\bm{k})$ is the Hamiltonian in the chiral basis
\begin{align}
  U_c\mathcal{H}_{\text{BdG}}(\bm{k})U^\dagger_c=\begin{pmatrix}0&q(\bm{k})\\q(\bm{k})^\dagger&0\end{pmatrix}. 
\end{align}
The unitary matrix $U_{\rm c}$ is chosen so as to diagonalize the chiral operator $\Gamma$,
\begin{align}
  U_c\Gamma U_c^\dagger=\begin{pmatrix}1_{2\times2}&0\\0&-1_{2\times2}\end{pmatrix},\quad\Gamma\equiv\begin{pmatrix}0&\sigma_y\\\sigma_y&0\end{pmatrix}.
\end{align}
The chiral operator is obtained by combining the time-reversal symmetry with the particle-hole symmetry and, satisfies the chiral symmetry $\{\Gamma,\mathcal{H}_{\text{BdG}}(\bm{k})\}=0$.
It turns out that in our model, the winding number is given by the sign reversal of the gap function along the Fermi surface:~\cite{Daido2016,Daido2017}
\begin{equation}
W(\bm{k}_0)=-\sgn\left[\frac{\partial\left(\psi(\bm{k})\pm\bm{d}(\bm{k})\cdot\hat{g}(\bm{k})\right)}{\partial k_\parallel}\right]_{\bm{k}=\bm{k}_0},
\label{eq:node_winding}
\end{equation}
for a node $\bm{k}_0$ on the Fermi surface with helicity $\pm$ defined in Eq.~\eqref{eq:node_cond}.
Here, $\hat{k}_\parallel\equiv \hat{z}\times\nabla_{\bm{k}}E_\pm(\bm{k})/|\hat{z}\times\nabla_{\bm{k}}E_\pm(\bm{k})|$ is the direction along the Fermi surface (See Fig.~\ref{fig:FermiSurface_D+p}), and $\partial_{k_\parallel}\equiv\hat{k}_\parallel\cdot\nabla_{\bm{k}}$ is the corresponding derivative.
This result is used later to identify TSC.

Because the chiral symmetry is preserved in time-reversal symmetric superconductors, the nodal excitation is robust against inversion-symmetry breaking alone. 
However, the gap may open when we additionally break time-reversal symmetry. Indeed, the Zeeman field $H\neq0$ changes the quasiparticle's energy to~\cite{Daido2016} 
\begin{gather}
  \mathcal{E}_\gamma(\bm{k})=-\gamma\muB\bm{H}\cdot\hat{g}(\bm{k})\pm\sqrt{E_\gamma(\bm{k})^2+\left|\Delta_\gamma(\bm{k})\right|^2},
  \label{eq:lambdas}
\end{gather}
with $\gamma=\pm$ and
\begin{align}
  \Delta_\pm(\bm{k})&= \psi(\bm{k})\pm\bm{d}(\bm{k})\cdot\hat{g}(\bm{k})\notag\\
  &\qquad\qquad+i\left[\frac{\muB\bm{H}\cdot\hat{g}(\bm{k})\times\bm{d}(\bm{k})}{\alpha g(\bm{k})}\right].
\label{eq:chiral_gapfunc}
\end{align}
A brief sketch of the derivation is given in Appendix~\ref{app:magenergy}.
When the energy shift term $\mp\muB\bm{H}\cdot\hat{g}$ vanishes for a direction of $\bm H$, the excitation energy is given by the gap function in the band basis $\Delta_\pm(\bm{k})$.
Because the gap function of each band contains both real and imaginary components,
the system has properties analogous to the chiral superconductivity.
As in the chiral $p_x\pm ip_y$-wave and $d_{x^2-y^2}\pm id_{xy}$-wave states~\cite{Kallin2016}, the superconducting state may be fully gapped, and may also carry a nontrivial Chern number as discussed in the next section. 
The condition for the gap opening is 
\begin{align}
M(\bm{k}_0)\equiv
\frac{\mu_B\bm{H}\cdot\hat{g}(\bm{k_0})\times\bm{d}(\bm{k_0})}{\alpha g(\bm{k}_0)} \ne 0, 
\label{eq:condition_gap}
\end{align}
at the nodal point ${\bm k}={\bm k}_0$.

For the condition~\eqref{eq:condition_gap} to be satisfied, the $d$-vector must have a component perpendicular to the $g$-vector at the nodes so that $\hat{g}\times\bm{d}\neq0$. This component is expected to be finite in general. 
It is true that the $d$-vector in noncentrosymmetric superconductors tends to be parallel to the $g$-vector, as it is thermodynamically favored by the spin-orbit coupling.~\cite{Frigeri2004,NCSC_book}
However, the relation $\hat{g} \parallel \bm{d}$ is not imposed by any point group symmetry, and hence is not a rigorous constraint.
Accordingly, $\hat{g}\parallel\bm{d}$ is not satisfied in situations, for example, where the $\bm{k}$-dependence of the pairing interaction for the dominant spin-triplet channel is mismatched with that of the $g$-vector~\cite{Yanase2012}.
Actually, a study on a noncentrosymmetric superconductor CePt$_3$Si points to a $d$-vector far from parallel to the $g$-vector~\cite{Yanase2008,Yanase2012}:
The $d$-vector has a simple $\bm{k}$ dependence corresponding to the short-ranged Cooper pairs,
while the $g$-vector has a complicated one as a result of the orbital degeneracy~\cite{Yanase2013,Harima2015}.
Furthermore, in unconventional superconductors, the $g$- and $d$-vectors belong to different irreducible representations of the point group.
In this case, the mismatch of the symmetries of the $g$- and $d$-vectors may also lead to the violation of $\hat{g}\parallel \bm{d}$, as we see in details for the case of the $d$-wave superconductivity.
Thus, the perpendicular component does not vanish in general. 

Let us consider a Rashba superconductor on the square lattice with dominant $d_{x^2-y^2}$-wave pairing.
We set 
\begin{subequations}
\begin{align}
\xi(\bm{k})&=-2t(\cos k_x+\cos k_y)+4t'\cos k_x\cos k_y-\mu, 
\label{eq:dispersion_sec1}
\\
\bm{g}(\bm{k})&=(-\sin k_y,\sin k_x,0)^T, 
\label{eq:g-vector_sec1}
\\
\psi(\bm{k})&=\psi_0(\cos k_x-\cos k_y), 
\label{eq:d-wave_gap_sec1}
\\
\bm{d}(\bm{k})&=d_0(\sin k_y,\sin k_x,0)^T,
\label{eq:d-vector_sec1}
\end{align}
\label{eq:model_dwaveSC}
\end{subequations}
and $\muB\bm{H}=\muB H\hat{z}$.
The dominant $d_{x^2-y^2}$-wave order parameter $\psi({\bm k})$ belongs to the $B_1$ irreducible representation of the $C_{4v}$ point group. Thus, the admixed spin-triplet order parameter naturally belongs to the $B_1$ representation, and the basis function is given by Eq.~\eqref{eq:d-vector_sec1} 
when we assume Cooper pairs on nearest-neighbor bonds. 
Note that a typical $g$-vector of the Rashba-type, Eq~\eqref{eq:g-vector_sec1}, is not parallel but perpendicular to the $d$-vector at the nodal points, ${\bm k}_0$ $\parallel [110]$ or $\parallel [1\bar{1}0]$.
From the symmetry points of view, this is because the ASOC must belong to
 the identity ($A_1$) representation, not to the $B_1$ representation.
The different symmetry properties of $\hat{g}(\bm{k})$ and $\bm{d}(\bm{k})$ lead to such an orthogonality (in this case, the mirror planes $(1\bar{1}0)$ and $(110)$ play the vital role).
Note also that $\hat{g}\parallel\bm{d}$ is satisfied for antinodal directions $\bm{k}\parallel[100]$ and $[010]$, where a dominant contribution to the condensation energy is gained.
The microscopic analysis of the Rashba-Hubbard model by random-phase approximation~\cite{Shigeta2013}, cluster dynamical mean-field theory~\cite{Lu2018}, and fluctuation-exchange approximation~\cite{Nogaki2020}
supports a finite admixture of the $d$-vector component for the nodal $\bm{k}$ directions.
Thus, the condition \eqref{eq:condition_gap} is satisfied, predicting the gap opening under a Zeeman field.

\begin{figure}[htbp]
  \centering
  \begin{tabular}{ll}
(a) $H=0$&(b) $H\neq 0$ \\
   \includegraphics[width=0.225\textwidth]{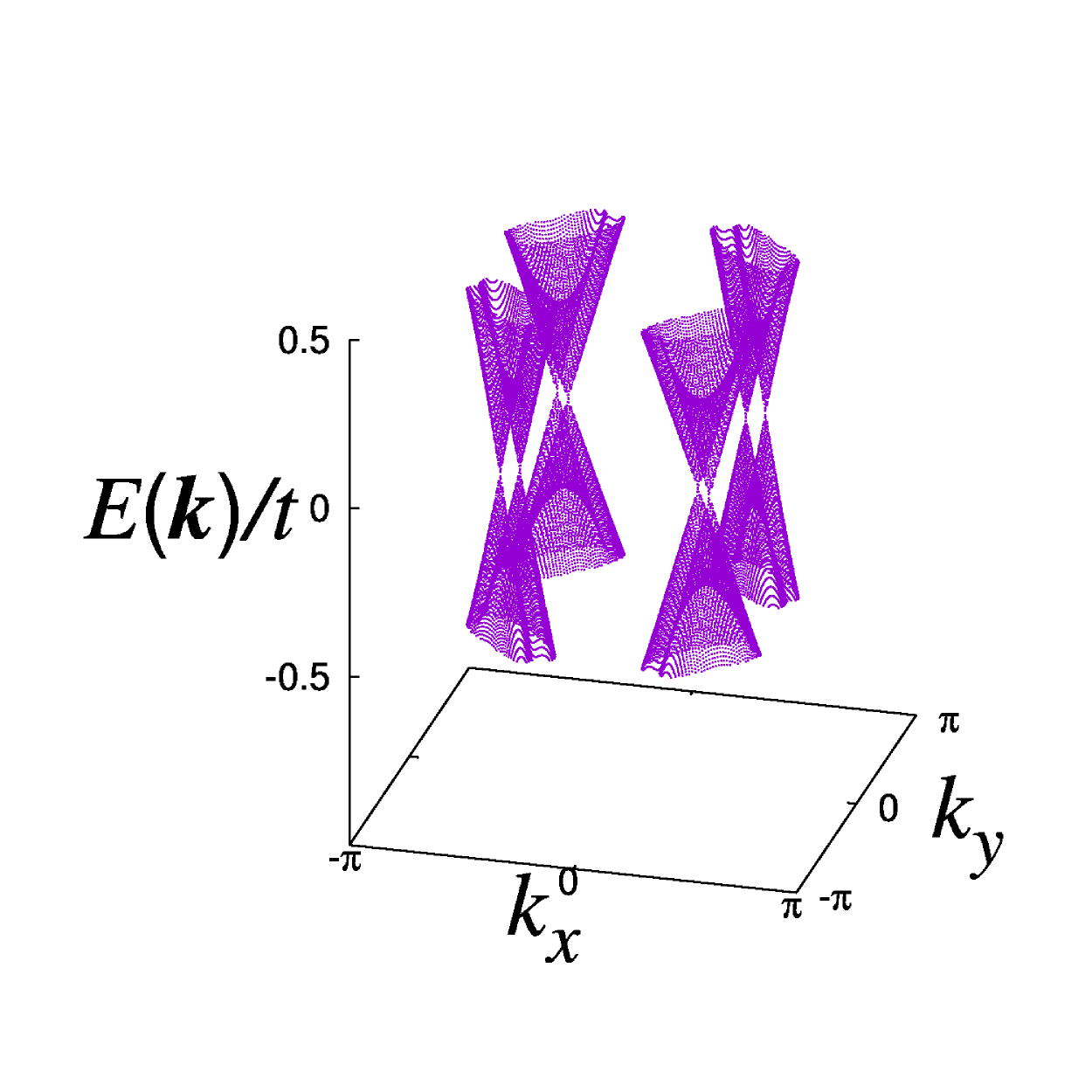}&
    \includegraphics[width=0.225\textwidth]{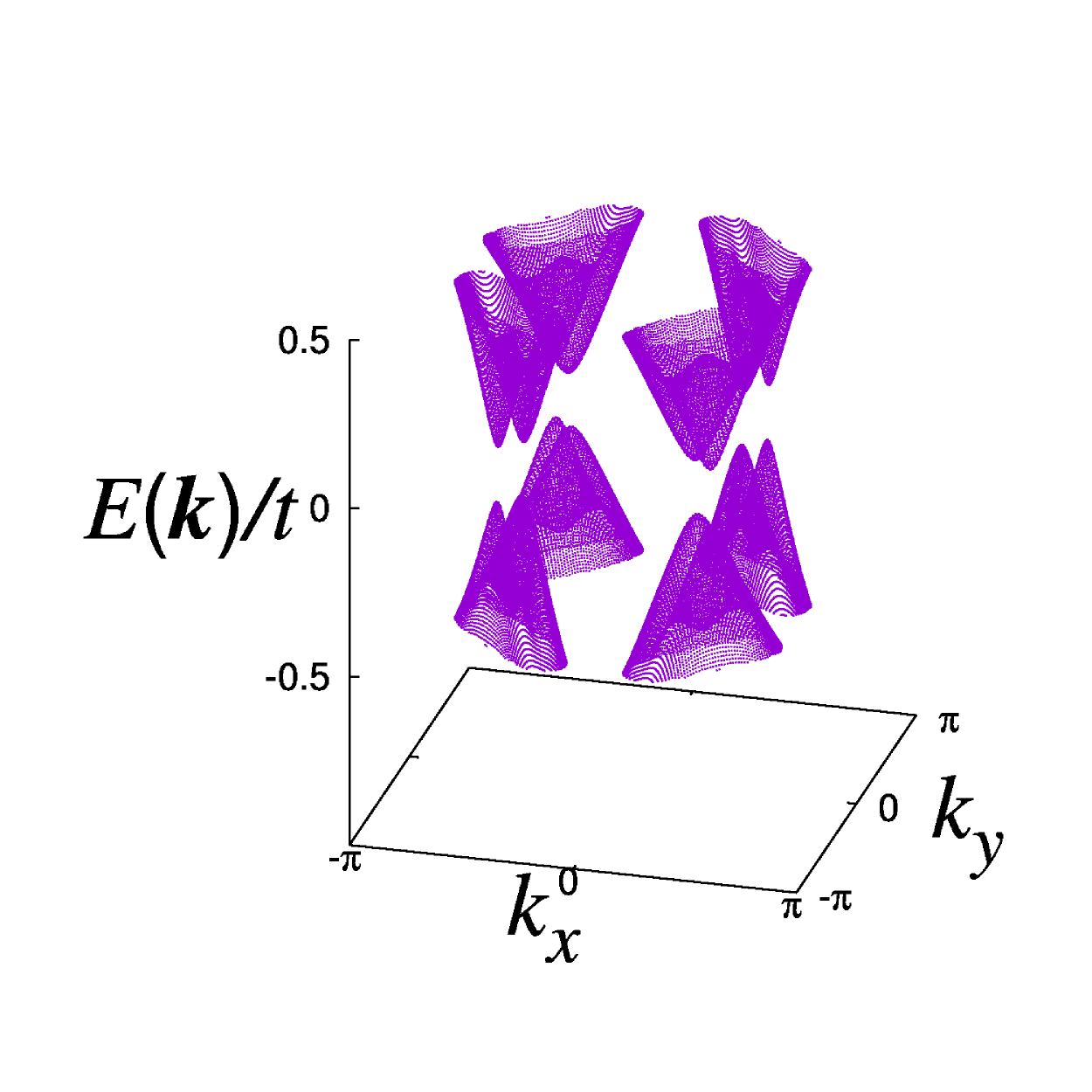}
    \end{tabular}
\caption{(a) and (b) Energy spectrum of Bogoliubov quasiparticles in a Rashba $d_{x^2-y^2}$-wave superconductor with a $p$-wave component admixed. The spectrum around $E\simeq0$ is shown for (a) $\muB H=0$, and (b) $\muB H= 0.3$.
Large values of $\psi_0$, $d_0$, and $\muB H$ are adopted for visibility.
  }
  \label{fig_D+p}
\end{figure}    

To verify the above discussions we show the numerically obtained energy spectrum of Bogoliubov quasiparticles. 
We adopt Eqs.~\eqref{eq:model_dwaveSC} and take $t=1$, $t'=0.2$, $\alpha=0.3$, $\mu=-0.79$, $\psi_0=0.5$, and $d_0=0.1$. 
While Fig.~\ref{fig_D+p}(a) shows the topologically protected nodal excitation at $H=0$, Fig.~\ref{fig_D+p}(b) reveals the gap generation due to the finite Zeeman term, $H \ne 0$. 
The appearance of the excitation gap can also be understood by analyzing $\Delta_\pm(\bm{k})$.
Equation~\eqref{eq:chiral_gapfunc} for the model~\eqref{eq:model_dwaveSC} is given by
\begin{equation}
    \Delta_\pm(\bm{k})=-\left(\psi_0\mp \frac{d_0}{|\bm{k}|}\right)(k_x^2-k_y^2)-i\frac{\mu_BHd_0 }{\alpha|\bm{k}|^2}k_xk_y.
\end{equation}
Here, the expression for small wave numbers is shown for simplicity.
Thus, we can interpret the Rashba $d$-wave superconductor under Zeeman field as the effective realization of the chiral $d_{x^2-y^2}+id_{xy}$-wave state, and thus a gapful spectrum naturally follows.

\subsection{Topological superconductivity}
\label{subsec:2C}

When a finite gap is generated by the paramagnetic effect as in Fig.~\ref{fig_D+p}(b), we can define the bulk topological invariant. This case belongs to the class D in the Altland-Zirnbauer classification~\cite{Schnyder_classification_free_2008,Kitaev_classification_free_2009,Ryu_classification_free_2010}.
Therefore, the topological property is specified by the Chern number $\nu \in  \mathbb{Z}$, defined by~\cite{Kohmoto1985}
\begin{gather}
  \nu\equiv\sum_{i,j;\,n:\text{occ.}}\frac{1}{2\pi i}\int_{\bm{k}\in[\text{2D BZ}]}d^2k\,\varepsilon_{ij}\partial_{k_i}\braket{u_n(\bm{k})|\partial_{k_j}|u_n(\bm{k})}.
  \label{Chn}
\end{gather}
The summation for $n$ is taken over all the occupied bands, and $\ket{u_n(\bm{k})}$ is the quasiparticle wave function of the $n$-th energy band. 
$\varepsilon_{ij}$ ($i,j=x,y$) is the antisymmetric tensor satisfying $\varepsilon_{xy}=1$.
To avoid confusion, note that $-\nu$ is sometimes adopted as the Chern number in the literature.

Here we show a simple formula for the Chern number 
in the gapped noncentrosymmetric superconductors. 
We again consider the general case represented by the model, Eq.~\eqref{eq:model_sec1}.
The idea for evaluating the Chern number is based on the original formula for the quantum anomalous Hall effect~\cite{Haldane1988}. 
The point is that the excitation in the nodal superconductor shows a linear dispersion similar to the Dirac electrons. 
The Chern number of a massive Dirac electron is known to be $-\sgn[WM]/2$, where $W$ and $M$ are the chirality and the mass of the Dirac electron, respectively.
In the same way, each node, {\it i.e.} each Bogoliubov-Dirac quasiparticle has the Chern number $-\sgn[W(\bm{k}_0)M(\bm{k}_0)]/2$ when gapped out by the Zeeman field.
Here, the winding number $W(\bm{k}_0)$ and the induced gap $M(\bm{k}_0)$ are given in Eqs.~\eqref{eq:node_winding} and \eqref{eq:condition_gap}, respectively.
Thus, the Chern number of the system is given by summing up the contribution from all the Bogoliubov-Dirac quasiparticles:~\cite{Daido2016,Daido2017}
\begin{align}
  \nu=&\sum_{(\pm,\ \bm{k}_0)}\frac{1}{2}\sgn
  \left[\frac{\partial\left(\psi\pm\bm{d}\cdot\hat{g}\right)/\partial k_\parallel}{\muB\bm{H}\cdot\hat{g}\times\bm{d}/\alpha}\right]_{\bm{k}=\bm{k}_0}\notag\\
  =&\sum_{(\pm,\ \bm{k}_0)}\frac{1}{2}\sgn
  \left[\frac{\left(\hat{z}\times\nabla_{\bm{k}}E_\pm\right)\cdot\nabla_{\bm{k}}\left(\psi\pm\bm{d}\cdot\hat{g}\right)}{\muB\bm{H}\cdot\hat{g}\times\bm{d}/\alpha}\right]_{\bm{k}=\bm{k}_0}.
  \label{TSCchern}
\end{align}
Here, $(\pm,\bm{k}_0)$ labels the gap node at $\bm{k}=\bm{k}_0$ on the $E_\pm$-Fermi surface, defined by Eq.~\eqref{eq:node_cond}.
Since the number of linear nodes must be even,  Eq.~\eqref{TSCchern} is quantized to be integer.
For more details, see Refs.~\cite{Daido2016,Daido2017}.

Let us apply the formula \eqref{TSCchern} to the Rashba $d$-wave superconductivity modeled by Eq.~\eqref{eq:model_sec1} with Eqs.~\eqref{eq:model_dwaveSC}.
Note that four nodes are located on each helicity band [Fig.~\ref{fig:FermiSurface_D+p}], all of which are crystallographically equivalent owing to the fourfold-rotation symmetry.
It is easy to check that the summand of Eq.~\eqref{TSCchern} is invariant under fourfold rotation.
Thus, we obtain
\begin{equation}
\nu=2(s_++s_-).
\end{equation}
Here, we defined $s_\pm$ by the sign in Eq.~\eqref{TSCchern} evaluated for the nodes in the first quadrant $k_x,k_y>0$.
Note that
\begin{align}
    s_+\cdot s_-&=\sgn[\partial_{k_\parallel}(\psi+\bm{d}\cdot\hat{g})]\sgn[\partial_{k_\parallel}(\psi-\bm{d}\cdot\hat{g})]\notag\\
    &=\sgn[\partial_{k_\parallel}\psi]^2\notag\\
    &=1,
\end{align}
because $\partial_{k_\parallel}(\psi\pm\bm{d}\cdot\hat{g})\simeq\partial_{k_\parallel}\psi$ for the dominant $d$-wave pairing.
Thus, we obtain a nontrivial Chern number
\begin{equation}
    \nu=4s_+=-4,
\end{equation}
by using $s_+=-1$ in this model.
In summary, we have shown that \textit{all the Bogoliubov-Dirac quasiparticles in the Rashba $d$-wave superconductivity make additive contributions to the Chern number}.
Topological $d$-wave superconductivity is realized in this way.
The result $\nu=-4$ is supported by the numerical calculation~\cite{Yoshida2016,Daido2016} using the Fukui-Hatsugai-Suzuki method~\cite{Fukui_Hatsugai_05}, as well as microscopic analysis of the Rashba-Hubbard model under the Zeeman field~\cite{Lu2018}.
The robustness of the topological $d$-wave superconductivity has been illustrated for an additional inclusion of the Dresselhaus spin-orbit coupling~\cite{Xiancong2020}.

Following the above discussion, we can obtain a simplified version of the general formula~\eqref{TSCchern}:
\begin{gather}
  \nu=\frac{s_+N_+ + s_-N_-}{2}.
  \label{reduced_formula}
\end{gather}
We denote the number of nodes on the Fermi surfaces with helicity $\pm$ by $N_\pm$, and the corresponding sign of Eq.~\eqref{TSCchern} by $s_\pm$.
This formula is obtained by showing the fact that crystallographically equivalent nodes contribute to the Chern number with the same sign, $s_+$ or $s_-$, when the order parameter belongs to a certain 1D irreducible representation of the point group and the Zeeman field is perpendicular to the system~\cite{Daido2016}.
Note that we have $N_+=N_-$, unless the spin-orbit coupling is so large as to cause the Lifshitz transition.

We can draw general conclusions from the formula~\Eq{reduced_formula}, by assuming either spin-singlet or -triplet pairing is dominant.
For spin-singlet-dominant superconductors, we have $\partial_{k_\parallel}(\psi\pm\bm{d}\cdot\hat{g})\simeq\partial_{k_\parallel}\psi$ and obtain $s_+=s_-$.
Thus, we generally conclude that the Chern number is nontrivial,
\begin{equation}
    \nu=s_+N_+\neq0.\label{eq:reduced_formula_singlet}
\end{equation}
Considering that almost all of the superconductors in nature are spin-singlet superconductors, our result implies the existence of various candidate materials. 
On the other hand, when the spin-triplet pairing is dominant, the contributions to the Chern number from the spin-split bands are canceled out. 
This is because $\partial_{k_\parallel}(\psi\pm\bm{d}\cdot\hat{g})\simeq\pm\partial_{k_\parallel}\bm{d}\cdot\hat{g}$ and therefore $s_+=-s_-$.
Thus, we conclude $\nu=0$ for the spin-triplet dominant pairings.
The result is in sharp contrast to the fact that the spin-triplet superconductors are promising candidates for time-reversal-invariant TSC~\cite{Tanaka2012,Sato2016_review,Sato2017_review}.

By Eq.~\eqref{eq:reduced_formula_singlet}, nontrivial Chern numbers are obtained for the spin-singlet-dominant pairings other than the $d$-wave state as well.
As an example, we discuss a dominant extended $s$-wave state admixed with the $p$-wave component~\cite{Daido2016}.
We consider
\begin{equation}
\psi(\bm{k})=\psi_0(\delta_1+\cos k_x+\cos k_y),
\end{equation}
and take $\delta_1$ to be small enough to possess accidental nodes.
We adopt the model~\eqref{eq:model_dwaveSC} for the normal state, and use the $d$-vector
\begin{equation}
    \bm{d}(\bm{k})=d_0(-\sin 2k_y,\sin 2k_x,0).\label{eq:S+p_dvec}
\end{equation}
Here, Eq.~\eqref{eq:S+p_dvec} is adopted to satisfy Eq.~\eqref{eq:condition_gap}, although a natural choice might be $\bm{d}(\bm{k})=d_0(-\sin k_y,\sin k_x,0)$.
Note again that such a component of the $d$-vector generally exists, as $\bm{d}(\bm{k})\parallel\bm{g}(\bm{k})$ is not required by any symmetry.
For $\delta_1=-0.1$ and $\mu=-0.8$, there exist eight point nodes on each one of the helicity $\pm$ Fermi surfaces, all of which are related by crystalline symmetries~\cite{Daido2016}.
Thus, we expect the Chern number $|\nu|=8$ according to the formula~\eqref{eq:reduced_formula_singlet}.
The numerical calculation revealed the Chern number $\nu=8$, and also revealed nontrivial values of the Chern number for $\mu$ and $\delta_1$ with which the system possesses gap nodes~\cite{Daido2016}.

We call the above pairing symmetry the $S+p$-wave state.
Here and hereafter, the capitalized pairing symmetry represents the dominant one and, hence, the model~\eqref{eq:model_dwaveSC} for the topological $d$-wave superconductivity is called the $D+p$-wave state.
The $S+p$-wave state has been proposed~\cite{Tada2008,Yanase2012} for noncentrosymmetric superconductors CeRhSi$_3$~\cite{Kimura2005} and CeIrSi$_3$~\cite{Sugitani2006}.
The above result suggests that a nontrivial Chern number is obtained for the effective 2D system
given by a slice of the three-dimensional (3D) Brillouin zone of these compounds with a fixed value of $k_z$.
We apply the results obtained here to 3D systems in Sec.~\ref{subsec:2E}.
In particular, we also discuss the spin-triplet-dominant $s+P$-wave state, which has been proposed~\cite{Yanase2012,Yanase2007} for a noncentrosymmetric superconductor CePt$_3$Si~\cite{Bauer2004}.

\subsection{Majorana edge states of topological $d$-wave superconductor}
\label{subsec:2D}

Nontrivial bulk topology is reflected in the appearance of gapless edge states, according to the bulk-edge correspondence.
The edge states of topological superconductors are conventionally called the Majorana edge states.
As the quantum (anomalous) Hall insulators accompany chiral edge states, \textit{chiral Majorana edge states} are predicted by the finite Chern number $\nu$ of the BdG Hamiltonian~\cite{Tanaka2012,Sato2016_review,Sato2017_review}.
To be specific, we focus on the case of topological $d$-wave superconductivity in this section.
The following numerical results are obtained by the model~\eqref{eq:model_sec1} with Eqs.~\eqref{eq:model_dwaveSC} for several values of $\bm{H}$.

\begin{figure}
    \centering
    \includegraphics[width=0.35\textwidth]{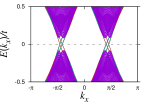}
    \caption{Chiral Majorana edge states of topological $d$-wave superconductivity.
    The purple lines correspond to the bulk quasiparticles gapped out by the Zeeman field.
    The green and red spectrum show the Majorana edge states on the $(010)$ and $(0\bar{1}0)$ edges, respectively. 
    }
    \label{fig:edge_perp}
\end{figure}

To obtain the edge spectrum, let us consider the periodic and open boundary conditions for the $x$ and $y$ directions, respectively.
In Fig.~\ref{fig:edge_perp}, we show the corresponding energy spectrum.
Note that $k_x$ is still a good quantum number since translational invariance is preserved for the $x$ direction.
The purple lines correspond to the bulk quasiparticles since almost the same dispersion can be reproduced by redrawing the bulk energy spectrum Fig.~\ref{fig_D+p}(b) in terms of $k_x$ vs. quasiparticle energy.
On the other hand, the gapless modes highlighted with red and green are absent in the bulk spectrum.
Indeed, the weight of wave functions of the red and green modes are localized around the $(010)$ and $(0\bar{1}0)$ edges, respectively, while delocalized in the $x$ direction.
Thus, we identify the red and green modes with the chiral Majorana edge states, which have negative and positive group velocities on the $(010)$ and $(0\bar{1}0)$ edges, respectively.
There are four modes on each edge, which is consistent with the Chern number $|\nu|=4$.

\begin{figure}
    \centering
    \begin{tabular}{ll}
    (a)&(b)\\
\includegraphics[width=0.24\textwidth]{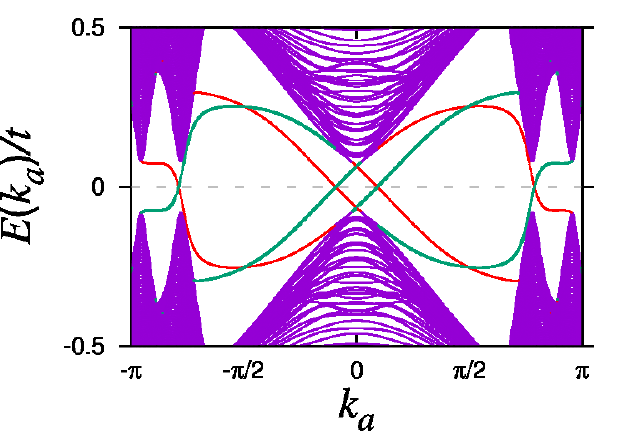}&
    \includegraphics[width=0.24\textwidth]{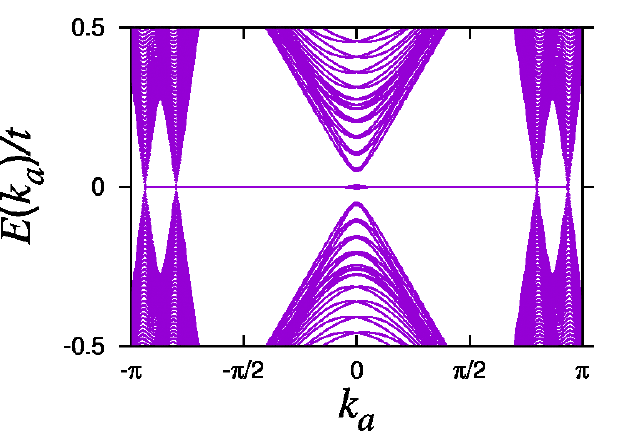}
    \end{tabular}
    \caption{(a) Chiral Majorana edge states of topological $d$-wave superconductivity with different boundary directions.
    The green and red spectrum show the Majorana edge states on the $(\bar{1}10)$ and $(1\bar{1}0)$ edges, respectively, while purple lines show bulk spectrum.
    The gray dashed line indicates the zero energy.
    (b) The spectrum of the same system with $H=0$.
    As decreasing $H$, the chiral edge states in the panel (a) are transformed to the Majorana flat bands.
    A minigap of the bulk quasiparticles at $k_a=0$ is a finite-size effect, and vanishes in the thermodynamic limit.}
    \label{fig:edge_para}
\end{figure}

The appearance of chiral Majorana edge states is independent of the boundary directions.
Indeed, we obtain chiral Majorana edge states also on the $(110)$ and $(1\bar{1}0)$ edges, as shown in Fig.~\ref{fig:edge_para}(a).
Here, $k_a=k_x+k_y$ is the good quantum number, and the bulk quasiparticles have a dispersion similar to Fig.~\ref{fig_D+p}(b) replotted in terms of $k_a$.
Although the dispersion of the edge states is a little complicated, we still find four gapless chiral modes with positive and negative velocities on each one of the edges.
By switching off the Zeeman field, the dispersion of the chiral Majorana edge states becomes flat as shown in Fig.~\ref{fig:edge_para}(b), connecting the gap nodes of the bulk $d$-wave superconductor.
These edge modes are the Andreev bound states~\cite{Tanaka2012} associated with the sign reversal of the $d$-wave gap function.
They are called \textit{the Majorana flat bands}, and are protected by the $k_a$-dependent winding number defined by an equation similar to Eq.~\eqref{winddef}~\cite{Sato2011,Yada2011,Schnyder2011,Schnyder2015_review}.
Thus, the chiral Majorana edge states on the $(110)$ and $(1\bar{1}0)$ edges can be regarded as reminiscent of Majorana flat bands, and \textit{vice versa}.
Note that Majorana flat bands do not appear on the $(100)$ and $(010)$ edges, in contrast to the chiral Majorana edge states.

\begin{figure}
    \centering
    \includegraphics[width=0.3\textwidth]{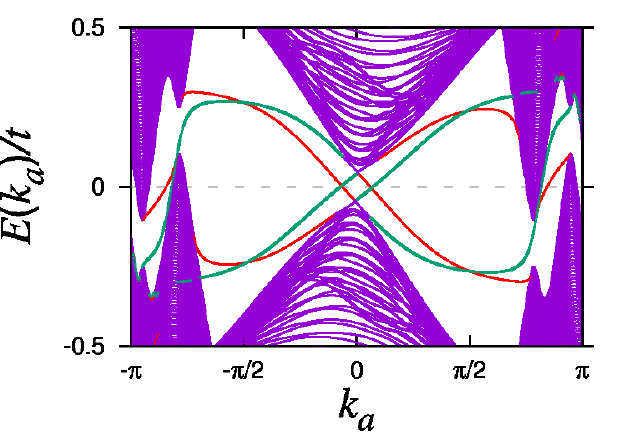}
    \caption{Unidirectional Majorana edge states of topological $d$-wave superconductivity.
    The purple lines correspond to the bulk quasiparticles.
    The green and red spectrum show the Majorana edge states on the $(\bar{1}10)$ and $(1\bar{1}0)$ edges, respectively. 
    }
    \label{fig:edge_unid}
\end{figure}

An advantage of the chiral Majorana edge states is their robustness against a broader class of perturbations.
Chiral Majorana edge states are robust against impurity scatterings, in contrast to the Majorana flat bands~\cite{Schnyder_classification_free_2008,Kitaev_classification_free_2009,Ryu_classification_free_2010}.
Furthermore, chiral Majorana edge states are robust against the correlation effects beyond the present analysis~\cite{Morimoto_2015} (See also the discussion in Sec.~\ref{sec:Yoshida} for interaction effects on the topological classification).
Thus, chiral Majorana edge states would be more easily accessible in experiments.

It is interesting to see the evolution of the energy spectrum as the Zeeman field $\bm{H}$ is tilted from the direction perpendicular to the $g$-vector.
According to Eqs.~\eqref{eq:lambdas}, there appears a paramagnetic shift $\mp\bm{H}\cdot\hat{g}(\bm{k}_0)$ in energy around the nodes.
As long as the tilting is small and the excitation gap remains finite, the Chern number $\nu$ is well-defined, and thus chiral edge states survive.
On the other hand, the energy spectrum becomes gapless when the inequality
\begin{equation}
    |\mu_B\bm{H}\cdot\hat{g}(\bm{k}_0)|>|M(\bm{k}_0)|,
\end{equation}
is satisfied, although the band gap remains open.
We show in Fig.~\ref{fig:edge_unid} the edge spectrum corresponding to Fig.~\ref{fig:edge_para}(a) but with a tilted Zeeman field
\begin{gather}
    \bm{H}=H(\sin\theta\cos\phi,\sin\theta\sin\phi,\cos\theta),\notag\\
    \theta=\pi/4,\quad,\phi=3\pi/4.
\end{gather}
Interestingly, some of the edge states are \textit{unidirectional} rather than chiral:
Compared with Fig.~\ref{fig:edge_para}(a), the two edge modes around $k_a=\pm\pi$ highlighted with red have the different sign of the group velocity, while the green ones and the edge modes around $k_a=0$ keep the original sign.
This indicates a unidirectional edge flow, where the quasiparticle currents on both edges do not completely cancel with each other.
Such edge states are called \textit{the unidirectional Majorana edge states}~\cite{Wong2013}.
Since a finite total quasiparticle current in the ground state is prohibited, a counterpropagating flow of the bulk gapless quasiparticles is expected
\footnote{
Here, the quasiparticle current refers to $\partial_{\bm{k}}\mathcal{H}_{\mathrm{BdG}}(\bm{k})$ rather than the electric current.
The total electric current should vanish, but needs a remark.
It is known that the thermodynamically stable state of noncentrosymmetric superconductors with a small parallel Zeeman field
has a tiny center-of-mass momentum of the Cooper pair (called the helical superconductivity)~\cite{Barzykin_Gorkov2002,Dimitrova_Feigelman2003,Agterberg2007,Agterberg2012,Smidman2017_review}.
The vanishing total electric current is ensured for this state, rather than the present model where a spatially-uniform order parameter is phenomenologically introduced.
It is expected that all the qualitative results do not change even when a tiny center-of-mass momentum is taken into account, and unidirectional Majorana edge states also appear in the helical superconducting state. Summing up the contributions from the bulk and edge states, the system with unidirectional Majorana edge states has the vanishing total electric current.
}.
Note that it has been recently shown that similar unidirectional edge modes can be induced with applying an external supercurrent~\cite{Takasan2021}.

\begin{figure}
    \centering
    \includegraphics[width=0.45\textwidth]{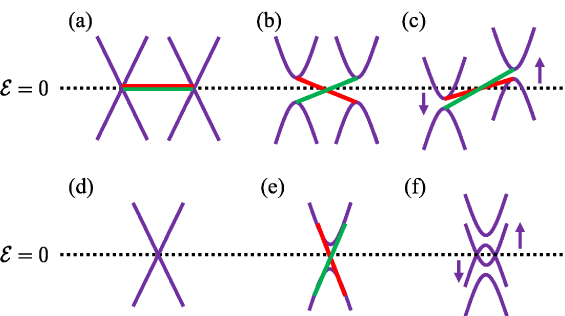}
    \caption{Schematic picture for the evolution of energy dispersion under $\bm{H}$.
    The purple lines correspond to the bulk spectrum, while the red and green lines represent edge modes.}
    \label{fig:schematics}
\end{figure}

The evolution of the edge states under $\bm{H}$ is summarized in  Figs.~\ref{fig:schematics}(a)-(c).
Here, the purple lines represent bulk massless/massive Bogoliubov-Dirac quasiparticles, while the red and green lines represent the dispersion of the edge states.
In the absence of the Zeeman field, there appear the Majorana flat bands for appropriate boundary directions (Fig.~\ref{fig:schematics}(a)).
By applying a Zeeman field perpendicular to the $g$-vector, the Majorana flat bands are lifted to form the chiral Majorana edge states (Fig.~\ref{fig:schematics}(b)).
After the bulk Bogoliubov-Dirac quasiparticles cross the zero energy by tilting the Zeeman field, these chiral Majorana edge states are transformed into the unidirectional Majorana edge states (Fig.~\ref{fig:schematics}(c)).
When the Zeeman field is further tilted and $\bm{H}$ becomes parallel to the $g$-vector, the mass gap of Bogoliubov-Dirac quasiparticles vanishes.

We also summarize in Figs.~\ref{fig:schematics}(d)-(e) the evolution of the edge states with a different choice of the boundary directions, e.g. those for Fig.~\ref{fig:edge_perp}.
Although there are no Majorana flat bands, chiral Majorana edge states appear by applying the Zeeman field perpendicular to the $g$-vector, as ensured by the bulk-edge correspondence of the Chern number.
When the Zeeman field is tilted and the dispersion becomes gapless, these edge states overlap with the bulk quasiparticles and unidirectional Majorana edge states do not appear.
The numerical results corresponding to Fig.~\ref{fig:schematics}(f) is available in Ref.~\cite{Daido2017}.

In summary, chiral Majorana edge states, Majorana flat bands, and unidirectional Majorana edge states appear in nodal noncentrosymmetric superconductors with and without Zeeman field.
The former appears irrespective of the boundary directions, while the latter two appear for appropriate boundary choices.
The appearance of these Majorana edge states is not limited to topological $d$-wave superconductivity, and is
a general feature of the paramagnetically-induced TSC~\cite{Daido2017}.

\subsection{Extension to three dimensions: Weyl superconductivity}
\label{subsec:2E}

We have seen that 
2D noncentrosymmetric superconductors with point nodes are gapped by the Zeeman field, and TSC is realized.
Here, we discuss the paramagnetic effect on 3D noncentrosymmetric superconductors with line nodes.
It turns out that such superconductors become point-nodal superconductivity under the Zeeman field.
These superconducting states are an example of the so-called \textit{Weyl superconductivity}~\cite{Volovik2003,Meng2012}, which has topologically-protected point nodes (the Weyl points).
The known candidates of Weyl superconductivity have almost been limited to (intrinsic) chiral superconductors~\cite{Sato2016_review,Sato2017_review,Kallin2016,Schnyder2015_review} such as URu$_2$Si$_2$~\cite{Kasahara2007,Kittaka2016,Sato2016_review}, SrPtAs~\cite{Biswas2013,Fischer2014}, UPt$_3$~\cite{Joynt2002,Yanase2016}, and ferromagnetic superconductors including UCoGe, URhGe, and UGe$_2$~\cite{Aoki2019,Sau2012,Sato2016_review}.
The following discussion reveals that Weyl superconductivity is ubiquitous in the low-Zeeman-field phases of noncentrosymmetric line-nodal superconductors for both spin-singlet and spin-triplet dominant states.

As we have already seen, the condition~\eqref{eq:condition_gap} must be satisfied for a node to be gapped by the Zeeman field.
In other words, if $\bm{k}_0$ remains to be a nodal point even in the presence of $\bm{H}$, the three constraints
\begin{align}
    E_\pm(\bm{k}_0)=\psi(\bm{k}_0)\pm\bm{d}(\bm{k})\cdot\hat{g}(\bm{k}_0)=M(\bm{k}_0)=0,
    \label{eq:constraints}
\end{align}
must be simultaneously satisfied.
For a 2D wave number $\bm{k}_0=(k_{0x},k_{0y})$,
this is possible only for accidental cases by fine-tuning the parameters of the system.
When we turn to three dimensions, on the other hand, we have an additional wave number $k_{0z}$, whose variation in the Brillouin zone may play the same role as the fine-tuning.
Thus, some discrete points on the line nodes may remain gapless even under the Zeeman field, satisfying Eq.~\eqref{eq:constraints},
although most of the nodal points are gapped out.
Accordingly, it is expected that noncentrosymmetric line-nodal superconductors become point-nodal superconductivity under the Zeeman field.

From the viewpoint of topology, the obtained point nodes are the so-called Weyl points with a nontrivial monopole charge,
\begin{gather}
C(\bm{k}_W)=\int_{S_2(\bm{k}_W)} \frac{dS_k}{2\pi i}\,\hat{n}(\bm{k})\cdot \bm{B}(\bm{k}),\label{eq:WeylCharge}\\
B_i(\bm{k})=\sum_{n:occ.}\epsilon_{ijl}\braket{\partial_ju_n(\bm{k})|\partial_lu_n(\bm{k})}.
\end{gather}
Here, $S_2(\bm{k}_W)$ is a small spherical surface enclosing the Weyl point $\bm{k}_W$, while $dS_k$ and $\hat{n}(\bm{k})$ are its areal element and unit normal vector, respectively.
The Weyl charge~\eqref{eq:WeylCharge} is analogous to the magnetic monopole: The difference is that it resides in the momentum space and, instead of the magnetic field, there appears the Berry curvature $\bm{B}(\bm{k})$, the integrand of the Chern number (Eq.~\eqref{Chn}).
The Weyl points are topologically protected by the Weyl charge and are robust against perturbation of the system.
Superconductivity with Weyl points is referred to as the {Weyl superconductivity}, in analogy with the Weyl semimetals~\cite{Armitage2018}.
Corresponding to the Fermi arcs in Weyl semimetals, Weyl superconductors host the Majorana arcs on the surface Brillouin zone~\cite{Sato2016_review,Schnyder2015_review,Sato2017_review}.
It has been clarified that Weyl quasiparticles show an anomalous thermal Hall effect depending on the nodal positions~\cite{Sato2016_review} and intriguing phenomena induced by an emergent pseudo-magnetic field caused by the lattice distortion~\cite{Massarelli2017,Liu2017,Matsushita2018,Pacholski2018}.

To be specific, let us consider the case of CeRhSi$_3$ and CeIrSi$_3$.
It has been shown that both of them have quasi-2D Fermi surfaces with spin splitting due to the Rashba ~AOSC\cite{Yanase2012,Terashima2007,Onuki2012}.
We adopt the model~\eqref{eq:model_dwaveSC} with an additional $k_z$-dependent term in $\xi(\bm{k})$:
\begin{align}
\xi(\bm{k})&=-2t(\cos k_x+\cos k_y)+4t'\cos k_x\cos k_y\notag\\
&\quad -8\tilde{t}\cos \frac{k_x}{2}\cos \frac{k_y}{2}\cos k_z-\mu,
\end{align}
following Ref.~\cite{Tada2008}.
In the theoretical analysis of a noncentrosymmetric Hubbard model~\cite{Tada2008}, $S+p$-wave state with $\psi(\bm{k})\sim \cos 2k_z$ has been pointed out.
Thus, we adopt 
\begin{equation}
    \psi(\bm{k})=\psi_0 \left[\cos 2k_z+\delta_2(\cos k_x+\cos k_y)\right].
\end{equation}
Here, we additionally take into account a small inplane $\bm{k}$ dependence with $\delta_2$, which is allowed by the $A_1$ symmetry of the order parameter.
Owing to a finite $\delta_2$, the line nodes originally at $k_z=\pm\pi/4,\,\pm3\pi/4$ become not completely horizontal, and extend in a small region of $k_z$ around $k_z=\pm\pi/4,\,\pm3\pi/4$.

\begin{figure}
    \centering
    \includegraphics[width=0.45\textwidth]{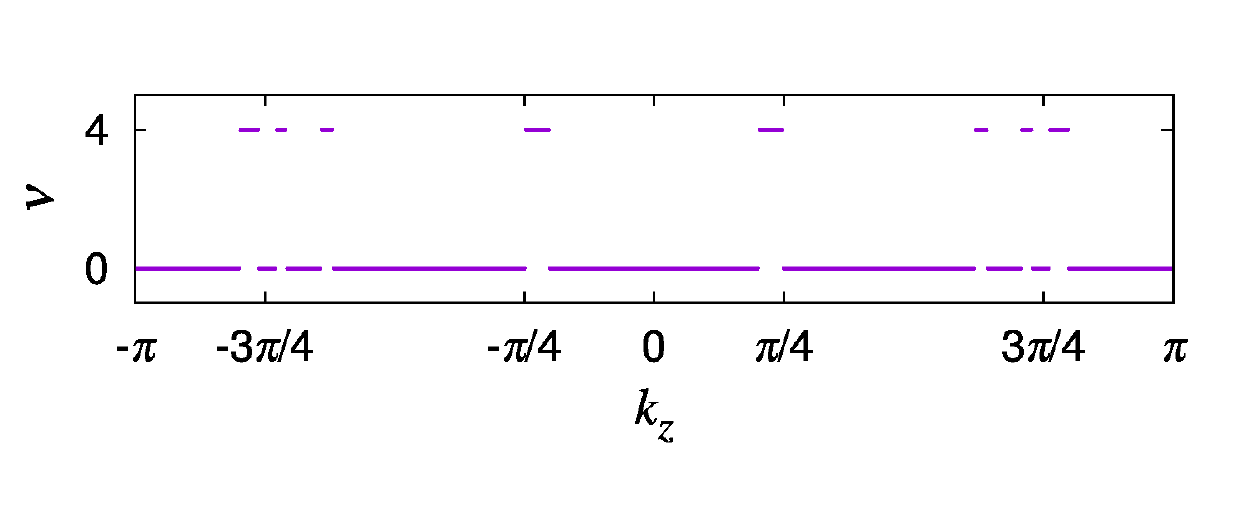}
    \caption{$k_z$-dependent Chern number $\nu(k_z)$ of the model for CeRhSi$_3$ and CeIrSi$_3$.
    We use $t=1$, $t'=0.475$, and $\tilde{t}=0.3$, following Ref.~\cite{Tada2008}.
    We adopt $\mu=-0.9$ to make charge density near the half filling.
    The other parameters are $\alpha=0.3$, $\psi_0=0.05$, $\delta_2=0.3$, and $d_0=0.01$.}
    \label{fig:Weyl}
\end{figure}

To identify the topology and Weyl points, we consider a 2D slice of the 3D Brillouin zone, fixing a value of $k_z$.
The Chern number $\nu(k_z)$ of the effective 2D system is well defined as long as the spectrum under $\bm{H}$ is gapful there.
Figure~\ref{fig:Weyl} shows the numerical result of $\nu(k_z)$ under finite Zeeman field $\bm{H}=H\hat{z}$.
Nontrivial Chern number $\nu(k_z)=4$ is obtained around the line nodes, where the effective 2D system is nodal for $H=0$ and is 2D TSC for $H\neq0$.
On the other hand, at $k_z$ away from the nodes the Chern number is trivial $\nu(k_z)=0$.
This is because the effective 2D system of such $k_z$ is gapful even in the absence of Zeeman field and, therefore, is topologically equivalent with the time-reversal symmetric $s$-wave superconductivity with the vanishing Chern number.

In Fig.~\ref{fig:Weyl}, the change of the $k_z$-dependent Chern number is seen for some $k_z$ around the line nodes.
Since the Chern number does not change as long as the band gap is open, such a change of $\nu(k_z)$ necessarily accompanies gapless points, {\it i.e.} the Weyl points.
Indeed, we have the relation
\begin{equation}
    \nu(k_{z}')-\nu(k_{z})=\sum_{k_{z}\le k_{Wz}\le k_{z}'}C(\bm{k}_W),
\end{equation}
according to the Stokes' theorem.
The left-hand side is the flux of Berry curvature threading the planes with constant $k_z$ and $k_z'$.
Its source is the Weyl points in between, as given on the right-hand side.
Each Weyl point carries $C(\bm{k}_W)=\pm1$ in this model as is usual and, therefore, four Weyl points are associated with each jump of $\nu(k_z)$. 
Thus, we have 64 Weyl nodes in this model.

We note that noncentrosymmetric $d$-wave superconductors with 3D Fermi surfaces become Weyl superconductivity under Zeeman field, by the same mechanism as discussed above~\cite{Yoshida2016}.
Actually, we obtain $\nu(k_z)=-4$ and $\nu(k_z)=0$ for $k_z$ with and without Fermi surfaces, respectively.
In reality, we also obtain $\nu(k_z)=-1$ and $-2$ for intermediate values of $k_z$.
For details, see Ref.~\cite{Yoshida2016}.

As another example, we consider 
a noncentrosymmetric superconductor CePt$_3$Si~\cite{Bauer2004,NCSC_book}.
Two pairing symmetries, namely, the $p+D+f$-wave~\cite{Yanase2007,Yanase2012} and $s+P$-wave~\cite{Frigeri2004,Yanase2007,Yanase2012} states, have been discussed as the candidate order parameter in accordance with experiments.
The first one is a spin-singlet dominant state and, therefore, Weyl superconductivity is naturally expected under the Zeeman field (See Ref.~\cite{Daido2016}).
Here we focus on the latter one,
the dominant $p$-wave state.
Such a state can be described by
\begin{equation}
\psi(\bm{k})=\psi_0,\quad    \bm{d}(\bm{k})=d_0(-\sin k_y,\sin k_x,0),
\end{equation}
with $d_0>\psi_0$.
The $s+P$-wave superconductivity with 3D Fermi surfaces hosts a line node, since one of the gap functions of the two bands with different helicities,
\begin{equation}
    \psi(\bm{k})\pm\bm{d}(\bm{k})\cdot\hat{g}(\bm{k})\sim \psi_0\pm d_0 |\bm{k}|,
\end{equation}
vanishes for a small wave number $\bm{k}\sim \psi_0 /d_0$~\cite{Hayashi2006_SFD,Hayashi2006_NMR,Smidman2017_review}.
The line node is protected by a finite winding number~\eqref{winddef}, in contrast to centrosymmetric odd-parity superconductors, where the Kramers degeneracy trivializes $W(\bm{k}_0)$ and line nodes are usually unstable (known as the Blount's theorem~\cite{Blount1985,SigristUeda_91,Kobayashi2014})
\footnote{
It has been revealed that the stable line nodes may exist at the Brillouin-zone faces of nonsymmorphic odd-parity superconductors~\cite{Norman1995,Micklitz-Norman2009,Kobayashi2016,Nomoto2017, Sumita_Sr2IrO42017, Sumita-Yanase2018,Sumita2019,Yanase2016}.
}.
By considering the cross section of the line node with a constant $k_z$ plane, we obtain an effective 2D model with nodal points.
According to the general considerations in Secs.~\ref{subsec:2B}~and~\ref{subsec:2C}, the nodal triplet-dominant state is gapped by the Zeeman field but should give a trivial Chern number.
However, the present case is exceptional, since the nodes are located on only one of the Fermi surfaces and the cancellation between the different helicity bands does not occur.
Thus, we also obtain a finite $k_z$-dependent Chern number for this type of line nodes and, Weyl superconductivity is realized under the Zeeman field.
For more details, see Ref.~\cite{Daido2016}.


Note that the models adopted here do not completely reproduce the Fermi surfaces of CeRhSi$_3$~\cite{Terashima2007,Onuki2012}, CeIrSi$_3$~\cite{Onuki2012}, and CePt$_3$Si~\cite{Onuki2012,Hashimoto2004_CePt3Si,Samokhin2004}.
However, the mechanism to realize Weyl superconductivity is generally applicable to line-nodal noncentrosymmetric superconductors.
In particular, the appearance of the Weyl points is independent of the details of the dispersion, although its details, e.g. where and how many Weyl points appear, may depend.
Thus, Weyl nodes are expected for realistic Fermi surfaces as well.

\subsection{Experimental platform}
\label{subsec:2F}

We have studied TSC induced by the paramagnetic effect.
The necessary ingredients are (1) thin films of nodal spin-singlet superconductors, (2) inversion-symmetry breaking, and (3) the Zeeman field.
In the following, we discuss the ingredients (1)-(3) order by order.

\begin{enumerate}
\renewcommand{\labelenumi}{(\arabic{enumi})}
\item The 2D $d$-wave superconductors naturally offer a promising platform to realize the topological $d$-wave superconductivity.
Thin films and superlattices of cuprate superconductors~\cite{Bollinger2011, Leng2011, Nojima2011} and a heavy-fermion superconductor CeCoIn$_5$~\cite{Izaki2007, Mizukami2011, Shimozawa_superlattice_RPP2016,naritsuka2021} have already been fabricated.
Recently, a twisted bilayer cuprate superconductor has also been fabricated~\cite{zhao2021emergent}. 

Thin films of anisotropic $s$-wave superconductors with accidental nodes
are also the candidate.
For example, implications of the line nodes have been reported for the bulk FeSe~\cite{Song2011, Kasahara2014, Shibauchi2020}.
Unfortunately, thin films of FeSe have been reported to show a gapful superconducting state~\cite{Zhang2017_FeSeARPES, Wang2017_FeSeFilm, Huang2017_FeSe}.
If nodal superconductivity is realized by changing system parameters and the environment, FeSe thin films can be a good platform for 2D TSC.
\item Inversion-symmetry breaking can be introduced by the structural asymmetry.
Thus, the presence of the substrate naturally makes the thin film superconductors noncentrosymmetric.
Since we have assumed the conditions given by Eq.~\eqref{eq:weak_coupling}, the strength of the ASOC $\alpha$ should at least amount to $\alpha\sim T_c$.
Although it is unclear that this condition is satisfied in the thin film superconductors mentioned above, $\alpha$ can be controlled to some extent by the gating techniques including the electric double-layer transistor~\cite{Liang2012_EDLT, Takase2017_EDLT, Premasiri2018}.
Artificial superlattices of CeCoIn$_5$~\cite{Shimozawa_superlattice_RPP2016} are also promising, where the Rashba spin-orbit coupling can be introduced by the tricolor stacking and is tunable by the layer thickness~\cite{Naritsuka2017,naritsuka2021}.
These noncentrosymmetric $d$-wave superconductors offer a fertile ground to explore TSC.

\item A natural way to introduce the Zeeman field is to apply an external magnetic field.
A concern is the orbital depairing effect, which is not taken into account in the present analysis.
A least orbital depairing effect is expected in (a) the heterostructure with a ferromagnet and (b) laser-irradiated cuprate thin films.
In the system (a), the Zeeman field is introduced by an exchange coupling.
Therefore, a sizable Zeeman field can be induced.
The heterostructures of the cuprate superconductors YBa$_2$Cu$_3$O$_{7}$~\cite{Chakhalian2006,Satapathy2012,Uribe2014,Sen2016} and La$_{2-x}$Sr$_x$CuO$_4$~\cite{Das2014,Uribe2014} with a ferromagnet La$_{2/3}$Ca$_{1/3}$MnO$_3$
have already been realized.
In the system (b), it is shown by the Floquet theory that a Zeeman field is introduced to the cuprate thin films by applying a high-frequency circularly-polarized laser light.
The details of this proposal are discussed in Sec.~\ref{sec:Takasan}.
The systems (a) and (b) free of the orbital depairing are the promising platforms of topological $d$-wave superconductivity.

We note that an external magnetic field may also be used to realize TSC, although we should take care of the existence of the vortices.
Cuprate superconductors and CeCoIn$_5$ are believed to have a large Maki parameter $\sqrt{2}H_0^c/H_0^P$,~\cite{Sigrist_LNCSreview2014}
meaning a small vortex density near the Pauli limiting field $H=H_0^P$.
Here,  $H_0^c$ and $H_0^P$ are the critical fields due to the orbital and Pauli depairing effects, which are estimated for the centrosymmetric cases.
On the other hand, the Pauli limiting field is absent in noncentrosymmetric superconductors for perpendicular magnetic fields, while $H_0^c$ and thus the coherence length would be of the same order of magnitude as the centrosymmetric cases.
Thus, we may obtain a superconducting state with sparse vortices for $H^P_0\lesssim H\ll H^c_0$, where a mean inter-vortex distance is considerably larger than the coherence length.
The region away from vortices, which spans most of the entire system, can be viewed as the system with a uniform Zeeman field.
To be precise, the vortices cause the Doppler shift due to the supercurrent, by which many experimental results are fitted well~\cite{Matsuda2006}.
However, its primary effect is to simply shift the quasiparticle spectrum and, therefore, topological properties are expected to be preserved.
Thus, it is feasible to observe gap opening and Majorana quasiparticles under an external magnetic field.
\end{enumerate}

In addition to the ingredients discussed above, it is desirable to obtain a large gap.
This allows an easier observation of the induced gap and Majorana edge states by {\it e.g.} scanning-tunneling microscopy (STM) and angle-resolved photoemission spectroscopy (ARPES).
Let us estimate the size of the mass gap induced by the paramagnetic effect.
For the evaluation of Eq.~\eqref{eq:condition_gap}, we have to know the ratio $r=d_0/\psi_0$ by which the spin-triplet component is admixed 
as a result of the inversion-symmetry breaking.
Here, $\psi_0$ and $d_0$ are the typical strength of the spin-singlet and -triplet components, respectively.
By using $r$, the induced band gap is given by
\begin{equation}
    \Delta E\equiv 2|M(\bm{k}_0)|\simeq\frac{2\mu_B H\psi_0}{\alpha}\,r.
\end{equation}

A rough estimation of $r$ is $r\sim\alpha/W$ with $W$ the energy scale of the normal state such as the Fermi energy and the bandwidth.
This is intuitive for small $\alpha/W$, but seems to be valid for relatively large $\alpha/W$ as well, as confirmed by a recent theoretical calculation of the Rashba-Hubbard model ($r\sim 0.5$ for $\alpha/W\sim0.5$)~\cite{Nogaki2020}.
We note that the ratio $r$ would be significantly enhanced even for a small $\alpha/W$, when the attractive interactions in the spin-singlet and spin-triplet channels are nearly degenerate.
Such a situation may also give a promising platform,
but here we limit ourselves to the simplest cases where $r\sim\alpha/W$ holds.
Thus, we obtain an estimate of the gap size,
\begin{equation}
    \Delta E\sim\frac{\psi_0}{W}\,\mu_B H.
\end{equation}
According to this expression, strong-coupling superconductors are suitable for obtaining a large induced gap.
Cuprate superconductors and CeCoIn$_5$ superlattices are promising candidates also for this reason.
Iron-based superconductors may also be a good candidate when nodal excitation is realized.
In particular, the parameter $\psi_0/W$ amounts to be $\lesssim 0.1$ in cuprate superconductors~\cite{Yanase2003}.
Thus, we obtain $\Delta E\sim 1\mathrm{K}$ under $\mu_B H\sim 10\mathrm{T}$.
The Zeeman field $\mu_B H\sim 10\mathrm{T}$ is not so small but would be accessible in experiments by external magnetic fields. The molecular field of ferromagnets may be much larger.
The gap $\Delta E\sim 1\mathrm{K}$ would also be within the precision of cutting-edge experimental techniques~\cite{Machida2019}.
Thus, we expect that the induced gap and Majorana edge states are observable in experiments.

\subsection{Summary of this section}
\label{subsec:2G}


In this section, we have discussed a general mechanism to realize 2D TSC with nontrivial Chern number.
The platform is nodal noncentrosymmetric superconductors.
The gapless Bogoliubov-Dirac quasiparticles become massive under the Zeeman field, owing to the interplay of the paramagnetic effect with the parity mixing of the order parameter.
The obtained gapful superconducting state has nontrivial Chern numbers for dominant spin-singlet pairings, while the dominantly spin-triplet pairing states usually have the trivial Chern number.

According to the bulk-boundary correspondence, chiral Majorana edge states appear in systems with open boundaries.
We have shown the evolution of the Majorana edge states under tilting of the Zeeman field and identified the unidirectional Majorana edge states with a peculiar unidirectional flow on the edges.
The relation of these edge states to the Majorana flat bands in the absence of the Zeeman field has also been revealed.
Furthermore, we have extended the theory to 3D systems.
3D line-nodal superconductors become Weyl superconductors under the Zeeman field.
We have discussed the possibility of the Weyl superconductivity in CeRhSi$_3$, CeIrSi$_3$, and CePt$_3$Si under the Zeeman field.

We have also discussed experimental platforms and estimated the induced gap size.
Thin films and superlattices of $d$-wave superconductors are the promising candidates of TSC.
The Zeeman field can be introduced by applying the external magnetic field, making heterostructure with ferromagnet, or irradiating a circularly polarized laser light.
We expect the excitation gap and Majorana edge states 
would be observable in future experiments.

\section{Laser-induced topological d-wave superconductivity}
\label{sec:Takasan}

As mentioned in the previous section, the application of laser light is one of the promising routes for realizing topological $d$-wave superconductivity. We proposed that thin films of $d$-wave superconductors under laser light can be topologically nontrivial and showed chiral Majorana edge modes~\cite{Takasan2017a}. In this section, we explain this proposal in detail.

The realization of TSC has been an intriguing research subject in condensed matter physics during the last decade. 
Thus, it is an important task for theorists to propose a promising new scenario for the realization of TSC in natural materials. For this purpose, in this section we focus on the idea of Floquet engineering, which is a scheme for controlling the states of matter with a periodic driving and gathering great attention in recent years~\cite{BukovReview2015, EckardtRMP2017, OkaReview2019, Rudner2020}. 
Based on this idea, we searched superconductors which become topologically nontrivial under the periodic driving with laser light. Then, we found that the 2D $d$-wave superconductors fabricated on a substrate can be a topological superconductor with application of circularly polarized laser light~\cite{Takasan2017a}. One of the strength of this proposal is its wide applicability. Our scheme can be applied to any $d$-wave superconductors typically realized in strongly correlated materials such as high-$T_c$ cuprate superconductors. Another strong point is its high controllability. Changing the properties of laser light ({\it e.g.}, intensity, frequency, polarization and so on), we can control various topological phases. Furthermore, using a short laser pulse, we can control the topological nature in an ultrafast time scale such as a picosecond or a femtosecond order. 

This section is organized as follows. First, we briefly review Floquet engineering and Floquet theory in Sec.~\ref{Sec:IIIA}. Then, we introduce our model describing the laser-irradiated $d$-wave superconductors in Sec.~\ref{Sec:IIIB} and derive the effective static model with the Floquet theory in Sec.~\ref{Sec:IIIC}. Using the model, we show the topological properties of $d$-wave superconductors under laser light in Sec.~\ref{Sec:IIID}. Finally, we discuss the experimental setup in Sec.~\ref{Sec:IIIE} and address the conclusion and the outlook in Sec.~\ref{Sec:IIIF}.

\subsection{Floquet engineering and Floquet theory}\label{Sec:IIIA}
Thanks to the recent developments of experimental techniques (e.g., realization of strong laser light applicable to solids and highly controllable quantum systems such as ultracold atoms), it becomes possible to realize periodically-driven quantum systems in experiments~\cite{BukovReview2015, EckardtRMP2017, OkaReview2019, Rudner2020}. Periodically-driven systems show nonequilibrium states different from equilibrium states. Choosing the adequate driving protocol, we can realize the desired quantum states in nonequilibrium states. This approach for controlling quantum states is called \textit{Floquet engineering}~\cite{BukovReview2015, EckardtRMP2017, OkaReview2019, Rudner2020}, which is named after Floquet theory. This is a theoretical framework for periodically-driven systems described with a time-periodic Hamiltonian $H(t)=H(t+T)$ where $T$ is a time period and based on the Floquet's theorem which is the analogue of the Bloch's theorem in the time direction. In general, time-dependent quantum many-body problems are difficult to solve, but the application of the Floquet's theorem makes the problem tractable even in strongly correlated systems. Floquet engineering schemes for various phases have been proposed~\cite{Oka2009, Lindner2011, Kitagawa2011, Mikami2016, Ezawa2015, Claassen2019, Sato2016, Claassen2017, Kitamura2017, Takasan2017b}, and the most widely-studied one is the engineering of band topology. A pioneering work by Oka and Aoki showed that a periodic driving with circularly polarized laser makes Dirac cones of graphene gapped out and induces topologically nontrivial states showing chiral edge modes~\cite{Oka2009}. This finding stimulated many researchers to study this kind of topological states called Floquet topological phases. 
However, it took a long time to realize them in experiments, especially in solids. The first experimental realization was achieved in ultracold atoms which are loaded in a periodically-shaken optical lattice~\cite{Jotzu2014}. In solid states, a similar phenomenon on the surface of the topological insulator Bi$_2$Se$_3$ has been reported~\cite{Wang2013, Mahmood2016}. Although the experiment in graphene had not been reported for a while, it was reported very recently that graphene irradiated by a short laser pulse shows a nearly-quantized transient Hall current, which should be a signature of the Floquet topological phases~\cite{McIver2020}. Floquet engineering in solids is becoming a possible way to control the topological states of matter in experiments and gathering great attention.

\begin{figure}[t]
\includegraphics[width=8.5cm]{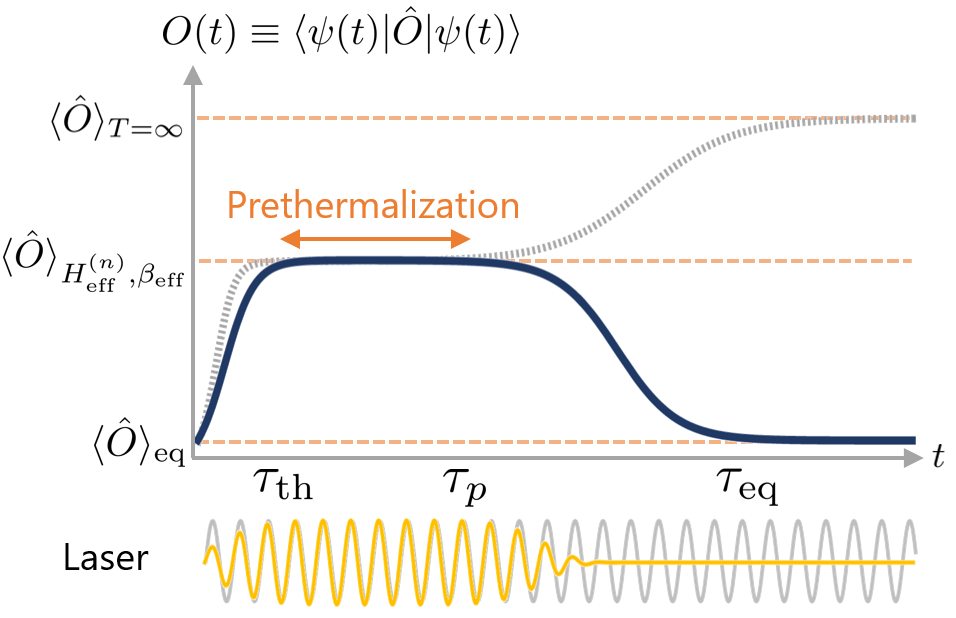}
\caption{Typical real-time evolution of an observable $\hat{O}$ under laser light. The gray broken and blue solid curve represent the time evolution induced by a continuous wave (gray curve) and a pulse wave (yellow curve), respectively. Prethermalization occurs within the time scale $\tau$ satisfying $\tau_\mathrm{th} \lesssim \tau \lesssim \tau_p$.}
\label{Fig:FP}
\end{figure}

In the rest of this subsection, we introduce the basic theoretical concepts in the Floquet theory needed for explaining our proposal. For the detail of the Floquet theory, see other review papers~\cite{BukovReview2015, EckardtRMP2017, OkaReview2019, Rudner2020}. As we mentioned above, we would like to consider periodically-driven $d$-wave superconductors. The $d$-wave superconductivity is typically realized in strongly correlated electron systems, but there is a crucial problem for periodically-driven interacting closed~\footnote{You might think electrons in solids are not a closed system since they interact with the other degrees of freedoms such as phonons. However, they are still considered to be approximated with closed systems in a short time scale where the typical experiments with strong laser light are performed~\cite{OkaReview2019}.} systems that they thermalize to the infinite temperature state in the long time limit~\cite{DAlessio2014}. Thus, it is naively expected that driven $d$-wave superconductors only show a topologically trivial phase. However, owing to the recent development of the Floquet theory~\cite{Abanin2015, Kuwahara2016}, it was proved that there appears a nontrivial prethermalized state before thermalizing to the infinite temperature state~\footnote{Strictly speaking, we additionally need the off-resonant condition to achieve the prethermalized state. We mention this point in Sec.~\ref{Sec:IIIE}.} and the prethermalized state's lifetime $\tau_p$ becomes exponentially longer when the driving frequency $\omega \equiv 2\pi/T$ increases. This means that the expectation value $O(t) \equiv \bra{\psi(t)}\hat{O} \ket{\psi(t)}$ of an observable $\hat{O}$ approaches $\langle \hat{O} \rangle_{H^{(n)}_\eff, \beta_\eff} \equiv \mathrm{Tr}[\hat{O} \exp(- \beta_\eff H^{(n)}_\eff)]$ up to the lifetime $\tau_p$ and finally reaches $\langle \hat{O} \rangle_{T=\infty} \equiv \mathrm{Tr}[\hat{O}]$. 
Here, $\beta_\eff$ and $H^{(n)}_\eff$ are the effective inverse temperature and the truncated effective Hamiltonian, respectively (the definition of $H^{(n)}_\eff$ is given below). 
The dynamics is schematically shown with the gray broken curve in Fig.~\ref{Fig:FP}, suggesting that a topologically nontrivial phase may appear in driven $d$-wave superconductors transiently in the prethermalized state. Such a transient topological phase is described with the thermal state of the static (time-independent) Hamiltonian $H^{(n)}_\eff$. To define $H^{(n)}_\eff$, we introduce the original effective Hamiltonian $H_\eff \equiv (i/T) \log U (T)$ where $U(T)$ is a time-evolution operator for one period. This Hamiltonian plays an essential role in the Floquet theory, but it is difficult to directly calculate it. To avoid this difficulty, the perturbative expansion in powers of $(1/\omega)$ is widely used~\cite{BukovReview2015, EckardtRMP2017, OkaReview2019, Rudner2020}. The effective Hamiltonian is expanded as
\begin{align}
H_{\mathrm{eff}} = \calH_0 + \sum_{n=1}^\infty \frac{[H_{+n},H_{-n}]}{n \omega} + \calO\left[\left(\frac{1}{\omega}\right)^{2}\right], \label{eq:Heff_expansion}
\end{align}
where $H_n \equiv\frac{1}{T} \int^{+T/2}_{-T/2}dt H(t) e^{-i n \omega t}$ is the Fourier components of the time-dependent Hamiltonian~\footnote{There are several schemes for the high-frequency expansion, and we take the van Vleck expansion. The term depending on the initial time does not appear in the van Vleck expansion while it appears in other expansion scheme (e.g. Floquet-Magnus expansion). For detail, see a detailed paper about this point~\cite{Mikami2016} or the review articles~\cite{OkaReview2019, EckardtRMP2017}.}. This formula provides a way to obtain the effective Hamiltonian, but the expansion is known not to be a convergent series in a large system size in general~\cite{Kuwahara2016}. Therefore, we typically adopt a truncated Hamiltonian $H^{(n)}_\eff$ which only contains terms up to $n$-th order of $(1/\omega)$. In the following, we use the $H^{(1)}_\eff$ as the effective Hamiltonian for simplicity. In other words, we redefine $H_\eff \equiv H^{(1)}_\eff$ here.

Finally, we mention the case when we use a pulse laser with a finite pulse width. To pump the electrons in solids, we need strong intensity of laser light in most cases. To gain it, we can use a short laser pulse. However, driving by a single pulse (shown at the bottom of Fig.~\ref{Fig:FP} with the yellow curve) is obviously not time-periodic and the Floquet theory is not applicable in the strict sense. 
On the other hand, signatures predicted by the Floquet theory are observed in solid-state experiments even when the pulse laser is used~\cite{Wang2013, Mahmood2016, McIver2020}. Also, it was theoretically reported that the Floquet-like behavior is reproduced with a pulse laser~\cite{Kalthoff2018}. These suggest that we can approximately apply the Floquet theory in the time window where the electric field is strongly oscillating near the pulse's center. Thus, as shown in Fig.~\ref{Fig:FP}, the prethermalized state is expected to be realized in the intermediate time scale even with a pulse laser, and after the application of the laser pulse, the system goes back to the thermal equilibrium at $\tau_\mathrm{eq}$ where $O(t)$ thermalizes to $\langle \hat{O} \rangle_\mathrm{eq}\equiv \mathrm{Tr}[\hat{O} \exp(- \beta_0 H(0))]$ with the initial temperature $\beta_0$. Therefore, to realize the Floquet engineering with pulse laser, we have to choose a pulse width $w$  as it satisfies the condition $\tau_\mathrm{th} \lesssim w \lesssim \tau_p$, where $\tau_p$ is the lifetime of the prethermalized states explained before and $\tau_\mathrm{th}$ is the thermalization time needed for reaching the prethermalized state from the initial state. In most strongly correlated materials, $\tau_\mathrm{th}$ is the order of femtoseconds determined by the electron-electron interaction. In contrast, $\tau_p$ depends on the details of material and laser light~\footnote{From theoretical studies, it is shown that $\tau_p \sim e^{\calO(\omega)}$ generally holds~\cite{Kuwahara2016}. However, the actual lifetime can be modified in solids because there are various ingredients not taken into account, such as unoccupied bands, impurities, and phonon bath.}. Thus, we should choose as short as possible satisfying $w \gtrsim \tau_\mathrm{th}$ and $w$ should be the order of ten femtoseconds or sub-picoseconds typically. We note that $\tau_p$ can be effectively longer considering the dissipation to thermal baths such as phonons or substrates which are neglected here. 

\begin{figure}[t]
\includegraphics[width=7.5cm]{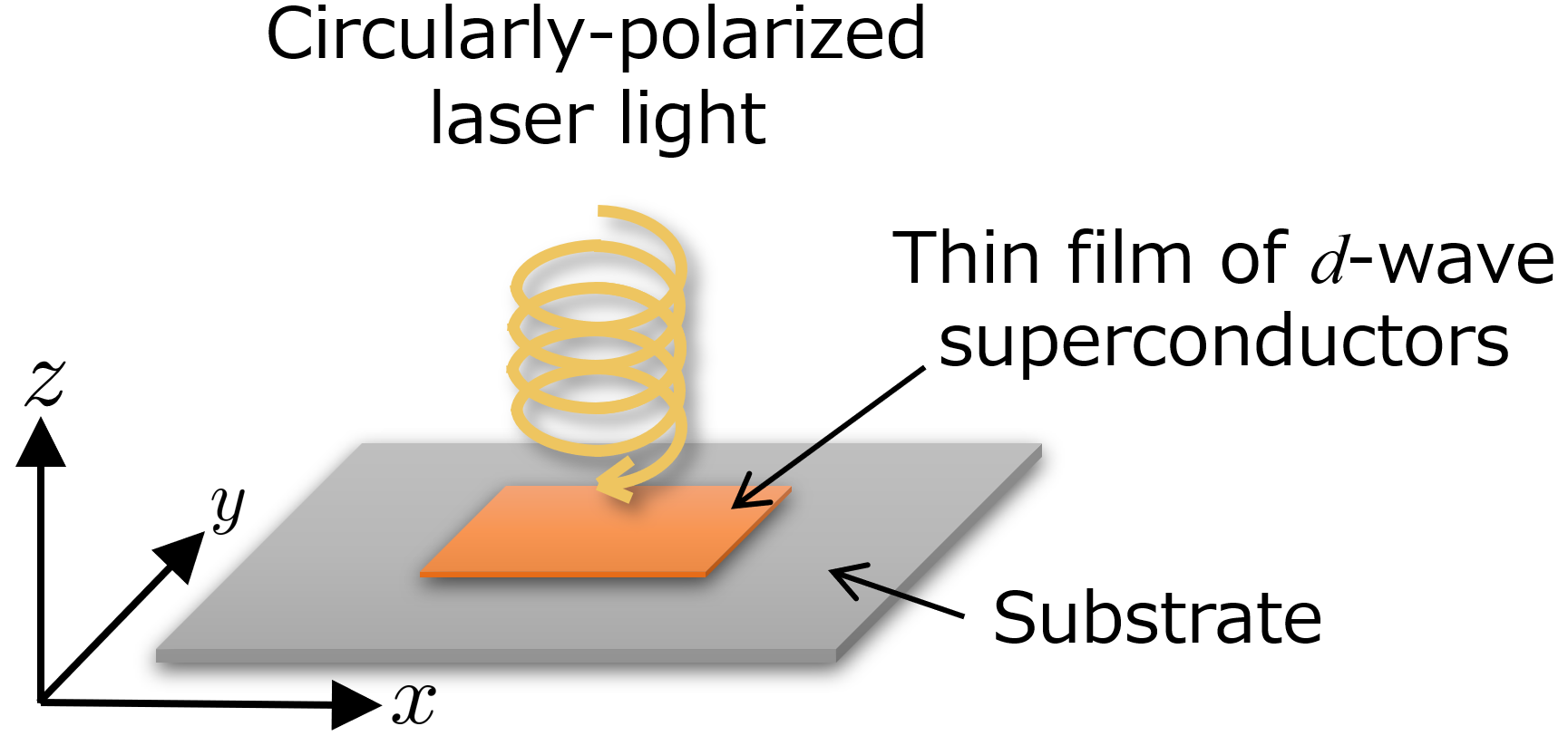}
\caption{Schematic picture of the setup. A $d$-wave superconductor thin film fabricated on a substrate is irradiated by circularly polarized laser light in the $z$-direction.}
\label{Fig:setup}
\end{figure}

\subsection{Model}\label{Sec:IIIB}
We consider the laser-irradiated thin film of $d$-wave superconductors (the setup is shown in Fig.~\ref{Fig:setup}) and discuss its topological properties. For this purpose, we first explain the model for the thin film of $d$-wave superconductors fabricated on a substrate. The potential induced by the substrate breaks the inversion symmetry, and thus a Rashba spin-orbit coupling appears in this system. To describe it, we introduce a Rashba-Hubbard model as 
\begin{align}
\calH&=\sum_{\bm k \sigma} \xi (\bm k) c_{\bm k \sigma} ^\dagger c_{\bm k \sigma} 
\nonumber \\
&+\sum_{\bm k \sigma \sigma'} (\alpha \bm g (\bm k) \cdot \bm \sigma)_{\sigma \sigma'}
c_{\bm k \sigma} ^\dagger c_{\bm k \sigma'} 
+U \sum_i n_{i \uparrow} n_{i \downarrow}, 
\end{align}
where the kinetic energy $\xi ({\bm k})$ and g-vector $\bm g (\bm k)$ have been given by 
Eqs.~\eqref{eq:dispersion_sec1} and \eqref{eq:g-vector_sec1}, respectively. 
We choose the form of $\xi ({\bm k})$ with the next-nearest neighbor hopping $t^\prime$ in a square lattice for reproducing the Fermi surface of typical cuprate materials well-known as $d$-wave superconductors. In addition, we incorporate the effect of laser light. The laser light is described by a classical electromagnetic field $\bm A (t)$ and introduced as a Peierls phase, which corresponds to the substitution $\bm k \to \bm k - \bm A (t)$ in the momentum space. Then, we obtain the time-dependent Hamiltonian as
\begin{align}
\calH(t)&=\sum_{\bm k \sigma} \xi (\bm k - \bm A (t)) c_{\bm k \sigma} ^\dagger c_{\bm k \sigma} 
\nonumber \\
&+\sum_{\bm k \sigma \sigma'} (\alpha \bm g (\bm k- \bm A (t)) \cdot \bm \sigma)_{\sigma \sigma'}
c_{\bm k \sigma} ^\dagger c_{\bm k \sigma'} 
+U \sum_i n_{i \uparrow} n_{i \downarrow}, \label{eq:t-dep}
\end{align}
with $\bm A(t) = (A_x \cos \omega t, A_y \sin \omega t ,0)$, which corresponds to the circularly ($A_x=A_y$) or elliptically ($A_x \neq A_y$) polarized laser light. We adopt the Hamiltonian~(\ref{eq:t-dep}) as a model describing laser-irradiated $d$-wave superconducting thin films.

\subsection{Effective Hamiltonian}\label{Sec:IIIC}
The model (\ref{eq:t-dep}) is a many-body and time-dependent Hamiltonian and thus not easy to treat directly. Here we focus on the time-periodicity of the Hamiltonian (\ref{eq:t-dep}) and apply the Floquet theory, which enable us to understand the properties of this model via the effective static Hamiltonian. To derive the effective Hamiltonian, we use the formula~(\ref{eq:Heff_expansion}) and then obtain 
\begin{align}
\calH_\eff &= \calH_0 + \sum_{n = 1}^\infty \frac{[\calH_{+n},\calH_{-n}]}{ n \omega} \nonumber \\
&=\sum_{\bm k \sigma} \tilde{\xi}_0 (\bm k) c_{\bm k \sigma} ^\dagger c_{\bm k \sigma} \nonumber\\
&+\sum_{\bm k \sigma \sigma'} (\alpha \bm \tilde{g}_0 (\bm k) \cdot \bm \sigma)_{\sigma \sigma'}
c_{\bm k \sigma} ^\dagger c_{\bm k \sigma'}+U \sum_i n_{i \uparrow} n_{i \downarrow}  \nonumber \\
&- \sum_{\bm k \sigma \sigma'} \mu_B \tilde{H}(\bm k)\sigma_z  c_{\bm k \sigma} ^\dagger c_{\bm k \sigma'},  \label{eq:eff_model}
\end{align}
where
\begin{align}
\tilde{\xi}_0 ({\bm k})&=-2 t(J_0(A_x)\cos k_x+J_0(A_y)\cos k_y) \nonumber \\
&\qquad + 4t^\prime J_0 \left(\sqrt{A_x^2+A_y^2}\right)  \cos k_x \cos k_y - \mu, \label{eq:eff_xi}\\
\tilde{\bm g}_0 (\bm k)&= (-J_0(A_y)\sin k_y, J_0(A_x)\sin k_x, 0), \label{eq:eff_SOC}\\
\tilde{H}(\bm k) &= - \frac{ 4 \alpha ^2 \calJ^2(A_x,A_y)}{\mu_B \omega}\cos k_x \cos k_y, \label{eq:eff_mag}\\
\calJ^2(A_x,A_y)&= \sum_{m = 0}^\infty  \frac{(-1)^{m}J_{2m+1} (A_x) J_{2m+1} (A_y)}{2m+1},
\end{align}
and $J_n(x)$ represents the $n$-th Bessel function. 
There appear two laser-induced effects in the effective model (\ref{eq:eff_model}). One is  reduction of the hopping and the spin-orbit coupling seen in Eqs.~(\ref{eq:eff_xi}) and (\ref{eq:eff_SOC}). This effect is known as dynamical localization which has been experimentally observed in ultracold atoms~\cite{Lignier2007} and solids~\cite{Ishikawa2014} and leads to a change in the shape of the Fermi surface. The other is the laser-induced magnetic field (\ref{eq:eff_mag}) which reflects the time-reversal symmetry breaking due to the circularly polarized laser light. This term is closely related to the inverse Faraday effect~\footnote{L. P. Pitaevskii, Sov. Phys. JETP {\bf 12}, 1008–1013 (1961).}, meaning the magnetization induced by the circularly polarized light.  

\begin{figure}[tbp]
\includegraphics[width=9cm]{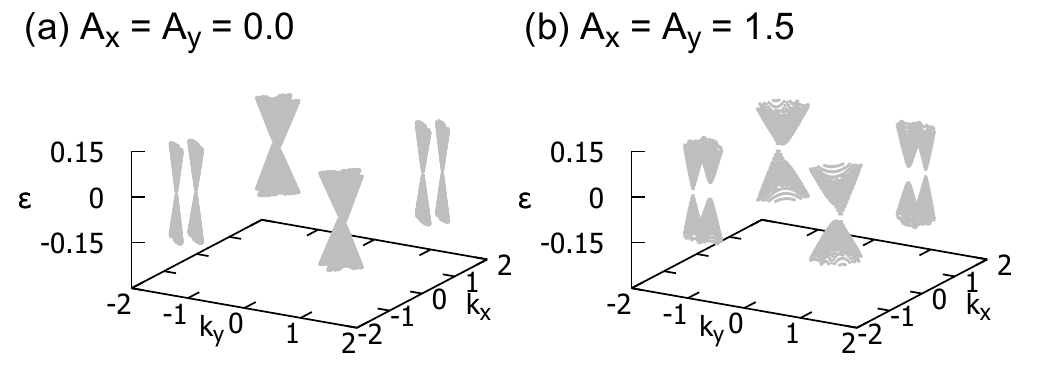}
\caption{Quasiparticle spectrum of the effective BdG Hamiltonian [Eq.~(\ref{eq:BdG})] (a) without and (b) with laser light. The parameters are set as $t=1.0$, $t^\prime=0.2$, $\alpha=0.3$, $\omega=0.4$, $\Delta_d = 0.4$, and $\Delta_p = 0.08$.}
\label{Fig:QP}
\end{figure}

Next, we take into account the superconducting order. We refer to the results of many-body calculations~\cite{Tada2009, Shigeta2013, Yanase2007, Yanase2008, Nogaki2020, Lu2018} and introduce the mean field of superconductivity. Due to the inversion symmetry breaking, there should be an additional odd-parity component, such as $p$-wave or $f$-wave pairing. To introduce this effect to our model, we adopt a simple form of $D$+$p$ wave order parameter $\Delta(\bm k)= i [\psi(\bm k)+ \bm d (\bm k) \cdot \bm \sigma]\sigma_y $ with $\psi(\bm k)= \Delta_d (\cos k_x - \cos k_y)$ and $\bm d (\bm k)= \Delta_p (\sin k_y, \sin k_x, 0)$, assuming that $|\Delta_d|$ is much larger than $|\Delta_p|$. With this order parameter, we write down the BdG Hamiltonian as $\calH_\mathrm{BdG} = \frac{1}{2} \sum_{\bm k} \Psi^\dagger_{\bm k} \calH(\bm k) \Psi_{\bm k}$,
where
\begin{align}
\calH(\bm k) &=
\begin{pmatrix}
\calH_N(\bm k) & \Delta (\bm k) \\
\Delta^\dagger(\bm k) & -\calH_N^T(- \bm k) 
\end{pmatrix}, \label{eq:BdG}\\
\calH_N(\bm k)&= \tilde{\xi}_0 (\bm k) \sigma_0 + \alpha \bm \tilde{g}_0 (\bm k) \cdot \bm \sigma -\mu_B \tilde{H}(\bm k)\sigma_z, 
\end{align}
and $\Psi^\dagger_{\bm k} = (c^\dagger_{\bm k \uparrow}, c^\dagger_{\bm k \downarrow}, c_{-\bm k \uparrow}, c_{-\bm k \downarrow})$. Differences from Eq.~\eqref{eq:model_dwaveSC} are only the renormalization due to dynamical localization and the momentum dependence in an effective magnetic field $\tilde{H}(\bm k)$. Therefore, we expect qualitatively the same behaviors as we have seen in the previous section. Diagonalizing this BdG Hamiltonian, we obtain the quasiparticle spectrum shown in Fig.~\ref{Fig:QP}. Similarly to Fig.~\ref{fig_D+p}, the energy spectrum has nodal points before irradiating laser light, and these points are gapped out with applying laser light. As mentioned in the previous section, the nodes in $d$-wave superconductors are protected by combination of time-reversal symmetry and particle-hole symmetry. In addition to breaking inversion symmetry by the substrate, the laser-induced magnetic field~[Eq.~(\ref{eq:eff_mag})] breaks the time-reversal symmetry and then the nodes are made gapped. Thanks to this energy gap, the system can host robust topological phases and indeed show topologically nontrivial phases as explained below. 

\subsection{Topological properties}\label{Sec:IIID}
In this subsection, we investigate the topological properties of the laser-irradiated $d$-wave superconductor thin films. We focus on the weak intensity regime ($A_x, A_y \lesssim 1.5$) and explain there appears a topological phase. In the strong intensity regime ($A_x, A_y \gtrsim 1.5$), different topological phases can be realized. See the original paper~\cite{Takasan2017a} for detail of the strong intensity regime.

\subsubsection{Edge modes}
\begin{figure}
\includegraphics[width=8.5cm]{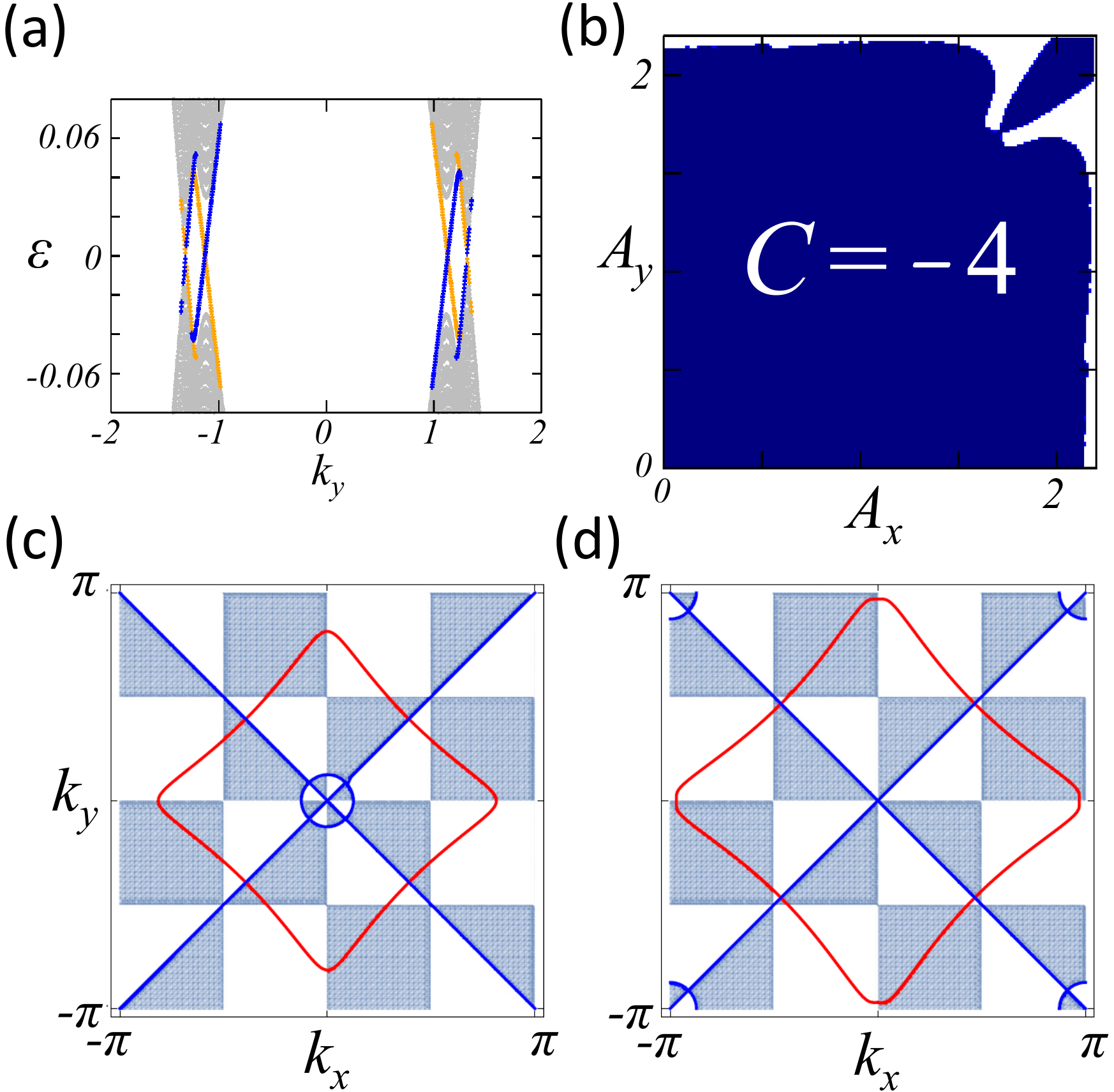}
\caption{(a) Energy spectrum of the effective Hamiltonian~[Eq.~(\ref{eq:BdG})] with the open boundary condition in the $x$-direction and the periodic boundary condition in the $y$-direction. The orange and blue dots represent the localized modes at each edges. (b) Topological phase diagram. Color plot shows numerically calculated Chern numbers for each $(A_x, A_y)$ point. The white area represents a topologically trivial phase ($C=0$) and the blue is a topologically nontrivial phase ($C=-4$). (c, d) Fermi surfaces (red lines) and zeros of the superconducting gap (blue lines) at $(A_x, A_y) = (0.1,0.1)$. The $E_+$ [$E_-$] band is shown in (c) [(d)], and the shaded (white) region represents $\psi \pm \bm d \cdot \hat{\bm g} / [\mu_B (\tilde{H} \hat{\bm z}) \cdot (\hat{\bm g} \times \bm d)/ \alpha] < 0~(>0)$. The parameters $(t, t^\prime, \alpha)$ are set as $(1.0, 0.2, 0.3)$ in all the figures. The others $(\omega, \Delta_d, \Delta_p)$ are set as $(0.4, 0.4, 0.08)$ in Fig.~\ref{Fig:topo}(a) and $(36.0, 0.05, 0.01)$ in Figs.~\ref{Fig:topo}(b-d).}
\label{Fig:topo}
\end{figure}

The most direct way to clarify if the topologically nontrivial phases are realized or not is to check the energy spectrum with the open boundary condition since there appear gapless modes localized at the boundary when the system is topologically nontrivial. For this purpose, we calculate the energy spectrum of the effective model [Eq.~(\ref{eq:BdG})] with the open boundary condition in the $x$-direction and the periodic boundary condition in the $y$-direction. The result is shown in Fig.~\ref{Fig:topo}~(a). There appear four chiral modes at each side of edges, and this fact suggests a topologically nontrivial phase based on the bulk-boundary correspondence. 

\subsubsection{Chern number and phase diagram}
Next, we study the topological index calculated from the bulk information. The above BdG Hamiltonian belongs to the class D in terms of the ten-fold classification~\cite{Schnyder_classification_free_2008,Kitaev_classification_free_2009,Ryu_classification_free_2010}, and it is known to be characterized by a $\mathbb{Z}$ topological index. Thus, we calculate the Chern number taking a value on $\mathbb{Z}$ defined for 2D systems. The definition has been given in Eq.~\eqref{Chn}. In this section, we denote the Chern number by $C$ instead of $\nu$ in Sec.~\ref{sec:Daido}.

To calculate the Chern number, we take two approaches. One is a numerical one called Fukui-Hatsugai-Suzuki method~\cite{Fukui_Hatsugai_05}, which is an efficient numerical way to compute the Chern number of the model defined on discretized momentum space. The numerical results are shown in Fig.~\ref{Fig:topo}~(b) as a phase diagram. It shows that the Chern number $C$ takes a non-zero value ($C=-4$) in a broad range of parameters including the infinitely weak intensity regime. Note that the Chern number on the $A_x$- and $A_y$-axes is ill-defined because the linearly polarized laser light preserving time-reversal symmetry does not gap out the nodal points. The absolute value of Chern number $|C|$ represents the number of chiral edge modes and the result is consistent with the edge spectrum in Fig.~\ref{Fig:topo}~(a).

The other approach is an analytical calculation following the formula explained in the previous section. 
From the same derivation as Eq.~\eqref{TSCchern}, the Chern number of our BdG Hamiltonian (\ref{eq:BdG}) can be calculated as
\begin{align}
C = \sum_{(\pm , \bm k_0)} \frac{1}{2} \mathrm{sgn} \left [ 
\frac{(\hat{\bm z} \times \nabla_{\bm k} E_\pm ) \cdot \nabla_{\bm k} (\psi \pm \bm d \cdot \hat{\bm g})}
{\mu_B (\tilde{H} \hat{\bm z}) \cdot (\hat{\bm g} \times \bm d)/ \alpha} \right]_{\bm k = \bm k_0}, \label{eq:ana_formula}
\end{align}
where $\hat{\bm z}$ is a unit vector in the $z$-direction, $E_\pm = \tilde{\xi_0}(\bm k) \pm \alpha |\tilde{\bm g} (\bm k)|$, $\hat{\bm g} = \tilde{\bm g} (\bm k) / | \tilde{\bm g} (\bm k)|$, and $\tilde{H}(\bm k)$ is ${\bm k}$ dependent~\cite{Daido2016}. The summation is taken over all the gapped nodes at $\bm k_0$ on the $E_\pm$ bands' Fermi surfaces defined as $E_\pm (\bm k)=0$. The analytic formula (\ref{eq:ana_formula}) enables us to evaluate the Chern number with counting the contribution from gapped nodes, which are intersections of a Fermi surface and zeros of the gap function $\psi (\bm k) \pm \bm d (\bm k) \cdot \tilde{\bm g (\bm k)}=0$. Each gapped node gives a contribution $+\frac{1}{2}$ or $-\frac{1}{2}$ and the sign of each contribution can be estimated as follows. First, we set the direction parallel to the Fermi surface of $E_\pm (\bm k)$ bands as $\hat{k}_{\pm}=\hat{\bm z} \times \nabla_{\bm k} E_\pm / |\hat{\bm z} \times \nabla_{\bm k} E_\pm|$. Next, we find the change in the sign of $\psi \pm \bm d \cdot \hat{\bm g} / (\mu_B (\tilde{H} \hat{\bm z}) \cdot (\hat{\bm g} \times \bm d)/ \alpha)$, which is in the argument of the function of Eq.~(\ref{eq:ana_formula}). When it changes from negative to positive (positive to negative) along the $\hat{k}_{\pm}$ direction at gapped nodes, the contribution is $+\frac{1}{2}$ ($-\frac{1}{2}$). Following this procedure, we can evaluate the Chern number analytically. Indeed, seeing Figs.~\ref{Fig:topo}(c)~and~(d), we can count each contribution from the gapped nodes, and the Chern number turns out to be $-4$, which coincides with the numerical result. 

\subsection{Experimental setups}\label{Sec:IIIE}
Our scheme has three advantages for realizing TSC in experiments. First one is that our theory can be applied to any $d$-wave superconductors. Second one is that the TSC can be induced by infinitesimally weak intensity of laser light. Third one is that the laser-induced magnetic field gives rise to only the paramagnetic effect and thus does not induce vortices which break superconductivity. In the following, we discuss the experimental setups for realizing TSC based on our proposal. 

\subsubsection{Material}
With a slight modification of the dispersion relation $\xi(\bm k)$, which is expected not to change the qualitative results, our results are basically applicable to any $d$-wave superconductor. In addition, for realizing stable TSC against perturbations, it is important to induce a sufficiently large Rashba spin-orbit coupling. For this purpose, making an atomically thin film on a substrate is one of the effective approaches. For these reasons, the most promising candidate material is a cuprate superconductor, of which thin films have already been fabricated~\cite{Bollinger2011, Leng2011}. Despite the 3d orbital character of electrons, a sizable spin-orbit coupling has been recently reported~\cite{Gotlieb2018}. Thanks to its high critical temperature, the cuprate superconductor is also a good candidate from other viewpoint that its superconducting state should be robust to irradiating laser light. Some of the heavy-fermion superconductors can also be good candidates because they show $d$-wave superconductivity. For instance, atomically-thin layers of $\mathrm{CeCoIn_5}$ have already been fabricated, and the spin-orbit coupling is controllable~\cite{Shimozawa_superlattice_RPP2016,naritsuka2021}. Thus, the heterostructure of $\mathrm{CeCoIn_5}$ is also a good platform to realize TSC based on our scenario.

\subsubsection{Frequency and intensity of laser light}
In this section, we used the high-frequency expansion in the Floquet theory and thus, strictly speaking, the frequency must be sufficiently high and off-resonant. Since the frequency has to be higher than the typical energy scale of the original Hamiltonian, which is the band width $D \sim 8t$ corresponding to the order of 1-10 eV. Thus, laser light should be visible or near-ultraviolet. As for the resonance, while there exist many unoccupied bands above Fermi energy, we have to choose an appropriate frequency so as to make it off-resonant. While it is not easy to strictly achieve all the above conditions in experiments, our result is expected to be approximately valid and the TSC should appear even out of these conditions. This is because the TSC in our model is known to universally appear in noncentrosymmetric systems without time-reversal symmetry~\cite{Daido2016}. Even if we change the frequency, the symmetry properties are unchanged. Indeed, the gap opening at the Dirac nodes predicted by the high-frequency expansion~\cite{Kitagawa2011} have been observed in experiments even when the frequency is much lower than the theory~\cite{Wang2013, McIver2020}. 

We mentioned above that the laser light opens the gap at the nodal point in superconducting gap and then TSC is realized with infinitesimally weak intensity of light. However, the weak intensity opens only a tiny gap which is fragile against perturbations and cannot be observed in experiments. To observe it, the gap must exceed the energy scale of thermal excitations at finite temperature. Thus, there exists the minimum intensity to observe the TSC in experiments. We estimate it from the formula representing the size of the energy gap 
\begin{align}
\left| \frac{\mu_B (\tilde{H}(\bm k) \hat{\bm z}) \cdot (\hat{\bm g}(\bm k) \times \bm d(\bm k))}{\alpha \tilde{\bm g}(\bm k)}\right|_{\bm k = \bm k_0} \sim \frac{4\alpha^2}{\omega} \frac{\Delta_p}{\tilde{\alpha}} J_1(A_x) J_1(A_y), \label{eq:intensity}
\end{align}
with $\tilde{\alpha}\equiv(J_0(A_x)^2+J_0(A_y)^2)^{1/2} \alpha$. The admixed $p$-wave component $\Delta_p$ is roughly estimated as $\Delta_p \sim \Delta_d \tilde{\alpha} / E_F$~\cite{NCSC_book, Fujimoto2007}. Assuming typical values as $\alpha=0.1~\mathrm{eV}$, $\omega=10~\mathrm{eV}$ and $\Delta_d / E_F= 0.1$, we need $A_x=A_y=1.21$ to induce the superconducting gap $0.1~\mathrm{meV} \sim 1 \mathrm{K}$. The corresponding electric field is almost $600~\mathrm{MV/cm}$. It is the minimum amplitude to observe TSC in the experiments at low temperatures around $1~\mathrm{K}$. 
On the other hand, the formula (\ref{eq:intensity}) implies that there is a realizable maximum gap size since the Bessel function $J_1(x)$ takes the maximum value($\sim~0.58$) at $x \sim 1.84$, which corresponds to the electric field amplitude $E~\sim~1~\mathrm{GV/cm}$. With this intensity, the energy gap becomes $0.4~\mathrm{meV}$ for the above parameter set. For realizing a larger gap, we need to prepare material with larger spin-orbit coupling or apply laser with lower frequency.

\subsubsection{Experimental method}
To obtain the strong intensity to realize the TSC, we have to use a short laser pulse whose time scale is typically the order of ten femtoseconds or sub-picoseconds. Therefore, the phenomena must be transient and they are called ultrafast phenomena. 
To observe the ultrafast phenomena, the methods must be time-resolved and achieve a good time resolution. Indeed, time-resolved optical measurements~\cite{Wang2013, Mahmood2016} and transport measurements~\cite{McIver2020} have been used to observe the Floquet states in solids. The optical and transport properties are expected to be changed transiently reflecting the TSC only when applying the laser light. By the transport measurement, it should be possible to probe the transient signature of the Majorana edge modes. With optical measurements, the gap opening at the Fermi level is a good signature to be probed. For this purpose, time-resolved ARPES would be most promising because the nodal structure of the superconducting gap in cuprates has already been observed in equilibrium ARPES measurements~\cite{Hashimoto2014}. Another approach is  time-resolved STM measurement~\cite{Terada2010, Yoshida2013, Pechenezhskiy2013, Cocker2016}. 
It can reveal the spatially-resolved information which can be direct evidence of the Majorana edge modes. Therefore, we believe this is an important direction to explore the laser-induced topological phases experimentally.

\subsection{Summary of this section}\label{Sec:IIIF}
We explained our proposal to realize $d$-wave TSC with laser light. Based on the concept of the Floquet engineering, we have considered the realization of TSC with a periodic driving by laser fields. To study the periodically driven system, we have used the Floquet theory and the high-frequency expansion. We have derived the effective model and discussed its topological properties. Then, we have found that TSC characterized by the Chern number is realized with infinitesimally small intensity of laser light. 
We have also discussed the experimental setup about materials and laser light (frequency and intensity) and experimental probes. 

There are various future directions left to be studied. One is to investigate different driving schemes. While we have studied the high frequency driving, there have been proposed the methods to obtain the effective Hamiltonian for different driving schemes such as low frequency driving or resonant frequency driving~\cite{Bukov2016, Mizuta2019}. It should be interesting to investigate how TSC is realized with these approaches. Whereas we can only study the prethermal steady states with these approaches since they are based on static effective Hamiltonian, it should also be of interest to calculate the real-time dynamics induced by laser irradiation and study how TSC emerges in the time evolution. Another direction is to study superconductors in other classes. 
For instance, we recently showed that transition metal dichalcogenide bilayers with circularly polarized laser light also become a topological superconductor while they are $s$-wave superconductors and do not have strong spin-orbit coupling~\cite{Chono2019}. 
We believe that the application of laser light provides a useful approach to change the quantum states of matter in a highly controllable way and open a new avenue to study various exotic states of matter including TSC.

While we study the external drive with AC electric fields, it is also interesting to consider a DC (static) drive. Recently, it has been shown that $d$-wave topological superconductivity can be induced with applying a DC supercurrent~\cite{Takasan2021}. The advantageous point is that this setup is free from the heating problem which is crucial for the Floquet engineering as explained above. We believe it is important to further explore a new pathway to realize TSCs for finding more useful and realizable ones.

\section{Topological superconductivity in locally noncentrosymmetric multilayers}\label{sec:Yanase}

While the superconductivity without inversion symmetry has been a topic of interest for a long time, attention has been paid to the global crystallographic symmetry of systems. This is mainly because the Kramers theorem for the degeneracy of electronic states relies on the global inversion symmetry in addition to the time-reversal symmetry.
To lift the Kramers degeneracy at a given momentum, either inversion symmetry or time-reversal symmetry is required to be broken. 
Actually, the spin splitting in the electronic states [see Eq.~\eqref{eq:spin_splitting}] is a characteristic property of the systems lacking global inversion symmetry. 
We have dealt with TSC in such globally noncentrosymmetric systems in Secs.~\ref{sec:Daido} and \ref{sec:Takasan}.

Here we switch the topic 
to locally noncentrosymmetric superconductivity.
Even when the global inversion symmetry in the crystal structure is preserved, the local site symmetry of atoms may be broken. The crystals with such symmetry are now called {\it locally noncentrosymmetric crystals}. Although less attention was paid to the locally noncentrosymmetric systems, recent studies shed light on unique properties of the systems which are attracting interest in a broad range of the fields from antiferromagnetic spintronics~\cite{Wadley2016} to exotic superconductivity. 

For the basic properties of superconductivity in the locally noncentrosymmetric systems we can refer to an early-stage review article~\cite{Sigrist_LNCSreview2014}. For instance, the selection rule for Cooper pairing~\cite{Fischer_LNCS2011}, anomalous paramagnetic effect~\cite{Maruyama_LISB_SpinOrb_JPSJ12}, field-induced odd-parity superconductivity~\cite{superlattice_Yanase_12}, and Fulde-Ferrell-Larkin-Ovchinnikov state with unusual phase modulation~\cite{Yoshida_complex_stripe2013} have been outlined. For one of the developments after Ref.~\onlinecite{Sigrist_LNCSreview2014}, 
in this article, we review ideas and results of the TSC in the field-induced odd-parity superconducting state. Recent experimental support in CeRh$_2$As$_2$~\cite{CeRh2As2} is also briefly discussed.

\subsection{Model and electronic states}

\begin{figure}[htbp]
\includegraphics[width=85mm]{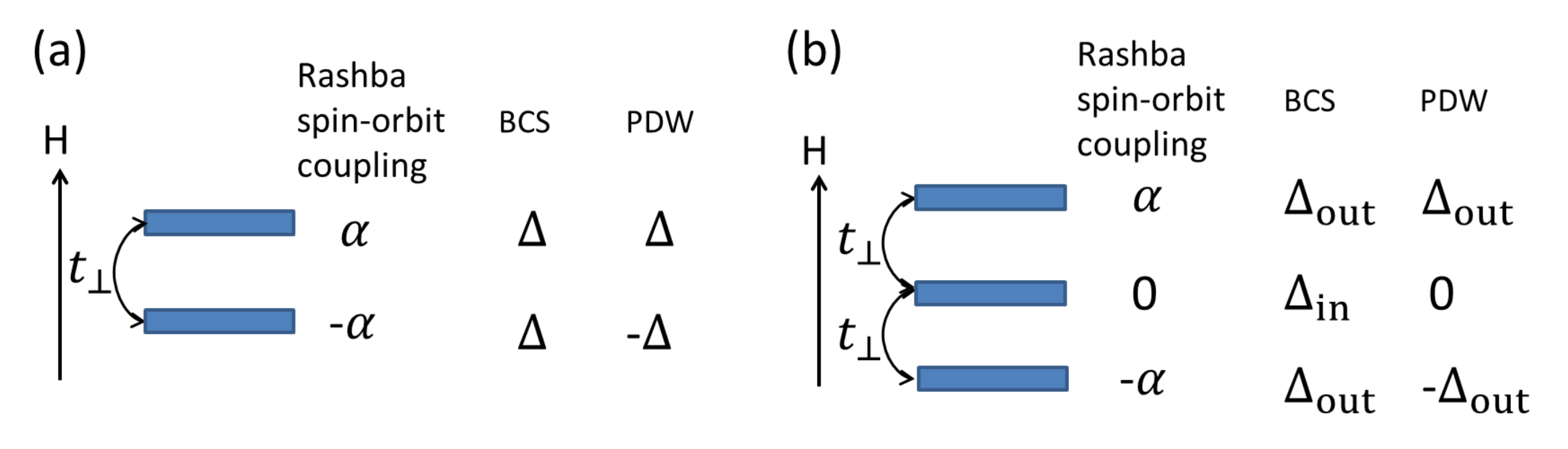}
\caption{Illustration of the multilayer crystal structure~\cite{superlattice_Yanase_12}. (a) Bilayers and (b) trilayers. Blue bars represent atomic layers. Layer dependence of the Rashba ASOC and order parameters in the BCS and PDW states are shown.
}
  \label{fig_multilayer}
\end{figure} 

Here we focus on a typical and realistic crystal structure lacking local inversion symmetry with keeping global inversion symmetry. An illustration of the multilayer crystalline structure is shown in Fig.~\ref{fig_multilayer}. 
Since an inversion center exists at the center of multilayers, global inversion symmetry is preserved. However, the local inversion symmetry on the outer layers is broken. 
Various compounds naturally crystallize in the multilayer structure. Examples of bulk crystals are found in a broad class of materials from high-$T_{\rm c}$ cuprate superconductors~\cite{Gotlieb2018} to a heavy-fermion superconductor CeRh$_2$As$_2$~\cite{CeRh2As2}.
Furthermore, recent advances in technology have provided a way to synthesize artificial 2D systems with multilayer structures. For examples, heavy-fermion superlattices~\cite{Mizukami2011,Goh_superlattice12,Shimozawa_superlattice_PRL14,Shimozawa_superlattice_RPP2016,naritsuka2021} and van der Waals heterostructures~\cite{Zheliuk2019} are current topics of interests. 
We will review a heavy fermion superlattice CeCoYb$_5$/YbCoIn$_5$ in Sec.~\ref{sec:Yoshida}.

Superconductivity in the locally noncentrosymmetric multilayers is modeled by the following minimal Hamiltonian, 
\begin{subequations}
\begin{align}
  {\cal H}&= {\cal H}_0 + {\cal H}_I, \label{eq1} \\
  {\cal H}_0&=\sum_{{\bm k},s,s',m}[\xi({\bm k})\sigma_0+\alpha_m{\bm g}({\bm k})\cdot{\bm \sigma}-\mu_{\rm B} {\bm H} \cdot {\bm \sigma}]_{ss'}
  c_{{\bm k}sm}^\dagger c_{{\bm k}s'm} \nonumber \\
  &+ \sum_{{\bm k},s,\langle m,m'\rangle} t_{\perp}({\bm k}) c_{{\bm k}sm}^\dagger c_{{\bm k}sm'}, \label{eq1'} \\
  {\cal H}_I&=\frac{1}{2N}\sum_{{\bm k},{\bm k}',s,s',m}V_{ss'}({\bm k},{\bm k}')c_{{\bm k}sm}^\dagger c_{-{\bm k}s'm}^\dagger c_{-{\bm k}'s'm}c_{{\bm k}sm},
  \label{eq1''}
\end{align}
\label{eq:model_LNCS}
\end{subequations}
where the index for layers $m$ runs from $1$ to $M$ ($M$-layer system). 
The first term in the single-particle part ${\cal H}_0$ includes the in-plane kinetic energy, ASOC, and Zeeman coupling. While the local inversion symmetry breaking gives rise to an ASOC at each layer, the global inversion symmetry constrains the layer-dependence of the coupling constant. The ASOC coupling constant, $\alpha_m$, must be layer-dependent and change the sign after the layer permutation by the inversion operation. 
Thus, it has the form, 
\begin{subequations}
\begin{align}
(\alpha_1, \alpha_2) = (\alpha, -\alpha)
\hspace{8.5mm} &{\rm in \hspace{1mm} bilayer \hspace{1mm} systems,} 
\\
(\alpha_1, \alpha_2, \alpha_3) = (\alpha, 0, -\alpha) \hspace{5mm} &{\rm in \hspace{1mm} trilayer \hspace{1mm} systems,}
\end{align}
\end{subequations}
as illustrated in Fig.~\ref{fig_multilayer}.
More generally, the relation $\alpha_m = -\alpha_{M+1-m}$ has to be satisfied. 
The second term in ${\cal H}_0$ represents the interlayer hopping with $t_\perp({\bm k})$, and ${\cal H}_I$ is an effective interaction term stabilizing superconductivity. 

\begin{figure}[htbp]
\includegraphics[width=75mm]{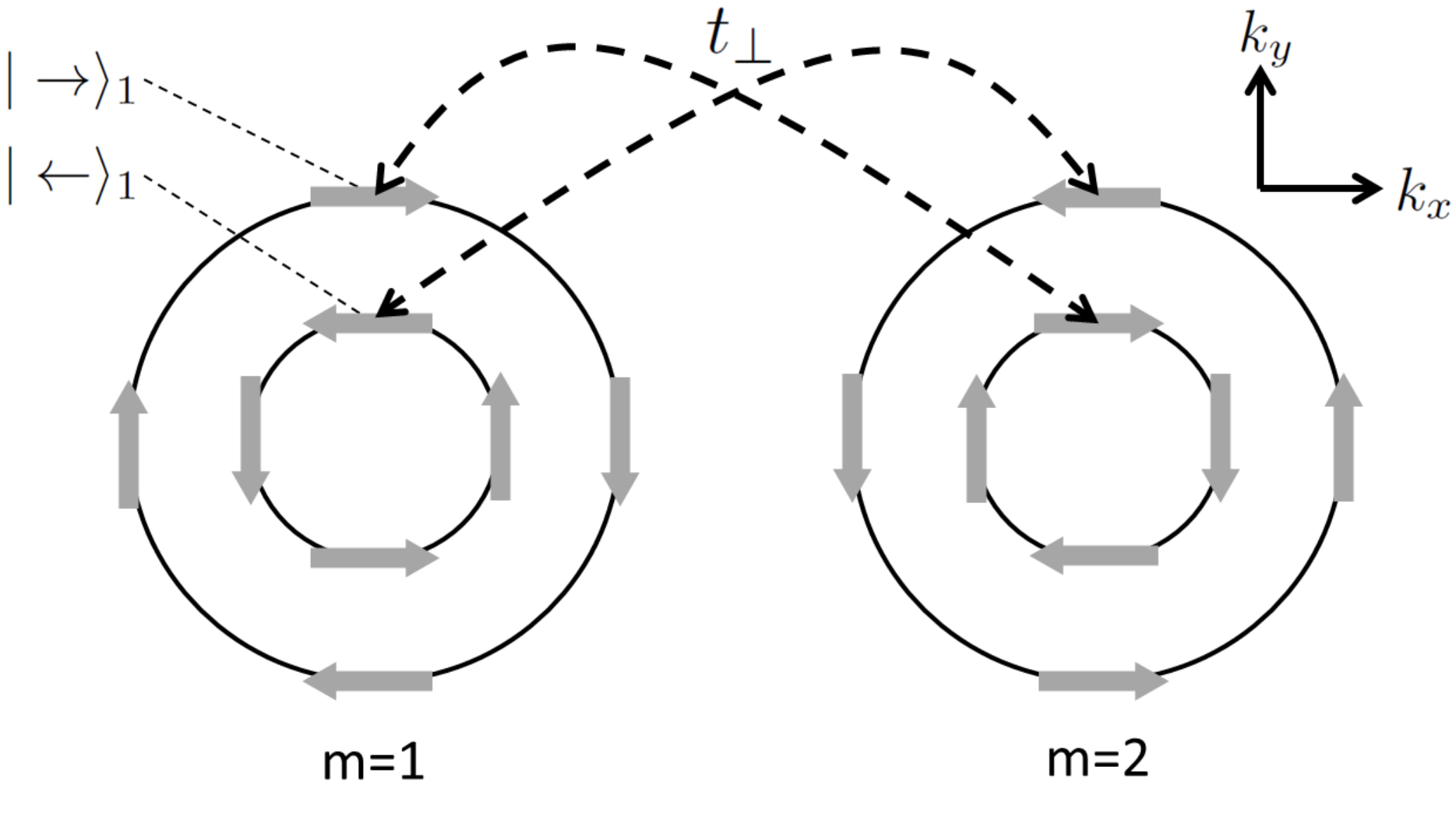}
\caption{Spin-momentum locking in the bilayer Rashba model~\cite{Maruyama_LISB_SpinOrb_JPSJ12}. Direction of spin orientation is opposite between the layers, because the Rashba spin-orbit coupling is opposite ($\alpha_1=-\alpha_2$). The interlayer hopping couples the states with different energy. 
}
  \label{fig_spin_momentum_locking}
\end{figure}

Electronic structures in the normal state are obtained by diagonalizing the single-particle part, ${\cal H}_0$. The energy band in the bilayer systems at ${\bm H}=0$ is 
\begin{align}
E_\pm(\bm{k}) = \xi(\bm{k}) \pm \sqrt{\alpha^2 g(\bm{k})^2 + t_\perp(\bm k)^2}, 
\label{eq:band_LNCS}
\end{align}
and all the bands are two-fold degenerate in accordance with the Kramers theorem. 
Comparing this with Eq.~\eqref{eq:spin_splitting}, we recognize that the layer-dependent ASOC is additive to the interlayer hopping. 
Therefore, it may be hard to extract contributions of the ASOC in the band dispersion. 
On the other hand, we can see a characteristic feature in the wave function. Analytic expressions of the wave function have been provided in Ref.~\onlinecite{Maruyama_LISB_SpinOrb_JPSJ12}, and we here discuss the limiting cases for bilayers. 
In the weak ASOC limit, $\alpha g(\bm k)/t_\perp(\bm k) \rightarrow 0$, the wave functions are conventional bonding and anti-bonding orbitals. 
In the strong ASOC limit, $t_\perp(\bm k)/\alpha g(\bm k) \rightarrow 0$, the Kramers pairs are formed by 
\begin{align}
\{ |1, \uparrow \rangle, \, |2, \downarrow \rangle \}, 
\hspace{2mm} {\rm and} \hspace{2mm}
\{|1, \downarrow \rangle, \, |2, \uparrow \rangle \}, 
\end{align}
in which spin ($s=\uparrow,\downarrow$) and sublattice (layer, $m=1,2$) degrees of freedom are entangled. 
A schematic illustration of the electronic states is shown in Fig.~\ref{fig_spin_momentum_locking}. The spin and sublattice entangled electronic states discussed above are recently called "hidden spin polarization"~\cite{Zhang_hidden_polarization2014}, because the feature is hidden in the band dispersion, Eq.~\eqref{eq:band_LNCS}. Such electronic structures have been observed in various compounds by the spin-resolved ARPES~\cite{Riley2014,Santos-Cottin2016,Gehlmann2016,Wu2017} and polarization-resolved optical measurements~\cite{Jones2014}.

It is expected from the above results that effects of the ASOC are significant when the ASOC is comparable or larger than the inter-sublattice hopping. This is true in most cases. Therefore, a strategy for uncovering features of locally noncentrosymmetric systems different from even-locally centrosymmetric systems is to study the case with a small $t_\perp(\bm k)$. Actually, intriguing phenomena, such as field-induced odd-parity superconductivity~\cite{Sumita_multipoleSC2016,Nakamura_bilayerTMD2017,cavanagh2021nonsymmorphic}, have been shown in the systems with vanishing inter-sublattice hopping $t_\perp(\bm k)=0$ at symmetric points in the Brillouin zone~\cite{Kane-Mele2005,Sumita_multipoleSC2016,Nakamura_bilayerTMD2017,Niu_CrAs2017,Ishizuka_Sr2IrO42018,cavanagh2021nonsymmorphic}. Disappearance of the inter-sublattice hybridyzation may be ensured by nonsymmorphic and/or rotation symmetry~\cite{Bradley_book,Akashi2017}. 
The nodal line excitation in nonsymmorphic odd-parity superconductors originates from this property~\cite{Norman1995,Micklitz-Norman2009,Yanase2016,Kobayashi2016,Sumita_Sr2IrO42017,Sumita-Yanase2018,Nomoto2017,Sumita2019}. 

\subsection{Odd-parity superconductivity}

In this section, we discuss the field-induced odd-parity superconductivity~\cite{superlattice_Yanase_12}.
Although the model \eqref{eq:model_LNCS} and its straightforward generalization are applicable to generic multi-sublattice systems, such as bilayer Ising superconductors~\cite{Nakamura_bilayerTMD2017,Zheliuk2019,Kanasugi_bilayerTMD,Chono2019}, we here focus on the multilayer Rashba system, supposing cuprate superconductors~\cite{Gotlieb2018}, heavy-fermion superlattices~\cite{Mizukami2011,Goh_superlattice12,Shimozawa_superlattice_PRL14,Shimozawa_superlattice_RPP2016,naritsuka2021}, and a recently discovered superconductor CeRh$_2$As$_2$~\cite{CeRh2As2}.

On the 2D square lattice, we have 
$\xi({\bm k})$ and ${\bm g}({\bm k})$ in Eqs.~\eqref{eq:dispersion_sec1} and \eqref{eq:g-vector_sec1} as in the previous sections. Hereafter, we take $t'=0$ for simplicity. Supposing weakly-coupled multilayers, we assume a small and momentum-independent interlayer hopping $t_\perp(\bm k)= t_\perp \ll W$, where $W$ is the band width. Accordingly, only the intralayer interaction is taken into account, which is given by 
$V_{ss'}({\bm k},{\bm k}')=-V_{\rm s} \psi_{\rm s}({\bm k}) \psi_{\rm s}({\bm k}') \delta_{s,-s'} 
- V_{\rm t} \left[\psi_{\rm t1}({\bm k}) \psi_{\rm t1}({\bm k}') + \psi_{\rm t2}({\bm k}) \psi_{\rm t2}({\bm k}') \right]$. 
The first term with the coupling constant $V_{\rm s}$ is the pairing interaction in the spin-singlet channel. We here consider either $s$-wave or $d$-wave pairing, namely, $\psi_{\rm s}({\bm k}) = 1$ or $\psi_{\rm s}({\bm k}) = \cos k_x - \cos k_y$. 
The second term is the interaction in the spin-triplet $p$-wave channel, $\psi_{\rm t1}({\bm k}) = \sqrt{2} \sin k_x $ and $\psi_{\rm t2}({\bm k}) = \sqrt{2} \sin k_y $. 
In the following part, we study dominantly spin-singlet pairing states and assume $V_{\rm s} > V_{\rm t}$, because the spin-singlet superconductors are much more ubiquitous than the spin-triplet ones. 


\begin{figure}[htbp]
\includegraphics[width=40mm]{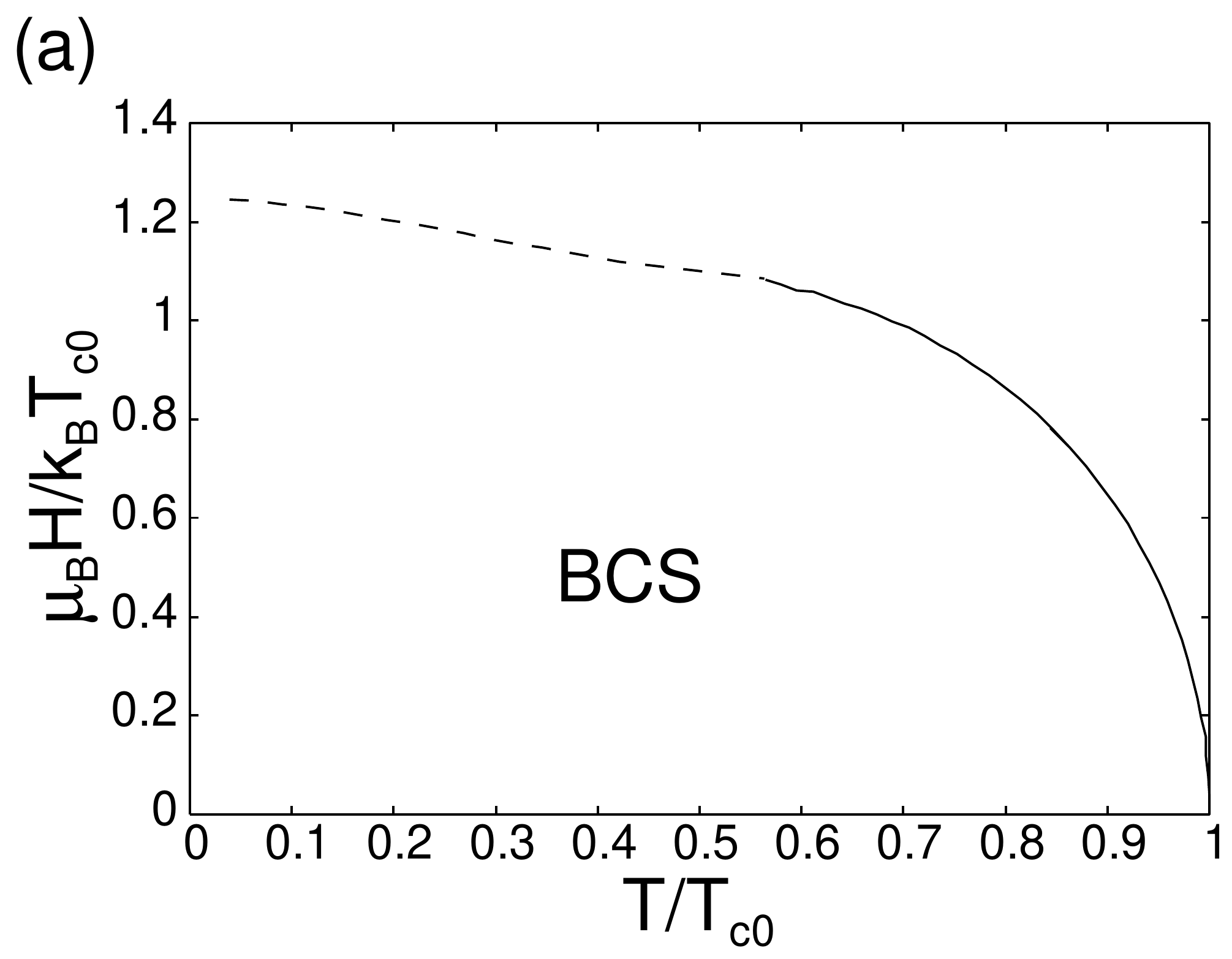} \hspace*{2mm}
\includegraphics[width=40mm]{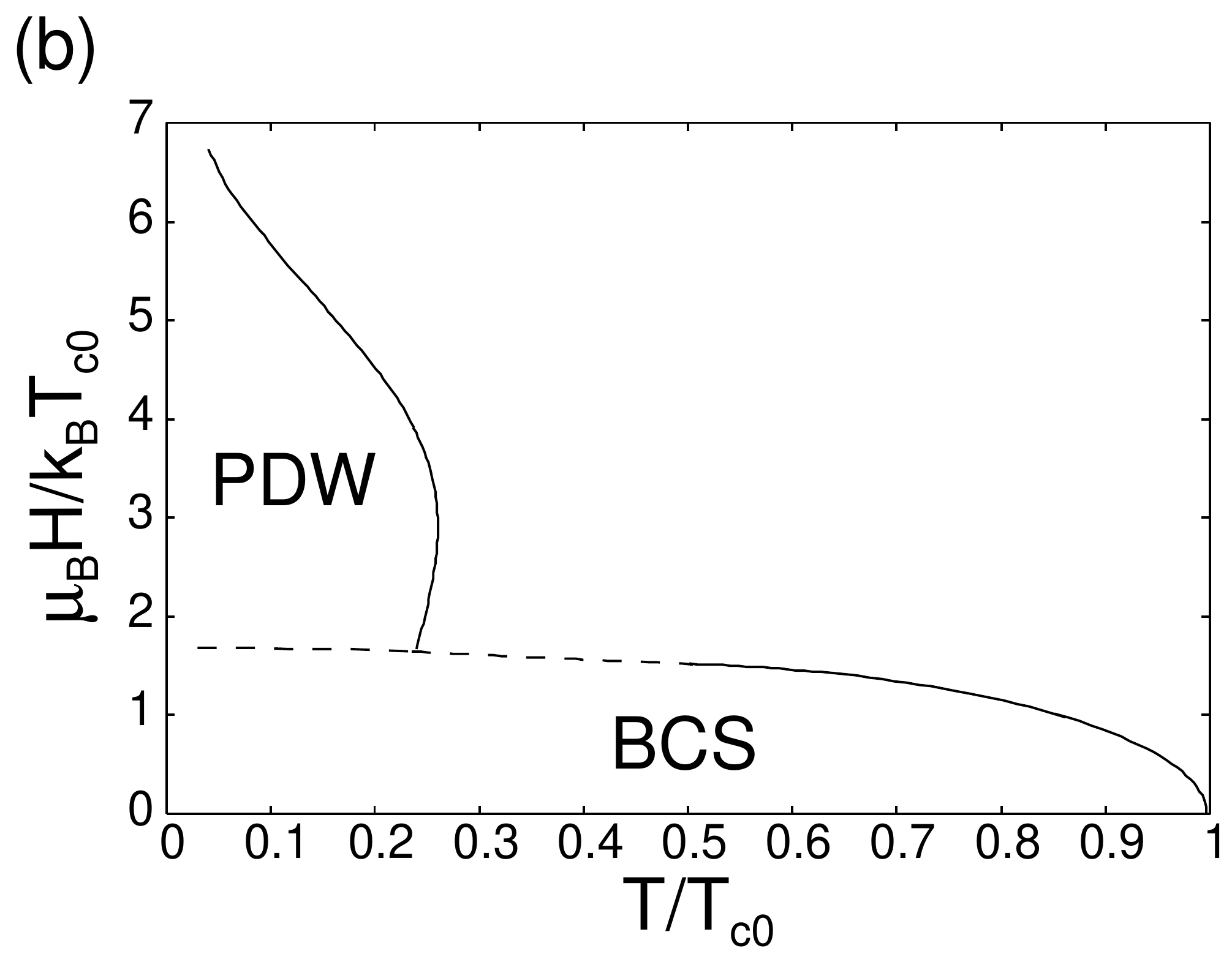} \\
\vspace*{3mm}
\includegraphics[width=40mm]{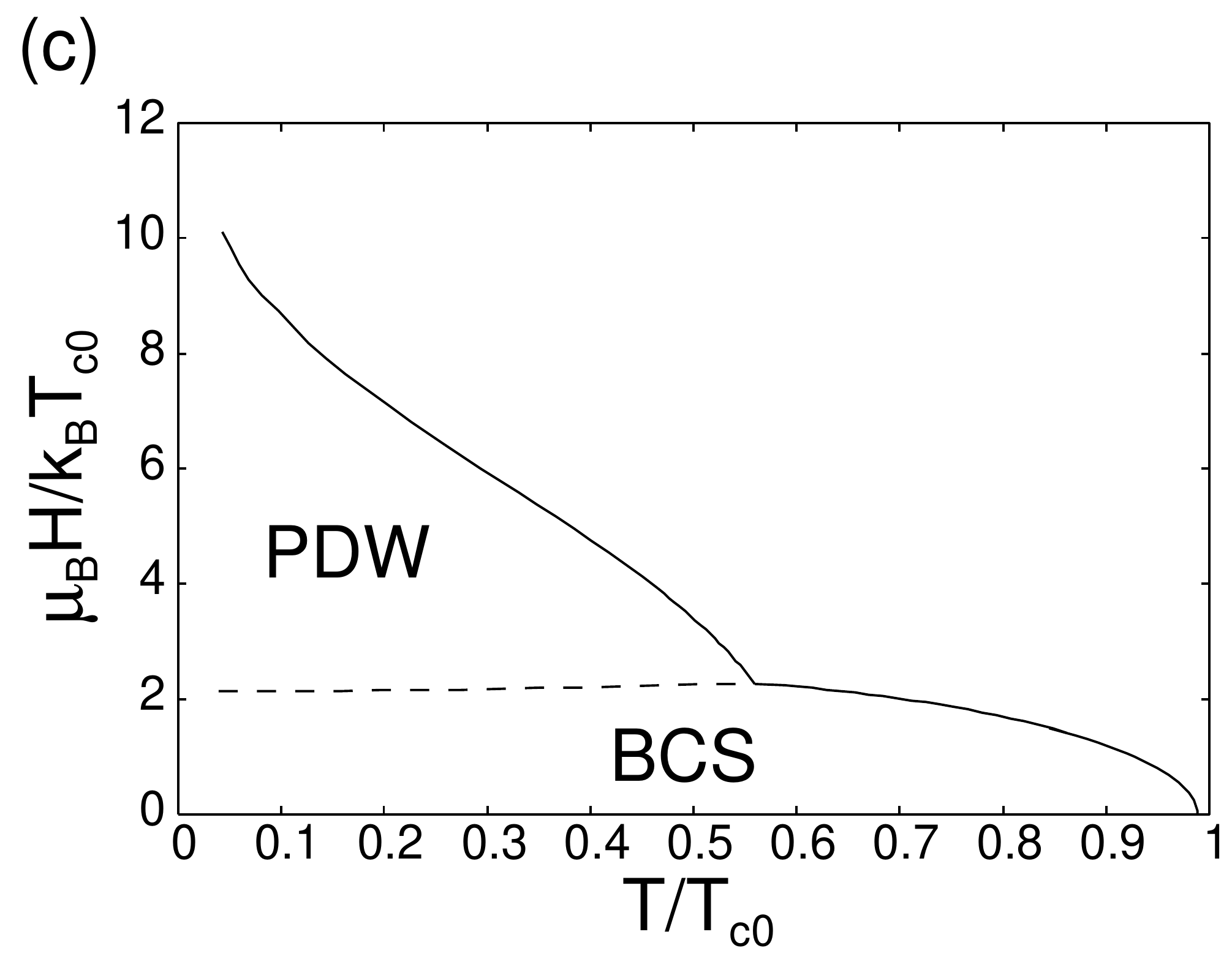} \hspace*{2mm}
\includegraphics[width=40mm]{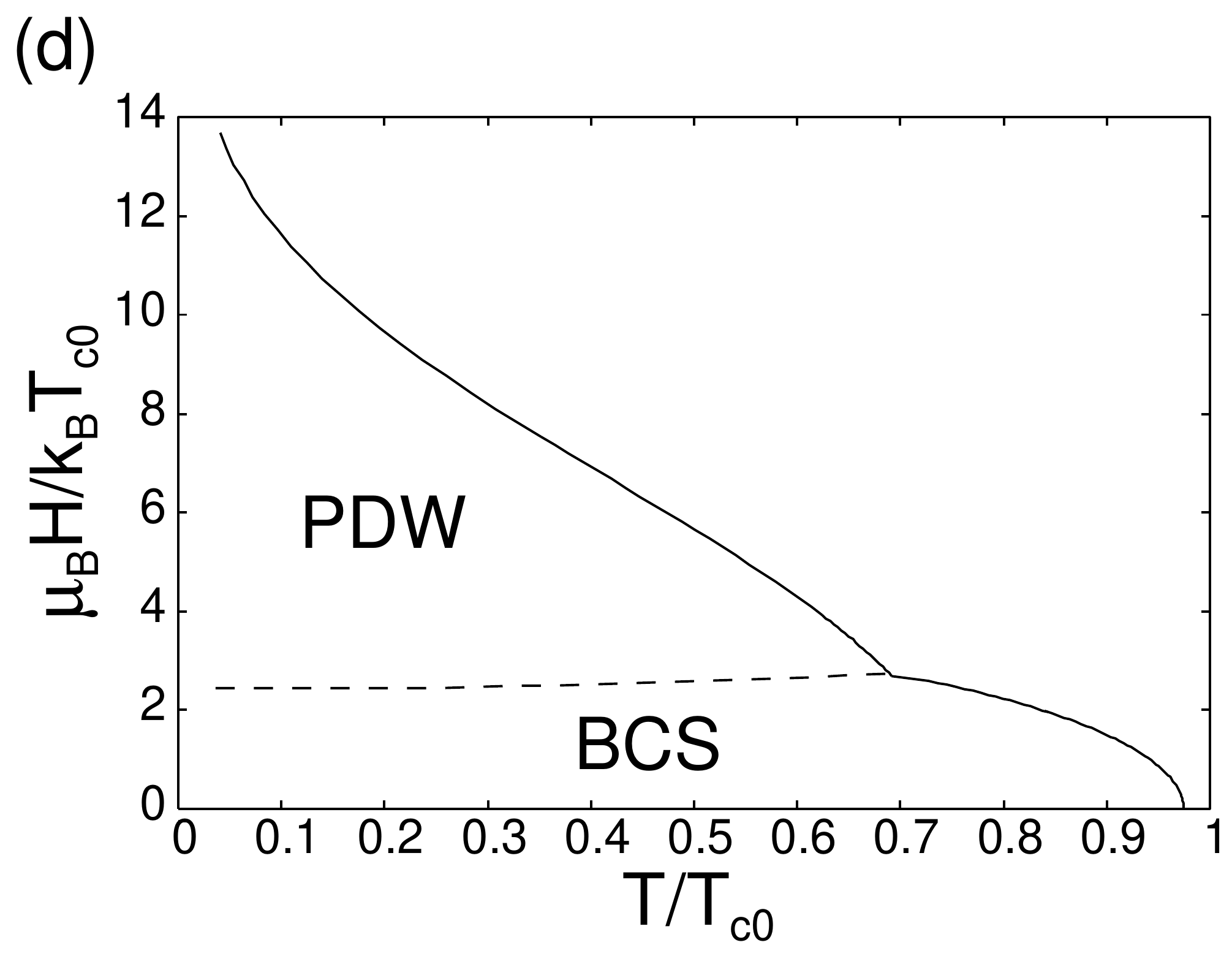}
\caption{Phase diagrams of the bilayer Rashba superconductors in the $H_z$-$T$ plane~\cite{superlattice_Yanase_12}. The BCS and PDW phases are shown. The dashed lines are the first-order phase transition lines. 
(a) $\alpha/t_\perp =0$, (b) $\alpha/t_\perp =1$, (c) $\alpha/t_\perp =2$, (d) $\alpha/t_\perp =3$. 
We set $t_\perp=0.1$ and $\mu=2$ in the unit $t=1$. 
  }
  \label{fig_PDW_bilayer}
\end{figure}

The BCS-type mean-field theory reveals the $H_z$-$T$ phase diagrams in Fig.~\ref{fig_PDW_bilayer}. Here, we consider the simplest case, namely, purely spin-singlet $s$-wave pairing state in bilayers. 
The superconducting order parameter has a simple form, 
$\Delta_{mss'} = \psi_m (i \sigma_y)_{ss'}$, 
and it can depend on layers. 
The two phases in Fig.~\ref{fig_PDW_bilayer} have distinct layer-dependence; $(\psi_1,\psi_2) = (\Delta,\Delta)$ in the BCS state while $(\psi_1,\psi_2) = (\Delta,-\Delta)$ in the {\it pair-density wave (PDW) state}. 
The BCS state is stable at zero magnetic field, $H=0$, so as to gain the Josephson coupling energy. 
On the other hand, the PDW state is stable in the high magnetic field region, when the ASOC strength is comparable or larger than the interlayer hopping.
As we emphasize later, the PDW state is an odd-parity superconducting state, although it is mainly stabilized by the spin-singlet pairing. Thus, Fig.~\ref{fig_PDW_bilayer} reveals {\it field-induced parity transition} in the superconducting state.

\begin{figure}[htbp]
\includegraphics[width=60mm]{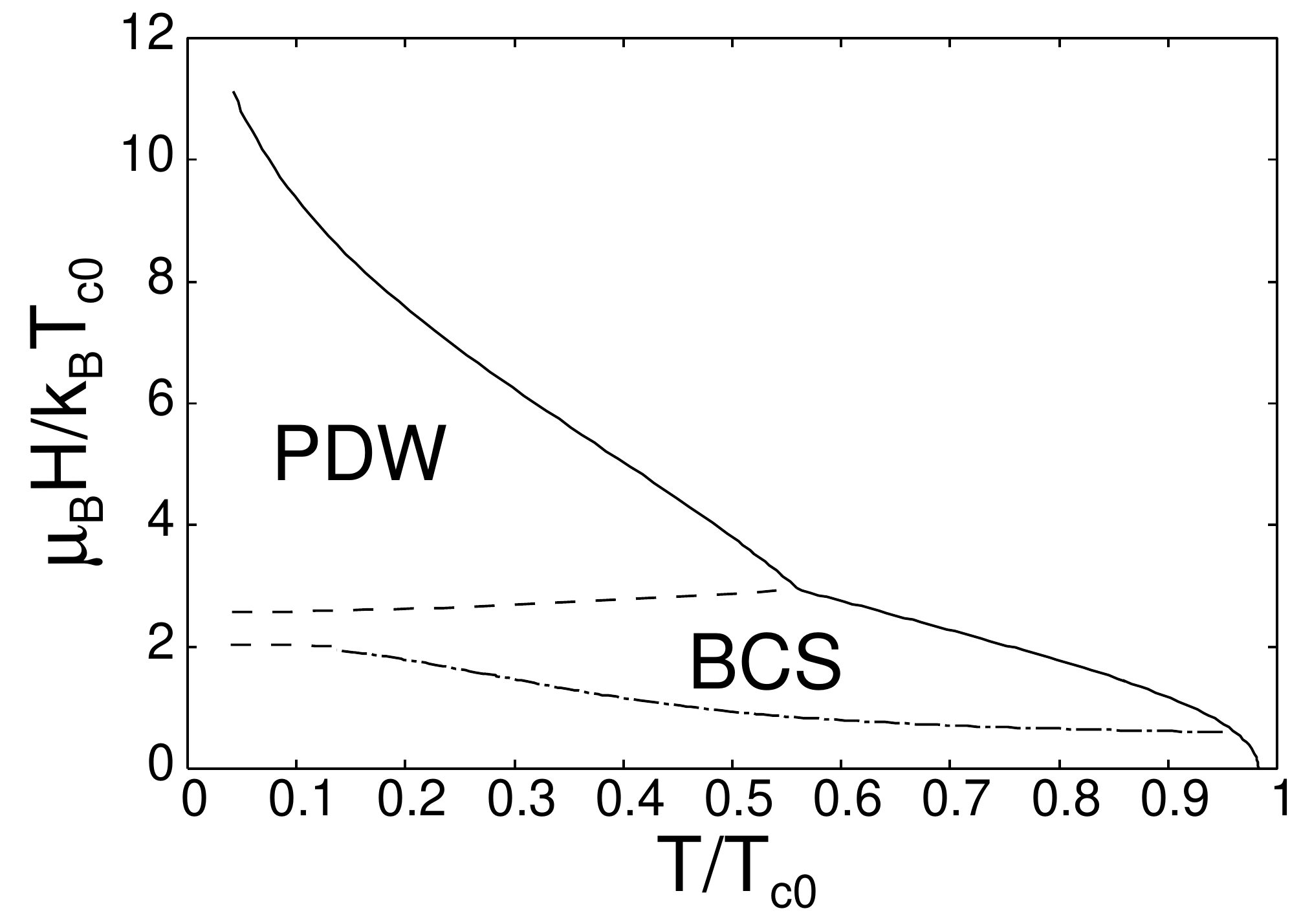} 
\caption{Phase diagram of the trilayer Rashba superconductors for $\alpha/t_\perp =3$. 
The dashed lines are the first-order phase transition lines. Dash-dotted and dashed lines in the BCS phase indicate the crossover and first-order transition, respectively. See Refs.~\onlinecite{superlattice_Yanase_12,Yoshida_parity_mixing2014} for details. 
}
  \label{fig_PDW_trilayer}
\end{figure}

The phase diagrams are understood by analyzing the gap function in the band basis. 
Since the bilayer model is a two-band model, the gap function is defined for each band and obtained by a unitary transformation of the order parameter in the sublattice basis. When the spin-triplet paring is neglected as in the calculations of Fig.~\ref{fig_PDW_bilayer}, 
the gap functions of the two bands are equivalent, and they are obtained as~\cite{Maruyama_LISB_SpinOrb_JPSJ12},  
\begin{align}
\Delta_{\rm BCS} = \Delta,
\label{eq:gap_BCS}
\end{align}
in the BCS state, while 
\begin{align}
\Delta_{\rm PDW}(\bm k) = \frac{\alpha g(\bm k)}{\sqrt{\alpha^2 g(\bm k)^2+t_\perp^2}} \Delta,  
\label{eq:gap_PDW}
\end{align}
in the PDW state. 
Equation~\eqref{eq:gap_PDW} indicates that intra-band Cooper pairs vanish in the PDW state when $\alpha=0$. Indeed, the pairs are formed between the bonding and anti-bonding orbitals with nonequivalent energy dispersion. Such pairing state is unstable as in the spin-polarized state beyond the Pauli-Chandrasekhar-Clogston limit. Therefore,  Fig.~\ref{fig_PDW_bilayer}(a) for $\alpha=0$ does not show the PDW phase. When we switch on the layer-dependent ASOC leading to the hidden spin polarization, the gap opens at the Fermi level, which makes the PDW state meta-stable. Although it is still less stable than the BCS state at $H=0$ because of $|\Delta_{\rm PDW}(\bm k)| < |\Delta_{\rm BCS}|$, the PDW state is thermodynamically stable in the high-field region. This is because the paramagnetic depairing effect is almost completely suppressed in the PDW state. The Pauli-Chandrasekhar-Clogston limit of the upper critical field is roughly estimated as
\begin{align}
H^{\rm P} = \frac{H^{\rm P}_0}{\sqrt{1-\chi_{\rm s}/\chi_{\rm n}}}, 
\end{align}
where $\chi_{\rm s}$ ($\chi_{\rm n}$) is the spin susceptibility in the superconducting (normal) state and $H^{\rm P}_0$ is the Pauli-Chandrasekhar-Clogston limit of spin-orbit coupling free systems. 
Both numerical and analytic calculations have shown the anomalous paramagnetic effect~\cite{Maruyama_LISB_SpinOrb_JPSJ12}, that is, $\chi_{\rm s}/\chi_{\rm n}=1$ in the PDW state while $0 < \chi_{\rm s}/\chi_{\rm n} < 1$ in the BCS state. Since the critical field is higher in the PDW state than the BCS state, field-induced phase transition occurs from the BCS state to the PDW state. The phase diagrams in Fig.~\ref{fig_PDW_bilayer} are determined by the competition between the paramagnetic depairing effect and decrease in the intra-band gap function.
The PDW phase is stabilized as the parameter $\alpha/t_\perp$ increases, as expected. 
When the interaction in the spin-triplet channel is taken into account, the PDW phase is furthermore stabilized~\cite{Yoshida_parity_mixing2014}. 
We have obtained similar phase diagrams for the $d$-wave superconducting state and for the trilayers~\cite{Yoshida_thesis}. An example for the trilayer system is shown in Fig.~\ref{fig_PDW_trilayer}, for instance~\cite{superlattice_Yanase_12}. We see the PDW phase as well as the additional crossover and first-order transition, at which the ratio of order parameters on the outer and inner layers changes with keeping the global symmetry. 

The sign-reversing order parameter of superconductivity was first proposed for bilayer systems at $H=0$ by Nakosai {\it et al.}~\cite{Nakosai2012}. However, fine-tuning of the band structure and interaction is required for the stable PDW state at $H=0$. On the other hand, the PDW state is ubiquitously stabilized at $H \ne 0$ in the locally noncentrosymmetric superconductors close to the Pauli limit. 
The phase diagrams in Figs.~\ref{fig_PDW_bilayer} and \ref{fig_PDW_trilayer} are obtained in the Pauli limit, namely, by neglecting the orbital effect. Stability of the PDW state against the orbital effect has been shown in Ref.~\onlinecite{Moecli_PDW2018}. 

Interestingly, the PDW state is an odd-parity superconducting state although the superconductivity is caused by the spin-singlet Cooper pairs. The odd-parity spin-singlet superconductivity looks incompatible with the text-book understanding of the BCS theory. However, the inversion operation accompanied by the layer permutation gives the additional negative sign in the representation of the PDW order parameter, which changes the parity of superconductivity. For the tetragonal $D_{\rm 4h}$ system, the irreducible representation of the $s$-wave PDW state is $A_{\rm u}$, which is the same as the spin-triplet 
$p_x \hat{y} - p_y \hat{x}$ state. 
Thus, Figs.~\ref{fig_PDW_bilayer} and ~\ref{fig_PDW_trilayer} show the field-induced transition from an even-parity superconducting state to an odd-parity one. 
This provides a way to realize odd-parity superconductivity without relying on the rare spin-triplet Cooper pairs. 
As we discuss in the next subsection, the field-induced odd-parity superconducting state is a candidate for topological crystalline superconductivity.

In the experiments, 
evidence for the PDW state and even-odd phase transition was recently obtained for the analogous bulk compound, CeRh$_2$As$_2$~\cite{CeRh2As2}. 
The phase transition in the superconducting state has been observed, and the phase diagram resembles the theoretical prediction in Fig.~\ref{fig_PDW_bilayer}.
After the experimental report of CeRh$_2$As$_2$, several theoretical studies investigated the effects of 3D stacking structure, interlayer pairing, and disorders on the anomalous paramagnetic effect and phase diagram~\cite{Schertenleib2021,moeckli2021,skurativska2021spin,cavanagh2021nonsymmorphic,Moeckli_disorder}. The present results basically support the parity transition in the superconducting state.

So far we considered the magnetic field along the $c$-axis. When the magnetic field is perpendicular to the $c$-axis, the Fulde-Ferrell-Larkin-Ovchinnikov state with unusual phase modulation~\cite{Yoshida_complex_stripe2013}, named complex stripe state, is stabilised in the Pauli limit. The PDW state may also be stabilized in this field direction when the paramagnetic effect is competing with the orbital effect~\cite{Watanabe_PDW2015}. 

Another representative of the ASOC in superconductors is the Zeeman-type ASOC in triclinic and hexagonal systems, where the $g$-vector is parallel to the {\it c}-axis~\cite{Saito2016,Lu_Ising_SC,Xi_Ising_SC}. With the choice of the Zeeman-type ASOC instead of the Rashba ASOC, the model corresponds to the bilayer Ising superconductors~\cite{Zheliuk2019}. 
In contrast to the multilayer Rashba superconductors, the PDW state may be stabilized in this case by the in-plane magnetic field. Realization in the bilayer MoS$_2$~\cite{Zheliuk2019} has been proposed~\cite{Nakamura_bilayerTMD2017}, and experimental progress is awaited.

\subsection{Topological mirror superconductivity}

In the search for TSC, the odd-parity superconducting state has been sought as a strong candidate. 
In the usual setup, odd-parity superconductivity is caused by spin-triplet Cooper pairs whose realization is established only in few compounds, mainly in Uranium-based heavy-fermion systems~\cite{Aoki2019,UTe2_review}.
On the other hand, the field-induced PDW phase is an odd-parity superconducting state due to the spin-singlet pairing, which occurs widely in nature. 
Therefore, we may expect TSC in the PDW phase, and its clarification may open a new route to realize TSC. 

As discussed in theoretical studies for the PDW phase~\cite{TomoYoshida_SupLatt_PRL15,Yoshida_thesis,Yoshida_ZxZtoZxZ8superlattice_PRL17}, it is hard to realize the TSC in terms of the so-called topological periodic table based on the Altland-Zirnbauer symmetry class~\cite{Schnyder_classification_free_2008,Kitaev_classification_free_2009,Ryu_classification_free_2010}. The symmetry class D implies the $\mathbb{Z}$ classification in two dimension, and the topological invariant is the Chern number. However, to obtain a finite Chern number, fine-tuning of the parameters is required to cause the Lifshitz transition as in the case of Rashba superconductors in magnetic fields~\cite{Sato-Fujimoto2009,Sato-Fujimoto2010,Sau2010,Alicea2010,Lutchyn2010}. Such fine-tuned parameters are hard to be realized in the intrinsic superconductors.

However, the above discussion does not mean that the PDW phases are topologically trivial. It is now widely known that the topological properties are enriched by crystalline symmetry. The topologically nontrivial insulators and superconductors protected by crystalline symmetry are named topological crystalline insulator/superconductor~\cite{Fu_TCI2011} and the classification theories have been extensively developed for them~\cite{Chiu2013,Morimoto2013,Shiozaki2013,Shiozaki2016,Shiozaki2017}.
Below we show that the PDW state may be a topological crystalline superconductor protected by the mirror symmetry~\cite{TomoYoshida_SupLatt_PRL15,Yoshida_thesis,Yoshida_ZxZtoZxZ8superlattice_PRL17}.

In the following, the discussions are based on the BdG Hamiltonian in which the interaction term Eq.~\eqref{eq1''} is approximated by the BCS-type mean field theory. 
The BdG Hamiltonian reads
\begin{align}
\label{eq: BdG Hami}
  {\cal H}&=\sum_{{\bm k},s,s',m}[\xi({\bm k})\sigma_0+\alpha_m{\bm g}({\bm k})\cdot{\bm \sigma}-\mu_{\rm B} {\bm H} \cdot {\bm \sigma}]_{ss'}
  c_{{\bm k}sm}^\dagger c_{{\bm k}s'm} \nonumber \\
  &+ \sum_{{\bm k},s,\langle m,m'\rangle} t_{\perp} c_{{\bm k}sm}^\dagger c_{{\bm k}sm'} \notag \\
  &+ \sum_{\bm{k},m,s,s'} \Delta_{m s s'}(\bm{k}) c^\dagger_{\bm{k}ms}c^\dagger_{-\bm{k}ms'} +\mathrm{H.c.}.
\end{align}
The layer-dependent order parameter can then be parameterized by 
$\hat{\Delta}_m({\bm k})=[\psi_m({\bm k})+{\bm d}_m({\bm k})\cdot{\bm \sigma}]i\sigma_y$,
where $\psi_m({\bm k})$ and ${\bm d}_m({\bm k})$ represent the spin-singlet and spin-triplet components of order parameters 
on the layer $m$, respectively. 
For the $d$-wave ($s$-wave) superconductor we adopt 
$\psi_m(\bm{k})=\Delta_m (\cos k_x -\cos k_y )$ ($\psi_m(\bm{k})=\Delta_m$). 
The $p$-wave component is induced by the spin-orbit coupling and pairing interaction in the spin-triplet channel, that is ${\bm d}_m({\bm k})=a_m(\sin k_y,\sin k_x,0)+ib_m(-\sin k_x,\sin k_y,0)$ for the $d$-wave superconductor, while
${\bm d}_m({\bm k})=a_m(-\sin k_y,\sin k_x,0)+ib_m(\sin k_x,\sin k_y,0)$ for the $s$-wave superconductor. For simplicity, we hereafter ignore the magnetic-field-induced component, $b_m=0$.
Definition of the BCS and PDW states is straightforwardly extended to more-than-two-layer systems: 
\begin{subequations}
\begin{align}
    {\rm BCS \,\,state:}\,\, 
    &\psi_m({\bm k}) = \psi_{M+1-m}({\bm k}), \\
    &{\bm d}_m({\bm k}) = - {\bm d}_{M+1-m}({\bm k}), 
\end{align}
\end{subequations}
\begin{subequations}
\begin{align}
      {\rm PDW \,\,state:}\,\, 
     &\psi_m({\bm k}) = -\psi_{M+1-m}({\bm k}), \\
     &{\bm d}_m({\bm k}) = {\bm d}_{M+1-m}({\bm k}). 
\end{align}
\end{subequations}
The BdG Hamiltonian is represented as 
${\cal H}=\frac{1}{2}\sum_{{\bm k}}\Psi^\dagger_{\bm k}{\cal H}({\bm k})\Psi_{\bm k} $ 
with use of Nambu operators $\Psi^\dagger_{\bm k}=(c^\dagger_{{\bm k}sm},c_{-{\bm k}sm})$ in $ 4 \times M $ dimension. 

The multilayer structures illustrated in Fig.~\ref{fig_multilayer} preserve the mirror symmetry with respect to the central {\it ab}-plane. 
Thus, the superconducting states have a well-defined mirror parity. The BCS state is mirror-even, while the PDW state is mirror-odd. Accordingly, the BdG Hamiltonian obeys the mirror symmetry.
\begin{eqnarray}
  {\cal M}_{xy}^\pm{\cal H}({\bm k}){\cal M}_{xy}^{\pm\dagger}={\cal H}({\bm k}), 
  \label{eq2}
\end{eqnarray}
with the mirror reflection operator in the particle-hole space, ${\cal M}^+_{xy}$ (${\cal M}^-_{xy}$) for the mirror-even (mirror-odd) state. 
Equation (\ref{eq2}) guarantees that the BdG Hamiltonian can be block-diagonalized in the eigenbasis of ${\cal M}_{xy}^\pm$. Thus, the system is divided into the two subsectors corresponding to the sector Hamiltonian ${\cal H}_\lambda^\pm({\bm k})$ $(\lambda=\pm 1)$ with $ \lambda i $ as eigenvalues of ${\cal M}_{xy}^\pm$. 

Here, let us focus on the PDW state. 
The mirror reflection operator for the mirror-odd PDW state is written as 
${\cal M}_{xy}^-=i\sigma^z\tau^0\mathcal{P}_z$ 
with $[\mathcal{P}_z]_{mm'}=\delta_{m M-m'}$ being the layer permutation operator. 
Noticing $({\cal M}_{xy}^-)^2=-1$ and the anti-commutation relation $\{ {\cal M}_{xy}^-, \mathcal{C} \}=0$, which ensure
the particle-hole symmetry closed in each mirror sector,
we recognize that the block-diagonalized Hamiltonian for each sector belongs to class D~\cite{TomoYoshida_SupLatt_PRL15}. 
Note the the time-reversal symmetry is broken in each mirror sector even at zero magnetic field.
Therefore, the classification of 2D systems is $\mathbb{Z}$ characterized by the Chern number, 
\begin{subequations}
\begin{eqnarray}
\nu_{\lambda} &=& \int_{\rm BZ} \! \frac{ d^2 \bm{k}}{2\pi} \; F^\lambda(\bm{k}),
\end{eqnarray}
with 
\begin{eqnarray}
 F^\lambda(\bm{k})&=& \partial_{k_x}A^{\lambda}_y(\bm{k})-\partial_{k_y}A^{\lambda}_x(\bm{k}), \\
 A^{\lambda}_\mu(\bm{k})&=& i \sum_{n:{\rm occ}}\langle u_{n,\lambda}(\bm{k})| \partial_\mu | u_{n,\lambda}(\bm{k}) \rangle.
\end{eqnarray}
\end{subequations}
Here, $| u_{n,\lambda}(\bm{k}) \rangle$ denotes the eigenvector of the sector Hamiltonian $\mathcal{H}_\lambda(\bm{k})$; $\mathcal{H}_\lambda(\bm{k})| u_{n,\lambda}(\bm{k}) \rangle = | u_{n,\lambda}(\bm{k}) \rangle E_{n\lambda}$ ($n=1,\cdots,\mathrm{dim} \mathcal{H}_\lambda$).
Different from previous sections for noncentrosymmetric superconductors, the Chern number is defined for each mirror sector. 
Because of the particle-hole symmetry in each sector, the stable Majorana state appears on the surface, when the Chern number is odd.

%

Because each mirror sector follows the $\mathbb{Z}$-classification, the total system follows the $\mathbb{Z}\times\mathbb{Z}$-classification.
For convenience, we rewrite the topological invariants 
\begin{eqnarray}
\nu_{\mathrm{M}} = \frac{\nu_+-\nu_-}{2}, &\quad& \nu_{\mathrm{tot}} = \nu_+ + \nu_-.
\end{eqnarray}
where the mirror Chern number $\nu_{\mathrm{M}}$~\cite{Teo_MCN_PRB08} is introduced, while the total Chern number is $\nu_{\mathrm{tot}}$. When the total Chern number is zero as the cases we consider here, the mirror Chern number is an integer. 
In accordance with the bulk-boundary correspondence, $\nu_{\mathrm{M}}$ and $\nu_{\mathrm{tot}}$ predict the numbers of helical Majorana edge modes and chiral edge modes, respectively.

Although the mirror Chern number depends on the parameters of the model~\cite{TomoYoshida_SupLatt_PRL15,Yoshida_thesis}, below we discuss the representative results for the cases where the magnetic field and interlayer hopping do not cause the Lifshitz transition. In the weakly coupled multilayer systems, this condition is satisfied in a wide range of parameters. The results are summarized in Table~\ref{tab:mirror_Chern}, and we discuss one by one below.

\begin{table}[htbp]
    \centering
    \begin{tabular}{c|cc}
         & s-wave SC & d-wave SC \\ \hline   
    Bilayer     & 0 & 4 \\
    Trilayer    & 1 & 1 \\
    Quad-layer   & 0 & 8 
    \end{tabular}
    \caption{Representative values of the mirror Chern number $\nu_{\rm M}$ in the multilayer PDW state~\cite{TomoYoshida_SupLatt_PRL15,Yoshida_thesis,Yoshida_ZxZtoZxZ8superlattice_PRL17}.}
    \label{tab:mirror_Chern}
\end{table}

First, we discuss the bilayer system~\cite{TomoYoshida_SupLatt_PRL15,Yoshida_thesis}. 
The layer-dependent order parameter is represented by 
$\left[\psi_{1}({\bm k}),\psi_{2}({\bm k})\right] = \psi({\bm k}) \left[1,-1\right]$ 
and $\left[{\bm d}_{1}({\bm k}),{\bm d}_{2}({\bm k})\right] ={\bm d}({\bm k}) \left[1,1\right] $. 
Unitary transformation diagonalizing the mirror reflection operator leads to the sector Hamiltonian   
\begin{align}
 & {\cal H}_{\pm}^-({\bm k}) =
\nonumber \\
& \left(
  \begin{array}{cc}
    {\cal H}'({\bm k}) \pm t_\perp\sigma_z &  -[\psi({\bm k})-{\bm d}({\bm k})\cdot{\bm \sigma}]i\sigma_y \\
    i\sigma_y[\psi({\bm k})^\ast-{\bm d}^\ast({\bm k})\cdot{\bm \sigma}] & -{\cal H}^{'T}(-{\bm k}) \mp t_\perp\sigma_z
  \end{array}
  \right), \nonumber \\
  \label{eq8}
\end{align}
where ${\cal H}'({\bm k})=\xi({\bm k})\sigma_0-\mu_{\rm B} H_z \sigma_z-\alpha{\bm g}({\bm k})\cdot {\bm \sigma}$. 
Interestingly, the sector Hamiltonian is equivalent to the BdG Hamiltonian of the noncentrosymmetric single-layer 
superconductors (see Sec.~\ref{sec:Daido}),  
when we regard $\mu_{\rm B} H_z'(\lambda=\pm) = \mu_{\rm B} H_z \mp t_\perp$ as fictitious Zeeman fields. From this correspondence, topological properties are clarified based on the results in Sec.~\ref{sec:Daido}. For the $S+p$-wave superconductor, the Chern numbers are trivial, $\nu_\pm =0$, unless the fictitious magnetic field causes the Lifshitz transition~\cite{Sato-Fujimoto2009,Sato-Fujimoto2010}. On the other hand, for the $D+p$-wave superconductor, we can apply the results in Sec.~\ref{sec:Daido} and obtain $\nu_\pm = \pm 4$. Thus, the field-induced $D+p$-wave PDW state is identified as a topological mirror superconductor specified by the mirror Chern number, $\nu_{\rm M}=4$.

Next, we discuss the trilayer PDW state~\cite{TomoYoshida_SupLatt_PRL15,Yoshida_thesis}. 
The layer-dependent order parameter is represented by 
$\left[\psi_{1}({\bm k}),\psi_{2}({\bm k}),\psi_{3}({\bm k})\right] = \left[\psi_{\rm out}({\bm k}),0,-\psi_{\rm out}({\bm k})\right]$ 
and $\left[{\bm d}_{1}({\bm k}),{\bm d}_{2}({\bm k}),{\bm d}_{3}({\bm k})\right] = 
\left[{\bm d}_{\rm out}({\bm k}),{\bm d}_{\rm in}({\bm k}),{\bm d}_{\rm out}({\bm k})\right] $. 
As a result of the block diagonalization of the BdG Hamiltonian, we obtain the sector Hamiltonian:
\begin{widetext}
  \begin{eqnarray}
    {\cal H}_{+}^-({\bm k})=\left(
    \begin{array}{cccccc}
      \xi_{+}({\bm k}) & \alpha k_+ & \sqrt{2}t_\perp & 0 & -d_{{\rm out}-}({\bm k}) & -\psi_{\rm out}({\bm k}) \\
      \alpha k_- & \xi_{-}({\bm k}) & 0 & 0 & \psi_{\rm out}({\bm k}) & d_{{\rm out}+}({\bm k}) \\
      \sqrt{2}t_\perp & 0 & \xi_{+}({\bm k}) & -d_{{\rm in}-}({\bm k}) & 0 & 0 \\
      0 & 0 & -d_{{\rm in}-}^\ast({\bm k}) & -\xi_{+}({\bm k}) & -\sqrt{2}t_\perp & 0 \\
      -d_{{\rm out}-}^\ast({\bm k}) & \psi_{\rm out}({\bm k})^\ast & 0 & -\sqrt{2}t_\perp & -\xi_{+}({\bm k}) & \alpha k_- \\
      -\psi_{\rm out}({\bm k})^\ast & d_{{\rm out}+}^\ast({\bm k}) & 0 & 0 & \alpha k_+ & -\xi_{-}({\bm k})
    \end{array}
    \right),  
    \label{eq9}
  \end{eqnarray}
  \begin{eqnarray}
    {\cal H}_{-}^-({\bm k})=\left(
    \begin{array}{cccccc}
      \xi_{+}({\bm k}) & \alpha k_+ & 0 & 0 & -d_{{\rm out}-}({\bm k}) & -\psi_{\rm out}({\bm k}) \\
      \alpha k_- & \xi_{-}({\bm k}) & \sqrt{2}t_\perp & 0 & \psi_{\rm out}({\bm k}) & d_{{\rm out}+}({\bm k}) \\
      0 & \sqrt{2}t_\perp & \xi_{-}({\bm k}) & d_{{\rm in}+}({\bm k}) & 0 & 0 \\
      0 & 0 & d_{{\rm in}+}^\ast({\bm k}) & -\xi_{-}({\bm k}) & 0 & -\sqrt{2}t_\perp \\
      -d_{{\rm out}-}^\ast({\bm k}) & \psi_{\rm out}({\bm k})^\ast & 0 & 0 & -\xi_{+}({\bm k}) & \alpha k_- \\
      -\psi_{\rm out}({\bm k})^\ast & d_{{\rm out}+}^\ast({\bm k}) & 0 & -\sqrt{2}t_\perp & \alpha k_+ & -\xi_{-}({\bm k})
    \end{array}
    \right),
    \label{eq10}
  \end{eqnarray}
\end{widetext}
where $\xi_{\pm}({\bm k}) = \xi({\bm k}) \mp \muB H_z$,  $k_\pm = \sin k_y \pm i \sin k_x$, 
and $d_{{\rm out/in}\pm}({\bm k}) = d_{{\rm out/in}}^{\, x}({\bm k}) \pm i d_{{\rm out/in}}^{\, y}({\bm k})$.
Although we do not know their counterpart in existing systems, we have shown that both $S+p$-wave and $D+p$-wave PDW states are topological mirror superconductors with the mirror Chern number, $\nu_{\rm M}=1$. 
Instead of discussing the details of the calculation, we show an intuitive understanding of the $S+p$-wave PDW state. In this case, the gap does not close when we adiabatically change the parameter as $t_\perp \rightarrow 0$. In the limit, the layers are isolated. The two outer layers are noncentrosymmetric $S+p$-wave superconductors which are mostly trivial as we discussed for bilayers. 
On the other hand, the sector Hamiltonian of the inner layer is equivalent to the spinless chiral $p$-wave superconductor, known as a topological superconductor hosting Majorana fermions~\cite{Read-Green,Ivanov}. Indeed, this part gives nontrivial Chern numbers $\nu_\pm = \pm 1$ leading to $\nu_{\rm M}=1$. Although such an intuitive understanding has not been obtained for the $D+p$-wave state, we numerically obtained the same topological invariant, $\nu_{\rm M}=1$.

\begin{figure}[htbp]
\begin{center}
\includegraphics[width=90mm]{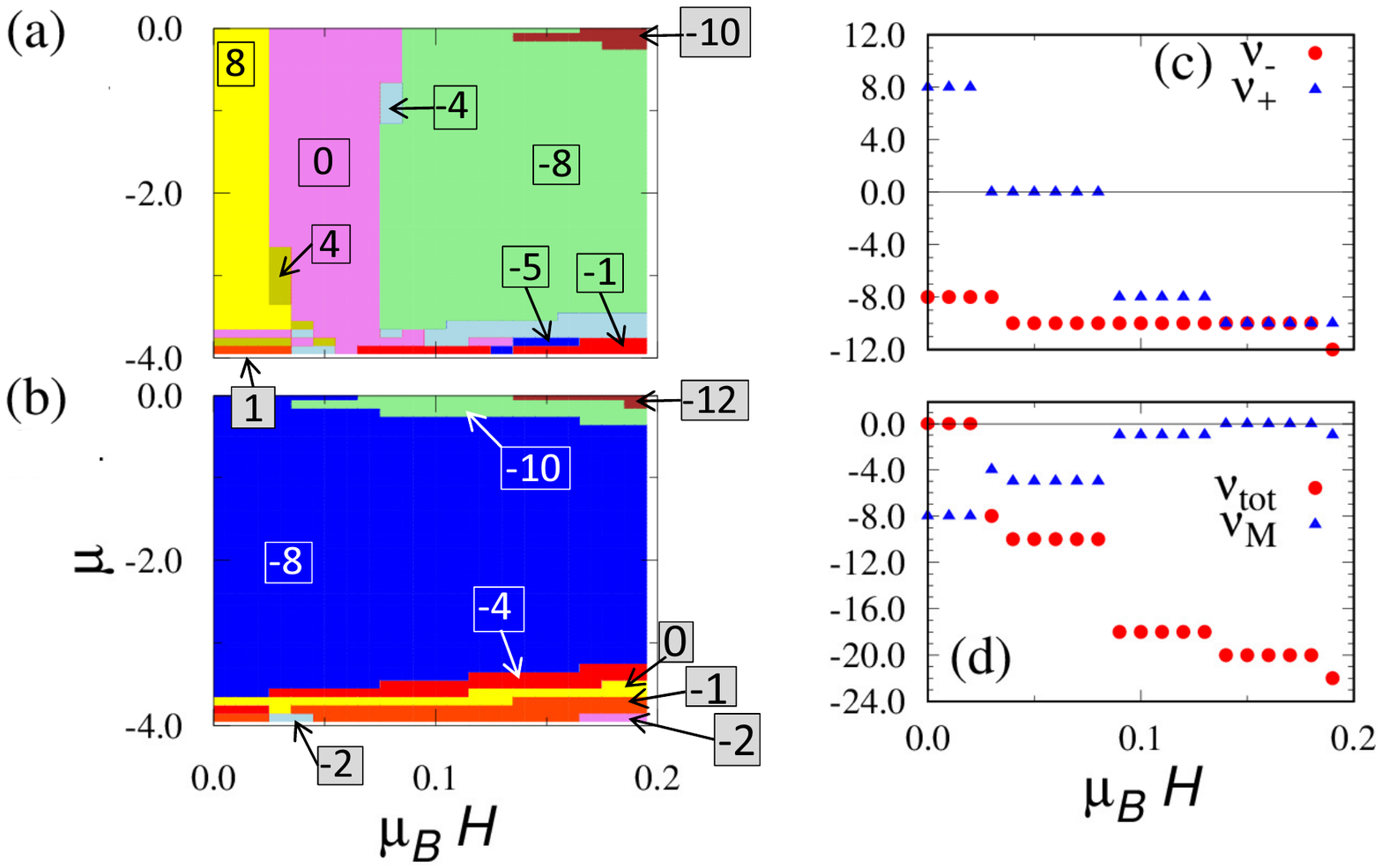}
\end{center}
\caption{
(a) and (b): Chern numbers of mirror sectors $\nu_+$ and $\nu_-$ against the magnetic field $\mu_BH$ and the chemical potential $\mu$, respectively~\cite{Yoshida_ZxZtoZxZ8superlattice_PRL17}. 
The Chern numbers are shown by numbers enclosed with solid lines.
(c): Chern numbers as functions of the magnetic field. 
(d) The mirror Chern number and the total Chern number as functions of the magnetic field.
These data are obtained for $t=1.0$, $t_\perp=0.1$, $\alpha=0.3$, $\alpha'=0.2$, $\Delta_d=\Delta'_d=0.05$, 
$a=-0.01$, and $a'=-0.0067$.
In panels (c) and (d), the chemical potential is fixed to $\mu=-0.1$.
}
\label{fig: ch}
\end{figure}

Finally, we discuss the quad-layer PDW state~\cite{Yoshida_ZxZtoZxZ8superlattice_PRL17}. By symmetry, the Rashba ASOC, spin-singlet gap function, and spin-triplet one follow the layer dependence, 
\begin{subequations}
\label{eq: alpha, psi, d}
\begin{eqnarray}
{}[\alpha_1,\psi_1(\bm{k}),\bm{d}_1(\bm{k})] &=& [\alpha,\psi(\bm{k}),\bm{d}(\bm{k})], \\
{}[\alpha_2,\psi_2(\bm{k}),\bm{d}_2(\bm{k})] &=& [\alpha',\psi'(\bm{k}),\bm{d}'(\bm{k})], \\
{}[\alpha_3,\psi_3(\bm{k}),\bm{d}_3(\bm{k})] &=& [-\alpha',-\psi'(\bm{k}),\bm{d}'(\bm{k})], \\
{}[\alpha_4,\psi_4(\bm{k}),\bm{d}_4(\bm{k})] &=& [-\alpha,\psi(\bm{k}),\bm{d}(\bm{k})].
\end{eqnarray}
\end{subequations}
Setting $\psi(\bm{k}):= \Delta_d (\cos k_x -\cos k_y )$ and $\bm{d}(\bm{k}):= a \left(\sin k_y,\sin k_x,0 \right)^T$ with real numbers $\Delta_d$ and $a$ in accordance with the $B$-representation of the $C_4$ point group, we numerically obtain the phase diagram in Fig.~\ref{fig: ch}.
In Fig.~\ref{fig: ch}, we plot the Chern numbers $\nu_{\pm}$ against the Zeeman field $\mu_BH$ and the chemical potential $\mu$.
This figure is obtained by employing the Fukui-Hatsugai-Suzuki method~\cite{Fukui_Hatsugai_05}.
In Figs.~\ref{fig: ch}(a)~and~(b), we can see that the mirror Chern number and the total Chern number take $(\nu_{\mathrm{M}},\nu_{\mathrm{tot}})=(8,0)$ when the magnetic field is weak, predicting eight pairs of helical Majorana edge modes.
The presence of these modes has been confirmed by the computation of energy spectrum~\cite{Yoshida_ZxZtoZxZ8superlattice_PRL17} (see Fig.~\ref{fig: OBC} in the next section).

\subsection{Summary and outlook of this section}

In this section, we have shown the field-induced transition from the even-parity superconducting state to the odd-parity state, which arises from the unique electronic structure in locally noncentrosymmetric multilayer systems. Based on this finding, we propose a way to realize odd-parity TSC by spin-singlet Cooper pairing, which are ubiquitous in materials. 

The field-induced odd-parity superconducting state, named PDW state, was identified as a topological crystalline superconductor protected by mirror reflection symmetry. The topological invariant, mirror Chern number, shows non-monotonic dependence on the number of layers (see Table.~\ref{tab:mirror_Chern}). 
For the $d$-wave superconductors, the representative mirror Chern numbers are $\nu_{\rm M}=4$, $1$, and $8$ for bilayers, trilayers, and quad-layers, respectively. All of these topological phases have intriguing properties. The bilayer PDW state is an analog of topological $d$-wave superconductivity discussed in Secs.~\ref{sec:Daido} and \ref{sec:Takasan}. The trilayer PDW state may host stable Majorana fermion on boundaries because $\nu_{\rm M} \in 2\mathbb{Z}+1$. As we show in the next section, the quad-layer PDW state is a testbed for the reduction of TSC by interactions, because $\nu_{\rm M} \in 8\mathbb{Z}$. 
These results shed light on the strongly correlated electron systems with multilayer structures as a fascinating platform of the TSC. For example, cuprate superconductors naturally form the multilayer structure~\cite{Gotlieb2018}, while recent developments of the heavy-fermion superlattice enable artificial control of the multilayer structure~\cite{Shimozawa_superlattice_RPP2016,naritsuka2021}. 

Very recently, the field-induced odd-parity superconducting state has been reported in the bulk CeRh$_2$As$_2$ having the locally noncentrosymmetric bilayer structure~\cite{CeRh2As2}. 
Motivated by this discovery, interest on the locally noncentrosymmetric superconductivity and the PDW state is increasing~\cite{skurativska2021spin,moeckli2021,Schertenleib2021,Moeckli_disorder}. 
Thus, attention is naturally paid to the possible TSC in CeRh$_2$As$_2$.
Classification of topological phases is different from the case in this section, because not the mirror symmetry but the nonsymmorphic glide symmetry is preserved in CeRh$_2$As$_2$. 
To clafiry TSC in CeRh$_2$As$_2$, topological analysis combined with a first-principles band calculation has been carried out. It shows that the PDW state in CeRh$_2$As$_2$ is a topological crystalline superconductor protected by the nonsymmorphic glide symmetry and specified by the glide $\mathbb{Z}_2$ invariant~\cite{nogaki2021topological}. 

\section{
Reduction of topological classification by interactions 
}
\label{sec:Yoshida}

So far, we have supposed that electron correlations are negligible.
However, extensive theoretical studies have elucidated that the electron correlations may induce novel topological phenomena which have not been observed for free fermions.
In particular, Fidkowski and Kitaev~\cite{Z_to_Zn_Fidkowski_10} have pointed out that electron correlations may change the classification results which play an essential role for the material searching~\cite{Schnyder_classification_free_2008,Kitaev_classification_free_2009,Ryu_classification_free_2010}. 
Namely, they have analyzed the one-dimensional topological superconductors which follow the $\mathbb{Z}$-classification at the non-interacting level and have found that one can gap out the Majorana edge modes without breaking the relevant symmetry when the number of the gapless Majorana modes is multiple of eight.
This fact indicates that the electron correlations change the topological classification from $\mathbb{Z}$ to $\mathbb{Z}_8$~\cite{Z_to_Zn_Fidkowski_10}.
In this review, we refer to this type of phenomena as reduction of the topological classification because the number of possible topological phases is reduced by electron correlations.
Further extensive theoretical studies have elucidated that the reduction occurs at any dimensions~\cite{Turner11,Fidkowski_1Dclassificatin_11,YaoRyu_Z_to_Z8_2013,Ryu_Z_to_Z8_2013,Qi_Z_to_Z8_2013,gu_supercohomology,kapustin_fermionic_cobordisms2014,Lu_CS_2011,Levin_CS_2012,Hsieh_CS_CPT_PRB14,Isobe_ZtoZ82015,Yoshida_ZtoZ8PRB15,Fidkowski_Z162013,Wang_Potter_Senthil2014,Metlitski_3Dinteraction2014,Wang_Senthil2014,You_Cenke2014,Morimoto_2015}, enhancing its significance.

In spite of the above theoretical progress, there are few experimental studies addressing this issue because of the absence of candidate platforms.
Therefore, it would make variable advance to theoretically propose a possible testbed.
In this section, we point out the possibility that a $\mathrm{CeCoIn_5}/\mathrm{YbCoIn_5}$ superlattice system is a feasible platform of the reduction of the topological classification from $\mathbb{Z}\times \mathbb{Z}$ to $\mathbb{Z}\times \mathbb{Z}_8$~\cite{Yoshida_ZxZtoZxZ8superlattice_PRL17}.

In the absence of correlations, the heavy fermions confined in the layer of $\mathrm{CeCoIn_5}$ can show the topological crystalline superconductivity following $\mathbb{Z}\times\mathbb{Z}$, which we have already discussed in Sec.~\ref{sec:Yanase}. For quad-layer (bilayer) of $\mathrm{CeCoIn_5}$, the eight (four) pairs of helical Majorana modes emerge.
In this section, we see that electron correlations can gap out the helical Majorana modes without breaking the symmetry when the number of the helical modes are multiple of eight. Thus, quad-layer $\mathrm{CeCoIn_5}$ can be a testbed of the reduction of the topological classification from $\mathbb{Z}\times\mathbb{Z}$ to $\mathbb{Z}\times\mathbb{Z}_8$.


\subsection{
$\mathrm{CeCoIn_5}/\mathrm{YbCoIn_5}$ superlattice
}

The $\mathrm{CeCoIn_5}/\mathrm{YbCoIn_5}$ superlattice system is a typical example of experimentally realizable 2D heavy-fermion superconductors~\cite{Mizukami2011,Goh_superlattice12,Shimozawa_superlattice_PRL14,Shimozawa_superlattice_RPP2016,naritsuka2021}.
In this system, the thickness of $\mathrm{CeCoIn_5}$-layers can be tuned at the atomic level.
%
%
Proximity effects between $\mathrm{CeCoIn_5}$ and $\mathrm{YbCoIn_5}$ are suppressed due to large mismatch of the Fermi velocity~\cite{superlattice_proximity_12,superlattice_Yanase_12,Yamashita_NMR_confinement_PRB15}. 
Thus, 2D heavy fermions emerge in $\mathrm{Ce}$-layers which show superconductivity around $1\mathrm{K}$.
In the following, we discuss the 2D superconducting phase in the subsystem composed of $\mathrm{CeCoIn_5}$. 

The CeCoIn$_5$ multilayers may be the platform of the model Eq.~\eqref{eq:model_LNCS}, because of the following facts. 
(i) Mirror reflection symmetry is locally broken owing to the presence of $\mathrm{YbCoIn}_5$-layers~\cite{Goh_superlattice12,Shimozawa_superlattice_PRL14,Shimozawa_superlattice_RPP2016,Maruyama_LISB_SpinOrb_JPSJ12}.
%
(ii) Bulk CeCoIn$_5$ is a $d_{x^2-y^2}$-wave superconductor~\cite{Matsuda_AngResTherm_JPCM06}.
(iii) The system is close to the Pauli limit~\cite{Tayama_CeCoIn5PauliLimPRB02}. 
(iv) The CeCoIn$_5$ superlattice is affected by the strong spin-orbit coupling~\cite{Goh_superlattice12,Maruyama_LISB_SpinOrb_JPSJ12,Shimozawa_superlattice_RPP2016,Shimozawa_superlattice_PRL14}.
Thus, we may expect that the field-induced odd-parity superconducting state, namely, the PDW state is realizable in the CeCoIn$_5$ superlattice. The PDW phase of the quad-layer CeCoIn$_5$ can be 
described by the BdG Hamiltonian Eq.~\eqref{eq: BdG Hami} with Eq.~\eqref{eq: alpha, psi, d}.

Topological phase diagram of the model has been shown in Fig.~\ref{fig: ch}. 
For the numerical calculation, we set parameters for quad-layer CeCoIn$_5$ as follows.
(i) Intralayer hopping: taking into account mass renormalization, we assume the intralayer hopping approximately $3\mathrm{meV}$. 
(ii) Pairing potentials: we adopt the pairing potential, $\Delta_d=0.05t$, in accordance with the experimental observation~\cite{Mizukami2011} showing the transition temperature approximately $T_{\rm c} \simeq 1\mathrm{K}$.
(iii) Rashba spin-orbit coupling: we set $\alpha=0.3t$. First-principles calculations indicate typical spin-splitting in the heavy fermions as  $1000\mathrm{K}$~\cite{Samokhin2004}. Here, we have taken into account the renormalization factor.
(iv) Interlayer hopping: we set the interlayer hopping as $t_\perp=0.1t$. This is consistent with the angular dependence of the upper critical field, which indicates the quasi-2D electronic structure and the interlayer hopping weaker than the Rashba spin-orbit coupling~\cite{Goh_superlattice12,Shimozawa_superlattice_RPP2016}.

\begin{figure}[ht]
\begin{center}
\includegraphics[width=80mm,clip]{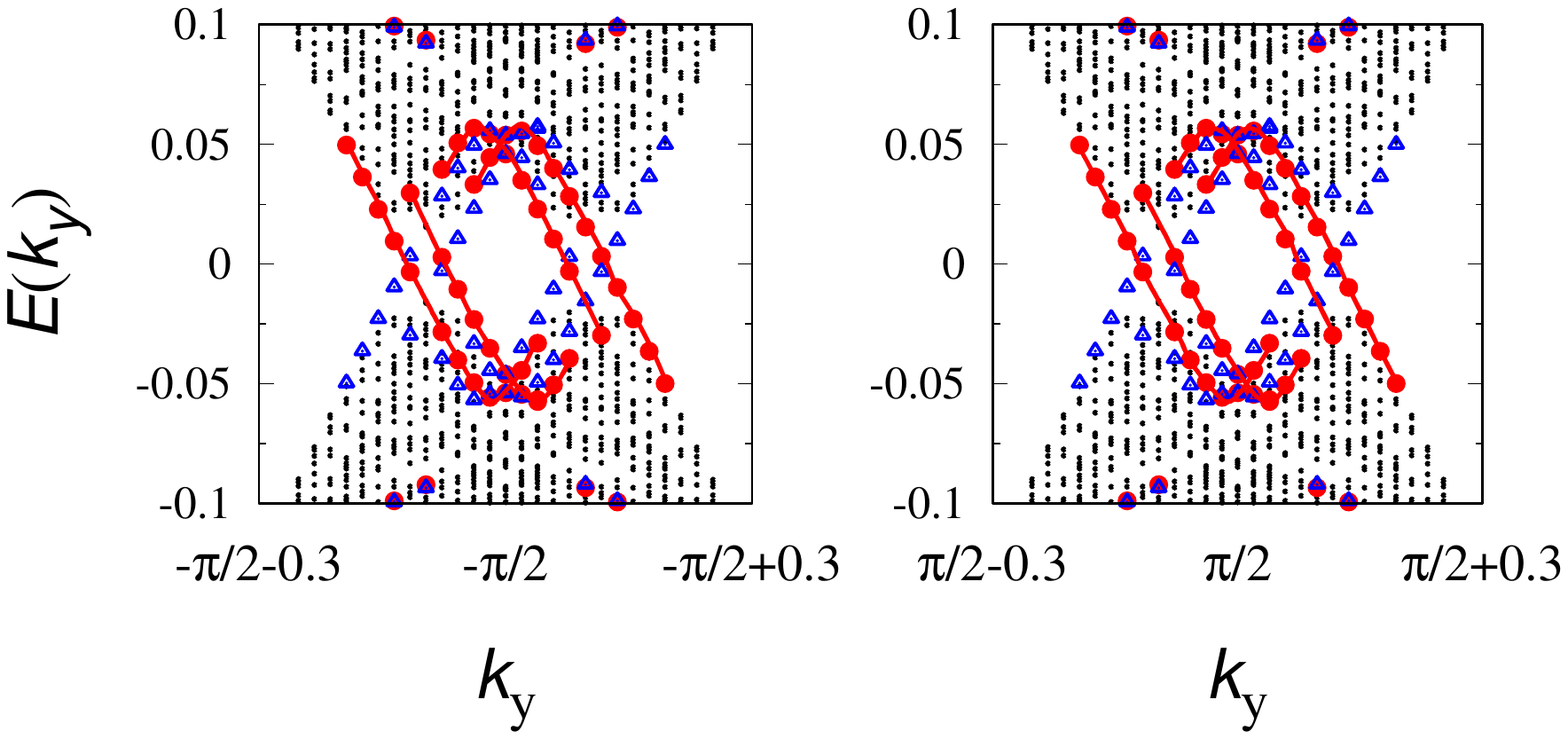}
\end{center}
\caption{
Energy spectrum of the mirror sector Hamiltonian $\mathcal{H}_+$ under the open (periodic) boundary condition for the $x$- ($y$-) direction. Here, we plot the data around $k_y=\pm \pi/2$.
The edge modes localized around the edge $x=1$ ($x=L$) are shown with red (blue) symbols. The red lines are for the guide of the eyes.
The data are obtained for the following parameter set in order to suppress the finite size effects: $t=1.0$, $t_\perp=0.1$, $\alpha=\alpha'=0.3$, 
$\Delta_d=\Delta'_d=0.4$, $\mu_{\rm B}H=\mu=0$,
and $L=300$.
At this parameter set, the topology of the bulk is characterized by $(\nu_{\mathrm{M}}, \nu_{\mathrm{tot}})=(8,0)$.
}
\label{fig: OBC}
\end{figure}

Figures~\ref{fig: ch}(a)~and~\ref{fig: ch}(b) reveal the mirror Chern number and the total Chern number $(\nu_{\mathrm{M}},\nu_{\mathrm{tot}})=(8,0)$ in the weak magnetic field region, predicting the eight pairs of helical Majorana edge modes localized at each edge. 
In order to verify this bulk-edge correspondence, we calculate the energy spectrum $E(k_y)$ by imposing the open (periodic) boundary condition for the $x$- ($y$-) direction. In Fig.~\ref{fig: OBC}, we observe eight Majorana edge modes in the $\lambda=+1$ mirror sector which is consistent with the Chern number $\nu_+=8$ in the bulk. For the $\lambda=-1$ mirror sector, the Chern number $\nu_-=-8$ predicts the eight Majorana edge modes propagating in the opposite direction, which has also been verified by numerics. 

With the above data, we expect that eight pairs of helical Majorana modes emerge in the quad-layer $\mathrm{CeCoIn_5}$ of CeCoIn$_5$/YbCoIn$_5$ superlattices.
We stress that the number of the helical Majorana modes can be tuned in experiments; {\it e.g.}, for the bilayer $\mathrm{CeCoIn_5}$, four pairs of helical Majorana modes may emerge (see Sec.~\ref{sec:Yanase}).

\subsection{
Gappability of the helical Majorana modes
}
Here, we show that the helical Majorana modes can be gapped out without breaking the mirror reflection symmetry when the number of helical edge modes is multiple of eight.

The effective Hamiltonian for the helical edge modes is written as
\begin{align}
H_\mathrm{edge}= \sum_{\lambda\alpha} \int \! dx\; \left[ \mathrm{sgn}(\lambda)\eta_{\lambda,\alpha}(x)(-iv \partial_x)\eta_{\lambda,\alpha}(x) \right],
\end{align}
where the summation is taken over $\lambda=\pm$ and $\alpha=1,\cdots,8$, and $\mathrm{sgn}(\lambda)$ takes $1$ ($-1$) for $\lambda=+$ $(-)$.
We note that the single pair of helical Majorana modes cannot be gapped out without breaking the symmetry~\cite{Z_to_Zn_Fidkowski_10,YaoRyu_Z_to_Z8_2013,Ryu_Z_to_Z8_2013,Qi_Z_to_Z8_2013,Hsieh_CS_CPT_PRB14,You_Cenke2014,Morimoto_2015}.
Before analyzing gappability of the multiple helical Majorana modes, we discuss the symmetry beyond the quadratic Hamiltonian.
The particle-hole symmetry changes to the symmetry of fermion number parity $P_f=(-1)^{N_f}$ with the fermion number operator $N_f$ when we go beyond the quadratic Hamiltonian~\cite{Z_to_Zn_Fidkowski_10}.
Therefore, the relevant symmetry to the helical gapless edge modes is the symmetry of mirror reflection and fermion number parity.

Now, by employing bosonization approach, let us see that one can gap out the eight pairs of helical Majorana modes without breaking the relevant symmetry.
To simplify the analysis, we rewrite the two pairs of helical Majorana fermions $\eta_{\lambda,\alpha}(x)$ ($\alpha=1,\cdots,8$) with a pair of helical complex fermions $f^\dagger_{\lambda,\alpha}:=\eta_{\lambda,2\alpha-1}(x)-i\eta_{\lambda,2\alpha}(x)$ ($\alpha=1,\cdots,4$),
\begin{eqnarray}
\mathcal{L}_\mathrm{edge}=\int \! dx \; \left[ K_{IJ}\partial_\tau \phi_I \partial_x \phi_J -V_{IJ} \partial_x\phi_I\partial_x\phi_J \right].
\end{eqnarray}
Here, $\phi_I(x)$ ($I=1,\cdots,N$) denotes the bosonic field defined as $f_I(x):=\kappa_I e^{-i\phi_I(x)}/\sqrt{2\pi\alpha}$ with the fermion operator $\bm{f}:=(f_{+,1},f_{-,1},f_{+,2},f_{-,2},\cdots,f_{+,N},f_{-,N})$.
The matrix $K:=\sigma^z\otimes \1_{N\times N}$ describes the chirality ({\it i.e.}, whether the modes propagate to left or right). $V=v\1_{2N\times2N}$ denotes the velocity.
With transformation laws of fermions $R\eta_{\lambda,\alpha}(x)R^{-1}=-\lambda \eta_{\lambda,\alpha}(x)$ and $P_f\eta_{\lambda,\alpha}(x) P_f^{-1}=-\eta_{\lambda,\alpha}(x)$, we obtain
\begin{subequations}
\begin{eqnarray}
R {\bm{\phi}} R^{-1}&=&\bm{\phi}+\pi (1,0,1,0,\cdots,1,0), \\
P_f{\bm{\phi}}P^{-1}_f&=&\bm{\phi}+\pi(1,1,\cdots,1,1).
\end{eqnarray}
\end{subequations}
Here, the mirror reflection operator is defined as $R=e^{-i\frac{\pi}{2}N_f}e^{-i\pi S^z}P_z$~\cite{Prefac_R_op_footnote} with $P_z$ denoting the exchange of the layers and the $z$-component of the total spin operator $S^z$.
We note that introducing a set of cosine terms $\cos(\bm{l}^T_i\cdot\bm{\phi})$ with $n$-independent integral vectors $\bm{l}$'s gaps out $n$-helical edge modes where $\bm{l}$'s satisfy the Haldane's criteria $\bm{l}^T_i K \bm{l}_j=0$ for all pair of $i,j=1,\cdots,n$~\cite{Haldane_nullvector_PRL95}.

Examining the presence/absence of the symmetry-protected gapless edge modes elucidates whether the system is topological/trivial.
In terms of the bosonization approach, it can be found by analyzing whether there exist cosine terms gapping out all of the edge modes without symmetry breaking.
We note that there are two ways to break symmetry: (i) introducing a cosine term which is not invariant under the transformation; (ii) spontaneous symmetry breaking.

As a first step, we discuss the case of $N=1$ where we can see a pair of helical Majorana modes which are symmetry-protected.
Firstly, we note that vectors $\bm{l}$'s satisfying the Haldane's criteria are written as
\begin{eqnarray}
\bm{l}=l(1,-1),\ l(1,1) && \quad l\in \mathbb{Z}.
\end{eqnarray}
If $l$ is odd, the cosine term $\cos(\bm{l}^T\cdot\bm{\phi})$ is not invariant under applying $R$. 
If $l$ is even, the cosine term $\cos(\bm{l}^T\cdot\bm{\phi})$ preserves the symmetry. 
We note, however, that reflection symmetry is spontaneously broken because applying the operator $R$ shifts $\langle \bm{l}^T\cdot \phi \rangle/l \to \langle \bm{l}^T\cdot \phi \rangle/l +\pi$~(mod $2\pi$).
Thus, there is no cosine term gapping out the edge mode without symmetry breaking.
Therefore, the single helical Majorana mode is symmetry-protected.

In a similar way, we can discuss the case of $N=2$. In this case, two linearly independent vectors $\bm{l}$'s describing cosine terms are required in order to gap out all of the edge modes.
Namely, at least a pair of helical Majorana modes survives even in the presence of cosine terms preserving the symmetry~\cite{Yoshida_TMI2D_PRB2016,Zhang_TMI2D_PRB2016}.

In the case of $N=4$, we can see that the eight pairs of helical Majorana modes are no longer symmetry-protected; all of these modes can be gapped out without symmetry breaking.
Specifically, introducing the cosine terms described by the following four independent vectors $\bm{l}$'s gaps out all of edge modes:
\begin{eqnarray}
{}\bm{l}^T_1= (1,0|1,0|0,-1|0,-1), &\quad& \bm{l}^T_2= (0,1|0,1|-1,0|-1,0), \nonumber \\
{}\bm{l}^T_3= (1,1|-1,-1|0,0|0,0), &\quad& \bm{l}^T_4= (0,0|0,0|1,1|-1,-1). \nonumber \\
\end{eqnarray}
%
Introducing these cosine terms do not break the symmetry because the following relations hold for arbitrary $i$ ($i=1,\cdots,4$):
\begin{subequations}
\begin{eqnarray}
R \bm{l}^T_i\cdot \bm{\phi} R^{-1} &=& \bm{l}^T_i\cdot \bm{\phi} \quad (\mathrm{mod} \,\, 2\pi),\\
P_f \bm{l}^T_i\cdot \bm{\phi} P^{-1}_f &=& \bm{l}^T_i\cdot \bm{\phi} \quad (\mathrm{mod} \,\, 2\pi).
\end{eqnarray}
\end{subequations}
These cosine potentials describe two-body interactions which can be induced by intralayer antiferromagnetic interactions.
Therefore, we can conclude that for $N=4$, the system is topologically trivial because of the electron correlations.
This means that the topological classification is reduced from $\mathbb{Z}\times \mathbb{Z}$ to $\mathbb{Z}\times \mathbb{Z}_8$~\cite{Yoshida_ZxZtoZxZ8superlattice_PRL17}.

We note that the problem for $N=3$ is reduced to the one for $N=-1$ because we know that the system is topologically trivial for $N=4$. In a similar way as the $N=1$ case, we can see that the Majorana modes are symmetry-protected for $N=-1$.

Therefore, we end up with the conclusion that the system characterized by $(\nu_{\mathrm{M}},\nu_{\mathrm{tot}})=(8,0)$ is topologically trivial in the presence of electron correlations, indicating that the $\mathrm{CeCoIn_5}/\mathrm{YbCoIn_5}$ superlattice system may serve as a testbed of the reduction of topological classification.

\subsection{
Summary of this section
}
In this section, we have pointed out that the superlattice system of $\mathrm{CeCoIn_5}/\mathrm{YbCoIn_5}$ can be a testbed of the reduction of topological classification, $\mathbb{Z}\times\mathbb{Z}\to \mathbb{Z}\times\mathbb{Z}_8$.
We have shown that the system may host eight pairs of helical edge modes protected by the mirror reflection symmetry. In addition, we have shown that these eight pairs of helical Majorana modes can be gapped out by interactions without symmetry breaking.
For experimental observation of the reduction, the STM measurement is a promising possibility. So far, the STM measurement has been carried out to detect the Majorana modes localized around the edge of the one-dimensional topological superconductors. In addition, recently, it became possible to apply the STM measurement to the heterostructures of $\mathrm{CeCoIn_5}$~\cite{Haze2018}. 
Thus, we consider that the STM measurement may experimentally support the reduction of topological classification by observing the following behaviors: the system hosts Majorana modes for the bilayer or trilayer of $\mathrm{CeCoIn_5}$ while it does not for the quad-layer systems.

We note that after the proposal for the $\mathrm{CeCoIn_5}/\mathrm{YbCoIn_5}$ superlattice~\cite{Yoshida_ZxZtoZxZ8superlattice_PRL17} another platform of the reduction was also proposed; a one-dimensional system of cold atoms showing $\mathbb{Z}\to \mathbb{Z}_4$~\cite{Yoshida_ZtoZ4ColdAtom_PRL18}. For this system, quantitative calculations are available by the density-matrix renormalization group (DMRG) analysis.
For the 2D testbed, further numerical analysis, which takes into account the details of electronic structure and electron correlations, are awaited. This issue is left as a future work.

\section{Summary and Outlook}

In this review, we have discussed the TSC based on the non-chiral $d$-wave superconductors. Our proposals rely on the globally or locally noncentrosymmetric crystal structure, where the spin-orbit coupling may affect superconducting properties. 
For noncentrosymmetric $d$-wave superconductors, the Zeeman term in combination with the spin-orbit coupling induces effectively chiral gap function and makes the bulk topological invariant nontrivial. The mechanism of TSC was explained in analogy with the quantum anomalous Hall insulator in the Haldane model. 
In locally noncentrosymmetric superconductors of either $s$-wave or $d$-wave symmetry, the paramagnetic pair-breaking effect causes the even-odd parity transition. The resulting topological crystalline superconductivity was clarified. 

Different from most proposals for topological $s$-wave superconductors, the topological $d$-wave superconductivity discussed here is realizable without fine-tuning of parameters, such as spin-orbit coupling, Zeeman term, and chemical potential. The platform is the 2D heterostructures of strongly correlated electron systems, and the controllability of recently developed $d$-wave superconductor heterostructures may enable various topological superconducting phases. For example, the quad-layer $d$-wave superconductor specified by the mirror Chern number $8$ may be a testbed for the reduction topological classification, which has not been demonstrated so far. 

Previous studies mainly focused on $s$-wave and spin-triplet superconductors. However, both of them have disadvantages. 
Although the $s$-wave superconductors are ubiquitous, fine-tuning of parameters is usually required for realizing TSC. The spin-triplet superconductors are possibly topological as it is. However, candidates are limited in nature. In addition to these directions, the topological $d$-wave superconductivity clarified here may open a new way for searching TSC. 

Progress in the topological $d$-wave superconductivity is further on-going recently. For instance, chiral $d$-wave superconductivity in twisted bilayer cuprate superconductors has been proposed~\cite{Can2021,can2021probing}, and experimental efforts are being conducted for realization~\cite{zhao2021emergent}. Furthermore, a newly discovered superconductor CeRh$_2$As$_2$~\cite{CeRh2As2} may be a bulk candidate of locally noncentrosymmetric topological superconductors~\cite{nogaki2021topological}. 
Identification of TSC and Majorana fermions and clarification of the unique properties are awaited for future studies.

\section*{
Acknowledgements
}
The authors are grateful to Tomohiro Yoshida, Daisuke Maruyama, Tatsuya Watanabe, Manfred Sigrist, and Norio Kawakami for collaboration in the original works.
The authors would like to thank Swee Kuan Goh, Yuji Matsuda, Takasada Shibauchi, Masaaki Shimozawa for fruitful discussions on the experiments.
This work was partly supported by JSPS KAKENHI Grant No.~JP21102506, JP23102709, JP24740230, JP25103711, JP15H05855, JP15H05884, JP16H00991, JP18H01140, JP18H05842, JP18H04225, JP18H05227, JP18H01178, JP19H01838, JP20H0515, JP20H04627, JP21K13880, and JP21K13850 and by SPIRITS 2020 of Kyoto University. 
K.T. was supported by the U.S. Department of Energy (DOE), Office of Science, Basic Energy Sciences (BES), under Contract No. AC02-05CH11231 within the Ultrafast Materials Science Program (KC2203).
The numerical calculations were performed on the supercomputer at the Institute for Solid State Physics in the University of Tokyo.

\appendix
\section{Derivation of energy dispersion}
\label{app:magenergy}
We show the outline of the derivation of Eq.~\eqref{eq:lambdas}.
It is convenient to work with the Nambu spinor ${\Psi'_{\bm{k}}}^\dagger=(c^\dagger_{\bm{k}\uparrow},c_{\bm{k}\downarrow}^\dagger,c_{-\bm{k}\downarrow},-c_{-\bm{k}\uparrow})$ rather than $\Psi^\dagger_{\bm{k}}$.
Then, we obtain the BdG Hamiltonian corresponding to ${\Psi'_{\bm{k}}}^\dagger$,
\begin{align}
\mathcal{H}'_{\mathrm{BdG}}(\bm{k})
&=(\xi(\bm{k})+\bm{g}(\bm{k})\cdot\bm{\sigma})\tau_z-\bm{h}\cdot\bm{\sigma}\notag\\
&\quad+\{(\psi(\bm{k})+\bm{d}(\bm{k})\cdot\bm{\sigma})\tau_++\mathrm{H.c.}\},\label{eq:model_BdG_appendix}
\end{align}
with $\bm{h}=\mu_B\bm{H}$ and $\alpha\bm{g}(\bm{k})\to\bm{g}(\bm{k})$ for simplicity.
The symbol $\tau_\mu$ ($\mu=0,1,2,3$) represents the Pauli matrices in the Nambu space, with $\tau_\pm\equiv(\tau_x\pm i\tau_y)/2$.
The essential procedure of the derivation is a somewhat technical unitary transformation $U_s=\exp(-i\pi\hat{g}(\bm{k})\cdot\bm{\sigma}\tau_z/4)$.
The transformed BdG Hamiltonian $U_s\mathcal{H}'_{\mathrm{BdG}}U_s^\dagger$
has the same form as \Eq{eq:model_BdG_appendix} with 
\begin{gather}
\bm{g}(\bm{k})\to\bm{g}_s(\bm{k})=\bm{g}(\bm{k})-\hat{g}(\bm{k})\times\bm{h},\\
\bm{h}\to\bm{h}_s(\bm{k})=(\hat{g}(\bm{k})\cdot\bm{h})\,\hat{g}(\bm{k}),\\
\psi(\bm{k})\to\psi_s(\bm{k})=-i\bm{d}(\bm{k})\cdot\hat{g}(\bm{k}),\\
\bm{d}(\bm{k})\to\bm{d}_s(\bm{k})=-i\psi(\bm{k})\hat{g}(\bm{k})+\hat{g}(\bm{k})\times(\bm{d}(\bm{k})\times\hat{g}(\bm{k})).
\end{gather}
When we consider the case $\bm{h}\perp\bm{g}(\bm{k})$ (such as for Rashba systems under $\bm{h}\parallel\hat{z}$), $\bm{h}_s=0$ and $U_s\mathcal{H}'_{\mathrm{BdG}}U_s^\dagger$ is formally equivalent with noncentrosymmetric superconductors in the absence of Zeeman field.
Thus, Eqs.~\eqref{eq:lambdas} are obtained by using the formula \eqref{zeromagspectrum} (Eq.~\eqref{zeromagspectrum} is valid for complex-valued $\psi(\bm{k})$ and $\bm{d}(\bm{k})$ as well).
In the presence of a parallel Zeeman field, it causes Pauli pair-breaking effect by $\pm\bm{h}_s(\bm{k})\cdot\hat{g}_s(\bm{k})$, leading to the full formula Eq.~\eqref{eq:lambdas}.
For more details, see Ref.~\onlinecite{Daido2016}.

\bibliographystyle{apsrev4-1}
\bibliography{TDSC2D.bib}

\begin{thebibliography}{245}%
\makeatletter
\providecommand \@ifxundefined [1]{%
 \@ifx{#1\undefined}
}%
\providecommand \@ifnum [1]{%
 \ifnum #1\expandafter \@firstoftwo
 \else \expandafter \@secondoftwo
 \fi
}%
\providecommand \@ifx [1]{%
 \ifx #1\expandafter \@firstoftwo
 \else \expandafter \@secondoftwo
 \fi
}%
\providecommand \natexlab [1]{#1}%
\providecommand \enquote  [1]{``#1''}%
\providecommand \bibnamefont  [1]{#1}%
\providecommand \bibfnamefont [1]{#1}%
\providecommand \citenamefont [1]{#1}%
\providecommand \href@noop [0]{\@secondoftwo}%
\providecommand \href [0]{\begingroup \@sanitize@url \@href}%
\providecommand \@href[1]{\@@startlink{#1}\@@href}%
\providecommand \@@href[1]{\endgroup#1\@@endlink}%
\providecommand \@sanitize@url [0]{\catcode `\\12\catcode `\$12\catcode
  `\&12\catcode `\#12\catcode `\^12\catcode `\_12\catcode `\%12\relax}%
\providecommand \@@startlink[1]{}%
\providecommand \@@endlink[0]{}%
\providecommand \url  [0]{\begingroup\@sanitize@url \@url }%
\providecommand \@url [1]{\endgroup\@href {#1}{\urlprefix }}%
\providecommand \urlprefix  [0]{URL }%
\providecommand \Eprint [0]{\href }%
\providecommand \doibase [0]{http://dx.doi.org/}%
\providecommand \selectlanguage [0]{\@gobble}%
\providecommand \bibinfo  [0]{\@secondoftwo}%
\providecommand \bibfield  [0]{\@secondoftwo}%
\providecommand \translation [1]{[#1]}%
\providecommand \BibitemOpen [0]{}%
\providecommand \bibitemStop [0]{}%
\providecommand \bibitemNoStop [0]{.\EOS\space}%
\providecommand \EOS [0]{\spacefactor3000\relax}%
\providecommand \BibitemShut  [1]{\csname bibitem#1\endcsname}%
\let\auto@bib@innerbib\@empty
\bibitem [{\citenamefont {Schnyder}\ \emph {et~al.}(2008)\citenamefont
  {Schnyder}, \citenamefont {Ryu}, \citenamefont {Furusaki},\ and\
  \citenamefont {Ludwig}}]{Schnyder_classification_free_2008}%
  \BibitemOpen
  \bibfield  {author} {\bibinfo {author} {\bibfnamefont {A.~P.}\ \bibnamefont
  {Schnyder}}, \bibinfo {author} {\bibfnamefont {S.}~\bibnamefont {Ryu}},
  \bibinfo {author} {\bibfnamefont {A.}~\bibnamefont {Furusaki}}, \ and\
  \bibinfo {author} {\bibfnamefont {A.~W.~W.}\ \bibnamefont {Ludwig}},\ }\href
  {\doibase 10.1103/PhysRevB.78.195125} {\bibfield  {journal} {\bibinfo
  {journal} {Phys. Rev. B}\ }\textbf {\bibinfo {volume} {78}},\ \bibinfo
  {pages} {195125} (\bibinfo {year} {2008})}\BibitemShut {NoStop}%
\bibitem [{\citenamefont {Kitaev}(2009)}]{Kitaev_classification_free_2009}%
  \BibitemOpen
  \bibfield  {author} {\bibinfo {author} {\bibfnamefont {A.}~\bibnamefont
  {Kitaev}},\ }\href {\doibase 10.1063/1.3149495} {\bibfield  {journal}
  {\bibinfo  {journal} {AIP Conf. Proc.}\ }\textbf {\bibinfo {volume} {1134}},\
  \bibinfo {pages} {22} (\bibinfo {year} {2009})}\BibitemShut {NoStop}%
\bibitem [{\citenamefont {Ryu}\ \emph {et~al.}(2010)\citenamefont {Ryu},
  \citenamefont {Schnyder}, \citenamefont {Furusaki},\ and\ \citenamefont
  {Ludwig}}]{Ryu_classification_free_2010}%
  \BibitemOpen
  \bibfield  {author} {\bibinfo {author} {\bibfnamefont {S.}~\bibnamefont
  {Ryu}}, \bibinfo {author} {\bibfnamefont {A.~P.}\ \bibnamefont {Schnyder}},
  \bibinfo {author} {\bibfnamefont {A.}~\bibnamefont {Furusaki}}, \ and\
  \bibinfo {author} {\bibfnamefont {A.~W.~W.}\ \bibnamefont {Ludwig}},\ }\href
  {http://stacks.iop.org/1367-2630/12/i=6/a=065010} {\bibfield  {journal}
  {\bibinfo  {journal} {New J. Phys.}\ }\textbf {\bibinfo {volume} {12}},\
  \bibinfo {pages} {065010} (\bibinfo {year} {2010})}\BibitemShut {NoStop}%
\bibitem [{\citenamefont {Kitaev}(2001)}]{Kitaev2001}%
  \BibitemOpen
  \bibfield  {author} {\bibinfo {author} {\bibfnamefont {A.~Y.}\ \bibnamefont
  {Kitaev}},\ }\href {\doibase 10.1070/1063-7869/44/10s/s29} {\bibfield
  {journal} {\bibinfo  {journal} {Physics-Uspekhi}\ }\textbf {\bibinfo {volume}
  {44}},\ \bibinfo {pages} {131} (\bibinfo {year} {2001})}\BibitemShut
  {NoStop}%
\bibitem [{\citenamefont {Kane}\ and\ \citenamefont
  {Mele}(2005)}]{Kane-Mele2005}%
  \BibitemOpen
  \bibfield  {author} {\bibinfo {author} {\bibfnamefont {C.~L.}\ \bibnamefont
  {Kane}}\ and\ \bibinfo {author} {\bibfnamefont {E.~J.}\ \bibnamefont
  {Mele}},\ }\href {\doibase 10.1103/PhysRevLett.95.226801} {\bibfield
  {journal} {\bibinfo  {journal} {Phys. Rev. Lett.}\ }\textbf {\bibinfo
  {volume} {95}},\ \bibinfo {pages} {226801} (\bibinfo {year}
  {2005})}\BibitemShut {NoStop}%
\bibitem [{\citenamefont {Qi}\ and\ \citenamefont
  {Zhang}(2011)}]{Qi-SCZ_review}%
  \BibitemOpen
  \bibfield  {author} {\bibinfo {author} {\bibfnamefont {X.-L.}\ \bibnamefont
  {Qi}}\ and\ \bibinfo {author} {\bibfnamefont {S.-C.}\ \bibnamefont {Zhang}},\
  }\href {\doibase 10.1103/RevModPhys.83.1057} {\bibfield  {journal} {\bibinfo
  {journal} {Rev. Mod. Phys.}\ }\textbf {\bibinfo {volume} {83}},\ \bibinfo
  {pages} {1057} (\bibinfo {year} {2011})}\BibitemShut {NoStop}%
\bibitem [{\citenamefont {Fu}(2011)}]{Fu_TCI2011}%
  \BibitemOpen
  \bibfield  {author} {\bibinfo {author} {\bibfnamefont {L.}~\bibnamefont
  {Fu}},\ }\href {\doibase 10.1103/PhysRevLett.106.106802} {\bibfield
  {journal} {\bibinfo  {journal} {Phys. Rev. Lett.}\ }\textbf {\bibinfo
  {volume} {106}},\ \bibinfo {pages} {106802} (\bibinfo {year}
  {2011})}\BibitemShut {NoStop}%
\bibitem [{\citenamefont {Sato}\ and\ \citenamefont
  {Fujimoto}(2016)}]{Sato2016_review}%
  \BibitemOpen
  \bibfield  {author} {\bibinfo {author} {\bibfnamefont {M.}~\bibnamefont
  {Sato}}\ and\ \bibinfo {author} {\bibfnamefont {S.}~\bibnamefont
  {Fujimoto}},\ }\href {\doibase 10.7566/JPSJ.85.072001} {\bibfield  {journal}
  {\bibinfo  {journal} {Journal of the Physical Society of Japan}\ }\textbf
  {\bibinfo {volume} {85}},\ \bibinfo {pages} {072001} (\bibinfo {year}
  {2016})},\ \Eprint
  {http://arxiv.org/abs/https://doi.org/10.7566/JPSJ.85.072001}
  {https://doi.org/10.7566/JPSJ.85.072001} \BibitemShut {NoStop}%
\bibitem [{\citenamefont {Sato}\ and\ \citenamefont
  {Ando}(2017)}]{Sato2017_review}%
  \BibitemOpen
  \bibfield  {author} {\bibinfo {author} {\bibfnamefont {M.}~\bibnamefont
  {Sato}}\ and\ \bibinfo {author} {\bibfnamefont {Y.}~\bibnamefont {Ando}},\
  }\href {\doibase 10.1088/1361-6633/aa6ac7} {\bibfield  {journal} {\bibinfo
  {journal} {Reports on Progress in Physics}\ }\textbf {\bibinfo {volume}
  {80}},\ \bibinfo {pages} {076501} (\bibinfo {year} {2017})}\BibitemShut
  {NoStop}%
\bibitem [{\citenamefont {Benalcazar}\ \emph {et~al.}(2017)\citenamefont
  {Benalcazar}, \citenamefont {Bernevig},\ and\ \citenamefont
  {Hughes}}]{BBH_science}%
  \BibitemOpen
  \bibfield  {author} {\bibinfo {author} {\bibfnamefont {W.~A.}\ \bibnamefont
  {Benalcazar}}, \bibinfo {author} {\bibfnamefont {B.~A.}\ \bibnamefont
  {Bernevig}}, \ and\ \bibinfo {author} {\bibfnamefont {T.~L.}\ \bibnamefont
  {Hughes}},\ }\href {\doibase 10.1126/science.aah6442} {\bibfield  {journal}
  {\bibinfo  {journal} {Science}\ }\textbf {\bibinfo {volume} {357}},\ \bibinfo
  {pages} {61} (\bibinfo {year} {2017})},\ \Eprint
  {http://arxiv.org/abs/https://science.sciencemag.org/content/357/6346/61.full.pdf}
  {https://science.sciencemag.org/content/357/6346/61.full.pdf} \BibitemShut
  {NoStop}%
\bibitem [{\citenamefont {Schindler}\ \emph {et~al.}(2018)\citenamefont
  {Schindler}, \citenamefont {Cook}, \citenamefont {Vergniory}, \citenamefont
  {Wang}, \citenamefont {Parkin}, \citenamefont {Bernevig},\ and\ \citenamefont
  {Neupert}}]{Schindler2018}%
  \BibitemOpen
  \bibfield  {author} {\bibinfo {author} {\bibfnamefont {F.}~\bibnamefont
  {Schindler}}, \bibinfo {author} {\bibfnamefont {A.~M.}\ \bibnamefont {Cook}},
  \bibinfo {author} {\bibfnamefont {M.~G.}\ \bibnamefont {Vergniory}}, \bibinfo
  {author} {\bibfnamefont {Z.}~\bibnamefont {Wang}}, \bibinfo {author}
  {\bibfnamefont {S.~S.~P.}\ \bibnamefont {Parkin}}, \bibinfo {author}
  {\bibfnamefont {B.~A.}\ \bibnamefont {Bernevig}}, \ and\ \bibinfo {author}
  {\bibfnamefont {T.}~\bibnamefont {Neupert}},\ }\href {\doibase
  10.1126/sciadv.aat0346} {\bibfield  {journal} {\bibinfo  {journal} {Science
  Advances}\ }\textbf {\bibinfo {volume} {4}} (\bibinfo {year} {2018}),\
  10.1126/sciadv.aat0346},\ \Eprint
  {http://arxiv.org/abs/https://advances.sciencemag.org/content/4/6/eaat0346.full.pdf}
  {https://advances.sciencemag.org/content/4/6/eaat0346.full.pdf} \BibitemShut
  {NoStop}%
\bibitem [{\citenamefont {Read}\ and\ \citenamefont
  {Green}(2000)}]{Read-Green}%
  \BibitemOpen
  \bibfield  {author} {\bibinfo {author} {\bibfnamefont {N.}~\bibnamefont
  {Read}}\ and\ \bibinfo {author} {\bibfnamefont {D.}~\bibnamefont {Green}},\
  }\href {\doibase 10.1103/PhysRevB.61.10267} {\bibfield  {journal} {\bibinfo
  {journal} {Phys. Rev. B}\ }\textbf {\bibinfo {volume} {61}},\ \bibinfo
  {pages} {10267} (\bibinfo {year} {2000})}\BibitemShut {NoStop}%
\bibitem [{\citenamefont {Ivanov}(2001)}]{Ivanov}%
  \BibitemOpen
  \bibfield  {author} {\bibinfo {author} {\bibfnamefont {D.~A.}\ \bibnamefont
  {Ivanov}},\ }\href {\doibase 10.1103/PhysRevLett.86.268} {\bibfield
  {journal} {\bibinfo  {journal} {Phys. Rev. Lett.}\ }\textbf {\bibinfo
  {volume} {86}},\ \bibinfo {pages} {268} (\bibinfo {year} {2001})}\BibitemShut
  {NoStop}%
\bibitem [{\citenamefont {Kitaev}(2003)}]{Kitaev2003}%
  \BibitemOpen
  \bibfield  {author} {\bibinfo {author} {\bibfnamefont {A.}~\bibnamefont
  {Kitaev}},\ }\href {\doibase https://doi.org/10.1016/S0003-4916(02)00018-0}
  {\bibfield  {journal} {\bibinfo  {journal} {Annals of Physics}\ }\textbf
  {\bibinfo {volume} {303}},\ \bibinfo {pages} {2} (\bibinfo {year}
  {2003})}\BibitemShut {NoStop}%
\bibitem [{\citenamefont {Nayak}\ \emph {et~al.}(2008)\citenamefont {Nayak},
  \citenamefont {Simon}, \citenamefont {Stern}, \citenamefont {Freedman},\ and\
  \citenamefont {Das~Sarma}}]{Nayak2008}%
  \BibitemOpen
  \bibfield  {author} {\bibinfo {author} {\bibfnamefont {C.}~\bibnamefont
  {Nayak}}, \bibinfo {author} {\bibfnamefont {S.~H.}\ \bibnamefont {Simon}},
  \bibinfo {author} {\bibfnamefont {A.}~\bibnamefont {Stern}}, \bibinfo
  {author} {\bibfnamefont {M.}~\bibnamefont {Freedman}}, \ and\ \bibinfo
  {author} {\bibfnamefont {S.}~\bibnamefont {Das~Sarma}},\ }\href {\doibase
  10.1103/RevModPhys.80.1083} {\bibfield  {journal} {\bibinfo  {journal} {Rev.
  Mod. Phys.}\ }\textbf {\bibinfo {volume} {80}},\ \bibinfo {pages} {1083}
  (\bibinfo {year} {2008})}\BibitemShut {NoStop}%
\bibitem [{\citenamefont {Leggett}(1975)}]{LeggettRMP}%
  \BibitemOpen
  \bibfield  {author} {\bibinfo {author} {\bibfnamefont {A.~J.}\ \bibnamefont
  {Leggett}},\ }\href {\doibase 10.1103/RevModPhys.47.331} {\bibfield
  {journal} {\bibinfo  {journal} {Rev. Mod. Phys.}\ }\textbf {\bibinfo {volume}
  {47}},\ \bibinfo {pages} {331} (\bibinfo {year} {1975})}\BibitemShut
  {NoStop}%
\bibitem [{\citenamefont {Sato}\ \emph {et~al.}(2009)\citenamefont {Sato},
  \citenamefont {Takahashi},\ and\ \citenamefont
  {Fujimoto}}]{Sato-Fujimoto2009}%
  \BibitemOpen
  \bibfield  {author} {\bibinfo {author} {\bibfnamefont {M.}~\bibnamefont
  {Sato}}, \bibinfo {author} {\bibfnamefont {Y.}~\bibnamefont {Takahashi}}, \
  and\ \bibinfo {author} {\bibfnamefont {S.}~\bibnamefont {Fujimoto}},\ }\href
  {\doibase 10.1103/PhysRevLett.103.020401} {\bibfield  {journal} {\bibinfo
  {journal} {Phys. Rev. Lett.}\ }\textbf {\bibinfo {volume} {103}},\ \bibinfo
  {pages} {020401} (\bibinfo {year} {2009})}\BibitemShut {NoStop}%
\bibitem [{\citenamefont {Sato}\ \emph {et~al.}(2010)\citenamefont {Sato},
  \citenamefont {Takahashi},\ and\ \citenamefont
  {Fujimoto}}]{Sato-Fujimoto2010}%
  \BibitemOpen
  \bibfield  {author} {\bibinfo {author} {\bibfnamefont {M.}~\bibnamefont
  {Sato}}, \bibinfo {author} {\bibfnamefont {Y.}~\bibnamefont {Takahashi}}, \
  and\ \bibinfo {author} {\bibfnamefont {S.}~\bibnamefont {Fujimoto}},\ }\href
  {\doibase 10.1103/PhysRevB.82.134521} {\bibfield  {journal} {\bibinfo
  {journal} {Phys. Rev. B}\ }\textbf {\bibinfo {volume} {82}},\ \bibinfo
  {pages} {134521} (\bibinfo {year} {2010})}\BibitemShut {NoStop}%
\bibitem [{\citenamefont {Sau}\ \emph {et~al.}(2010)\citenamefont {Sau},
  \citenamefont {Lutchyn}, \citenamefont {Tewari},\ and\ \citenamefont
  {Das~Sarma}}]{Sau2010}%
  \BibitemOpen
  \bibfield  {author} {\bibinfo {author} {\bibfnamefont {J.~D.}\ \bibnamefont
  {Sau}}, \bibinfo {author} {\bibfnamefont {R.~M.}\ \bibnamefont {Lutchyn}},
  \bibinfo {author} {\bibfnamefont {S.}~\bibnamefont {Tewari}}, \ and\ \bibinfo
  {author} {\bibfnamefont {S.}~\bibnamefont {Das~Sarma}},\ }\href {\doibase
  10.1103/PhysRevLett.104.040502} {\bibfield  {journal} {\bibinfo  {journal}
  {Phys. Rev. Lett.}\ }\textbf {\bibinfo {volume} {104}},\ \bibinfo {pages}
  {040502} (\bibinfo {year} {2010})}\BibitemShut {NoStop}%
\bibitem [{\citenamefont {Lutchyn}\ \emph {et~al.}(2010)\citenamefont
  {Lutchyn}, \citenamefont {Sau},\ and\ \citenamefont
  {Das~Sarma}}]{Lutchyn2010}%
  \BibitemOpen
  \bibfield  {author} {\bibinfo {author} {\bibfnamefont {R.~M.}\ \bibnamefont
  {Lutchyn}}, \bibinfo {author} {\bibfnamefont {J.~D.}\ \bibnamefont {Sau}}, \
  and\ \bibinfo {author} {\bibfnamefont {S.}~\bibnamefont {Das~Sarma}},\ }\href
  {\doibase 10.1103/PhysRevLett.105.077001} {\bibfield  {journal} {\bibinfo
  {journal} {Phys. Rev. Lett.}\ }\textbf {\bibinfo {volume} {105}},\ \bibinfo
  {pages} {077001} (\bibinfo {year} {2010})}\BibitemShut {NoStop}%
\bibitem [{\citenamefont {Alicea}(2010)}]{Alicea2010}%
  \BibitemOpen
  \bibfield  {author} {\bibinfo {author} {\bibfnamefont {J.}~\bibnamefont
  {Alicea}},\ }\href {\doibase 10.1103/PhysRevB.81.125318} {\bibfield
  {journal} {\bibinfo  {journal} {Phys. Rev. B}\ }\textbf {\bibinfo {volume}
  {81}},\ \bibinfo {pages} {125318} (\bibinfo {year} {2010})}\BibitemShut
  {NoStop}%
\bibitem [{\citenamefont {Sato}(2010)}]{Sato2010}%
  \BibitemOpen
  \bibfield  {author} {\bibinfo {author} {\bibfnamefont {M.}~\bibnamefont
  {Sato}},\ }\href {\doibase 10.1103/PhysRevB.81.220504} {\bibfield  {journal}
  {\bibinfo  {journal} {Phys. Rev. B}\ }\textbf {\bibinfo {volume} {81}},\
  \bibinfo {pages} {220504(R)} (\bibinfo {year} {2010})}\BibitemShut {NoStop}%
\bibitem [{\citenamefont {Fu}\ and\ \citenamefont {Berg}(2010)}]{Fu2010}%
  \BibitemOpen
  \bibfield  {author} {\bibinfo {author} {\bibfnamefont {L.}~\bibnamefont
  {Fu}}\ and\ \bibinfo {author} {\bibfnamefont {E.}~\bibnamefont {Berg}},\
  }\href {\doibase 10.1103/PhysRevLett.105.097001} {\bibfield  {journal}
  {\bibinfo  {journal} {Phys. Rev. Lett.}\ }\textbf {\bibinfo {volume} {105}},\
  \bibinfo {pages} {097001} (\bibinfo {year} {2010})}\BibitemShut {NoStop}%
\bibitem [{\citenamefont {Mourik}\ \emph {et~al.}(2012)\citenamefont {Mourik},
  \citenamefont {Zuo}, \citenamefont {Frolov}, \citenamefont {Plissard},
  \citenamefont {Bakkers},\ and\ \citenamefont {Kouwenhoven}}]{Mourik2012}%
  \BibitemOpen
  \bibfield  {author} {\bibinfo {author} {\bibfnamefont {V.}~\bibnamefont
  {Mourik}}, \bibinfo {author} {\bibfnamefont {K.}~\bibnamefont {Zuo}},
  \bibinfo {author} {\bibfnamefont {S.~M.}\ \bibnamefont {Frolov}}, \bibinfo
  {author} {\bibfnamefont {S.~R.}\ \bibnamefont {Plissard}}, \bibinfo {author}
  {\bibfnamefont {E.~P. A.~M.}\ \bibnamefont {Bakkers}}, \ and\ \bibinfo
  {author} {\bibfnamefont {L.~P.}\ \bibnamefont {Kouwenhoven}},\ }\href
  {\doibase 10.1126/science.1222360} {\bibfield  {journal} {\bibinfo  {journal}
  {Science}\ }\textbf {\bibinfo {volume} {336}},\ \bibinfo {pages} {1003}
  (\bibinfo {year} {2012})},\ \Eprint
  {http://arxiv.org/abs/https://science.sciencemag.org/content/336/6084/1003.full.pdf}
  {https://science.sciencemag.org/content/336/6084/1003.full.pdf} \BibitemShut
  {NoStop}%
\bibitem [{\citenamefont {Nadj-Perge}\ \emph {et~al.}(2014)\citenamefont
  {Nadj-Perge}, \citenamefont {Drozdov}, \citenamefont {Li}, \citenamefont
  {Chen}, \citenamefont {Jeon}, \citenamefont {Seo}, \citenamefont {MacDonald},
  \citenamefont {Bernevig},\ and\ \citenamefont {Yazdani}}]{Nadj-Perge2014}%
  \BibitemOpen
  \bibfield  {author} {\bibinfo {author} {\bibfnamefont {S.}~\bibnamefont
  {Nadj-Perge}}, \bibinfo {author} {\bibfnamefont {I.~K.}\ \bibnamefont
  {Drozdov}}, \bibinfo {author} {\bibfnamefont {J.}~\bibnamefont {Li}},
  \bibinfo {author} {\bibfnamefont {H.}~\bibnamefont {Chen}}, \bibinfo {author}
  {\bibfnamefont {S.}~\bibnamefont {Jeon}}, \bibinfo {author} {\bibfnamefont
  {J.}~\bibnamefont {Seo}}, \bibinfo {author} {\bibfnamefont {A.~H.}\
  \bibnamefont {MacDonald}}, \bibinfo {author} {\bibfnamefont {B.~A.}\
  \bibnamefont {Bernevig}}, \ and\ \bibinfo {author} {\bibfnamefont
  {A.}~\bibnamefont {Yazdani}},\ }\href {\doibase 10.1126/science.1259327}
  {\bibfield  {journal} {\bibinfo  {journal} {Science}\ }\textbf {\bibinfo
  {volume} {346}},\ \bibinfo {pages} {602} (\bibinfo {year} {2014})},\ \Eprint
  {http://arxiv.org/abs/https://science.sciencemag.org/content/346/6209/602.full.pdf}
  {https://science.sciencemag.org/content/346/6209/602.full.pdf} \BibitemShut
  {NoStop}%
\bibitem [{\citenamefont {He}\ \emph {et~al.}(2017)\citenamefont {He},
  \citenamefont {Pan}, \citenamefont {Stern}, \citenamefont {Burks},
  \citenamefont {Che}, \citenamefont {Yin}, \citenamefont {Wang}, \citenamefont
  {Lian}, \citenamefont {Zhou}, \citenamefont {Choi}, \citenamefont {Murata},
  \citenamefont {Kou}, \citenamefont {Chen}, \citenamefont {Nie}, \citenamefont
  {Shao}, \citenamefont {Fan}, \citenamefont {Zhang}, \citenamefont {Liu},
  \citenamefont {Xia},\ and\ \citenamefont {Wang}}]{He2017}%
  \BibitemOpen
  \bibfield  {author} {\bibinfo {author} {\bibfnamefont {Q.~L.}\ \bibnamefont
  {He}}, \bibinfo {author} {\bibfnamefont {L.}~\bibnamefont {Pan}}, \bibinfo
  {author} {\bibfnamefont {A.~L.}\ \bibnamefont {Stern}}, \bibinfo {author}
  {\bibfnamefont {E.~C.}\ \bibnamefont {Burks}}, \bibinfo {author}
  {\bibfnamefont {X.}~\bibnamefont {Che}}, \bibinfo {author} {\bibfnamefont
  {G.}~\bibnamefont {Yin}}, \bibinfo {author} {\bibfnamefont {J.}~\bibnamefont
  {Wang}}, \bibinfo {author} {\bibfnamefont {B.}~\bibnamefont {Lian}}, \bibinfo
  {author} {\bibfnamefont {Q.}~\bibnamefont {Zhou}}, \bibinfo {author}
  {\bibfnamefont {E.~S.}\ \bibnamefont {Choi}}, \bibinfo {author}
  {\bibfnamefont {K.}~\bibnamefont {Murata}}, \bibinfo {author} {\bibfnamefont
  {X.}~\bibnamefont {Kou}}, \bibinfo {author} {\bibfnamefont {Z.}~\bibnamefont
  {Chen}}, \bibinfo {author} {\bibfnamefont {T.}~\bibnamefont {Nie}}, \bibinfo
  {author} {\bibfnamefont {Q.}~\bibnamefont {Shao}}, \bibinfo {author}
  {\bibfnamefont {Y.}~\bibnamefont {Fan}}, \bibinfo {author} {\bibfnamefont
  {S.-C.}\ \bibnamefont {Zhang}}, \bibinfo {author} {\bibfnamefont
  {K.}~\bibnamefont {Liu}}, \bibinfo {author} {\bibfnamefont {J.}~\bibnamefont
  {Xia}}, \ and\ \bibinfo {author} {\bibfnamefont {K.~L.}\ \bibnamefont
  {Wang}},\ }\href {\doibase 10.1126/science.aag2792} {\bibfield  {journal}
  {\bibinfo  {journal} {Science}\ }\textbf {\bibinfo {volume} {357}},\ \bibinfo
  {pages} {294} (\bibinfo {year} {2017})},\ \Eprint
  {http://arxiv.org/abs/https://science.sciencemag.org/content/357/6348/294.full.pdf}
  {https://science.sciencemag.org/content/357/6348/294.full.pdf} \BibitemShut
  {NoStop}%
\bibitem [{\citenamefont {Yanase}\ and\ \citenamefont
  {Shiozaki}(2017)}]{Yanase2017}%
  \BibitemOpen
  \bibfield  {author} {\bibinfo {author} {\bibfnamefont {Y.}~\bibnamefont
  {Yanase}}\ and\ \bibinfo {author} {\bibfnamefont {K.}~\bibnamefont
  {Shiozaki}},\ }\href {\doibase 10.1103/PhysRevB.95.224514} {\bibfield
  {journal} {\bibinfo  {journal} {Phys. Rev. B}\ }\textbf {\bibinfo {volume}
  {95}},\ \bibinfo {pages} {224514} (\bibinfo {year} {2017})}\BibitemShut
  {NoStop}%
\bibitem [{\citenamefont {Daido}\ \emph {et~al.}(2019)\citenamefont {Daido},
  \citenamefont {Yoshida},\ and\ \citenamefont {Yanase}}]{Daido2019}%
  \BibitemOpen
  \bibfield  {author} {\bibinfo {author} {\bibfnamefont {A.}~\bibnamefont
  {Daido}}, \bibinfo {author} {\bibfnamefont {T.}~\bibnamefont {Yoshida}}, \
  and\ \bibinfo {author} {\bibfnamefont {Y.}~\bibnamefont {Yanase}},\ }\href
  {\doibase 10.1103/PhysRevLett.122.227001} {\bibfield  {journal} {\bibinfo
  {journal} {Phys. Rev. Lett.}\ }\textbf {\bibinfo {volume} {122}},\ \bibinfo
  {pages} {227001} (\bibinfo {year} {2019})}\BibitemShut {NoStop}%
\bibitem [{\citenamefont {Ran}\ \emph {et~al.}(2019)\citenamefont {Ran},
  \citenamefont {Eckberg}, \citenamefont {Ding}, \citenamefont {Furukawa},
  \citenamefont {Metz}, \citenamefont {Saha}, \citenamefont {Liu},
  \citenamefont {Zic}, \citenamefont {Kim}, \citenamefont {Paglione},\ and\
  \citenamefont {Butch}}]{Ran_UTe2_2019}%
  \BibitemOpen
  \bibfield  {author} {\bibinfo {author} {\bibfnamefont {S.}~\bibnamefont
  {Ran}}, \bibinfo {author} {\bibfnamefont {C.}~\bibnamefont {Eckberg}},
  \bibinfo {author} {\bibfnamefont {Q.-P.}\ \bibnamefont {Ding}}, \bibinfo
  {author} {\bibfnamefont {Y.}~\bibnamefont {Furukawa}}, \bibinfo {author}
  {\bibfnamefont {T.}~\bibnamefont {Metz}}, \bibinfo {author} {\bibfnamefont
  {S.~R.}\ \bibnamefont {Saha}}, \bibinfo {author} {\bibfnamefont {I.-L.}\
  \bibnamefont {Liu}}, \bibinfo {author} {\bibfnamefont {M.}~\bibnamefont
  {Zic}}, \bibinfo {author} {\bibfnamefont {H.}~\bibnamefont {Kim}}, \bibinfo
  {author} {\bibfnamefont {J.}~\bibnamefont {Paglione}}, \ and\ \bibinfo
  {author} {\bibfnamefont {N.~P.}\ \bibnamefont {Butch}},\ }\href {\doibase
  10.1126/science.aav8645} {\bibfield  {journal} {\bibinfo  {journal}
  {Science}\ }\textbf {\bibinfo {volume} {365}},\ \bibinfo {pages} {684}
  (\bibinfo {year} {2019})}\BibitemShut {NoStop}%
\bibitem [{\citenamefont {Ishizuka}\ \emph {et~al.}(2019)\citenamefont
  {Ishizuka}, \citenamefont {Sumita}, \citenamefont {Daido},\ and\
  \citenamefont {Yanase}}]{Ishizuka_UTe22019}%
  \BibitemOpen
  \bibfield  {author} {\bibinfo {author} {\bibfnamefont {J.}~\bibnamefont
  {Ishizuka}}, \bibinfo {author} {\bibfnamefont {S.}~\bibnamefont {Sumita}},
  \bibinfo {author} {\bibfnamefont {A.}~\bibnamefont {Daido}}, \ and\ \bibinfo
  {author} {\bibfnamefont {Y.}~\bibnamefont {Yanase}},\ }\href {\doibase
  10.1103/PhysRevLett.123.217001} {\bibfield  {journal} {\bibinfo  {journal}
  {Phys. Rev. Lett.}\ }\textbf {\bibinfo {volume} {123}},\ \bibinfo {pages}
  {217001} (\bibinfo {year} {2019})}\BibitemShut {NoStop}%
\bibitem [{\citenamefont {Sasaki}\ \emph {et~al.}(2011)\citenamefont {Sasaki},
  \citenamefont {Kriener}, \citenamefont {Segawa}, \citenamefont {Yada},
  \citenamefont {Tanaka}, \citenamefont {Sato},\ and\ \citenamefont
  {Ando}}]{Sasaki2011}%
  \BibitemOpen
  \bibfield  {author} {\bibinfo {author} {\bibfnamefont {S.}~\bibnamefont
  {Sasaki}}, \bibinfo {author} {\bibfnamefont {M.}~\bibnamefont {Kriener}},
  \bibinfo {author} {\bibfnamefont {K.}~\bibnamefont {Segawa}}, \bibinfo
  {author} {\bibfnamefont {K.}~\bibnamefont {Yada}}, \bibinfo {author}
  {\bibfnamefont {Y.}~\bibnamefont {Tanaka}}, \bibinfo {author} {\bibfnamefont
  {M.}~\bibnamefont {Sato}}, \ and\ \bibinfo {author} {\bibfnamefont
  {Y.}~\bibnamefont {Ando}},\ }\href {\doibase 10.1103/PhysRevLett.107.217001}
  {\bibfield  {journal} {\bibinfo  {journal} {Phys. Rev. Lett.}\ }\textbf
  {\bibinfo {volume} {107}},\ \bibinfo {pages} {217001} (\bibinfo {year}
  {2011})}\BibitemShut {NoStop}%
\bibitem [{\citenamefont {Levy}\ \emph {et~al.}(2013)\citenamefont {Levy},
  \citenamefont {Zhang}, \citenamefont {Ha}, \citenamefont {Sharifi},
  \citenamefont {Talin}, \citenamefont {Kuk},\ and\ \citenamefont
  {Stroscio}}]{Levy2013}%
  \BibitemOpen
  \bibfield  {author} {\bibinfo {author} {\bibfnamefont {N.}~\bibnamefont
  {Levy}}, \bibinfo {author} {\bibfnamefont {T.}~\bibnamefont {Zhang}},
  \bibinfo {author} {\bibfnamefont {J.}~\bibnamefont {Ha}}, \bibinfo {author}
  {\bibfnamefont {F.}~\bibnamefont {Sharifi}}, \bibinfo {author} {\bibfnamefont
  {A.~A.}\ \bibnamefont {Talin}}, \bibinfo {author} {\bibfnamefont
  {Y.}~\bibnamefont {Kuk}}, \ and\ \bibinfo {author} {\bibfnamefont {J.~A.}\
  \bibnamefont {Stroscio}},\ }\href {\doibase 10.1103/PhysRevLett.110.117001}
  {\bibfield  {journal} {\bibinfo  {journal} {Phys. Rev. Lett.}\ }\textbf
  {\bibinfo {volume} {110}},\ \bibinfo {pages} {117001} (\bibinfo {year}
  {2013})}\BibitemShut {NoStop}%
\bibitem [{\citenamefont {Yanase}\ \emph {et~al.}(2003)\citenamefont {Yanase},
  \citenamefont {Jujo}, \citenamefont {Nomura}, \citenamefont {Ikeda},
  \citenamefont {Hotta},\ and\ \citenamefont {Yamada}}]{Yanase2003}%
  \BibitemOpen
  \bibfield  {author} {\bibinfo {author} {\bibfnamefont {Y.}~\bibnamefont
  {Yanase}}, \bibinfo {author} {\bibfnamefont {T.}~\bibnamefont {Jujo}},
  \bibinfo {author} {\bibfnamefont {T.}~\bibnamefont {Nomura}}, \bibinfo
  {author} {\bibfnamefont {H.}~\bibnamefont {Ikeda}}, \bibinfo {author}
  {\bibfnamefont {T.}~\bibnamefont {Hotta}}, \ and\ \bibinfo {author}
  {\bibfnamefont {K.}~\bibnamefont {Yamada}},\ }\href {\doibase
  https://doi.org/10.1016/j.physrep.2003.07.002} {\bibfield  {journal}
  {\bibinfo  {journal} {Physics Reports}\ }\textbf {\bibinfo {volume} {387}},\
  \bibinfo {pages} {1 } (\bibinfo {year} {2003})}\BibitemShut {NoStop}%
\bibitem [{\citenamefont {Nandkishore}\ \emph {et~al.}(2012)\citenamefont
  {Nandkishore}, \citenamefont {Levitov},\ and\ \citenamefont
  {Chubukov}}]{Nandkishore2012}%
  \BibitemOpen
  \bibfield  {author} {\bibinfo {author} {\bibfnamefont {R.}~\bibnamefont
  {Nandkishore}}, \bibinfo {author} {\bibfnamefont {L.~S.}\ \bibnamefont
  {Levitov}}, \ and\ \bibinfo {author} {\bibfnamefont {A.~V.}\ \bibnamefont
  {Chubukov}},\ }\href {\doibase 10.1038/nphys2208} {\bibfield  {journal}
  {\bibinfo  {journal} {Nature Physics}\ }\textbf {\bibinfo {volume} {8}},\
  \bibinfo {pages} {158} (\bibinfo {year} {2012})}\BibitemShut {NoStop}%
\bibitem [{\citenamefont {Fischer}\ \emph {et~al.}(2014)\citenamefont
  {Fischer}, \citenamefont {Neupert}, \citenamefont {Platt}, \citenamefont
  {Schnyder}, \citenamefont {Hanke}, \citenamefont {Goryo}, \citenamefont
  {Thomale},\ and\ \citenamefont {Sigrist}}]{Fischer2014}%
  \BibitemOpen
  \bibfield  {author} {\bibinfo {author} {\bibfnamefont {M.~H.}\ \bibnamefont
  {Fischer}}, \bibinfo {author} {\bibfnamefont {T.}~\bibnamefont {Neupert}},
  \bibinfo {author} {\bibfnamefont {C.}~\bibnamefont {Platt}}, \bibinfo
  {author} {\bibfnamefont {A.~P.}\ \bibnamefont {Schnyder}}, \bibinfo {author}
  {\bibfnamefont {W.}~\bibnamefont {Hanke}}, \bibinfo {author} {\bibfnamefont
  {J.}~\bibnamefont {Goryo}}, \bibinfo {author} {\bibfnamefont
  {R.}~\bibnamefont {Thomale}}, \ and\ \bibinfo {author} {\bibfnamefont
  {M.}~\bibnamefont {Sigrist}},\ }\href {\doibase 10.1103/PhysRevB.89.020509}
  {\bibfield  {journal} {\bibinfo  {journal} {Phys. Rev. B}\ }\textbf {\bibinfo
  {volume} {89}},\ \bibinfo {pages} {020509} (\bibinfo {year}
  {2014})}\BibitemShut {NoStop}%
\bibitem [{\citenamefont {Sigrist}\ and\ \citenamefont
  {Ueda}(1991)}]{SigristUeda_91}%
  \BibitemOpen
  \bibfield  {author} {\bibinfo {author} {\bibfnamefont {M.}~\bibnamefont
  {Sigrist}}\ and\ \bibinfo {author} {\bibfnamefont {K.}~\bibnamefont {Ueda}},\
  }\href {\doibase 10.1103/RevModPhys.63.239} {\bibfield  {journal} {\bibinfo
  {journal} {Rev. Mod. Phys.}\ }\textbf {\bibinfo {volume} {63}},\ \bibinfo
  {pages} {239} (\bibinfo {year} {1991})}\BibitemShut {NoStop}%
\bibitem [{\citenamefont {Frigeri}(2005)}]{Frigeri_Dthesis}%
  \BibitemOpen
  \bibfield  {author} {\bibinfo {author} {\bibfnamefont {P.~A.}\ \bibnamefont
  {Frigeri}},\ }\emph {\bibinfo {title} {Superconductivity in crystals without
  an inversion center}},\ \href@noop {} {Ph.D. thesis},\ \bibinfo  {school}
  {ETH Zurich} (\bibinfo {year} {2005})\BibitemShut {NoStop}%
\bibitem [{\citenamefont {Smidman}\ \emph {et~al.}(2017)\citenamefont
  {Smidman}, \citenamefont {Salamon}, \citenamefont {Yuan},\ and\ \citenamefont
  {Agterberg}}]{Smidman2017_review}%
  \BibitemOpen
  \bibfield  {author} {\bibinfo {author} {\bibfnamefont {M.}~\bibnamefont
  {Smidman}}, \bibinfo {author} {\bibfnamefont {M.}~\bibnamefont {Salamon}},
  \bibinfo {author} {\bibfnamefont {H.}~\bibnamefont {Yuan}}, \ and\ \bibinfo
  {author} {\bibfnamefont {D.}~\bibnamefont {Agterberg}},\ }\href@noop {}
  {\bibfield  {journal} {\bibinfo  {journal} {Reports on Progress in Physics}\
  }\textbf {\bibinfo {volume} {80}},\ \bibinfo {pages} {036501} (\bibinfo
  {year} {2017})}\BibitemShut {NoStop}%
\bibitem [{\citenamefont {Bauer}\ and\ \citenamefont
  {(eds.)}(2012)}]{NCSC_book}%
  \BibitemOpen
  \bibfield  {author} {\bibinfo {author} {\bibfnamefont {E.}~\bibnamefont
  {Bauer}}\ and\ \bibinfo {author} {\bibfnamefont {M.~S.}\ \bibnamefont
  {(eds.)}},\ }\href@noop {} {\emph {\bibinfo {title} {Non-Centrosymmetric
  Superconductors: Introduction and Overview}}}\ (\bibinfo  {publisher}
  {Springer, Berlin},\ \bibinfo {year} {2012})\BibitemShut {NoStop}%
\bibitem [{\citenamefont {Daido}\ and\ \citenamefont
  {Yanase}(2016)}]{Daido2016}%
  \BibitemOpen
  \bibfield  {author} {\bibinfo {author} {\bibfnamefont {A.}~\bibnamefont
  {Daido}}\ and\ \bibinfo {author} {\bibfnamefont {Y.}~\bibnamefont {Yanase}},\
  }\href {\doibase 10.1103/PhysRevB.94.054519} {\bibfield  {journal} {\bibinfo
  {journal} {Phys. Rev. B}\ }\textbf {\bibinfo {volume} {94}},\ \bibinfo
  {pages} {054519} (\bibinfo {year} {2016})}\BibitemShut {NoStop}%
\bibitem [{\citenamefont {Schnyder}\ and\ \citenamefont
  {Ryu}(2011)}]{Schnyder2011}%
  \BibitemOpen
  \bibfield  {author} {\bibinfo {author} {\bibfnamefont {A.~P.}\ \bibnamefont
  {Schnyder}}\ and\ \bibinfo {author} {\bibfnamefont {S.}~\bibnamefont {Ryu}},\
  }\href {\doibase 10.1103/PhysRevB.84.060504} {\bibfield  {journal} {\bibinfo
  {journal} {Phys. Rev. B}\ }\textbf {\bibinfo {volume} {84}},\ \bibinfo
  {pages} {060504} (\bibinfo {year} {2011})}\BibitemShut {NoStop}%
\bibitem [{\citenamefont {Sato}\ \emph {et~al.}(2011)\citenamefont {Sato},
  \citenamefont {Tanaka}, \citenamefont {Yada},\ and\ \citenamefont
  {Yokoyama}}]{Sato2011}%
  \BibitemOpen
  \bibfield  {author} {\bibinfo {author} {\bibfnamefont {M.}~\bibnamefont
  {Sato}}, \bibinfo {author} {\bibfnamefont {Y.}~\bibnamefont {Tanaka}},
  \bibinfo {author} {\bibfnamefont {K.}~\bibnamefont {Yada}}, \ and\ \bibinfo
  {author} {\bibfnamefont {T.}~\bibnamefont {Yokoyama}},\ }\href {\doibase
  10.1103/PhysRevB.83.224511} {\bibfield  {journal} {\bibinfo  {journal} {Phys.
  Rev. B}\ }\textbf {\bibinfo {volume} {83}},\ \bibinfo {pages} {224511}
  (\bibinfo {year} {2011})}\BibitemShut {NoStop}%
\bibitem [{\citenamefont {Yada}\ \emph {et~al.}(2011)\citenamefont {Yada},
  \citenamefont {Sato}, \citenamefont {Tanaka},\ and\ \citenamefont
  {Yokoyama}}]{Yada2011}%
  \BibitemOpen
  \bibfield  {author} {\bibinfo {author} {\bibfnamefont {K.}~\bibnamefont
  {Yada}}, \bibinfo {author} {\bibfnamefont {M.}~\bibnamefont {Sato}}, \bibinfo
  {author} {\bibfnamefont {Y.}~\bibnamefont {Tanaka}}, \ and\ \bibinfo {author}
  {\bibfnamefont {T.}~\bibnamefont {Yokoyama}},\ }\href {\doibase
  10.1103/PhysRevB.83.064505} {\bibfield  {journal} {\bibinfo  {journal} {Phys.
  Rev. B}\ }\textbf {\bibinfo {volume} {83}},\ \bibinfo {pages} {064505}
  (\bibinfo {year} {2011})}\BibitemShut {NoStop}%
\bibitem [{\citenamefont {Daido}\ and\ \citenamefont
  {Yanase}(2017)}]{Daido2017}%
  \BibitemOpen
  \bibfield  {author} {\bibinfo {author} {\bibfnamefont {A.}~\bibnamefont
  {Daido}}\ and\ \bibinfo {author} {\bibfnamefont {Y.}~\bibnamefont {Yanase}},\
  }\href {\doibase 10.1103/PhysRevB.95.134507} {\bibfield  {journal} {\bibinfo
  {journal} {Phys. Rev. B}\ }\textbf {\bibinfo {volume} {95}},\ \bibinfo
  {pages} {134507} (\bibinfo {year} {2017})}\BibitemShut {NoStop}%
\bibitem [{\citenamefont {Kallin}\ and\ \citenamefont
  {Berlinsky}(2016)}]{Kallin2016}%
  \BibitemOpen
  \bibfield  {author} {\bibinfo {author} {\bibfnamefont {C.}~\bibnamefont
  {Kallin}}\ and\ \bibinfo {author} {\bibfnamefont {J.}~\bibnamefont
  {Berlinsky}},\ }\href {\doibase 10.1088/0034-4885/79/5/054502} {\bibfield
  {journal} {\bibinfo  {journal} {Reports on Progress in Physics}\ }\textbf
  {\bibinfo {volume} {79}},\ \bibinfo {pages} {054502} (\bibinfo {year}
  {2016})}\BibitemShut {NoStop}%
\bibitem [{\citenamefont {Frigeri}\ \emph {et~al.}(2004)\citenamefont
  {Frigeri}, \citenamefont {Agterberg}, \citenamefont {Koga},\ and\
  \citenamefont {Sigrist}}]{Frigeri2004}%
  \BibitemOpen
  \bibfield  {author} {\bibinfo {author} {\bibfnamefont {P.~A.}\ \bibnamefont
  {Frigeri}}, \bibinfo {author} {\bibfnamefont {D.~F.}\ \bibnamefont
  {Agterberg}}, \bibinfo {author} {\bibfnamefont {A.}~\bibnamefont {Koga}}, \
  and\ \bibinfo {author} {\bibfnamefont {M.}~\bibnamefont {Sigrist}},\ }\href
  {\doibase 10.1103/PhysRevLett.92.097001} {\bibfield  {journal} {\bibinfo
  {journal} {Phys. Rev. Lett.}\ }\textbf {\bibinfo {volume} {92}},\ \bibinfo
  {pages} {097001} (\bibinfo {year} {2004})}\BibitemShut {NoStop}%
\bibitem [{\citenamefont {Yanase}\ and\ \citenamefont
  {Fujimoto}(2012)}]{Yanase2012}%
  \BibitemOpen
  \bibfield  {author} {\bibinfo {author} {\bibfnamefont {Y.}~\bibnamefont
  {Yanase}}\ and\ \bibinfo {author} {\bibfnamefont {S.}~\bibnamefont
  {Fujimoto}},\ }\enquote {\bibinfo {title} {Microscopic theory of pairing
  mechanisms},}\ in\ \href {\doibase 10.1007/978-3-642-24624-1_6} {\emph
  {\bibinfo {booktitle} {Non-Centrosymmetric Superconductors: Introduction and
  Overview}}},\ \bibinfo {editor} {edited by\ \bibinfo {editor} {\bibfnamefont
  {E.}~\bibnamefont {Bauer}}\ and\ \bibinfo {editor} {\bibfnamefont
  {M.}~\bibnamefont {Sigrist}}}\ (\bibinfo  {publisher} {Springer Berlin
  Heidelberg},\ \bibinfo {address} {Berlin, Heidelberg},\ \bibinfo {year}
  {2012})\ pp.\ \bibinfo {pages} {171--210}\BibitemShut {NoStop}%
\bibitem [{\citenamefont {Yanase}\ and\ \citenamefont
  {Sigrist}(2008)}]{Yanase2008}%
  \BibitemOpen
  \bibfield  {author} {\bibinfo {author} {\bibfnamefont {Y.}~\bibnamefont
  {Yanase}}\ and\ \bibinfo {author} {\bibfnamefont {M.}~\bibnamefont
  {Sigrist}},\ }\href {\doibase 10.1143/JPSJ.77.124711} {\bibfield  {journal}
  {\bibinfo  {journal} {Journal of the Physical Society of Japan}\ }\textbf
  {\bibinfo {volume} {77}},\ \bibinfo {pages} {124711} (\bibinfo {year}
  {2008})},\ \Eprint
  {http://arxiv.org/abs/https://doi.org/10.1143/JPSJ.77.124711}
  {https://doi.org/10.1143/JPSJ.77.124711} \BibitemShut {NoStop}%
\bibitem [{\citenamefont {Yanase}(2013)}]{Yanase2013}%
  \BibitemOpen
  \bibfield  {author} {\bibinfo {author} {\bibfnamefont {Y.}~\bibnamefont
  {Yanase}},\ }\href {\doibase 10.7566/JPSJ.82.044711} {\bibfield  {journal}
  {\bibinfo  {journal} {Journal of the Physical Society of Japan}\ }\textbf
  {\bibinfo {volume} {82}},\ \bibinfo {pages} {044711} (\bibinfo {year}
  {2013})},\ \Eprint
  {http://arxiv.org/abs/https://doi.org/10.7566/JPSJ.82.044711}
  {https://doi.org/10.7566/JPSJ.82.044711} \BibitemShut {NoStop}%
\bibitem [{\citenamefont {Harima}\ \emph {et~al.}(2015)\citenamefont {Harima},
  \citenamefont {Goho},\ and\ \citenamefont {Tomi}}]{Harima2015}%
  \BibitemOpen
  \bibfield  {author} {\bibinfo {author} {\bibfnamefont {H.}~\bibnamefont
  {Harima}}, \bibinfo {author} {\bibfnamefont {T.}~\bibnamefont {Goho}}, \ and\
  \bibinfo {author} {\bibfnamefont {T.}~\bibnamefont {Tomi}},\ }\href {\doibase
  10.1088/1742-6596/592/1/012040} {\bibfield  {journal} {\bibinfo  {journal}
  {Journal of Physics: Conference Series}\ }\textbf {\bibinfo {volume} {592}},\
  \bibinfo {pages} {012040} (\bibinfo {year} {2015})}\BibitemShut {NoStop}%
\bibitem [{\citenamefont {Shigeta}\ \emph {et~al.}(2013)\citenamefont
  {Shigeta}, \citenamefont {Onari},\ and\ \citenamefont
  {Tanaka}}]{Shigeta2013}%
  \BibitemOpen
  \bibfield  {author} {\bibinfo {author} {\bibfnamefont {K.}~\bibnamefont
  {Shigeta}}, \bibinfo {author} {\bibfnamefont {S.}~\bibnamefont {Onari}}, \
  and\ \bibinfo {author} {\bibfnamefont {Y.}~\bibnamefont {Tanaka}},\ }\href
  {\doibase 10.7566/JPSJ.82.014702} {\bibfield  {journal} {\bibinfo  {journal}
  {Journal of the Physical Society of Japan}\ }\textbf {\bibinfo {volume}
  {82}},\ \bibinfo {pages} {014702} (\bibinfo {year} {2013})},\ \Eprint
  {http://arxiv.org/abs/https://doi.org/10.7566/JPSJ.82.014702}
  {https://doi.org/10.7566/JPSJ.82.014702} \BibitemShut {NoStop}%
\bibitem [{\citenamefont {Lu}\ and\ \citenamefont
  {S\'en\'echal}(2018)}]{Lu2018}%
  \BibitemOpen
  \bibfield  {author} {\bibinfo {author} {\bibfnamefont {X.}~\bibnamefont
  {Lu}}\ and\ \bibinfo {author} {\bibfnamefont {D.}~\bibnamefont
  {S\'en\'echal}},\ }\href {\doibase 10.1103/PhysRevB.98.245118} {\bibfield
  {journal} {\bibinfo  {journal} {Phys. Rev. B}\ }\textbf {\bibinfo {volume}
  {98}},\ \bibinfo {pages} {245118} (\bibinfo {year} {2018})}\BibitemShut
  {NoStop}%
\bibitem [{\citenamefont {Nogaki}\ and\ \citenamefont
  {Yanase}(2020)}]{Nogaki2020}%
  \BibitemOpen
  \bibfield  {author} {\bibinfo {author} {\bibfnamefont {K.}~\bibnamefont
  {Nogaki}}\ and\ \bibinfo {author} {\bibfnamefont {Y.}~\bibnamefont
  {Yanase}},\ }\href {\doibase 10.1103/PhysRevB.102.165114} {\bibfield
  {journal} {\bibinfo  {journal} {Phys. Rev. B}\ }\textbf {\bibinfo {volume}
  {102}},\ \bibinfo {pages} {165114} (\bibinfo {year} {2020})}\BibitemShut
  {NoStop}%
\bibitem [{\citenamefont {Kohmoto}(1985)}]{Kohmoto1985}%
  \BibitemOpen
  \bibfield  {author} {\bibinfo {author} {\bibfnamefont {M.}~\bibnamefont
  {Kohmoto}},\ }\href@noop {} {\bibfield  {journal} {\bibinfo  {journal}
  {Annals of Physics}\ }\textbf {\bibinfo {volume} {160}},\ \bibinfo {pages}
  {343} (\bibinfo {year} {1985})}\BibitemShut {NoStop}%
\bibitem [{\citenamefont {Haldane}(1988)}]{Haldane1988}%
  \BibitemOpen
  \bibfield  {author} {\bibinfo {author} {\bibfnamefont {F.~D.~M.}\
  \bibnamefont {Haldane}},\ }\href {\doibase 10.1103/PhysRevLett.61.2015}
  {\bibfield  {journal} {\bibinfo  {journal} {Phys. Rev. Lett.}\ }\textbf
  {\bibinfo {volume} {61}},\ \bibinfo {pages} {2015} (\bibinfo {year}
  {1988})}\BibitemShut {NoStop}%
\bibitem [{\citenamefont {Yoshida}\ and\ \citenamefont
  {Yanase}(2016)}]{Yoshida2016}%
  \BibitemOpen
  \bibfield  {author} {\bibinfo {author} {\bibfnamefont {T.}~\bibnamefont
  {Yoshida}}\ and\ \bibinfo {author} {\bibfnamefont {Y.}~\bibnamefont
  {Yanase}},\ }\href {\doibase 10.1103/PhysRevB.93.054504} {\bibfield
  {journal} {\bibinfo  {journal} {Phys. Rev. B}\ }\textbf {\bibinfo {volume}
  {93}},\ \bibinfo {pages} {054504} (\bibinfo {year} {2016})}\BibitemShut
  {NoStop}%
\bibitem [{\citenamefont {Fukui}\ \emph {et~al.}(2005)\citenamefont {Fukui},
  \citenamefont {Hatsugai},\ and\ \citenamefont {Suzuki}}]{Fukui_Hatsugai_05}%
  \BibitemOpen
  \bibfield  {author} {\bibinfo {author} {\bibfnamefont {T.}~\bibnamefont
  {Fukui}}, \bibinfo {author} {\bibfnamefont {Y.}~\bibnamefont {Hatsugai}}, \
  and\ \bibinfo {author} {\bibfnamefont {H.}~\bibnamefont {Suzuki}},\ }\href
  {\doibase 10.1143/JPSJ.74.1674} {\bibfield  {journal} {\bibinfo  {journal}
  {Journal of the Physical Society of Japan}\ }\textbf {\bibinfo {volume}
  {74}},\ \bibinfo {pages} {1674} (\bibinfo {year} {2005})},\ \Eprint
  {http://arxiv.org/abs/http://dx.doi.org/10.1143/JPSJ.74.1674}
  {http://dx.doi.org/10.1143/JPSJ.74.1674} \BibitemShut {NoStop}%
\bibitem [{\citenamefont {Lu}\ and\ \citenamefont {Liu}(2020)}]{Xiancong2020}%
  \BibitemOpen
  \bibfield  {author} {\bibinfo {author} {\bibfnamefont {X.}~\bibnamefont
  {Lu}}\ and\ \bibinfo {author} {\bibfnamefont {H.}~\bibnamefont {Liu}},\
  }\href {\doibase 10.1088/1361-648x/aba980} {\bibfield  {journal} {\bibinfo
  {journal} {J. Phys.: Condens. Matter}\ }\textbf {\bibinfo {volume} {32}},\
  \bibinfo {pages} {455601} (\bibinfo {year} {2020})}\BibitemShut {NoStop}%
\bibitem [{\citenamefont {Tanaka}\ \emph {et~al.}(2012)\citenamefont {Tanaka},
  \citenamefont {Sato},\ and\ \citenamefont {Nagaosa}}]{Tanaka2012}%
  \BibitemOpen
  \bibfield  {author} {\bibinfo {author} {\bibfnamefont {Y.}~\bibnamefont
  {Tanaka}}, \bibinfo {author} {\bibfnamefont {M.}~\bibnamefont {Sato}}, \ and\
  \bibinfo {author} {\bibfnamefont {N.}~\bibnamefont {Nagaosa}},\ }\href
  {\doibase 10.1143/JPSJ.81.011013} {\bibfield  {journal} {\bibinfo  {journal}
  {Journal of the Physical Society of Japan}\ }\textbf {\bibinfo {volume}
  {81}},\ \bibinfo {pages} {011013} (\bibinfo {year} {2012})},\ \Eprint
  {http://arxiv.org/abs/https://doi.org/10.1143/JPSJ.81.011013}
  {https://doi.org/10.1143/JPSJ.81.011013} \BibitemShut {NoStop}%
\bibitem [{\citenamefont {Tada}\ \emph {et~al.}(2008)\citenamefont {Tada},
  \citenamefont {Kawakami},\ and\ \citenamefont {Fujimoto}}]{Tada2008}%
  \BibitemOpen
  \bibfield  {author} {\bibinfo {author} {\bibfnamefont {Y.}~\bibnamefont
  {Tada}}, \bibinfo {author} {\bibfnamefont {N.}~\bibnamefont {Kawakami}}, \
  and\ \bibinfo {author} {\bibfnamefont {S.}~\bibnamefont {Fujimoto}},\ }\href
  {\doibase 10.1143/JPSJ.77.054707} {\bibfield  {journal} {\bibinfo  {journal}
  {Journal of the Physical Society of Japan}\ }\textbf {\bibinfo {volume}
  {77}},\ \bibinfo {pages} {054707} (\bibinfo {year} {2008})},\ \Eprint
  {http://arxiv.org/abs/https://doi.org/10.1143/JPSJ.77.054707}
  {https://doi.org/10.1143/JPSJ.77.054707} \BibitemShut {NoStop}%
\bibitem [{\citenamefont {Kimura}\ \emph {et~al.}(2005)\citenamefont {Kimura},
  \citenamefont {Ito}, \citenamefont {Saitoh}, \citenamefont {Umeda},
  \citenamefont {Aoki},\ and\ \citenamefont {Terashima}}]{Kimura2005}%
  \BibitemOpen
  \bibfield  {author} {\bibinfo {author} {\bibfnamefont {N.}~\bibnamefont
  {Kimura}}, \bibinfo {author} {\bibfnamefont {K.}~\bibnamefont {Ito}},
  \bibinfo {author} {\bibfnamefont {K.}~\bibnamefont {Saitoh}}, \bibinfo
  {author} {\bibfnamefont {Y.}~\bibnamefont {Umeda}}, \bibinfo {author}
  {\bibfnamefont {H.}~\bibnamefont {Aoki}}, \ and\ \bibinfo {author}
  {\bibfnamefont {T.}~\bibnamefont {Terashima}},\ }\href {\doibase
  10.1103/PhysRevLett.95.247004} {\bibfield  {journal} {\bibinfo  {journal}
  {Phys. Rev. Lett.}\ }\textbf {\bibinfo {volume} {95}},\ \bibinfo {pages}
  {247004} (\bibinfo {year} {2005})}\BibitemShut {NoStop}%
\bibitem [{\citenamefont {Sugitani}\ \emph {et~al.}(2006)\citenamefont
  {Sugitani}, \citenamefont {Okuda}, \citenamefont {Shishido}, \citenamefont
  {Yamada}, \citenamefont {Thamizhavel}, \citenamefont {Yamamoto},
  \citenamefont {D.~Matsuda}, \citenamefont {Haga}, \citenamefont {Takeuchi},
  \citenamefont {Settai},\ and\ \citenamefont {Ōnuki}}]{Sugitani2006}%
  \BibitemOpen
  \bibfield  {author} {\bibinfo {author} {\bibfnamefont {I.}~\bibnamefont
  {Sugitani}}, \bibinfo {author} {\bibfnamefont {Y.}~\bibnamefont {Okuda}},
  \bibinfo {author} {\bibfnamefont {H.}~\bibnamefont {Shishido}}, \bibinfo
  {author} {\bibfnamefont {T.}~\bibnamefont {Yamada}}, \bibinfo {author}
  {\bibfnamefont {A.}~\bibnamefont {Thamizhavel}}, \bibinfo {author}
  {\bibfnamefont {E.}~\bibnamefont {Yamamoto}}, \bibinfo {author}
  {\bibfnamefont {T.}~\bibnamefont {D.~Matsuda}}, \bibinfo {author}
  {\bibfnamefont {Y.}~\bibnamefont {Haga}}, \bibinfo {author} {\bibfnamefont
  {T.}~\bibnamefont {Takeuchi}}, \bibinfo {author} {\bibfnamefont
  {R.}~\bibnamefont {Settai}}, \ and\ \bibinfo {author} {\bibfnamefont
  {Y.}~\bibnamefont {Ōnuki}},\ }\href {\doibase 10.1143/JPSJ.75.043703}
  {\bibfield  {journal} {\bibinfo  {journal} {Journal of the Physical Society
  of Japan}\ }\textbf {\bibinfo {volume} {75}},\ \bibinfo {pages} {043703}
  (\bibinfo {year} {2006})},\ \Eprint
  {http://arxiv.org/abs/https://doi.org/10.1143/JPSJ.75.043703}
  {https://doi.org/10.1143/JPSJ.75.043703} \BibitemShut {NoStop}%
\bibitem [{\citenamefont {Yanase}\ and\ \citenamefont
  {Sigrist}(2007)}]{Yanase2007}%
  \BibitemOpen
  \bibfield  {author} {\bibinfo {author} {\bibfnamefont {Y.}~\bibnamefont
  {Yanase}}\ and\ \bibinfo {author} {\bibfnamefont {M.}~\bibnamefont
  {Sigrist}},\ }\href {\doibase 10.1143/JPSJ.76.043712} {\bibfield  {journal}
  {\bibinfo  {journal} {Journal of the Physical Society of Japan}\ }\textbf
  {\bibinfo {volume} {76}},\ \bibinfo {pages} {043712} (\bibinfo {year}
  {2007})},\ \Eprint
  {http://arxiv.org/abs/https://doi.org/10.1143/JPSJ.76.043712}
  {https://doi.org/10.1143/JPSJ.76.043712} \BibitemShut {NoStop}%
\bibitem [{\citenamefont {Bauer}\ \emph {et~al.}(2004)\citenamefont {Bauer},
  \citenamefont {Hilscher}, \citenamefont {Michor}, \citenamefont {Paul},
  \citenamefont {Scheidt}, \citenamefont {Gribanov}, \citenamefont {Seropegin},
  \citenamefont {No\"el}, \citenamefont {Sigrist},\ and\ \citenamefont
  {Rogl}}]{Bauer2004}%
  \BibitemOpen
  \bibfield  {author} {\bibinfo {author} {\bibfnamefont {E.}~\bibnamefont
  {Bauer}}, \bibinfo {author} {\bibfnamefont {G.}~\bibnamefont {Hilscher}},
  \bibinfo {author} {\bibfnamefont {H.}~\bibnamefont {Michor}}, \bibinfo
  {author} {\bibfnamefont {C.}~\bibnamefont {Paul}}, \bibinfo {author}
  {\bibfnamefont {E.~W.}\ \bibnamefont {Scheidt}}, \bibinfo {author}
  {\bibfnamefont {A.}~\bibnamefont {Gribanov}}, \bibinfo {author}
  {\bibfnamefont {Y.}~\bibnamefont {Seropegin}}, \bibinfo {author}
  {\bibfnamefont {H.}~\bibnamefont {No\"el}}, \bibinfo {author} {\bibfnamefont
  {M.}~\bibnamefont {Sigrist}}, \ and\ \bibinfo {author} {\bibfnamefont
  {P.}~\bibnamefont {Rogl}},\ }\href {\doibase 10.1103/PhysRevLett.92.027003}
  {\bibfield  {journal} {\bibinfo  {journal} {Phys. Rev. Lett.}\ }\textbf
  {\bibinfo {volume} {92}},\ \bibinfo {pages} {027003} (\bibinfo {year}
  {2004})}\BibitemShut {NoStop}%
\bibitem [{\citenamefont {Schnyder}\ and\ \citenamefont
  {Brydon}(2015)}]{Schnyder2015_review}%
  \BibitemOpen
  \bibfield  {author} {\bibinfo {author} {\bibfnamefont {A.~P.}\ \bibnamefont
  {Schnyder}}\ and\ \bibinfo {author} {\bibfnamefont {P.~M.~R.}\ \bibnamefont
  {Brydon}},\ }\href {\doibase 10.1088/0953-8984/27/24/243201} {\bibfield
  {journal} {\bibinfo  {journal} {Journal of Physics: Condensed Matter}\
  }\textbf {\bibinfo {volume} {27}},\ \bibinfo {pages} {243201} (\bibinfo
  {year} {2015})}\BibitemShut {NoStop}%
\bibitem [{\citenamefont {Morimoto}\ \emph {et~al.}(2015)\citenamefont
  {Morimoto}, \citenamefont {Furusaki},\ and\ \citenamefont
  {Mudry}}]{Morimoto_2015}%
  \BibitemOpen
  \bibfield  {author} {\bibinfo {author} {\bibfnamefont {T.}~\bibnamefont
  {Morimoto}}, \bibinfo {author} {\bibfnamefont {A.}~\bibnamefont {Furusaki}},
  \ and\ \bibinfo {author} {\bibfnamefont {C.}~\bibnamefont {Mudry}},\ }\href
  {\doibase 10.1103/PhysRevB.92.125104} {\bibfield  {journal} {\bibinfo
  {journal} {Phys. Rev. B}\ }\textbf {\bibinfo {volume} {92}},\ \bibinfo
  {pages} {125104} (\bibinfo {year} {2015})}\BibitemShut {NoStop}%
\bibitem [{\citenamefont {Wong}\ \emph {et~al.}(2013)\citenamefont {Wong},
  \citenamefont {Liu}, \citenamefont {Law},\ and\ \citenamefont
  {Lee}}]{Wong2013}%
  \BibitemOpen
  \bibfield  {author} {\bibinfo {author} {\bibfnamefont {C.~L.~M.}\
  \bibnamefont {Wong}}, \bibinfo {author} {\bibfnamefont {J.}~\bibnamefont
  {Liu}}, \bibinfo {author} {\bibfnamefont {K.~T.}\ \bibnamefont {Law}}, \ and\
  \bibinfo {author} {\bibfnamefont {P.~A.}\ \bibnamefont {Lee}},\ }\href
  {\doibase 10.1103/PhysRevB.88.060504} {\bibfield  {journal} {\bibinfo
  {journal} {Phys. Rev. B}\ }\textbf {\bibinfo {volume} {88}},\ \bibinfo
  {pages} {060504} (\bibinfo {year} {2013})}\BibitemShut {NoStop}%
\bibitem [{Note1()}]{Note1}%
  \BibitemOpen
  \bibinfo {note} {Here, the quasiparticle current refers to $\partial
  _{\protect \bm {k}}\protect \mathcal {H}_{\protect \mathrm {BdG}}(\protect
  \bm {k})$ rather than the electric current. The total electric current should
  vanish, but needs a remark. It is known that the thermodynamically stable
  state of noncentrosymmetric superconductors with a small parallel Zeeman
  field has a tiny center-of-mass momentum of the Cooper pair (called the
  helical superconductivity)~\cite
  {Barzykin_Gorkov2002,Dimitrova_Feigelman2003,Agterberg2007,Agterberg2012,Smidman2017_review}.
  The vanishing total electric current is ensured for this state, rather than
  the present model where a spatially-uniform order parameter is
  phenomenologically introduced. It is expected that all the qualitative
  results do not change even when a tiny center-of-mass momentum is taken into
  account, and unidirectional Majorana edge states also appear in the helical
  superconducting state. Summing up the contributions from the bulk and edge
  states, the system with unidirectional Majorana edge states has the vanishing
  total electric current.}\BibitemShut {Stop}%
\bibitem [{\citenamefont {Takasan}\ \emph {et~al.}(2021)\citenamefont
  {Takasan}, \citenamefont {Sumita},\ and\ \citenamefont
  {Yanase}}]{Takasan2021}%
  \BibitemOpen
  \bibfield  {author} {\bibinfo {author} {\bibfnamefont {K.}~\bibnamefont
  {Takasan}}, \bibinfo {author} {\bibfnamefont {S.}~\bibnamefont {Sumita}}, \
  and\ \bibinfo {author} {\bibfnamefont {Y.}~\bibnamefont {Yanase}},\
  }\href@noop {} {} (\bibinfo {year} {2021}),\ \Eprint
  {http://arxiv.org/abs/2110.06959} {arXiv:2110.06959} \BibitemShut {NoStop}%
\bibitem [{\citenamefont {Volovik}(2003)}]{Volovik2003}%
  \BibitemOpen
  \bibfield  {author} {\bibinfo {author} {\bibfnamefont {G.~E.}\ \bibnamefont
  {Volovik}},\ }\href@noop {} {\emph {\bibinfo {title} {The universe in a
  helium droplet}}}\ (\bibinfo  {publisher} {Oxford University Press on
  Demand},\ \bibinfo {year} {2003})\BibitemShut {NoStop}%
\bibitem [{\citenamefont {Meng}\ and\ \citenamefont
  {Balents}(2012)}]{Meng2012}%
  \BibitemOpen
  \bibfield  {author} {\bibinfo {author} {\bibfnamefont {T.}~\bibnamefont
  {Meng}}\ and\ \bibinfo {author} {\bibfnamefont {L.}~\bibnamefont {Balents}},\
  }\href {\doibase 10.1103/PhysRevB.86.054504} {\bibfield  {journal} {\bibinfo
  {journal} {Phys. Rev. B}\ }\textbf {\bibinfo {volume} {86}},\ \bibinfo
  {pages} {054504} (\bibinfo {year} {2012})}\BibitemShut {NoStop}%
\bibitem [{\citenamefont {Kasahara}\ \emph {et~al.}(2007)\citenamefont
  {Kasahara}, \citenamefont {Iwasawa}, \citenamefont {Shishido}, \citenamefont
  {Shibauchi}, \citenamefont {Behnia}, \citenamefont {Haga}, \citenamefont
  {Matsuda}, \citenamefont {Onuki}, \citenamefont {Sigrist},\ and\
  \citenamefont {Matsuda}}]{Kasahara2007}%
  \BibitemOpen
  \bibfield  {author} {\bibinfo {author} {\bibfnamefont {Y.}~\bibnamefont
  {Kasahara}}, \bibinfo {author} {\bibfnamefont {T.}~\bibnamefont {Iwasawa}},
  \bibinfo {author} {\bibfnamefont {H.}~\bibnamefont {Shishido}}, \bibinfo
  {author} {\bibfnamefont {T.}~\bibnamefont {Shibauchi}}, \bibinfo {author}
  {\bibfnamefont {K.}~\bibnamefont {Behnia}}, \bibinfo {author} {\bibfnamefont
  {Y.}~\bibnamefont {Haga}}, \bibinfo {author} {\bibfnamefont {T.~D.}\
  \bibnamefont {Matsuda}}, \bibinfo {author} {\bibfnamefont {Y.}~\bibnamefont
  {Onuki}}, \bibinfo {author} {\bibfnamefont {M.}~\bibnamefont {Sigrist}}, \
  and\ \bibinfo {author} {\bibfnamefont {Y.}~\bibnamefont {Matsuda}},\ }\href
  {\doibase 10.1103/PhysRevLett.99.116402} {\bibfield  {journal} {\bibinfo
  {journal} {Phys. Rev. Lett.}\ }\textbf {\bibinfo {volume} {99}},\ \bibinfo
  {pages} {116402} (\bibinfo {year} {2007})}\BibitemShut {NoStop}%
\bibitem [{\citenamefont {Kittaka}\ \emph {et~al.}(2016)\citenamefont
  {Kittaka}, \citenamefont {Shimizu}, \citenamefont {Sakakibara}, \citenamefont
  {Haga}, \citenamefont {Yamamoto}, \citenamefont {Ōnuki}, \citenamefont
  {Tsutsumi}, \citenamefont {Nomoto}, \citenamefont {Ikeda},\ and\
  \citenamefont {Machida}}]{Kittaka2016}%
  \BibitemOpen
  \bibfield  {author} {\bibinfo {author} {\bibfnamefont {S.}~\bibnamefont
  {Kittaka}}, \bibinfo {author} {\bibfnamefont {Y.}~\bibnamefont {Shimizu}},
  \bibinfo {author} {\bibfnamefont {T.}~\bibnamefont {Sakakibara}}, \bibinfo
  {author} {\bibfnamefont {Y.}~\bibnamefont {Haga}}, \bibinfo {author}
  {\bibfnamefont {E.}~\bibnamefont {Yamamoto}}, \bibinfo {author}
  {\bibfnamefont {Y.}~\bibnamefont {Ōnuki}}, \bibinfo {author} {\bibfnamefont
  {Y.}~\bibnamefont {Tsutsumi}}, \bibinfo {author} {\bibfnamefont
  {T.}~\bibnamefont {Nomoto}}, \bibinfo {author} {\bibfnamefont
  {H.}~\bibnamefont {Ikeda}}, \ and\ \bibinfo {author} {\bibfnamefont
  {K.}~\bibnamefont {Machida}},\ }\href {\doibase 10.7566/JPSJ.85.033704}
  {\bibfield  {journal} {\bibinfo  {journal} {Journal of the Physical Society
  of Japan}\ }\textbf {\bibinfo {volume} {85}},\ \bibinfo {pages} {033704}
  (\bibinfo {year} {2016})},\ \Eprint
  {http://arxiv.org/abs/https://doi.org/10.7566/JPSJ.85.033704}
  {https://doi.org/10.7566/JPSJ.85.033704} \BibitemShut {NoStop}%
\bibitem [{\citenamefont {Biswas}\ \emph {et~al.}(2013)\citenamefont {Biswas},
  \citenamefont {Luetkens}, \citenamefont {Neupert}, \citenamefont {St\"urzer},
  \citenamefont {Baines}, \citenamefont {Pascua}, \citenamefont {Schnyder},
  \citenamefont {Fischer}, \citenamefont {Goryo}, \citenamefont {Lees},
  \citenamefont {Maeter}, \citenamefont {Br\"uckner}, \citenamefont {Klauss},
  \citenamefont {Nicklas}, \citenamefont {Baker}, \citenamefont {Hillier},
  \citenamefont {Sigrist}, \citenamefont {Amato},\ and\ \citenamefont
  {Johrendt}}]{Biswas2013}%
  \BibitemOpen
  \bibfield  {author} {\bibinfo {author} {\bibfnamefont {P.~K.}\ \bibnamefont
  {Biswas}}, \bibinfo {author} {\bibfnamefont {H.}~\bibnamefont {Luetkens}},
  \bibinfo {author} {\bibfnamefont {T.}~\bibnamefont {Neupert}}, \bibinfo
  {author} {\bibfnamefont {T.}~\bibnamefont {St\"urzer}}, \bibinfo {author}
  {\bibfnamefont {C.}~\bibnamefont {Baines}}, \bibinfo {author} {\bibfnamefont
  {G.}~\bibnamefont {Pascua}}, \bibinfo {author} {\bibfnamefont {A.~P.}\
  \bibnamefont {Schnyder}}, \bibinfo {author} {\bibfnamefont {M.~H.}\
  \bibnamefont {Fischer}}, \bibinfo {author} {\bibfnamefont {J.}~\bibnamefont
  {Goryo}}, \bibinfo {author} {\bibfnamefont {M.~R.}\ \bibnamefont {Lees}},
  \bibinfo {author} {\bibfnamefont {H.}~\bibnamefont {Maeter}}, \bibinfo
  {author} {\bibfnamefont {F.}~\bibnamefont {Br\"uckner}}, \bibinfo {author}
  {\bibfnamefont {H.-H.}\ \bibnamefont {Klauss}}, \bibinfo {author}
  {\bibfnamefont {M.}~\bibnamefont {Nicklas}}, \bibinfo {author} {\bibfnamefont
  {P.~J.}\ \bibnamefont {Baker}}, \bibinfo {author} {\bibfnamefont {A.~D.}\
  \bibnamefont {Hillier}}, \bibinfo {author} {\bibfnamefont {M.}~\bibnamefont
  {Sigrist}}, \bibinfo {author} {\bibfnamefont {A.}~\bibnamefont {Amato}}, \
  and\ \bibinfo {author} {\bibfnamefont {D.}~\bibnamefont {Johrendt}},\ }\href
  {\doibase 10.1103/PhysRevB.87.180503} {\bibfield  {journal} {\bibinfo
  {journal} {Phys. Rev. B}\ }\textbf {\bibinfo {volume} {87}},\ \bibinfo
  {pages} {180503} (\bibinfo {year} {2013})}\BibitemShut {NoStop}%
\bibitem [{\citenamefont {Joynt}\ and\ \citenamefont
  {Taillefer}(2002)}]{Joynt2002}%
  \BibitemOpen
  \bibfield  {author} {\bibinfo {author} {\bibfnamefont {R.}~\bibnamefont
  {Joynt}}\ and\ \bibinfo {author} {\bibfnamefont {L.}~\bibnamefont
  {Taillefer}},\ }\href {\doibase 10.1103/RevModPhys.74.235} {\bibfield
  {journal} {\bibinfo  {journal} {Rev. Mod. Phys.}\ }\textbf {\bibinfo {volume}
  {74}},\ \bibinfo {pages} {235} (\bibinfo {year} {2002})}\BibitemShut
  {NoStop}%
\bibitem [{\citenamefont {Yanase}(2016)}]{Yanase2016}%
  \BibitemOpen
  \bibfield  {author} {\bibinfo {author} {\bibfnamefont {Y.}~\bibnamefont
  {Yanase}},\ }\href {\doibase 10.1103/PhysRevB.94.174502} {\bibfield
  {journal} {\bibinfo  {journal} {Phys. Rev. B}\ }\textbf {\bibinfo {volume}
  {94}},\ \bibinfo {pages} {174502} (\bibinfo {year} {2016})}\BibitemShut
  {NoStop}%
\bibitem [{\citenamefont {Aoki}\ \emph {et~al.}(2019)\citenamefont {Aoki},
  \citenamefont {Ishida},\ and\ \citenamefont {Flouquet}}]{Aoki2019}%
  \BibitemOpen
  \bibfield  {author} {\bibinfo {author} {\bibfnamefont {D.}~\bibnamefont
  {Aoki}}, \bibinfo {author} {\bibfnamefont {K.}~\bibnamefont {Ishida}}, \ and\
  \bibinfo {author} {\bibfnamefont {J.}~\bibnamefont {Flouquet}},\ }\href
  {\doibase 10.7566/JPSJ.88.022001} {\bibfield  {journal} {\bibinfo  {journal}
  {Journal of the Physical Society of Japan}\ }\textbf {\bibinfo {volume}
  {88}},\ \bibinfo {pages} {022001} (\bibinfo {year} {2019})},\ \Eprint
  {http://arxiv.org/abs/https://doi.org/10.7566/JPSJ.88.022001}
  {https://doi.org/10.7566/JPSJ.88.022001} \BibitemShut {NoStop}%
\bibitem [{\citenamefont {Sau}\ and\ \citenamefont {Tewari}(2012)}]{Sau2012}%
  \BibitemOpen
  \bibfield  {author} {\bibinfo {author} {\bibfnamefont {J.~D.}\ \bibnamefont
  {Sau}}\ and\ \bibinfo {author} {\bibfnamefont {S.}~\bibnamefont {Tewari}},\
  }\href {\doibase 10.1103/PhysRevB.86.104509} {\bibfield  {journal} {\bibinfo
  {journal} {Phys. Rev. B}\ }\textbf {\bibinfo {volume} {86}},\ \bibinfo
  {pages} {104509} (\bibinfo {year} {2012})}\BibitemShut {NoStop}%
\bibitem [{\citenamefont {Armitage}\ \emph {et~al.}(2018)\citenamefont
  {Armitage}, \citenamefont {Mele},\ and\ \citenamefont
  {Vishwanath}}]{Armitage2018}%
  \BibitemOpen
  \bibfield  {author} {\bibinfo {author} {\bibfnamefont {N.~P.}\ \bibnamefont
  {Armitage}}, \bibinfo {author} {\bibfnamefont {E.~J.}\ \bibnamefont {Mele}},
  \ and\ \bibinfo {author} {\bibfnamefont {A.}~\bibnamefont {Vishwanath}},\
  }\href {\doibase 10.1103/RevModPhys.90.015001} {\bibfield  {journal}
  {\bibinfo  {journal} {Rev. Mod. Phys.}\ }\textbf {\bibinfo {volume} {90}},\
  \bibinfo {pages} {015001} (\bibinfo {year} {2018})}\BibitemShut {NoStop}%
\bibitem [{\citenamefont {Massarelli}\ \emph {et~al.}(2017)\citenamefont
  {Massarelli}, \citenamefont {Wachtel}, \citenamefont {Wei},\ and\
  \citenamefont {Paramekanti}}]{Massarelli2017}%
  \BibitemOpen
  \bibfield  {author} {\bibinfo {author} {\bibfnamefont {G.}~\bibnamefont
  {Massarelli}}, \bibinfo {author} {\bibfnamefont {G.}~\bibnamefont {Wachtel}},
  \bibinfo {author} {\bibfnamefont {J.~Y.~T.}\ \bibnamefont {Wei}}, \ and\
  \bibinfo {author} {\bibfnamefont {A.}~\bibnamefont {Paramekanti}},\ }\href
  {\doibase 10.1103/PhysRevB.96.224516} {\bibfield  {journal} {\bibinfo
  {journal} {Phys. Rev. B}\ }\textbf {\bibinfo {volume} {96}},\ \bibinfo
  {pages} {224516} (\bibinfo {year} {2017})}\BibitemShut {NoStop}%
\bibitem [{\citenamefont {Liu}\ \emph {et~al.}(2017)\citenamefont {Liu},
  \citenamefont {Franz},\ and\ \citenamefont {Fujimoto}}]{Liu2017}%
  \BibitemOpen
  \bibfield  {author} {\bibinfo {author} {\bibfnamefont {T.}~\bibnamefont
  {Liu}}, \bibinfo {author} {\bibfnamefont {M.}~\bibnamefont {Franz}}, \ and\
  \bibinfo {author} {\bibfnamefont {S.}~\bibnamefont {Fujimoto}},\ }\href
  {\doibase 10.1103/PhysRevB.96.224518} {\bibfield  {journal} {\bibinfo
  {journal} {Phys. Rev. B}\ }\textbf {\bibinfo {volume} {96}},\ \bibinfo
  {pages} {224518} (\bibinfo {year} {2017})}\BibitemShut {NoStop}%
\bibitem [{\citenamefont {Matsushita}\ \emph {et~al.}(2018)\citenamefont
  {Matsushita}, \citenamefont {Liu}, \citenamefont {Mizushima},\ and\
  \citenamefont {Fujimoto}}]{Matsushita2018}%
  \BibitemOpen
  \bibfield  {author} {\bibinfo {author} {\bibfnamefont {T.}~\bibnamefont
  {Matsushita}}, \bibinfo {author} {\bibfnamefont {T.}~\bibnamefont {Liu}},
  \bibinfo {author} {\bibfnamefont {T.}~\bibnamefont {Mizushima}}, \ and\
  \bibinfo {author} {\bibfnamefont {S.}~\bibnamefont {Fujimoto}},\ }\href
  {\doibase 10.1103/PhysRevB.97.134519} {\bibfield  {journal} {\bibinfo
  {journal} {Phys. Rev. B}\ }\textbf {\bibinfo {volume} {97}},\ \bibinfo
  {pages} {134519} (\bibinfo {year} {2018})}\BibitemShut {NoStop}%
\bibitem [{\citenamefont {Pacholski}\ \emph {et~al.}(2018)\citenamefont
  {Pacholski}, \citenamefont {Beenakker},\ and\ \citenamefont
  {Adagideli}}]{Pacholski2018}%
  \BibitemOpen
  \bibfield  {author} {\bibinfo {author} {\bibfnamefont {M.~J.}\ \bibnamefont
  {Pacholski}}, \bibinfo {author} {\bibfnamefont {C.~W.~J.}\ \bibnamefont
  {Beenakker}}, \ and\ \bibinfo {author} {\bibfnamefont {i.~d.~I.}\
  \bibnamefont {Adagideli}},\ }\href {\doibase 10.1103/PhysRevLett.121.037701}
  {\bibfield  {journal} {\bibinfo  {journal} {Phys. Rev. Lett.}\ }\textbf
  {\bibinfo {volume} {121}},\ \bibinfo {pages} {037701} (\bibinfo {year}
  {2018})}\BibitemShut {NoStop}%
\bibitem [{\citenamefont {Terashima}\ \emph {et~al.}(2007)\citenamefont
  {Terashima}, \citenamefont {Takahide}, \citenamefont {Matsumoto},
  \citenamefont {Uji}, \citenamefont {Kimura}, \citenamefont {Aoki},\ and\
  \citenamefont {Harima}}]{Terashima2007}%
  \BibitemOpen
  \bibfield  {author} {\bibinfo {author} {\bibfnamefont {T.}~\bibnamefont
  {Terashima}}, \bibinfo {author} {\bibfnamefont {Y.}~\bibnamefont {Takahide}},
  \bibinfo {author} {\bibfnamefont {T.}~\bibnamefont {Matsumoto}}, \bibinfo
  {author} {\bibfnamefont {S.}~\bibnamefont {Uji}}, \bibinfo {author}
  {\bibfnamefont {N.}~\bibnamefont {Kimura}}, \bibinfo {author} {\bibfnamefont
  {H.}~\bibnamefont {Aoki}}, \ and\ \bibinfo {author} {\bibfnamefont
  {H.}~\bibnamefont {Harima}},\ }\href {\doibase 10.1103/PhysRevB.76.054506}
  {\bibfield  {journal} {\bibinfo  {journal} {Phys. Rev. B}\ }\textbf {\bibinfo
  {volume} {76}},\ \bibinfo {pages} {054506} (\bibinfo {year}
  {2007})}\BibitemShut {NoStop}%
\bibitem [{\citenamefont {Onuki}\ and\ \citenamefont
  {Settai}(2012)}]{Onuki2012}%
  \BibitemOpen
  \bibfield  {author} {\bibinfo {author} {\bibfnamefont {Y.}~\bibnamefont
  {Onuki}}\ and\ \bibinfo {author} {\bibfnamefont {R.}~\bibnamefont {Settai}},\
  }\enquote {\bibinfo {title} {Electronic states and superconducting properties
  of non-centrosymmetric rare earth compounds},}\ in\ \href {\doibase
  10.1007/978-3-642-24624-1_3} {\emph {\bibinfo {booktitle}
  {Non-Centrosymmetric Superconductors: Introduction and Overview}}},\ \bibinfo
  {editor} {edited by\ \bibinfo {editor} {\bibfnamefont {E.}~\bibnamefont
  {Bauer}}\ and\ \bibinfo {editor} {\bibfnamefont {M.}~\bibnamefont
  {Sigrist}}}\ (\bibinfo  {publisher} {Springer Berlin Heidelberg},\ \bibinfo
  {address} {Berlin, Heidelberg},\ \bibinfo {year} {2012})\ pp.\ \bibinfo
  {pages} {81--126}\BibitemShut {NoStop}%
\bibitem [{\citenamefont {Hayashi}\ \emph
  {et~al.}(2006{\natexlab{a}})\citenamefont {Hayashi}, \citenamefont
  {Wakabayashi}, \citenamefont {Frigeri},\ and\ \citenamefont
  {Sigrist}}]{Hayashi2006_SFD}%
  \BibitemOpen
  \bibfield  {author} {\bibinfo {author} {\bibfnamefont {N.}~\bibnamefont
  {Hayashi}}, \bibinfo {author} {\bibfnamefont {K.}~\bibnamefont
  {Wakabayashi}}, \bibinfo {author} {\bibfnamefont {P.~A.}\ \bibnamefont
  {Frigeri}}, \ and\ \bibinfo {author} {\bibfnamefont {M.}~\bibnamefont
  {Sigrist}},\ }\href {\doibase 10.1103/PhysRevB.73.024504} {\bibfield
  {journal} {\bibinfo  {journal} {Phys. Rev. B}\ }\textbf {\bibinfo {volume}
  {73}},\ \bibinfo {pages} {024504} (\bibinfo {year}
  {2006}{\natexlab{a}})}\BibitemShut {NoStop}%
\bibitem [{\citenamefont {Hayashi}\ \emph
  {et~al.}(2006{\natexlab{b}})\citenamefont {Hayashi}, \citenamefont
  {Wakabayashi}, \citenamefont {Frigeri},\ and\ \citenamefont
  {Sigrist}}]{Hayashi2006_NMR}%
  \BibitemOpen
  \bibfield  {author} {\bibinfo {author} {\bibfnamefont {N.}~\bibnamefont
  {Hayashi}}, \bibinfo {author} {\bibfnamefont {K.}~\bibnamefont
  {Wakabayashi}}, \bibinfo {author} {\bibfnamefont {P.~A.}\ \bibnamefont
  {Frigeri}}, \ and\ \bibinfo {author} {\bibfnamefont {M.}~\bibnamefont
  {Sigrist}},\ }\href {\doibase 10.1103/PhysRevB.73.092508} {\bibfield
  {journal} {\bibinfo  {journal} {Phys. Rev. B}\ }\textbf {\bibinfo {volume}
  {73}},\ \bibinfo {pages} {092508} (\bibinfo {year}
  {2006}{\natexlab{b}})}\BibitemShut {NoStop}%
\bibitem [{\citenamefont {Blount}(1985)}]{Blount1985}%
  \BibitemOpen
  \bibfield  {author} {\bibinfo {author} {\bibfnamefont {E.~I.}\ \bibnamefont
  {Blount}},\ }\href {\doibase 10.1103/PhysRevB.32.2935} {\bibfield  {journal}
  {\bibinfo  {journal} {Phys. Rev. B}\ }\textbf {\bibinfo {volume} {32}},\
  \bibinfo {pages} {2935} (\bibinfo {year} {1985})}\BibitemShut {NoStop}%
\bibitem [{\citenamefont {Kobayashi}\ \emph {et~al.}(2014)\citenamefont
  {Kobayashi}, \citenamefont {Shiozaki}, \citenamefont {Tanaka},\ and\
  \citenamefont {Sato}}]{Kobayashi2014}%
  \BibitemOpen
  \bibfield  {author} {\bibinfo {author} {\bibfnamefont {S.}~\bibnamefont
  {Kobayashi}}, \bibinfo {author} {\bibfnamefont {K.}~\bibnamefont {Shiozaki}},
  \bibinfo {author} {\bibfnamefont {Y.}~\bibnamefont {Tanaka}}, \ and\ \bibinfo
  {author} {\bibfnamefont {M.}~\bibnamefont {Sato}},\ }\href {\doibase
  10.1103/PhysRevB.90.024516} {\bibfield  {journal} {\bibinfo  {journal} {Phys.
  Rev. B}\ }\textbf {\bibinfo {volume} {90}},\ \bibinfo {pages} {024516}
  (\bibinfo {year} {2014})}\BibitemShut {NoStop}%
\bibitem [{Note2()}]{Note2}%
  \BibitemOpen
  \bibinfo {note} {It has been revealed that the stable line nodes may exist at
  the Brillouin-zone faces of nonsymmorphic odd-parity superconductors~\cite
  {Norman1995,Micklitz-Norman2009,Kobayashi2016,Nomoto2017, Sumita_Sr2IrO42017,
  Sumita-Yanase2018,Sumita2019,Yanase2016}.}\BibitemShut {Stop}%
\bibitem [{\citenamefont {Hashimoto}\ \emph {et~al.}(2004)\citenamefont
  {Hashimoto}, \citenamefont {Yasuda}, \citenamefont {Kubo}, \citenamefont
  {Shishido}, \citenamefont {Ueda}, \citenamefont {Settai}, \citenamefont
  {Matsuda}, \citenamefont {Haga}, \citenamefont {Harima},\ and\ \citenamefont
  {Onuki}}]{Hashimoto2004_CePt3Si}%
  \BibitemOpen
  \bibfield  {author} {\bibinfo {author} {\bibfnamefont {S.}~\bibnamefont
  {Hashimoto}}, \bibinfo {author} {\bibfnamefont {T.}~\bibnamefont {Yasuda}},
  \bibinfo {author} {\bibfnamefont {T.}~\bibnamefont {Kubo}}, \bibinfo {author}
  {\bibfnamefont {H.}~\bibnamefont {Shishido}}, \bibinfo {author}
  {\bibfnamefont {T.}~\bibnamefont {Ueda}}, \bibinfo {author} {\bibfnamefont
  {R.}~\bibnamefont {Settai}}, \bibinfo {author} {\bibfnamefont {T.~D.}\
  \bibnamefont {Matsuda}}, \bibinfo {author} {\bibfnamefont {Y.}~\bibnamefont
  {Haga}}, \bibinfo {author} {\bibfnamefont {H.}~\bibnamefont {Harima}}, \ and\
  \bibinfo {author} {\bibfnamefont {Y.}~\bibnamefont {Onuki}},\ }\href
  {\doibase 10.1088/0953-8984/16/23/l02} {\bibfield  {journal} {\bibinfo
  {journal} {Journal of Physics: Condensed Matter}\ }\textbf {\bibinfo {volume}
  {16}},\ \bibinfo {pages} {L287} (\bibinfo {year} {2004})}\BibitemShut
  {NoStop}%
\bibitem [{\citenamefont {Samokhin}\ \emph {et~al.}(2004)\citenamefont
  {Samokhin}, \citenamefont {Zijlstra},\ and\ \citenamefont
  {Bose}}]{Samokhin2004}%
  \BibitemOpen
  \bibfield  {author} {\bibinfo {author} {\bibfnamefont {K.~V.}\ \bibnamefont
  {Samokhin}}, \bibinfo {author} {\bibfnamefont {E.~S.}\ \bibnamefont
  {Zijlstra}}, \ and\ \bibinfo {author} {\bibfnamefont {S.~K.}\ \bibnamefont
  {Bose}},\ }\href {\doibase 10.1103/PhysRevB.69.094514} {\bibfield  {journal}
  {\bibinfo  {journal} {Phys. Rev. B}\ }\textbf {\bibinfo {volume} {69}},\
  \bibinfo {pages} {094514} (\bibinfo {year} {2004})}\BibitemShut {NoStop}%
\bibitem [{\citenamefont {Bollinger}\ \emph {et~al.}(2011)\citenamefont
  {Bollinger}, \citenamefont {Dubuis}, \citenamefont {Yoon}, \citenamefont
  {Pavuna}, \citenamefont {Misewich},\ and\ \citenamefont
  {Bo{\v{z}}ovi{\'{c}}}}]{Bollinger2011}%
  \BibitemOpen
  \bibfield  {author} {\bibinfo {author} {\bibfnamefont {A.~T.}\ \bibnamefont
  {Bollinger}}, \bibinfo {author} {\bibfnamefont {G.}~\bibnamefont {Dubuis}},
  \bibinfo {author} {\bibfnamefont {J.}~\bibnamefont {Yoon}}, \bibinfo {author}
  {\bibfnamefont {D.}~\bibnamefont {Pavuna}}, \bibinfo {author} {\bibfnamefont
  {J.}~\bibnamefont {Misewich}}, \ and\ \bibinfo {author} {\bibfnamefont
  {I.}~\bibnamefont {Bo{\v{z}}ovi{\'{c}}}},\ }\href {\doibase
  10.1038/nature09998} {\bibfield  {journal} {\bibinfo  {journal} {Nature}\
  }\textbf {\bibinfo {volume} {472}},\ \bibinfo {pages} {458} (\bibinfo {year}
  {2011})}\BibitemShut {NoStop}%
\bibitem [{\citenamefont {Leng}\ \emph {et~al.}(2011)\citenamefont {Leng},
  \citenamefont {Garcia-Barriocanal}, \citenamefont {Bose}, \citenamefont
  {Lee},\ and\ \citenamefont {Goldman}}]{Leng2011}%
  \BibitemOpen
  \bibfield  {author} {\bibinfo {author} {\bibfnamefont {X.}~\bibnamefont
  {Leng}}, \bibinfo {author} {\bibfnamefont {J.}~\bibnamefont
  {Garcia-Barriocanal}}, \bibinfo {author} {\bibfnamefont {S.}~\bibnamefont
  {Bose}}, \bibinfo {author} {\bibfnamefont {Y.}~\bibnamefont {Lee}}, \ and\
  \bibinfo {author} {\bibfnamefont {A.~M.}\ \bibnamefont {Goldman}},\ }\href
  {\doibase 10.1103/PhysRevLett.107.027001} {\bibfield  {journal} {\bibinfo
  {journal} {Phys. Rev. Lett.}\ }\textbf {\bibinfo {volume} {107}},\ \bibinfo
  {pages} {027001} (\bibinfo {year} {2011})}\BibitemShut {NoStop}%
\bibitem [{\citenamefont {Nojima}\ \emph {et~al.}(2011)\citenamefont {Nojima},
  \citenamefont {Tada}, \citenamefont {Nakamura}, \citenamefont {Kobayashi},
  \citenamefont {Shimotani},\ and\ \citenamefont {Iwasa}}]{Nojima2011}%
  \BibitemOpen
  \bibfield  {author} {\bibinfo {author} {\bibfnamefont {T.}~\bibnamefont
  {Nojima}}, \bibinfo {author} {\bibfnamefont {H.}~\bibnamefont {Tada}},
  \bibinfo {author} {\bibfnamefont {S.}~\bibnamefont {Nakamura}}, \bibinfo
  {author} {\bibfnamefont {N.}~\bibnamefont {Kobayashi}}, \bibinfo {author}
  {\bibfnamefont {H.}~\bibnamefont {Shimotani}}, \ and\ \bibinfo {author}
  {\bibfnamefont {Y.}~\bibnamefont {Iwasa}},\ }\href {\doibase
  10.1103/PhysRevB.84.020502} {\bibfield  {journal} {\bibinfo  {journal} {Phys.
  Rev. B}\ }\textbf {\bibinfo {volume} {84}},\ \bibinfo {pages} {020502}
  (\bibinfo {year} {2011})}\BibitemShut {NoStop}%
\bibitem [{\citenamefont {Izaki}\ \emph {et~al.}(2007)\citenamefont {Izaki},
  \citenamefont {Shishido}, \citenamefont {Kato}, \citenamefont {Shibauchi},
  \citenamefont {Matsuda},\ and\ \citenamefont {Terashima}}]{Izaki2007}%
  \BibitemOpen
  \bibfield  {author} {\bibinfo {author} {\bibfnamefont {M.}~\bibnamefont
  {Izaki}}, \bibinfo {author} {\bibfnamefont {H.}~\bibnamefont {Shishido}},
  \bibinfo {author} {\bibfnamefont {T.}~\bibnamefont {Kato}}, \bibinfo {author}
  {\bibfnamefont {T.}~\bibnamefont {Shibauchi}}, \bibinfo {author}
  {\bibfnamefont {Y.}~\bibnamefont {Matsuda}}, \ and\ \bibinfo {author}
  {\bibfnamefont {T.}~\bibnamefont {Terashima}},\ }\href {\doibase
  10.1063/1.2787969} {\bibfield  {journal} {\bibinfo  {journal} {Applied
  Physics Letters}\ }\textbf {\bibinfo {volume} {91}},\ \bibinfo {pages}
  {122507} (\bibinfo {year} {2007})},\ \Eprint
  {http://arxiv.org/abs/https://doi.org/10.1063/1.2787969}
  {https://doi.org/10.1063/1.2787969} \BibitemShut {NoStop}%
\bibitem [{\citenamefont {Mizukami}\ \emph {et~al.}(2011)\citenamefont
  {Mizukami}, \citenamefont {Shishido}, \citenamefont {Shibauchi},
  \citenamefont {Shimozawa}, \citenamefont {Yasumoto}, \citenamefont
  {Watanabe}, \citenamefont {Yamashita}, \citenamefont {Ikeda}, \citenamefont
  {Terashima}, \citenamefont {Kontani},\ and\ \citenamefont
  {Matsuda}}]{Mizukami2011}%
  \BibitemOpen
  \bibfield  {author} {\bibinfo {author} {\bibfnamefont {Y.}~\bibnamefont
  {Mizukami}}, \bibinfo {author} {\bibfnamefont {H.}~\bibnamefont {Shishido}},
  \bibinfo {author} {\bibfnamefont {T.}~\bibnamefont {Shibauchi}}, \bibinfo
  {author} {\bibfnamefont {M.}~\bibnamefont {Shimozawa}}, \bibinfo {author}
  {\bibfnamefont {S.}~\bibnamefont {Yasumoto}}, \bibinfo {author}
  {\bibfnamefont {D.}~\bibnamefont {Watanabe}}, \bibinfo {author}
  {\bibfnamefont {M.}~\bibnamefont {Yamashita}}, \bibinfo {author}
  {\bibfnamefont {H.}~\bibnamefont {Ikeda}}, \bibinfo {author} {\bibfnamefont
  {T.}~\bibnamefont {Terashima}}, \bibinfo {author} {\bibfnamefont
  {H.}~\bibnamefont {Kontani}}, \ and\ \bibinfo {author} {\bibfnamefont
  {Y.}~\bibnamefont {Matsuda}},\ }\href {\doibase 10.1038/nphys2112} {\bibfield
   {journal} {\bibinfo  {journal} {Nature Physics}\ }\textbf {\bibinfo {volume}
  {7}},\ \bibinfo {pages} {849} (\bibinfo {year} {2011})}\BibitemShut {NoStop}%
\bibitem [{\citenamefont {Shimozawa}\ \emph {et~al.}(2016)\citenamefont
  {Shimozawa}, \citenamefont {Goh}, \citenamefont {Shibauchi},\ and\
  \citenamefont {Matsuda}}]{Shimozawa_superlattice_RPP2016}%
  \BibitemOpen
  \bibfield  {author} {\bibinfo {author} {\bibfnamefont {M.}~\bibnamefont
  {Shimozawa}}, \bibinfo {author} {\bibfnamefont {S.~K.}\ \bibnamefont {Goh}},
  \bibinfo {author} {\bibfnamefont {T.}~\bibnamefont {Shibauchi}}, \ and\
  \bibinfo {author} {\bibfnamefont {Y.}~\bibnamefont {Matsuda}},\ }\href
  {http://stacks.iop.org/0034-4885/79/i=7/a=074503} {\bibfield  {journal}
  {\bibinfo  {journal} {Reports on Progress in Physics}\ }\textbf {\bibinfo
  {volume} {79}},\ \bibinfo {pages} {074503} (\bibinfo {year}
  {2016})}\BibitemShut {NoStop}%
\bibitem [{\citenamefont {Naritsuka}\ \emph {et~al.}(2021)\citenamefont
  {Naritsuka}, \citenamefont {Terashima},\ and\ \citenamefont
  {Matsuda}}]{naritsuka2021}%
  \BibitemOpen
  \bibfield  {author} {\bibinfo {author} {\bibfnamefont {M.}~\bibnamefont
  {Naritsuka}}, \bibinfo {author} {\bibfnamefont {T.}~\bibnamefont
  {Terashima}}, \ and\ \bibinfo {author} {\bibfnamefont {Y.}~\bibnamefont
  {Matsuda}},\ }\href {\doibase 10.1088/1361-648x/abfdf2} {\bibfield  {journal}
  {\bibinfo  {journal} {J. Phys.: Condens. Matter}\ }\textbf {\bibinfo {volume}
  {33}},\ \bibinfo {pages} {273001} (\bibinfo {year} {2021})}\BibitemShut
  {NoStop}%
\bibitem [{\citenamefont {Zhao}\ \emph {et~al.}(2021)\citenamefont {Zhao},
  \citenamefont {Poccia}, \citenamefont {Cui}, \citenamefont {Volkov},
  \citenamefont {Yoo}, \citenamefont {Engelke}, \citenamefont {Ronen},
  \citenamefont {Zhong}, \citenamefont {Gu}, \citenamefont {Plugge},
  \citenamefont {Tummuru}, \citenamefont {Franz}, \citenamefont {Pixley},\ and\
  \citenamefont {Kim}}]{zhao2021emergent}%
  \BibitemOpen
  \bibfield  {author} {\bibinfo {author} {\bibfnamefont {S.~Y.~F.}\
  \bibnamefont {Zhao}}, \bibinfo {author} {\bibfnamefont {N.}~\bibnamefont
  {Poccia}}, \bibinfo {author} {\bibfnamefont {X.}~\bibnamefont {Cui}},
  \bibinfo {author} {\bibfnamefont {P.~A.}\ \bibnamefont {Volkov}}, \bibinfo
  {author} {\bibfnamefont {H.}~\bibnamefont {Yoo}}, \bibinfo {author}
  {\bibfnamefont {R.}~\bibnamefont {Engelke}}, \bibinfo {author} {\bibfnamefont
  {Y.}~\bibnamefont {Ronen}}, \bibinfo {author} {\bibfnamefont
  {R.}~\bibnamefont {Zhong}}, \bibinfo {author} {\bibfnamefont
  {G.}~\bibnamefont {Gu}}, \bibinfo {author} {\bibfnamefont {S.}~\bibnamefont
  {Plugge}}, \bibinfo {author} {\bibfnamefont {T.}~\bibnamefont {Tummuru}},
  \bibinfo {author} {\bibfnamefont {M.}~\bibnamefont {Franz}}, \bibinfo
  {author} {\bibfnamefont {J.~H.}\ \bibnamefont {Pixley}}, \ and\ \bibinfo
  {author} {\bibfnamefont {P.}~\bibnamefont {Kim}},\ }\href@noop {} {\
  (\bibinfo {year} {2021})},\ \Eprint {http://arxiv.org/abs/2108.13455}
  {arXiv:2108.13455 [cond-mat.supr-con]} \BibitemShut {NoStop}%
\bibitem [{\citenamefont {Song}\ \emph {et~al.}(2011)\citenamefont {Song},
  \citenamefont {Wang}, \citenamefont {Cheng}, \citenamefont {Jiang},
  \citenamefont {Li}, \citenamefont {Zhang}, \citenamefont {Li}, \citenamefont
  {He}, \citenamefont {Wang}, \citenamefont {Jia}, \citenamefont {Hung},
  \citenamefont {Wu}, \citenamefont {Ma}, \citenamefont {Chen},\ and\
  \citenamefont {Xue}}]{Song2011}%
  \BibitemOpen
  \bibfield  {author} {\bibinfo {author} {\bibfnamefont {C.-L.}\ \bibnamefont
  {Song}}, \bibinfo {author} {\bibfnamefont {Y.-L.}\ \bibnamefont {Wang}},
  \bibinfo {author} {\bibfnamefont {P.}~\bibnamefont {Cheng}}, \bibinfo
  {author} {\bibfnamefont {Y.-P.}\ \bibnamefont {Jiang}}, \bibinfo {author}
  {\bibfnamefont {W.}~\bibnamefont {Li}}, \bibinfo {author} {\bibfnamefont
  {T.}~\bibnamefont {Zhang}}, \bibinfo {author} {\bibfnamefont
  {Z.}~\bibnamefont {Li}}, \bibinfo {author} {\bibfnamefont {K.}~\bibnamefont
  {He}}, \bibinfo {author} {\bibfnamefont {L.}~\bibnamefont {Wang}}, \bibinfo
  {author} {\bibfnamefont {J.-F.}\ \bibnamefont {Jia}}, \bibinfo {author}
  {\bibfnamefont {H.-H.}\ \bibnamefont {Hung}}, \bibinfo {author}
  {\bibfnamefont {C.}~\bibnamefont {Wu}}, \bibinfo {author} {\bibfnamefont
  {X.}~\bibnamefont {Ma}}, \bibinfo {author} {\bibfnamefont {X.}~\bibnamefont
  {Chen}}, \ and\ \bibinfo {author} {\bibfnamefont {Q.-K.}\ \bibnamefont
  {Xue}},\ }\href {\doibase 10.1126/science.1202226} {\bibfield  {journal}
  {\bibinfo  {journal} {Science}\ }\textbf {\bibinfo {volume} {332}},\ \bibinfo
  {pages} {1410} (\bibinfo {year} {2011})},\ \Eprint
  {http://arxiv.org/abs/https://science.sciencemag.org/content/332/6036/1410.full.pdf}
  {https://science.sciencemag.org/content/332/6036/1410.full.pdf} \BibitemShut
  {NoStop}%
\bibitem [{\citenamefont {Kasahara}\ \emph {et~al.}(2014)\citenamefont
  {Kasahara}, \citenamefont {Watashige}, \citenamefont {Hanaguri},
  \citenamefont {Kohsaka}, \citenamefont {Yamashita}, \citenamefont
  {Shimoyama}, \citenamefont {Mizukami}, \citenamefont {Endo}, \citenamefont
  {Ikeda}, \citenamefont {Aoyama}, \citenamefont {Terashima}, \citenamefont
  {Uji}, \citenamefont {Wolf}, \citenamefont {von L{\"o}hneysen}, \citenamefont
  {Shibauchi},\ and\ \citenamefont {Matsuda}}]{Kasahara2014}%
  \BibitemOpen
  \bibfield  {author} {\bibinfo {author} {\bibfnamefont {S.}~\bibnamefont
  {Kasahara}}, \bibinfo {author} {\bibfnamefont {T.}~\bibnamefont {Watashige}},
  \bibinfo {author} {\bibfnamefont {T.}~\bibnamefont {Hanaguri}}, \bibinfo
  {author} {\bibfnamefont {Y.}~\bibnamefont {Kohsaka}}, \bibinfo {author}
  {\bibfnamefont {T.}~\bibnamefont {Yamashita}}, \bibinfo {author}
  {\bibfnamefont {Y.}~\bibnamefont {Shimoyama}}, \bibinfo {author}
  {\bibfnamefont {Y.}~\bibnamefont {Mizukami}}, \bibinfo {author}
  {\bibfnamefont {R.}~\bibnamefont {Endo}}, \bibinfo {author} {\bibfnamefont
  {H.}~\bibnamefont {Ikeda}}, \bibinfo {author} {\bibfnamefont
  {K.}~\bibnamefont {Aoyama}}, \bibinfo {author} {\bibfnamefont
  {T.}~\bibnamefont {Terashima}}, \bibinfo {author} {\bibfnamefont
  {S.}~\bibnamefont {Uji}}, \bibinfo {author} {\bibfnamefont {T.}~\bibnamefont
  {Wolf}}, \bibinfo {author} {\bibfnamefont {H.}~\bibnamefont {von
  L{\"o}hneysen}}, \bibinfo {author} {\bibfnamefont {T.}~\bibnamefont
  {Shibauchi}}, \ and\ \bibinfo {author} {\bibfnamefont {Y.}~\bibnamefont
  {Matsuda}},\ }\href {\doibase 10.1073/pnas.1413477111} {\bibfield  {journal}
  {\bibinfo  {journal} {Proceedings of the National Academy of Sciences}\
  }\textbf {\bibinfo {volume} {111}},\ \bibinfo {pages} {16309} (\bibinfo
  {year} {2014})},\ \Eprint
  {http://arxiv.org/abs/https://www.pnas.org/content/111/46/16309.full.pdf}
  {https://www.pnas.org/content/111/46/16309.full.pdf} \BibitemShut {NoStop}%
\bibitem [{\citenamefont {Shibauchi}\ \emph {et~al.}(2020)\citenamefont
  {Shibauchi}, \citenamefont {Hanaguri},\ and\ \citenamefont
  {Matsuda}}]{Shibauchi2020}%
  \BibitemOpen
  \bibfield  {author} {\bibinfo {author} {\bibfnamefont {T.}~\bibnamefont
  {Shibauchi}}, \bibinfo {author} {\bibfnamefont {T.}~\bibnamefont {Hanaguri}},
  \ and\ \bibinfo {author} {\bibfnamefont {Y.}~\bibnamefont {Matsuda}},\ }\href
  {\doibase 10.7566/JPSJ.89.102002} {\bibfield  {journal} {\bibinfo  {journal}
  {Journal of the Physical Society of Japan}\ }\textbf {\bibinfo {volume}
  {89}},\ \bibinfo {pages} {102002} (\bibinfo {year} {2020})},\ \Eprint
  {http://arxiv.org/abs/https://doi.org/10.7566/JPSJ.89.102002}
  {https://doi.org/10.7566/JPSJ.89.102002} \BibitemShut {NoStop}%
\bibitem [{\citenamefont {Zhang}\ \emph
  {et~al.}(2016{\natexlab{a}})\citenamefont {Zhang}, \citenamefont {Lee},
  \citenamefont {Moore}, \citenamefont {Li}, \citenamefont {Yi}, \citenamefont
  {Hashimoto}, \citenamefont {Lu}, \citenamefont {Devereaux}, \citenamefont
  {Lee},\ and\ \citenamefont {Shen}}]{Zhang2017_FeSeARPES}%
  \BibitemOpen
  \bibfield  {author} {\bibinfo {author} {\bibfnamefont {Y.}~\bibnamefont
  {Zhang}}, \bibinfo {author} {\bibfnamefont {J.~J.}\ \bibnamefont {Lee}},
  \bibinfo {author} {\bibfnamefont {R.~G.}\ \bibnamefont {Moore}}, \bibinfo
  {author} {\bibfnamefont {W.}~\bibnamefont {Li}}, \bibinfo {author}
  {\bibfnamefont {M.}~\bibnamefont {Yi}}, \bibinfo {author} {\bibfnamefont
  {M.}~\bibnamefont {Hashimoto}}, \bibinfo {author} {\bibfnamefont {D.~H.}\
  \bibnamefont {Lu}}, \bibinfo {author} {\bibfnamefont {T.~P.}\ \bibnamefont
  {Devereaux}}, \bibinfo {author} {\bibfnamefont {D.-H.}\ \bibnamefont {Lee}},
  \ and\ \bibinfo {author} {\bibfnamefont {Z.-X.}\ \bibnamefont {Shen}},\
  }\href {\doibase 10.1103/PhysRevLett.117.117001} {\bibfield  {journal}
  {\bibinfo  {journal} {Phys. Rev. Lett.}\ }\textbf {\bibinfo {volume} {117}},\
  \bibinfo {pages} {117001} (\bibinfo {year} {2016}{\natexlab{a}})}\BibitemShut
  {NoStop}%
\bibitem [{\citenamefont {Wang}\ \emph {et~al.}(2017)\citenamefont {Wang},
  \citenamefont {Liu}, \citenamefont {Liu},\ and\ \citenamefont
  {Wang}}]{Wang2017_FeSeFilm}%
  \BibitemOpen
  \bibfield  {author} {\bibinfo {author} {\bibfnamefont {Z.}~\bibnamefont
  {Wang}}, \bibinfo {author} {\bibfnamefont {C.}~\bibnamefont {Liu}}, \bibinfo
  {author} {\bibfnamefont {Y.}~\bibnamefont {Liu}}, \ and\ \bibinfo {author}
  {\bibfnamefont {J.}~\bibnamefont {Wang}},\ }\href {\doibase
  10.1088/1361-648x/aa5f26} {\bibfield  {journal} {\bibinfo  {journal} {Journal
  of Physics: Condensed Matter}\ }\textbf {\bibinfo {volume} {29}},\ \bibinfo
  {pages} {153001} (\bibinfo {year} {2017})}\BibitemShut {NoStop}%
\bibitem [{\citenamefont {Huang}\ and\ \citenamefont
  {Hoffman}(2017)}]{Huang2017_FeSe}%
  \BibitemOpen
  \bibfield  {author} {\bibinfo {author} {\bibfnamefont {D.}~\bibnamefont
  {Huang}}\ and\ \bibinfo {author} {\bibfnamefont {J.~E.}\ \bibnamefont
  {Hoffman}},\ }\href {\doibase 10.1146/annurev-conmatphys-031016-025242}
  {\bibfield  {journal} {\bibinfo  {journal} {Annual Review of Condensed Matter
  Physics}\ }\textbf {\bibinfo {volume} {8}},\ \bibinfo {pages} {311} (\bibinfo
  {year} {2017})},\ \Eprint
  {http://arxiv.org/abs/https://doi.org/10.1146/annurev-conmatphys-031016-025242}
  {https://doi.org/10.1146/annurev-conmatphys-031016-025242} \BibitemShut
  {NoStop}%
\bibitem [{\citenamefont {Liang}\ and\ \citenamefont
  {Gao}(2012)}]{Liang2012_EDLT}%
  \BibitemOpen
  \bibfield  {author} {\bibinfo {author} {\bibfnamefont {D.}~\bibnamefont
  {Liang}}\ and\ \bibinfo {author} {\bibfnamefont {X.~P.}\ \bibnamefont
  {Gao}},\ }\href {\doibase 10.1021/nl301325h} {\bibfield  {journal} {\bibinfo
  {journal} {Nano Letters}\ }\textbf {\bibinfo {volume} {12}},\ \bibinfo
  {pages} {3263} (\bibinfo {year} {2012})}\BibitemShut {NoStop}%
\bibitem [{\citenamefont {Takase}\ \emph {et~al.}(2017)\citenamefont {Takase},
  \citenamefont {Ashikawa}, \citenamefont {Zhang}, \citenamefont {Tateno},\
  and\ \citenamefont {Sasaki}}]{Takase2017_EDLT}%
  \BibitemOpen
  \bibfield  {author} {\bibinfo {author} {\bibfnamefont {K.}~\bibnamefont
  {Takase}}, \bibinfo {author} {\bibfnamefont {Y.}~\bibnamefont {Ashikawa}},
  \bibinfo {author} {\bibfnamefont {G.}~\bibnamefont {Zhang}}, \bibinfo
  {author} {\bibfnamefont {K.}~\bibnamefont {Tateno}}, \ and\ \bibinfo {author}
  {\bibfnamefont {S.}~\bibnamefont {Sasaki}},\ }\href {\doibase
  10.1038/s41598-017-01080-0} {\bibfield  {journal} {\bibinfo  {journal}
  {Scientific Reports}\ }\textbf {\bibinfo {volume} {7}},\ \bibinfo {pages}
  {930} (\bibinfo {year} {2017})}\BibitemShut {NoStop}%
\bibitem [{\citenamefont {Premasiri}\ \emph {et~al.}(2018)\citenamefont
  {Premasiri}, \citenamefont {Radha}, \citenamefont {Sucharitakul},
  \citenamefont {Kumar}, \citenamefont {Sankar}, \citenamefont {Chou},
  \citenamefont {Chen},\ and\ \citenamefont {Gao}}]{Premasiri2018}%
  \BibitemOpen
  \bibfield  {author} {\bibinfo {author} {\bibfnamefont {K.}~\bibnamefont
  {Premasiri}}, \bibinfo {author} {\bibfnamefont {S.~K.}\ \bibnamefont
  {Radha}}, \bibinfo {author} {\bibfnamefont {S.}~\bibnamefont {Sucharitakul}},
  \bibinfo {author} {\bibfnamefont {U.~R.}\ \bibnamefont {Kumar}}, \bibinfo
  {author} {\bibfnamefont {R.}~\bibnamefont {Sankar}}, \bibinfo {author}
  {\bibfnamefont {F.-C.}\ \bibnamefont {Chou}}, \bibinfo {author}
  {\bibfnamefont {Y.-T.}\ \bibnamefont {Chen}}, \ and\ \bibinfo {author}
  {\bibfnamefont {X.~P.~A.}\ \bibnamefont {Gao}},\ }\href {\doibase
  10.1021/acs.nanolett.8b01462} {\bibfield  {journal} {\bibinfo  {journal}
  {Nano Letters}\ }\textbf {\bibinfo {volume} {18}},\ \bibinfo {pages} {4403}
  (\bibinfo {year} {2018})}\BibitemShut {NoStop}%
\bibitem [{\citenamefont {Naritsuka}\ \emph {et~al.}(2017)\citenamefont
  {Naritsuka}, \citenamefont {Ishii}, \citenamefont {Miyake}, \citenamefont
  {Tokiwa}, \citenamefont {Toda}, \citenamefont {Shimozawa}, \citenamefont
  {Terashima}, \citenamefont {Shibauchi}, \citenamefont {Matsuda},\ and\
  \citenamefont {Kasahara}}]{Naritsuka2017}%
  \BibitemOpen
  \bibfield  {author} {\bibinfo {author} {\bibfnamefont {M.}~\bibnamefont
  {Naritsuka}}, \bibinfo {author} {\bibfnamefont {T.}~\bibnamefont {Ishii}},
  \bibinfo {author} {\bibfnamefont {S.}~\bibnamefont {Miyake}}, \bibinfo
  {author} {\bibfnamefont {Y.}~\bibnamefont {Tokiwa}}, \bibinfo {author}
  {\bibfnamefont {R.}~\bibnamefont {Toda}}, \bibinfo {author} {\bibfnamefont
  {M.}~\bibnamefont {Shimozawa}}, \bibinfo {author} {\bibfnamefont
  {T.}~\bibnamefont {Terashima}}, \bibinfo {author} {\bibfnamefont
  {T.}~\bibnamefont {Shibauchi}}, \bibinfo {author} {\bibfnamefont
  {Y.}~\bibnamefont {Matsuda}}, \ and\ \bibinfo {author} {\bibfnamefont
  {Y.}~\bibnamefont {Kasahara}},\ }\href {\doibase 10.1103/PhysRevB.96.174512}
  {\bibfield  {journal} {\bibinfo  {journal} {Phys. Rev. B}\ }\textbf {\bibinfo
  {volume} {96}},\ \bibinfo {pages} {174512} (\bibinfo {year}
  {2017})}\BibitemShut {NoStop}%
\bibitem [{\citenamefont {Chakhalian}\ \emph {et~al.}(2006)\citenamefont
  {Chakhalian}, \citenamefont {Freeland}, \citenamefont {Srajer}, \citenamefont
  {Strempfer}, \citenamefont {Khaliullin}, \citenamefont {Cezar}, \citenamefont
  {Charlton}, \citenamefont {Dalgliesh}, \citenamefont {Bernhard},
  \citenamefont {Cristiani}, \citenamefont {Habermeier},\ and\ \citenamefont
  {Keimer}}]{Chakhalian2006}%
  \BibitemOpen
  \bibfield  {author} {\bibinfo {author} {\bibfnamefont {J.}~\bibnamefont
  {Chakhalian}}, \bibinfo {author} {\bibfnamefont {J.~W.}\ \bibnamefont
  {Freeland}}, \bibinfo {author} {\bibfnamefont {G.}~\bibnamefont {Srajer}},
  \bibinfo {author} {\bibfnamefont {J.}~\bibnamefont {Strempfer}}, \bibinfo
  {author} {\bibfnamefont {G.}~\bibnamefont {Khaliullin}}, \bibinfo {author}
  {\bibfnamefont {J.~C.}\ \bibnamefont {Cezar}}, \bibinfo {author}
  {\bibfnamefont {T.}~\bibnamefont {Charlton}}, \bibinfo {author}
  {\bibfnamefont {R.}~\bibnamefont {Dalgliesh}}, \bibinfo {author}
  {\bibfnamefont {C.}~\bibnamefont {Bernhard}}, \bibinfo {author}
  {\bibfnamefont {G.}~\bibnamefont {Cristiani}}, \bibinfo {author}
  {\bibfnamefont {H.-U.}\ \bibnamefont {Habermeier}}, \ and\ \bibinfo {author}
  {\bibfnamefont {B.}~\bibnamefont {Keimer}},\ }\href {\doibase
  10.1038/nphys272} {\bibfield  {journal} {\bibinfo  {journal} {Nature
  Physics}\ }\textbf {\bibinfo {volume} {2}},\ \bibinfo {pages} {244} (\bibinfo
  {year} {2006})}\BibitemShut {NoStop}%
\bibitem [{\citenamefont {Satapathy}\ \emph {et~al.}(2012)\citenamefont
  {Satapathy}, \citenamefont {Uribe-Laverde}, \citenamefont {Marozau},
  \citenamefont {Malik}, \citenamefont {Das}, \citenamefont {Wagner},
  \citenamefont {Marcelot}, \citenamefont {Stahn}, \citenamefont {Br\"uck},
  \citenamefont {R\"uhm}, \citenamefont {Macke}, \citenamefont {Tietze},
  \citenamefont {Goering}, \citenamefont {Fra\~n\'o}, \citenamefont {Kim},
  \citenamefont {Wu}, \citenamefont {Benckiser}, \citenamefont {Keimer},
  \citenamefont {Devishvili}, \citenamefont {Toperverg}, \citenamefont {Merz},
  \citenamefont {Nagel}, \citenamefont {Schuppler},\ and\ \citenamefont
  {Bernhard}}]{Satapathy2012}%
  \BibitemOpen
  \bibfield  {author} {\bibinfo {author} {\bibfnamefont {D.~K.}\ \bibnamefont
  {Satapathy}}, \bibinfo {author} {\bibfnamefont {M.~A.}\ \bibnamefont
  {Uribe-Laverde}}, \bibinfo {author} {\bibfnamefont {I.}~\bibnamefont
  {Marozau}}, \bibinfo {author} {\bibfnamefont {V.~K.}\ \bibnamefont {Malik}},
  \bibinfo {author} {\bibfnamefont {S.}~\bibnamefont {Das}}, \bibinfo {author}
  {\bibfnamefont {T.}~\bibnamefont {Wagner}}, \bibinfo {author} {\bibfnamefont
  {C.}~\bibnamefont {Marcelot}}, \bibinfo {author} {\bibfnamefont
  {J.}~\bibnamefont {Stahn}}, \bibinfo {author} {\bibfnamefont
  {S.}~\bibnamefont {Br\"uck}}, \bibinfo {author} {\bibfnamefont
  {A.}~\bibnamefont {R\"uhm}}, \bibinfo {author} {\bibfnamefont
  {S.}~\bibnamefont {Macke}}, \bibinfo {author} {\bibfnamefont
  {T.}~\bibnamefont {Tietze}}, \bibinfo {author} {\bibfnamefont
  {E.}~\bibnamefont {Goering}}, \bibinfo {author} {\bibfnamefont
  {A.}~\bibnamefont {Fra\~n\'o}}, \bibinfo {author} {\bibfnamefont {J.~H.}\
  \bibnamefont {Kim}}, \bibinfo {author} {\bibfnamefont {M.}~\bibnamefont
  {Wu}}, \bibinfo {author} {\bibfnamefont {E.}~\bibnamefont {Benckiser}},
  \bibinfo {author} {\bibfnamefont {B.}~\bibnamefont {Keimer}}, \bibinfo
  {author} {\bibfnamefont {A.}~\bibnamefont {Devishvili}}, \bibinfo {author}
  {\bibfnamefont {B.~P.}\ \bibnamefont {Toperverg}}, \bibinfo {author}
  {\bibfnamefont {M.}~\bibnamefont {Merz}}, \bibinfo {author} {\bibfnamefont
  {P.}~\bibnamefont {Nagel}}, \bibinfo {author} {\bibfnamefont
  {S.}~\bibnamefont {Schuppler}}, \ and\ \bibinfo {author} {\bibfnamefont
  {C.}~\bibnamefont {Bernhard}},\ }\href {\doibase
  10.1103/PhysRevLett.108.197201} {\bibfield  {journal} {\bibinfo  {journal}
  {Phys. Rev. Lett.}\ }\textbf {\bibinfo {volume} {108}},\ \bibinfo {pages}
  {197201} (\bibinfo {year} {2012})}\BibitemShut {NoStop}%
\bibitem [{\citenamefont {Uribe-Laverde}\ \emph {et~al.}(2014)\citenamefont
  {Uribe-Laverde}, \citenamefont {Das}, \citenamefont {Sen}, \citenamefont
  {Marozau}, \citenamefont {Perret}, \citenamefont {Alberca}, \citenamefont
  {Heidler}, \citenamefont {Piamonteze}, \citenamefont {Merz}, \citenamefont
  {Nagel}, \citenamefont {Schuppler}, \citenamefont {Munzar},\ and\
  \citenamefont {Bernhard}}]{Uribe2014}%
  \BibitemOpen
  \bibfield  {author} {\bibinfo {author} {\bibfnamefont {M.~A.}\ \bibnamefont
  {Uribe-Laverde}}, \bibinfo {author} {\bibfnamefont {S.}~\bibnamefont {Das}},
  \bibinfo {author} {\bibfnamefont {K.}~\bibnamefont {Sen}}, \bibinfo {author}
  {\bibfnamefont {I.}~\bibnamefont {Marozau}}, \bibinfo {author} {\bibfnamefont
  {E.}~\bibnamefont {Perret}}, \bibinfo {author} {\bibfnamefont
  {A.}~\bibnamefont {Alberca}}, \bibinfo {author} {\bibfnamefont
  {J.}~\bibnamefont {Heidler}}, \bibinfo {author} {\bibfnamefont
  {C.}~\bibnamefont {Piamonteze}}, \bibinfo {author} {\bibfnamefont
  {M.}~\bibnamefont {Merz}}, \bibinfo {author} {\bibfnamefont {P.}~\bibnamefont
  {Nagel}}, \bibinfo {author} {\bibfnamefont {S.}~\bibnamefont {Schuppler}},
  \bibinfo {author} {\bibfnamefont {D.}~\bibnamefont {Munzar}}, \ and\ \bibinfo
  {author} {\bibfnamefont {C.}~\bibnamefont {Bernhard}},\ }\href {\doibase
  10.1103/PhysRevB.90.205135} {\bibfield  {journal} {\bibinfo  {journal} {Phys.
  Rev. B}\ }\textbf {\bibinfo {volume} {90}},\ \bibinfo {pages} {205135}
  (\bibinfo {year} {2014})}\BibitemShut {NoStop}%
\bibitem [{\citenamefont {Sen}\ \emph {et~al.}(2016)\citenamefont {Sen},
  \citenamefont {Perret}, \citenamefont {Alberca}, \citenamefont
  {Uribe-Laverde}, \citenamefont {Marozau}, \citenamefont {Yazdi-Rizi},
  \citenamefont {Mallett}, \citenamefont {Marsik}, \citenamefont {Piamonteze},
  \citenamefont {Khaydukov}, \citenamefont {D\"obeli}, \citenamefont {Keller},
  \citenamefont {Bi\ifmmode~\check{s}\else \v{s}\fi{}kup}, \citenamefont
  {Varela}, \citenamefont {Va\ifmmode~\check{s}\else \v{s}\fi{}\'atko},
  \citenamefont {Munzar},\ and\ \citenamefont {Bernhard}}]{Sen2016}%
  \BibitemOpen
  \bibfield  {author} {\bibinfo {author} {\bibfnamefont {K.}~\bibnamefont
  {Sen}}, \bibinfo {author} {\bibfnamefont {E.}~\bibnamefont {Perret}},
  \bibinfo {author} {\bibfnamefont {A.}~\bibnamefont {Alberca}}, \bibinfo
  {author} {\bibfnamefont {M.~A.}\ \bibnamefont {Uribe-Laverde}}, \bibinfo
  {author} {\bibfnamefont {I.}~\bibnamefont {Marozau}}, \bibinfo {author}
  {\bibfnamefont {M.}~\bibnamefont {Yazdi-Rizi}}, \bibinfo {author}
  {\bibfnamefont {B.~P.~P.}\ \bibnamefont {Mallett}}, \bibinfo {author}
  {\bibfnamefont {P.}~\bibnamefont {Marsik}}, \bibinfo {author} {\bibfnamefont
  {C.}~\bibnamefont {Piamonteze}}, \bibinfo {author} {\bibfnamefont
  {Y.}~\bibnamefont {Khaydukov}}, \bibinfo {author} {\bibfnamefont
  {M.}~\bibnamefont {D\"obeli}}, \bibinfo {author} {\bibfnamefont
  {T.}~\bibnamefont {Keller}}, \bibinfo {author} {\bibfnamefont
  {N.}~\bibnamefont {Bi\ifmmode~\check{s}\else \v{s}\fi{}kup}}, \bibinfo
  {author} {\bibfnamefont {M.}~\bibnamefont {Varela}}, \bibinfo {author}
  {\bibfnamefont {J.}~\bibnamefont {Va\ifmmode~\check{s}\else
  \v{s}\fi{}\'atko}}, \bibinfo {author} {\bibfnamefont {D.}~\bibnamefont
  {Munzar}}, \ and\ \bibinfo {author} {\bibfnamefont {C.}~\bibnamefont
  {Bernhard}},\ }\href {\doibase 10.1103/PhysRevB.93.205131} {\bibfield
  {journal} {\bibinfo  {journal} {Phys. Rev. B}\ }\textbf {\bibinfo {volume}
  {93}},\ \bibinfo {pages} {205131} (\bibinfo {year} {2016})}\BibitemShut
  {NoStop}%
\bibitem [{\citenamefont {Das}\ \emph {et~al.}(2014)\citenamefont {Das},
  \citenamefont {Sen}, \citenamefont {Marozau}, \citenamefont {Uribe-Laverde},
  \citenamefont {Biskup}, \citenamefont {Varela}, \citenamefont {Khaydukov},
  \citenamefont {Soltwedel}, \citenamefont {Keller}, \citenamefont {D\"obeli},
  \citenamefont {Schneider},\ and\ \citenamefont {Bernhard}}]{Das2014}%
  \BibitemOpen
  \bibfield  {author} {\bibinfo {author} {\bibfnamefont {S.}~\bibnamefont
  {Das}}, \bibinfo {author} {\bibfnamefont {K.}~\bibnamefont {Sen}}, \bibinfo
  {author} {\bibfnamefont {I.}~\bibnamefont {Marozau}}, \bibinfo {author}
  {\bibfnamefont {M.~A.}\ \bibnamefont {Uribe-Laverde}}, \bibinfo {author}
  {\bibfnamefont {N.}~\bibnamefont {Biskup}}, \bibinfo {author} {\bibfnamefont
  {M.}~\bibnamefont {Varela}}, \bibinfo {author} {\bibfnamefont
  {Y.}~\bibnamefont {Khaydukov}}, \bibinfo {author} {\bibfnamefont
  {O.}~\bibnamefont {Soltwedel}}, \bibinfo {author} {\bibfnamefont
  {T.}~\bibnamefont {Keller}}, \bibinfo {author} {\bibfnamefont
  {M.}~\bibnamefont {D\"obeli}}, \bibinfo {author} {\bibfnamefont {C.~W.}\
  \bibnamefont {Schneider}}, \ and\ \bibinfo {author} {\bibfnamefont
  {C.}~\bibnamefont {Bernhard}},\ }\href {\doibase 10.1103/PhysRevB.89.094511}
  {\bibfield  {journal} {\bibinfo  {journal} {Phys. Rev. B}\ }\textbf {\bibinfo
  {volume} {89}},\ \bibinfo {pages} {094511} (\bibinfo {year}
  {2014})}\BibitemShut {NoStop}%
\bibitem [{\citenamefont {Sigrist}\ \emph {et~al.}(2014)\citenamefont
  {Sigrist}, \citenamefont {Agterberg}, \citenamefont {Fischer}, \citenamefont
  {Goryo}, \citenamefont {Loder}, \citenamefont {Rhim}, \citenamefont
  {Maruyama}, \citenamefont {Yanase}, \citenamefont {Yoshida},\ and\
  \citenamefont {Youn}}]{Sigrist_LNCSreview2014}%
  \BibitemOpen
  \bibfield  {author} {\bibinfo {author} {\bibfnamefont {M.}~\bibnamefont
  {Sigrist}}, \bibinfo {author} {\bibfnamefont {D.~F.}\ \bibnamefont
  {Agterberg}}, \bibinfo {author} {\bibfnamefont {M.~H.}\ \bibnamefont
  {Fischer}}, \bibinfo {author} {\bibfnamefont {J.}~\bibnamefont {Goryo}},
  \bibinfo {author} {\bibfnamefont {F.}~\bibnamefont {Loder}}, \bibinfo
  {author} {\bibfnamefont {S.-H.}\ \bibnamefont {Rhim}}, \bibinfo {author}
  {\bibfnamefont {D.}~\bibnamefont {Maruyama}}, \bibinfo {author}
  {\bibfnamefont {Y.}~\bibnamefont {Yanase}}, \bibinfo {author} {\bibfnamefont
  {T.}~\bibnamefont {Yoshida}}, \ and\ \bibinfo {author} {\bibfnamefont
  {S.~J.}\ \bibnamefont {Youn}},\ }\href {\doibase 10.7566/JPSJ.83.061014}
  {\bibfield  {journal} {\bibinfo  {journal} {Journal of the Physical Society
  of Japan}\ }\textbf {\bibinfo {volume} {83}},\ \bibinfo {pages} {061014}
  (\bibinfo {year} {2014})},\ \Eprint
  {http://arxiv.org/abs/https://doi.org/10.7566/JPSJ.83.061014}
  {https://doi.org/10.7566/JPSJ.83.061014} \BibitemShut {NoStop}%
\bibitem [{\citenamefont {Matsuda}\ \emph
  {et~al.}(2006{\natexlab{a}})\citenamefont {Matsuda}, \citenamefont {Izawa},\
  and\ \citenamefont {Vekhter}}]{Matsuda2006}%
  \BibitemOpen
  \bibfield  {author} {\bibinfo {author} {\bibfnamefont {Y.}~\bibnamefont
  {Matsuda}}, \bibinfo {author} {\bibfnamefont {K.}~\bibnamefont {Izawa}}, \
  and\ \bibinfo {author} {\bibfnamefont {I.}~\bibnamefont {Vekhter}},\ }\href
  {\doibase 10.1088/0953-8984/18/44/r01} {\bibfield  {journal} {\bibinfo
  {journal} {Journal of Physics: Condensed Matter}\ }\textbf {\bibinfo {volume}
  {18}},\ \bibinfo {pages} {R705} (\bibinfo {year}
  {2006}{\natexlab{a}})}\BibitemShut {NoStop}%
\bibitem [{\citenamefont {Machida}\ \emph {et~al.}(2019)\citenamefont
  {Machida}, \citenamefont {Sun}, \citenamefont {Pyon}, \citenamefont {Takeda},
  \citenamefont {Kohsaka}, \citenamefont {Hanaguri}, \citenamefont {Sasagawa},\
  and\ \citenamefont {Tamegai}}]{Machida2019}%
  \BibitemOpen
  \bibfield  {author} {\bibinfo {author} {\bibfnamefont {T.}~\bibnamefont
  {Machida}}, \bibinfo {author} {\bibfnamefont {Y.}~\bibnamefont {Sun}},
  \bibinfo {author} {\bibfnamefont {S.}~\bibnamefont {Pyon}}, \bibinfo {author}
  {\bibfnamefont {S.}~\bibnamefont {Takeda}}, \bibinfo {author} {\bibfnamefont
  {Y.}~\bibnamefont {Kohsaka}}, \bibinfo {author} {\bibfnamefont
  {T.}~\bibnamefont {Hanaguri}}, \bibinfo {author} {\bibfnamefont
  {T.}~\bibnamefont {Sasagawa}}, \ and\ \bibinfo {author} {\bibfnamefont
  {T.}~\bibnamefont {Tamegai}},\ }\href {\doibase 10.1038/s41563-019-0397-1}
  {\bibfield  {journal} {\bibinfo  {journal} {Nature Materials}\ }\textbf
  {\bibinfo {volume} {18}},\ \bibinfo {pages} {811} (\bibinfo {year}
  {2019})}\BibitemShut {NoStop}%
\bibitem [{\citenamefont {Takasan}\ \emph
  {et~al.}(2017{\natexlab{a}})\citenamefont {Takasan}, \citenamefont {Daido},
  \citenamefont {Kawakami},\ and\ \citenamefont {Yanase}}]{Takasan2017a}%
  \BibitemOpen
  \bibfield  {author} {\bibinfo {author} {\bibfnamefont {K.}~\bibnamefont
  {Takasan}}, \bibinfo {author} {\bibfnamefont {A.}~\bibnamefont {Daido}},
  \bibinfo {author} {\bibfnamefont {N.}~\bibnamefont {Kawakami}}, \ and\
  \bibinfo {author} {\bibfnamefont {Y.}~\bibnamefont {Yanase}},\ }\href
  {\doibase 10.1103/PhysRevB.95.134508} {\bibfield  {journal} {\bibinfo
  {journal} {Phys. Rev. B}\ }\textbf {\bibinfo {volume} {95}},\ \bibinfo
  {pages} {134508} (\bibinfo {year} {2017}{\natexlab{a}})}\BibitemShut
  {NoStop}%
\bibitem [{\citenamefont {Bukov}\ \emph {et~al.}(2015)\citenamefont {Bukov},
  \citenamefont {D'Alessio},\ and\ \citenamefont
  {Polkovnikov}}]{BukovReview2015}%
  \BibitemOpen
  \bibfield  {author} {\bibinfo {author} {\bibfnamefont {M.}~\bibnamefont
  {Bukov}}, \bibinfo {author} {\bibfnamefont {L.}~\bibnamefont {D'Alessio}}, \
  and\ \bibinfo {author} {\bibfnamefont {A.}~\bibnamefont {Polkovnikov}},\
  }\href {\doibase 10.1080/00018732.2015.1055918} {\bibfield  {journal}
  {\bibinfo  {journal} {Advances in Physics}\ }\textbf {\bibinfo {volume}
  {64}},\ \bibinfo {pages} {139} (\bibinfo {year} {2015})}\BibitemShut
  {NoStop}%
\bibitem [{\citenamefont {Eckardt}(2017)}]{EckardtRMP2017}%
  \BibitemOpen
  \bibfield  {author} {\bibinfo {author} {\bibfnamefont {A.}~\bibnamefont
  {Eckardt}},\ }\href {\doibase 10.1103/RevModPhys.89.011004} {\bibfield
  {journal} {\bibinfo  {journal} {Rev. Mod. Phys.}\ }\textbf {\bibinfo {volume}
  {89}},\ \bibinfo {pages} {011004} (\bibinfo {year} {2017})}\BibitemShut
  {NoStop}%
\bibitem [{\citenamefont {Oka}\ and\ \citenamefont
  {Kitamura}(2019)}]{OkaReview2019}%
  \BibitemOpen
  \bibfield  {author} {\bibinfo {author} {\bibfnamefont {T.}~\bibnamefont
  {Oka}}\ and\ \bibinfo {author} {\bibfnamefont {S.}~\bibnamefont {Kitamura}},\
  }\href {\doibase 10.1146/annurev-conmatphys-031218-013423} {\bibfield
  {journal} {\bibinfo  {journal} {Annual Review of Condensed Matter Physics}\
  }\textbf {\bibinfo {volume} {10}},\ \bibinfo {pages} {387} (\bibinfo {year}
  {2019})}\BibitemShut {NoStop}%
\bibitem [{\citenamefont {Rudner}\ and\ \citenamefont
  {Lindner}(2020)}]{Rudner2020}%
  \BibitemOpen
  \bibfield  {author} {\bibinfo {author} {\bibfnamefont {M.~S.}\ \bibnamefont
  {Rudner}}\ and\ \bibinfo {author} {\bibfnamefont {N.~H.}\ \bibnamefont
  {Lindner}},\ }\href {\doibase 10.1038/s42254-020-0170-z} {\bibfield
  {journal} {\bibinfo  {journal} {Nature Reviews Physics}\ }\textbf {\bibinfo
  {volume} {2}},\ \bibinfo {pages} {229} (\bibinfo {year} {2020})}\BibitemShut
  {NoStop}%
\bibitem [{\citenamefont {Oka}\ and\ \citenamefont {Aoki}(2009)}]{Oka2009}%
  \BibitemOpen
  \bibfield  {author} {\bibinfo {author} {\bibfnamefont {T.}~\bibnamefont
  {Oka}}\ and\ \bibinfo {author} {\bibfnamefont {H.}~\bibnamefont {Aoki}},\
  }\href {\doibase 10.1103/PhysRevB.79.081406} {\bibfield  {journal} {\bibinfo
  {journal} {Phys. Rev. B}\ }\textbf {\bibinfo {volume} {79}},\ \bibinfo
  {pages} {081406} (\bibinfo {year} {2009})}\BibitemShut {NoStop}%
\bibitem [{\citenamefont {Lindner}\ \emph {et~al.}(2011)\citenamefont
  {Lindner}, \citenamefont {Refael},\ and\ \citenamefont
  {Galitski}}]{Lindner2011}%
  \BibitemOpen
  \bibfield  {author} {\bibinfo {author} {\bibfnamefont {N.~H.}\ \bibnamefont
  {Lindner}}, \bibinfo {author} {\bibfnamefont {G.}~\bibnamefont {Refael}}, \
  and\ \bibinfo {author} {\bibfnamefont {V.}~\bibnamefont {Galitski}},\ }\href
  {https://doi.org/10.1038/nphys1926} {\bibfield  {journal} {\bibinfo
  {journal} {Nature Physics}\ }\textbf {\bibinfo {volume} {7}},\ \bibinfo
  {pages} {490 EP } (\bibinfo {year} {2011})}\BibitemShut {NoStop}%
\bibitem [{\citenamefont {Kitagawa}\ \emph {et~al.}(2011)\citenamefont
  {Kitagawa}, \citenamefont {Oka}, \citenamefont {Brataas}, \citenamefont
  {Fu},\ and\ \citenamefont {Demler}}]{Kitagawa2011}%
  \BibitemOpen
  \bibfield  {author} {\bibinfo {author} {\bibfnamefont {T.}~\bibnamefont
  {Kitagawa}}, \bibinfo {author} {\bibfnamefont {T.}~\bibnamefont {Oka}},
  \bibinfo {author} {\bibfnamefont {A.}~\bibnamefont {Brataas}}, \bibinfo
  {author} {\bibfnamefont {L.}~\bibnamefont {Fu}}, \ and\ \bibinfo {author}
  {\bibfnamefont {E.}~\bibnamefont {Demler}},\ }\href {\doibase
  10.1103/PhysRevB.84.235108} {\bibfield  {journal} {\bibinfo  {journal} {Phys.
  Rev. B}\ }\textbf {\bibinfo {volume} {84}},\ \bibinfo {pages} {235108}
  (\bibinfo {year} {2011})}\BibitemShut {NoStop}%
\bibitem [{\citenamefont {Mikami}\ \emph {et~al.}(2016)\citenamefont {Mikami},
  \citenamefont {Kitamura}, \citenamefont {Yasuda}, \citenamefont {Tsuji},
  \citenamefont {Oka},\ and\ \citenamefont {Aoki}}]{Mikami2016}%
  \BibitemOpen
  \bibfield  {author} {\bibinfo {author} {\bibfnamefont {T.}~\bibnamefont
  {Mikami}}, \bibinfo {author} {\bibfnamefont {S.}~\bibnamefont {Kitamura}},
  \bibinfo {author} {\bibfnamefont {K.}~\bibnamefont {Yasuda}}, \bibinfo
  {author} {\bibfnamefont {N.}~\bibnamefont {Tsuji}}, \bibinfo {author}
  {\bibfnamefont {T.}~\bibnamefont {Oka}}, \ and\ \bibinfo {author}
  {\bibfnamefont {H.}~\bibnamefont {Aoki}},\ }\href {\doibase
  10.1103/PhysRevB.93.144307} {\bibfield  {journal} {\bibinfo  {journal} {Phys.
  Rev. B}\ }\textbf {\bibinfo {volume} {93}},\ \bibinfo {pages} {144307}
  (\bibinfo {year} {2016})}\BibitemShut {NoStop}%
\bibitem [{\citenamefont {Ezawa}(2015)}]{Ezawa2015}%
  \BibitemOpen
  \bibfield  {author} {\bibinfo {author} {\bibfnamefont {M.}~\bibnamefont
  {Ezawa}},\ }\href {\doibase 10.1007/s10948-014-2900-x} {\bibfield  {journal}
  {\bibinfo  {journal} {Journal of Superconductivity and Novel Magnetism}\
  }\textbf {\bibinfo {volume} {28}},\ \bibinfo {pages} {1249} (\bibinfo {year}
  {2015})}\BibitemShut {NoStop}%
\bibitem [{\citenamefont {Claassen}\ \emph {et~al.}(2019)\citenamefont
  {Claassen}, \citenamefont {Kennes}, \citenamefont {Zingl}, \citenamefont
  {Sentef},\ and\ \citenamefont {Rubio}}]{Claassen2019}%
  \BibitemOpen
  \bibfield  {author} {\bibinfo {author} {\bibfnamefont {M.}~\bibnamefont
  {Claassen}}, \bibinfo {author} {\bibfnamefont {D.~M.}\ \bibnamefont
  {Kennes}}, \bibinfo {author} {\bibfnamefont {M.}~\bibnamefont {Zingl}},
  \bibinfo {author} {\bibfnamefont {M.~A.}\ \bibnamefont {Sentef}}, \ and\
  \bibinfo {author} {\bibfnamefont {A.}~\bibnamefont {Rubio}},\ }\href
  {\doibase 10.1038/s41567-019-0532-6} {\bibfield  {journal} {\bibinfo
  {journal} {Nature Physics}\ }\textbf {\bibinfo {volume} {15}},\ \bibinfo
  {pages} {766} (\bibinfo {year} {2019})}\BibitemShut {NoStop}%
\bibitem [{\citenamefont {Sato}\ \emph {et~al.}(2016)\citenamefont {Sato},
  \citenamefont {Takayoshi},\ and\ \citenamefont {Oka}}]{Sato2016}%
  \BibitemOpen
  \bibfield  {author} {\bibinfo {author} {\bibfnamefont {M.}~\bibnamefont
  {Sato}}, \bibinfo {author} {\bibfnamefont {S.}~\bibnamefont {Takayoshi}}, \
  and\ \bibinfo {author} {\bibfnamefont {T.}~\bibnamefont {Oka}},\ }\href
  {\doibase 10.1103/PhysRevLett.117.147202} {\bibfield  {journal} {\bibinfo
  {journal} {Phys. Rev. Lett.}\ }\textbf {\bibinfo {volume} {117}},\ \bibinfo
  {pages} {147202} (\bibinfo {year} {2016})}\BibitemShut {NoStop}%
\bibitem [{\citenamefont {Claassen}\ \emph {et~al.}(2017)\citenamefont
  {Claassen}, \citenamefont {Jiang}, \citenamefont {Moritz},\ and\
  \citenamefont {Devereaux}}]{Claassen2017}%
  \BibitemOpen
  \bibfield  {author} {\bibinfo {author} {\bibfnamefont {M.}~\bibnamefont
  {Claassen}}, \bibinfo {author} {\bibfnamefont {H.-C.}\ \bibnamefont {Jiang}},
  \bibinfo {author} {\bibfnamefont {B.}~\bibnamefont {Moritz}}, \ and\ \bibinfo
  {author} {\bibfnamefont {T.~P.}\ \bibnamefont {Devereaux}},\ }\href {\doibase
  10.1038/s41467-017-00876-y} {\bibfield  {journal} {\bibinfo  {journal}
  {Nature Communications}\ }\textbf {\bibinfo {volume} {8}},\ \bibinfo {pages}
  {1192} (\bibinfo {year} {2017})}\BibitemShut {NoStop}%
\bibitem [{\citenamefont {Kitamura}\ \emph {et~al.}(2017)\citenamefont
  {Kitamura}, \citenamefont {Oka},\ and\ \citenamefont {Aoki}}]{Kitamura2017}%
  \BibitemOpen
  \bibfield  {author} {\bibinfo {author} {\bibfnamefont {S.}~\bibnamefont
  {Kitamura}}, \bibinfo {author} {\bibfnamefont {T.}~\bibnamefont {Oka}}, \
  and\ \bibinfo {author} {\bibfnamefont {H.}~\bibnamefont {Aoki}},\ }\href
  {\doibase 10.1103/PhysRevB.96.014406} {\bibfield  {journal} {\bibinfo
  {journal} {Phys. Rev. B}\ }\textbf {\bibinfo {volume} {96}},\ \bibinfo
  {pages} {014406} (\bibinfo {year} {2017})}\BibitemShut {NoStop}%
\bibitem [{\citenamefont {Takasan}\ \emph
  {et~al.}(2017{\natexlab{b}})\citenamefont {Takasan}, \citenamefont
  {Nakagawa},\ and\ \citenamefont {Kawakami}}]{Takasan2017b}%
  \BibitemOpen
  \bibfield  {author} {\bibinfo {author} {\bibfnamefont {K.}~\bibnamefont
  {Takasan}}, \bibinfo {author} {\bibfnamefont {M.}~\bibnamefont {Nakagawa}}, \
  and\ \bibinfo {author} {\bibfnamefont {N.}~\bibnamefont {Kawakami}},\ }\href
  {\doibase 10.1103/PhysRevB.96.115120} {\bibfield  {journal} {\bibinfo
  {journal} {Phys. Rev. B}\ }\textbf {\bibinfo {volume} {96}},\ \bibinfo
  {pages} {115120} (\bibinfo {year} {2017}{\natexlab{b}})}\BibitemShut
  {NoStop}%
\bibitem [{\citenamefont {Jotzu}\ \emph {et~al.}(2014)\citenamefont {Jotzu},
  \citenamefont {Messer}, \citenamefont {Desbuquois}, \citenamefont {Lebrat},
  \citenamefont {Uehlinger}, \citenamefont {Greif},\ and\ \citenamefont
  {Esslinger}}]{Jotzu2014}%
  \BibitemOpen
  \bibfield  {author} {\bibinfo {author} {\bibfnamefont {G.}~\bibnamefont
  {Jotzu}}, \bibinfo {author} {\bibfnamefont {M.}~\bibnamefont {Messer}},
  \bibinfo {author} {\bibfnamefont {R.}~\bibnamefont {Desbuquois}}, \bibinfo
  {author} {\bibfnamefont {M.}~\bibnamefont {Lebrat}}, \bibinfo {author}
  {\bibfnamefont {T.}~\bibnamefont {Uehlinger}}, \bibinfo {author}
  {\bibfnamefont {D.}~\bibnamefont {Greif}}, \ and\ \bibinfo {author}
  {\bibfnamefont {T.}~\bibnamefont {Esslinger}},\ }\href
  {https://doi.org/10.1038/nature13915} {\bibfield  {journal} {\bibinfo
  {journal} {Nature}\ }\textbf {\bibinfo {volume} {515}},\ \bibinfo {pages}
  {237 EP } (\bibinfo {year} {2014})}\BibitemShut {NoStop}%
\bibitem [{\citenamefont {Wang}\ \emph {et~al.}(2013)\citenamefont {Wang},
  \citenamefont {Steinberg}, \citenamefont {Jarillo-Herrero},\ and\
  \citenamefont {Gedik}}]{Wang2013}%
  \BibitemOpen
  \bibfield  {author} {\bibinfo {author} {\bibfnamefont {Y.~H.}\ \bibnamefont
  {Wang}}, \bibinfo {author} {\bibfnamefont {H.}~\bibnamefont {Steinberg}},
  \bibinfo {author} {\bibfnamefont {P.}~\bibnamefont {Jarillo-Herrero}}, \ and\
  \bibinfo {author} {\bibfnamefont {N.}~\bibnamefont {Gedik}},\ }\href
  {\doibase 10.1126/science.1239834} {\bibfield  {journal} {\bibinfo  {journal}
  {Science}\ }\textbf {\bibinfo {volume} {342}},\ \bibinfo {pages} {453}
  (\bibinfo {year} {2013})}\BibitemShut {NoStop}%
\bibitem [{\citenamefont {Mahmood}\ \emph {et~al.}(2016)\citenamefont
  {Mahmood}, \citenamefont {Chan}, \citenamefont {Alpichshev}, \citenamefont
  {Gardner}, \citenamefont {Lee}, \citenamefont {Lee},\ and\ \citenamefont
  {Gedik}}]{Mahmood2016}%
  \BibitemOpen
  \bibfield  {author} {\bibinfo {author} {\bibfnamefont {F.}~\bibnamefont
  {Mahmood}}, \bibinfo {author} {\bibfnamefont {C.-K.}\ \bibnamefont {Chan}},
  \bibinfo {author} {\bibfnamefont {Z.}~\bibnamefont {Alpichshev}}, \bibinfo
  {author} {\bibfnamefont {D.}~\bibnamefont {Gardner}}, \bibinfo {author}
  {\bibfnamefont {Y.}~\bibnamefont {Lee}}, \bibinfo {author} {\bibfnamefont
  {P.~A.}\ \bibnamefont {Lee}}, \ and\ \bibinfo {author} {\bibfnamefont
  {N.}~\bibnamefont {Gedik}},\ }\href {https://doi.org/10.1038/nphys3609}
  {\bibfield  {journal} {\bibinfo  {journal} {Nature Physics}\ }\textbf
  {\bibinfo {volume} {12}},\ \bibinfo {pages} {306 EP } (\bibinfo {year}
  {2016})}\BibitemShut {NoStop}%
\bibitem [{\citenamefont {McIver}\ \emph {et~al.}(2020)\citenamefont {McIver},
  \citenamefont {Schulte}, \citenamefont {Stein}, \citenamefont {Matsuyama},
  \citenamefont {Jotzu}, \citenamefont {Meier},\ and\ \citenamefont
  {Cavalleri}}]{McIver2020}%
  \BibitemOpen
  \bibfield  {author} {\bibinfo {author} {\bibfnamefont {J.~W.}\ \bibnamefont
  {McIver}}, \bibinfo {author} {\bibfnamefont {B.}~\bibnamefont {Schulte}},
  \bibinfo {author} {\bibfnamefont {F.-U.}\ \bibnamefont {Stein}}, \bibinfo
  {author} {\bibfnamefont {T.}~\bibnamefont {Matsuyama}}, \bibinfo {author}
  {\bibfnamefont {G.}~\bibnamefont {Jotzu}}, \bibinfo {author} {\bibfnamefont
  {G.}~\bibnamefont {Meier}}, \ and\ \bibinfo {author} {\bibfnamefont
  {A.}~\bibnamefont {Cavalleri}},\ }\href {\doibase 10.1038/s41567-019-0698-y}
  {\bibfield  {journal} {\bibinfo  {journal} {Nature Physics}\ }\textbf
  {\bibinfo {volume} {16}},\ \bibinfo {pages} {38} (\bibinfo {year}
  {2020})}\BibitemShut {NoStop}%
\bibitem [{Note3()}]{Note3}%
  \BibitemOpen
  \bibinfo {note} {You might think electrons in solids are not a closed system
  since they interact with the other degrees of freedoms such as phonons.
  However, they are still considered to be approximated with closed systems in
  a short time scale where the typical experiments with strong laser light are
  performed~\cite {OkaReview2019}.}\BibitemShut {Stop}%
\bibitem [{\citenamefont {D'Alessio}\ and\ \citenamefont
  {Rigol}(2014)}]{DAlessio2014}%
  \BibitemOpen
  \bibfield  {author} {\bibinfo {author} {\bibfnamefont {L.}~\bibnamefont
  {D'Alessio}}\ and\ \bibinfo {author} {\bibfnamefont {M.}~\bibnamefont
  {Rigol}},\ }\href {\doibase 10.1103/PhysRevX.4.041048} {\bibfield  {journal}
  {\bibinfo  {journal} {Phys. Rev. X}\ }\textbf {\bibinfo {volume} {4}},\
  \bibinfo {pages} {041048} (\bibinfo {year} {2014})}\BibitemShut {NoStop}%
\bibitem [{\citenamefont {Abanin}\ \emph {et~al.}(2015)\citenamefont {Abanin},
  \citenamefont {De~Roeck},\ and\ \citenamefont {Huveneers}}]{Abanin2015}%
  \BibitemOpen
  \bibfield  {author} {\bibinfo {author} {\bibfnamefont {D.~A.}\ \bibnamefont
  {Abanin}}, \bibinfo {author} {\bibfnamefont {W.}~\bibnamefont {De~Roeck}}, \
  and\ \bibinfo {author} {\bibfnamefont {F.~m.~c.}\ \bibnamefont {Huveneers}},\
  }\href {\doibase 10.1103/PhysRevLett.115.256803} {\bibfield  {journal}
  {\bibinfo  {journal} {Phys. Rev. Lett.}\ }\textbf {\bibinfo {volume} {115}},\
  \bibinfo {pages} {256803} (\bibinfo {year} {2015})}\BibitemShut {NoStop}%
\bibitem [{\citenamefont {Kuwahara}\ \emph {et~al.}(2016)\citenamefont
  {Kuwahara}, \citenamefont {Mori},\ and\ \citenamefont
  {Saito}}]{Kuwahara2016}%
  \BibitemOpen
  \bibfield  {author} {\bibinfo {author} {\bibfnamefont {T.}~\bibnamefont
  {Kuwahara}}, \bibinfo {author} {\bibfnamefont {T.}~\bibnamefont {Mori}}, \
  and\ \bibinfo {author} {\bibfnamefont {K.}~\bibnamefont {Saito}},\ }\href
  {\doibase https://doi.org/10.1016/j.aop.2016.01.012} {\bibfield  {journal}
  {\bibinfo  {journal} {Annals of Physics}\ }\textbf {\bibinfo {volume}
  {367}},\ \bibinfo {pages} {96 } (\bibinfo {year} {2016})}\BibitemShut
  {NoStop}%
\bibitem [{Note4()}]{Note4}%
  \BibitemOpen
  \bibinfo {note} {Strictly speaking, we additionally need the off-resonant
  condition to achieve the prethermalized state. We mention this point in
  Sec.~\ref {Sec:IIIE}.}\BibitemShut {Stop}%
\bibitem [{Note5()}]{Note5}%
  \BibitemOpen
  \bibinfo {note} {There are several schemes for the high-frequency expansion,
  and we take the van Vleck expansion. The term depending on the initial time
  does not appear in the van Vleck expansion while it appears in other
  expansion scheme (e.g. Floquet-Magnus expansion). For detail, see a detailed
  paper about this point~\cite {Mikami2016} or the review articles~\cite
  {OkaReview2019, EckardtRMP2017}.}\BibitemShut {Stop}%
\bibitem [{\citenamefont {Kalthoff}\ \emph {et~al.}(2018)\citenamefont
  {Kalthoff}, \citenamefont {Uhrig},\ and\ \citenamefont
  {Freericks}}]{Kalthoff2018}%
  \BibitemOpen
  \bibfield  {author} {\bibinfo {author} {\bibfnamefont {M.~H.}\ \bibnamefont
  {Kalthoff}}, \bibinfo {author} {\bibfnamefont {G.~S.}\ \bibnamefont {Uhrig}},
  \ and\ \bibinfo {author} {\bibfnamefont {J.~K.}\ \bibnamefont {Freericks}},\
  }\href {\doibase 10.1103/PhysRevB.98.035138} {\bibfield  {journal} {\bibinfo
  {journal} {Phys. Rev. B}\ }\textbf {\bibinfo {volume} {98}},\ \bibinfo
  {pages} {035138} (\bibinfo {year} {2018})}\BibitemShut {NoStop}%
\bibitem [{Note6()}]{Note6}%
  \BibitemOpen
  \bibinfo {note} {From theoretical studies, it is shown that $\tau _p \sim
  e^{\protect \mathcal {O}(\omega )}$ generally holds~\cite {Kuwahara2016}.
  However, the actual lifetime can be modified in solids because there are
  various ingredients not taken into account, such as unoccupied bands,
  impurities, and phonon bath.}\BibitemShut {Stop}%
\bibitem [{\citenamefont {Lignier}\ \emph {et~al.}(2007)\citenamefont
  {Lignier}, \citenamefont {Sias}, \citenamefont {Ciampini}, \citenamefont
  {Singh}, \citenamefont {Zenesini}, \citenamefont {Morsch},\ and\
  \citenamefont {Arimondo}}]{Lignier2007}%
  \BibitemOpen
  \bibfield  {author} {\bibinfo {author} {\bibfnamefont {H.}~\bibnamefont
  {Lignier}}, \bibinfo {author} {\bibfnamefont {C.}~\bibnamefont {Sias}},
  \bibinfo {author} {\bibfnamefont {D.}~\bibnamefont {Ciampini}}, \bibinfo
  {author} {\bibfnamefont {Y.}~\bibnamefont {Singh}}, \bibinfo {author}
  {\bibfnamefont {A.}~\bibnamefont {Zenesini}}, \bibinfo {author}
  {\bibfnamefont {O.}~\bibnamefont {Morsch}}, \ and\ \bibinfo {author}
  {\bibfnamefont {E.}~\bibnamefont {Arimondo}},\ }\href {\doibase
  10.1103/PhysRevLett.99.220403} {\bibfield  {journal} {\bibinfo  {journal}
  {Phys. Rev. Lett.}\ }\textbf {\bibinfo {volume} {99}},\ \bibinfo {pages}
  {220403} (\bibinfo {year} {2007})}\BibitemShut {NoStop}%
\bibitem [{\citenamefont {Ishikawa}\ \emph {et~al.}(2014)\citenamefont
  {Ishikawa}, \citenamefont {Sagae}, \citenamefont {Naitoh}, \citenamefont
  {Kawakami}, \citenamefont {Itoh}, \citenamefont {Yamamoto}, \citenamefont
  {Yakushi}, \citenamefont {Kishida}, \citenamefont {Sasaki}, \citenamefont
  {Ishihara}, \citenamefont {Tanaka}, \citenamefont {Yonemitsu},\ and\
  \citenamefont {Iwai}}]{Ishikawa2014}%
  \BibitemOpen
  \bibfield  {author} {\bibinfo {author} {\bibfnamefont {T.}~\bibnamefont
  {Ishikawa}}, \bibinfo {author} {\bibfnamefont {Y.}~\bibnamefont {Sagae}},
  \bibinfo {author} {\bibfnamefont {Y.}~\bibnamefont {Naitoh}}, \bibinfo
  {author} {\bibfnamefont {Y.}~\bibnamefont {Kawakami}}, \bibinfo {author}
  {\bibfnamefont {H.}~\bibnamefont {Itoh}}, \bibinfo {author} {\bibfnamefont
  {K.}~\bibnamefont {Yamamoto}}, \bibinfo {author} {\bibfnamefont
  {K.}~\bibnamefont {Yakushi}}, \bibinfo {author} {\bibfnamefont
  {H.}~\bibnamefont {Kishida}}, \bibinfo {author} {\bibfnamefont
  {T.}~\bibnamefont {Sasaki}}, \bibinfo {author} {\bibfnamefont
  {S.}~\bibnamefont {Ishihara}}, \bibinfo {author} {\bibfnamefont
  {Y.}~\bibnamefont {Tanaka}}, \bibinfo {author} {\bibfnamefont
  {K.}~\bibnamefont {Yonemitsu}}, \ and\ \bibinfo {author} {\bibfnamefont
  {S.}~\bibnamefont {Iwai}},\ }\href {\doibase 10.1038/ncomms6528} {\bibfield
  {journal} {\bibinfo  {journal} {Nature Communications}\ }\textbf {\bibinfo
  {volume} {5}},\ \bibinfo {pages} {5528} (\bibinfo {year} {2014})}\BibitemShut
  {NoStop}%
\bibitem [{Note7()}]{Note7}%
  \BibitemOpen
  \bibinfo {note} {L. P. Pitaevskii, Sov. Phys. JETP {\protect \bf 12},
  1008–1013 (1961).}\BibitemShut {Stop}%
\bibitem [{\citenamefont {Tada}\ \emph {et~al.}(2009)\citenamefont {Tada},
  \citenamefont {Kawakami},\ and\ \citenamefont {Fujimoto}}]{Tada2009}%
  \BibitemOpen
  \bibfield  {author} {\bibinfo {author} {\bibfnamefont {Y.}~\bibnamefont
  {Tada}}, \bibinfo {author} {\bibfnamefont {N.}~\bibnamefont {Kawakami}}, \
  and\ \bibinfo {author} {\bibfnamefont {S.}~\bibnamefont {Fujimoto}},\ }\href
  {\doibase 10.1088/1367-2630/11/5/055070} {\bibfield  {journal} {\bibinfo
  {journal} {New Journal of Physics}\ }\textbf {\bibinfo {volume} {11}},\
  \bibinfo {pages} {055070} (\bibinfo {year} {2009})}\BibitemShut {NoStop}%
\bibitem [{\citenamefont {Gotlieb}\ \emph {et~al.}(2018)\citenamefont
  {Gotlieb}, \citenamefont {Lin}, \citenamefont {Serbyn}, \citenamefont
  {Zhang}, \citenamefont {Smallwood}, \citenamefont {Jozwiak}, \citenamefont
  {Eisaki}, \citenamefont {Hussain}, \citenamefont {Vishwanath},\ and\
  \citenamefont {Lanzara}}]{Gotlieb2018}%
  \BibitemOpen
  \bibfield  {author} {\bibinfo {author} {\bibfnamefont {K.}~\bibnamefont
  {Gotlieb}}, \bibinfo {author} {\bibfnamefont {C.-Y.}\ \bibnamefont {Lin}},
  \bibinfo {author} {\bibfnamefont {M.}~\bibnamefont {Serbyn}}, \bibinfo
  {author} {\bibfnamefont {W.}~\bibnamefont {Zhang}}, \bibinfo {author}
  {\bibfnamefont {C.~L.}\ \bibnamefont {Smallwood}}, \bibinfo {author}
  {\bibfnamefont {C.}~\bibnamefont {Jozwiak}}, \bibinfo {author} {\bibfnamefont
  {H.}~\bibnamefont {Eisaki}}, \bibinfo {author} {\bibfnamefont
  {Z.}~\bibnamefont {Hussain}}, \bibinfo {author} {\bibfnamefont
  {A.}~\bibnamefont {Vishwanath}}, \ and\ \bibinfo {author} {\bibfnamefont
  {A.}~\bibnamefont {Lanzara}},\ }\href {\doibase 10.1126/science.aao0980}
  {\bibfield  {journal} {\bibinfo  {journal} {Science}\ }\textbf {\bibinfo
  {volume} {362}},\ \bibinfo {pages} {1271} (\bibinfo {year} {2018})},\ \Eprint
  {http://arxiv.org/abs/https://science.sciencemag.org/content/362/6420/1271.full.pdf}
  {https://science.sciencemag.org/content/362/6420/1271.full.pdf} \BibitemShut
  {NoStop}%
\bibitem [{\citenamefont {Fujimoto}(2007)}]{Fujimoto2007}%
  \BibitemOpen
  \bibfield  {author} {\bibinfo {author} {\bibfnamefont {S.}~\bibnamefont
  {Fujimoto}},\ }\href {\doibase 10.1143/JPSJ.76.051008} {\bibfield  {journal}
  {\bibinfo  {journal} {J. Phys. Soc. Jpn.}\ }\textbf {\bibinfo {volume}
  {76}},\ \bibinfo {pages} {051008} (\bibinfo {year} {2007})}\BibitemShut
  {NoStop}%
\bibitem [{\citenamefont {Hashimoto}\ \emph {et~al.}(2014)\citenamefont
  {Hashimoto}, \citenamefont {Vishik}, \citenamefont {He}, \citenamefont
  {Devereaux},\ and\ \citenamefont {Shen}}]{Hashimoto2014}%
  \BibitemOpen
  \bibfield  {author} {\bibinfo {author} {\bibfnamefont {M.}~\bibnamefont
  {Hashimoto}}, \bibinfo {author} {\bibfnamefont {I.~M.}\ \bibnamefont
  {Vishik}}, \bibinfo {author} {\bibfnamefont {R.-H.}\ \bibnamefont {He}},
  \bibinfo {author} {\bibfnamefont {T.~P.}\ \bibnamefont {Devereaux}}, \ and\
  \bibinfo {author} {\bibfnamefont {Z.-X.}\ \bibnamefont {Shen}},\ }\href
  {https://doi.org/10.1038/nphys3009} {\bibfield  {journal} {\bibinfo
  {journal} {Nature Physics}\ }\textbf {\bibinfo {volume} {10}},\ \bibinfo
  {pages} {483} (\bibinfo {year} {2014})}\BibitemShut {NoStop}%
\bibitem [{\citenamefont {Terada}\ \emph {et~al.}(2010)\citenamefont {Terada},
  \citenamefont {Yoshida}, \citenamefont {Takeuchi},\ and\ \citenamefont
  {Shigekawa}}]{Terada2010}%
  \BibitemOpen
  \bibfield  {author} {\bibinfo {author} {\bibfnamefont {Y.}~\bibnamefont
  {Terada}}, \bibinfo {author} {\bibfnamefont {S.}~\bibnamefont {Yoshida}},
  \bibinfo {author} {\bibfnamefont {O.}~\bibnamefont {Takeuchi}}, \ and\
  \bibinfo {author} {\bibfnamefont {H.}~\bibnamefont {Shigekawa}},\ }\href
  {\doibase 10.1038/nphoton.2010.235} {\bibfield  {journal} {\bibinfo
  {journal} {Nature Photonics}\ }\textbf {\bibinfo {volume} {4}},\ \bibinfo
  {pages} {869} (\bibinfo {year} {2010})}\BibitemShut {NoStop}%
\bibitem [{\citenamefont {Yoshida}\ \emph
  {et~al.}(2013{\natexlab{a}})\citenamefont {Yoshida}, \citenamefont {Terada},
  \citenamefont {Yokota}, \citenamefont {Takeuchi}, \citenamefont {Mera},\ and\
  \citenamefont {Shigekawa}}]{Yoshida2013}%
  \BibitemOpen
  \bibfield  {author} {\bibinfo {author} {\bibfnamefont {S.}~\bibnamefont
  {Yoshida}}, \bibinfo {author} {\bibfnamefont {Y.}~\bibnamefont {Terada}},
  \bibinfo {author} {\bibfnamefont {M.}~\bibnamefont {Yokota}}, \bibinfo
  {author} {\bibfnamefont {O.}~\bibnamefont {Takeuchi}}, \bibinfo {author}
  {\bibfnamefont {Y.}~\bibnamefont {Mera}}, \ and\ \bibinfo {author}
  {\bibfnamefont {H.}~\bibnamefont {Shigekawa}},\ }\href {\doibase
  10.7567/apex.6.016601} {\bibfield  {journal} {\bibinfo  {journal} {Applied
  Physics Express}\ }\textbf {\bibinfo {volume} {6}},\ \bibinfo {pages}
  {016601} (\bibinfo {year} {2013}{\natexlab{a}})}\BibitemShut {NoStop}%
\bibitem [{\citenamefont {Pechenezhskiy}\ \emph {et~al.}(2013)\citenamefont
  {Pechenezhskiy}, \citenamefont {Hong}, \citenamefont {Nguyen}, \citenamefont
  {Dahl}, \citenamefont {Carlson}, \citenamefont {Wang},\ and\ \citenamefont
  {Crommie}}]{Pechenezhskiy2013}%
  \BibitemOpen
  \bibfield  {author} {\bibinfo {author} {\bibfnamefont {I.~V.}\ \bibnamefont
  {Pechenezhskiy}}, \bibinfo {author} {\bibfnamefont {X.}~\bibnamefont {Hong}},
  \bibinfo {author} {\bibfnamefont {G.~D.}\ \bibnamefont {Nguyen}}, \bibinfo
  {author} {\bibfnamefont {J.~E.~P.}\ \bibnamefont {Dahl}}, \bibinfo {author}
  {\bibfnamefont {R.~M.~K.}\ \bibnamefont {Carlson}}, \bibinfo {author}
  {\bibfnamefont {F.}~\bibnamefont {Wang}}, \ and\ \bibinfo {author}
  {\bibfnamefont {M.~F.}\ \bibnamefont {Crommie}},\ }\href {\doibase
  10.1103/PhysRevLett.111.126101} {\bibfield  {journal} {\bibinfo  {journal}
  {Phys. Rev. Lett.}\ }\textbf {\bibinfo {volume} {111}},\ \bibinfo {pages}
  {126101} (\bibinfo {year} {2013})}\BibitemShut {NoStop}%
\bibitem [{\citenamefont {Cocker}\ \emph {et~al.}(2016)\citenamefont {Cocker},
  \citenamefont {Peller}, \citenamefont {Yu}, \citenamefont {Repp},\ and\
  \citenamefont {Huber}}]{Cocker2016}%
  \BibitemOpen
  \bibfield  {author} {\bibinfo {author} {\bibfnamefont {T.~L.}\ \bibnamefont
  {Cocker}}, \bibinfo {author} {\bibfnamefont {D.}~\bibnamefont {Peller}},
  \bibinfo {author} {\bibfnamefont {P.}~\bibnamefont {Yu}}, \bibinfo {author}
  {\bibfnamefont {J.}~\bibnamefont {Repp}}, \ and\ \bibinfo {author}
  {\bibfnamefont {R.}~\bibnamefont {Huber}},\ }\href {\doibase
  10.1038/nature19816} {\bibfield  {journal} {\bibinfo  {journal} {Nature}\
  }\textbf {\bibinfo {volume} {539}},\ \bibinfo {pages} {263} (\bibinfo {year}
  {2016})}\BibitemShut {NoStop}%
\bibitem [{\citenamefont {Bukov}\ \emph {et~al.}(2016)\citenamefont {Bukov},
  \citenamefont {Kolodrubetz},\ and\ \citenamefont {Polkovnikov}}]{Bukov2016}%
  \BibitemOpen
  \bibfield  {author} {\bibinfo {author} {\bibfnamefont {M.}~\bibnamefont
  {Bukov}}, \bibinfo {author} {\bibfnamefont {M.}~\bibnamefont {Kolodrubetz}},
  \ and\ \bibinfo {author} {\bibfnamefont {A.}~\bibnamefont {Polkovnikov}},\
  }\href {\doibase 10.1103/PhysRevLett.116.125301} {\bibfield  {journal}
  {\bibinfo  {journal} {Phys. Rev. Lett.}\ }\textbf {\bibinfo {volume} {116}},\
  \bibinfo {pages} {125301} (\bibinfo {year} {2016})}\BibitemShut {NoStop}%
\bibitem [{\citenamefont {Mizuta}\ \emph {et~al.}(2019)\citenamefont {Mizuta},
  \citenamefont {Takasan},\ and\ \citenamefont {Kawakami}}]{Mizuta2019}%
  \BibitemOpen
  \bibfield  {author} {\bibinfo {author} {\bibfnamefont {K.}~\bibnamefont
  {Mizuta}}, \bibinfo {author} {\bibfnamefont {K.}~\bibnamefont {Takasan}}, \
  and\ \bibinfo {author} {\bibfnamefont {N.}~\bibnamefont {Kawakami}},\ }\href
  {\doibase 10.1103/PhysRevB.100.020301} {\bibfield  {journal} {\bibinfo
  {journal} {Phys. Rev. B}\ }\textbf {\bibinfo {volume} {100}},\ \bibinfo
  {pages} {020301} (\bibinfo {year} {2019})}\BibitemShut {NoStop}%
\bibitem [{\citenamefont {Chono}\ \emph {et~al.}(2020)\citenamefont {Chono},
  \citenamefont {Takasan},\ and\ \citenamefont {Yanase}}]{Chono2019}%
  \BibitemOpen
  \bibfield  {author} {\bibinfo {author} {\bibfnamefont {H.}~\bibnamefont
  {Chono}}, \bibinfo {author} {\bibfnamefont {K.}~\bibnamefont {Takasan}}, \
  and\ \bibinfo {author} {\bibfnamefont {Y.}~\bibnamefont {Yanase}},\ }\href
  {\doibase 10.1103/PhysRevB.102.174508} {\bibfield  {journal} {\bibinfo
  {journal} {Phys. Rev. B}\ }\textbf {\bibinfo {volume} {102}},\ \bibinfo
  {pages} {174508} (\bibinfo {year} {2020})}\BibitemShut {NoStop}%
\bibitem [{\citenamefont {Wadley}\ \emph {et~al.}(2016)\citenamefont {Wadley},
  \citenamefont {Howells}, \citenamefont {{\v Z}elezn{\'y}}, \citenamefont
  {Andrews}, \citenamefont {Hills}, \citenamefont {Campion}, \citenamefont
  {Nov{\'a}k}, \citenamefont {Olejn{\'\i}k}, \citenamefont {Maccherozzi},
  \citenamefont {Dhesi}, \citenamefont {Martin}, \citenamefont {Wagner},
  \citenamefont {Wunderlich}, \citenamefont {Freimuth}, \citenamefont
  {Mokrousov}, \citenamefont {Kune{\v s}}, \citenamefont {Chauhan},
  \citenamefont {Grzybowski}, \citenamefont {Rushforth}, \citenamefont
  {Edmonds}, \citenamefont {Gallagher},\ and\ \citenamefont
  {Jungwirth}}]{Wadley2016}%
  \BibitemOpen
  \bibfield  {author} {\bibinfo {author} {\bibfnamefont {P.}~\bibnamefont
  {Wadley}}, \bibinfo {author} {\bibfnamefont {B.}~\bibnamefont {Howells}},
  \bibinfo {author} {\bibfnamefont {J.}~\bibnamefont {{\v Z}elezn{\'y}}},
  \bibinfo {author} {\bibfnamefont {C.}~\bibnamefont {Andrews}}, \bibinfo
  {author} {\bibfnamefont {V.}~\bibnamefont {Hills}}, \bibinfo {author}
  {\bibfnamefont {R.~P.}\ \bibnamefont {Campion}}, \bibinfo {author}
  {\bibfnamefont {V.}~\bibnamefont {Nov{\'a}k}}, \bibinfo {author}
  {\bibfnamefont {K.}~\bibnamefont {Olejn{\'\i}k}}, \bibinfo {author}
  {\bibfnamefont {F.}~\bibnamefont {Maccherozzi}}, \bibinfo {author}
  {\bibfnamefont {S.~S.}\ \bibnamefont {Dhesi}}, \bibinfo {author}
  {\bibfnamefont {S.~Y.}\ \bibnamefont {Martin}}, \bibinfo {author}
  {\bibfnamefont {T.}~\bibnamefont {Wagner}}, \bibinfo {author} {\bibfnamefont
  {J.}~\bibnamefont {Wunderlich}}, \bibinfo {author} {\bibfnamefont
  {F.}~\bibnamefont {Freimuth}}, \bibinfo {author} {\bibfnamefont
  {Y.}~\bibnamefont {Mokrousov}}, \bibinfo {author} {\bibfnamefont
  {J.}~\bibnamefont {Kune{\v s}}}, \bibinfo {author} {\bibfnamefont {J.~S.}\
  \bibnamefont {Chauhan}}, \bibinfo {author} {\bibfnamefont {M.~J.}\
  \bibnamefont {Grzybowski}}, \bibinfo {author} {\bibfnamefont {A.~W.}\
  \bibnamefont {Rushforth}}, \bibinfo {author} {\bibfnamefont {K.~W.}\
  \bibnamefont {Edmonds}}, \bibinfo {author} {\bibfnamefont {B.~L.}\
  \bibnamefont {Gallagher}}, \ and\ \bibinfo {author} {\bibfnamefont
  {T.}~\bibnamefont {Jungwirth}},\ }\href {\doibase 10.1126/science.aab1031}
  {\bibfield  {journal} {\bibinfo  {journal} {Science}\ }\textbf {\bibinfo
  {volume} {351}},\ \bibinfo {pages} {587} (\bibinfo {year} {2016})},\ \Eprint
  {http://arxiv.org/abs/https://science.sciencemag.org/content/351/6273/587.full.pdf}
  {https://science.sciencemag.org/content/351/6273/587.full.pdf} \BibitemShut
  {NoStop}%
\bibitem [{\citenamefont {Fischer}\ \emph {et~al.}(2011)\citenamefont
  {Fischer}, \citenamefont {Loder},\ and\ \citenamefont
  {Sigrist}}]{Fischer_LNCS2011}%
  \BibitemOpen
  \bibfield  {author} {\bibinfo {author} {\bibfnamefont {M.~H.}\ \bibnamefont
  {Fischer}}, \bibinfo {author} {\bibfnamefont {F.}~\bibnamefont {Loder}}, \
  and\ \bibinfo {author} {\bibfnamefont {M.}~\bibnamefont {Sigrist}},\ }\href
  {\doibase 10.1103/PhysRevB.84.184533} {\bibfield  {journal} {\bibinfo
  {journal} {Phys. Rev. B}\ }\textbf {\bibinfo {volume} {84}},\ \bibinfo
  {pages} {184533} (\bibinfo {year} {2011})}\BibitemShut {NoStop}%
\bibitem [{\citenamefont {Maruyama}\ \emph {et~al.}(2012)\citenamefont
  {Maruyama}, \citenamefont {Sigrist},\ and\ \citenamefont
  {Yanase}}]{Maruyama_LISB_SpinOrb_JPSJ12}%
  \BibitemOpen
  \bibfield  {author} {\bibinfo {author} {\bibfnamefont {D.}~\bibnamefont
  {Maruyama}}, \bibinfo {author} {\bibfnamefont {M.}~\bibnamefont {Sigrist}}, \
  and\ \bibinfo {author} {\bibfnamefont {Y.}~\bibnamefont {Yanase}},\ }\href
  {\doibase 10.1143/JPSJ.81.034702} {\bibfield  {journal} {\bibinfo  {journal}
  {Journal of the Physical Society of Japan}\ }\textbf {\bibinfo {volume}
  {81}},\ \bibinfo {pages} {034702} (\bibinfo {year} {2012})},\ \Eprint
  {http://arxiv.org/abs/https://doi.org/10.1143/JPSJ.81.034702}
  {https://doi.org/10.1143/JPSJ.81.034702} \BibitemShut {NoStop}%
\bibitem [{\citenamefont {Yoshida}\ \emph {et~al.}(2012)\citenamefont
  {Yoshida}, \citenamefont {Sigrist},\ and\ \citenamefont
  {Yanase}}]{superlattice_Yanase_12}%
  \BibitemOpen
  \bibfield  {author} {\bibinfo {author} {\bibfnamefont {T.}~\bibnamefont
  {Yoshida}}, \bibinfo {author} {\bibfnamefont {M.}~\bibnamefont {Sigrist}}, \
  and\ \bibinfo {author} {\bibfnamefont {Y.}~\bibnamefont {Yanase}},\ }\href
  {\doibase 10.1103/PhysRevB.86.134514} {\bibfield  {journal} {\bibinfo
  {journal} {Phys. Rev. B}\ }\textbf {\bibinfo {volume} {86}},\ \bibinfo
  {pages} {134514} (\bibinfo {year} {2012})}\BibitemShut {NoStop}%
\bibitem [{\citenamefont {Yoshida}\ \emph
  {et~al.}(2013{\natexlab{b}})\citenamefont {Yoshida}, \citenamefont
  {Sigrist},\ and\ \citenamefont {Yanase}}]{Yoshida_complex_stripe2013}%
  \BibitemOpen
  \bibfield  {author} {\bibinfo {author} {\bibfnamefont {T.}~\bibnamefont
  {Yoshida}}, \bibinfo {author} {\bibfnamefont {M.}~\bibnamefont {Sigrist}}, \
  and\ \bibinfo {author} {\bibfnamefont {Y.}~\bibnamefont {Yanase}},\ }\href
  {\doibase 10.7566/JPSJ.82.074714} {\bibfield  {journal} {\bibinfo  {journal}
  {Journal of the Physical Society of Japan}\ }\textbf {\bibinfo {volume}
  {82}},\ \bibinfo {pages} {074714} (\bibinfo {year} {2013}{\natexlab{b}})},\
  \Eprint {http://arxiv.org/abs/https://doi.org/10.7566/JPSJ.82.074714}
  {https://doi.org/10.7566/JPSJ.82.074714} \BibitemShut {NoStop}%
\bibitem [{\citenamefont {Khim}\ \emph {et~al.}(2021)\citenamefont {Khim},
  \citenamefont {Landaeta}, \citenamefont {Banda}, \citenamefont {Bannor},
  \citenamefont {Brando}, \citenamefont {Brydon}, \citenamefont {Hafner},
  \citenamefont {Küchler}, \citenamefont {Cardoso-Gil}, \citenamefont
  {Stockert}, \citenamefont {Mackenzie}, \citenamefont {Agterberg},
  \citenamefont {Geibel},\ and\ \citenamefont {Hassinger}}]{CeRh2As2}%
  \BibitemOpen
  \bibfield  {author} {\bibinfo {author} {\bibfnamefont {S.}~\bibnamefont
  {Khim}}, \bibinfo {author} {\bibfnamefont {J.~F.}\ \bibnamefont {Landaeta}},
  \bibinfo {author} {\bibfnamefont {J.}~\bibnamefont {Banda}}, \bibinfo
  {author} {\bibfnamefont {N.}~\bibnamefont {Bannor}}, \bibinfo {author}
  {\bibfnamefont {M.}~\bibnamefont {Brando}}, \bibinfo {author} {\bibfnamefont
  {P.~M.~R.}\ \bibnamefont {Brydon}}, \bibinfo {author} {\bibfnamefont
  {D.}~\bibnamefont {Hafner}}, \bibinfo {author} {\bibfnamefont
  {R.}~\bibnamefont {Küchler}}, \bibinfo {author} {\bibfnamefont
  {R.}~\bibnamefont {Cardoso-Gil}}, \bibinfo {author} {\bibfnamefont
  {U.}~\bibnamefont {Stockert}}, \bibinfo {author} {\bibfnamefont {A.~P.}\
  \bibnamefont {Mackenzie}}, \bibinfo {author} {\bibfnamefont {D.~F.}\
  \bibnamefont {Agterberg}}, \bibinfo {author} {\bibfnamefont {C.}~\bibnamefont
  {Geibel}}, \ and\ \bibinfo {author} {\bibfnamefont {E.}~\bibnamefont
  {Hassinger}},\ }\href {\doibase 10.1126/science.abe7518} {\bibfield
  {journal} {\bibinfo  {journal} {Science}\ }\textbf {\bibinfo {volume}
  {373}},\ \bibinfo {pages} {1012} (\bibinfo {year} {2021})}\BibitemShut
  {NoStop}%
\bibitem [{\citenamefont {Goh}\ \emph {et~al.}(2012)\citenamefont {Goh},
  \citenamefont {Mizukami}, \citenamefont {Shishido}, \citenamefont {Watanabe},
  \citenamefont {Yasumoto}, \citenamefont {Shimozawa}, \citenamefont
  {Yamashita}, \citenamefont {Terashima}, \citenamefont {Yanase}, \citenamefont
  {Shibauchi}, \citenamefont {Buzdin},\ and\ \citenamefont
  {Matsuda}}]{Goh_superlattice12}%
  \BibitemOpen
  \bibfield  {author} {\bibinfo {author} {\bibfnamefont {S.~K.}\ \bibnamefont
  {Goh}}, \bibinfo {author} {\bibfnamefont {Y.}~\bibnamefont {Mizukami}},
  \bibinfo {author} {\bibfnamefont {H.}~\bibnamefont {Shishido}}, \bibinfo
  {author} {\bibfnamefont {D.}~\bibnamefont {Watanabe}}, \bibinfo {author}
  {\bibfnamefont {S.}~\bibnamefont {Yasumoto}}, \bibinfo {author}
  {\bibfnamefont {M.}~\bibnamefont {Shimozawa}}, \bibinfo {author}
  {\bibfnamefont {M.}~\bibnamefont {Yamashita}}, \bibinfo {author}
  {\bibfnamefont {T.}~\bibnamefont {Terashima}}, \bibinfo {author}
  {\bibfnamefont {Y.}~\bibnamefont {Yanase}}, \bibinfo {author} {\bibfnamefont
  {T.}~\bibnamefont {Shibauchi}}, \bibinfo {author} {\bibfnamefont {A.~I.}\
  \bibnamefont {Buzdin}}, \ and\ \bibinfo {author} {\bibfnamefont
  {Y.}~\bibnamefont {Matsuda}},\ }\href {\doibase
  10.1103/PhysRevLett.109.157006} {\bibfield  {journal} {\bibinfo  {journal}
  {Phys. Rev. Lett.}\ }\textbf {\bibinfo {volume} {109}},\ \bibinfo {pages}
  {157006} (\bibinfo {year} {2012})}\BibitemShut {NoStop}%
\bibitem [{\citenamefont {Shimozawa}\ \emph {et~al.}(2014)\citenamefont
  {Shimozawa}, \citenamefont {Goh}, \citenamefont {Endo}, \citenamefont
  {Kobayashi}, \citenamefont {Watashige}, \citenamefont {Mizukami},
  \citenamefont {Ikeda}, \citenamefont {Shishido}, \citenamefont {Yanase},
  \citenamefont {Terashima}, \citenamefont {Shibauchi},\ and\ \citenamefont
  {Matsuda}}]{Shimozawa_superlattice_PRL14}%
  \BibitemOpen
  \bibfield  {author} {\bibinfo {author} {\bibfnamefont {M.}~\bibnamefont
  {Shimozawa}}, \bibinfo {author} {\bibfnamefont {S.~K.}\ \bibnamefont {Goh}},
  \bibinfo {author} {\bibfnamefont {R.}~\bibnamefont {Endo}}, \bibinfo {author}
  {\bibfnamefont {R.}~\bibnamefont {Kobayashi}}, \bibinfo {author}
  {\bibfnamefont {T.}~\bibnamefont {Watashige}}, \bibinfo {author}
  {\bibfnamefont {Y.}~\bibnamefont {Mizukami}}, \bibinfo {author}
  {\bibfnamefont {H.}~\bibnamefont {Ikeda}}, \bibinfo {author} {\bibfnamefont
  {H.}~\bibnamefont {Shishido}}, \bibinfo {author} {\bibfnamefont
  {Y.}~\bibnamefont {Yanase}}, \bibinfo {author} {\bibfnamefont
  {T.}~\bibnamefont {Terashima}}, \bibinfo {author} {\bibfnamefont
  {T.}~\bibnamefont {Shibauchi}}, \ and\ \bibinfo {author} {\bibfnamefont
  {Y.}~\bibnamefont {Matsuda}},\ }\href {\doibase
  10.1103/PhysRevLett.112.156404} {\bibfield  {journal} {\bibinfo  {journal}
  {Phys. Rev. Lett.}\ }\textbf {\bibinfo {volume} {112}},\ \bibinfo {pages}
  {156404} (\bibinfo {year} {2014})}\BibitemShut {NoStop}%
\bibitem [{\citenamefont {Zheliuk}\ \emph {et~al.}(2019)\citenamefont
  {Zheliuk}, \citenamefont {Lu}, \citenamefont {Chen}, \citenamefont {Yumin},
  \citenamefont {Golightly},\ and\ \citenamefont {Ye}}]{Zheliuk2019}%
  \BibitemOpen
  \bibfield  {author} {\bibinfo {author} {\bibfnamefont {O.}~\bibnamefont
  {Zheliuk}}, \bibinfo {author} {\bibfnamefont {J.~M.}\ \bibnamefont {Lu}},
  \bibinfo {author} {\bibfnamefont {Q.~H.}\ \bibnamefont {Chen}}, \bibinfo
  {author} {\bibfnamefont {A.~A.~E.}\ \bibnamefont {Yumin}}, \bibinfo {author}
  {\bibfnamefont {S.}~\bibnamefont {Golightly}}, \ and\ \bibinfo {author}
  {\bibfnamefont {J.~T.}\ \bibnamefont {Ye}},\ }\href {\doibase
  10.1038/s41565-019-0564-1} {\bibfield  {journal} {\bibinfo  {journal} {Nature
  Nanotechnology}\ }\textbf {\bibinfo {volume} {14}},\ \bibinfo {pages} {1123}
  (\bibinfo {year} {2019})}\BibitemShut {NoStop}%
\bibitem [{\citenamefont {Zhang}\ \emph {et~al.}(2014)\citenamefont {Zhang},
  \citenamefont {Liu}, \citenamefont {Luo}, \citenamefont {Freeman},\ and\
  \citenamefont {Zunger}}]{Zhang_hidden_polarization2014}%
  \BibitemOpen
  \bibfield  {author} {\bibinfo {author} {\bibfnamefont {X.}~\bibnamefont
  {Zhang}}, \bibinfo {author} {\bibfnamefont {Q.}~\bibnamefont {Liu}}, \bibinfo
  {author} {\bibfnamefont {J.-W.}\ \bibnamefont {Luo}}, \bibinfo {author}
  {\bibfnamefont {A.~J.}\ \bibnamefont {Freeman}}, \ and\ \bibinfo {author}
  {\bibfnamefont {A.}~\bibnamefont {Zunger}},\ }\href {\doibase
  10.1038/nphys2933} {\bibfield  {journal} {\bibinfo  {journal} {Nature
  Physics}\ }\textbf {\bibinfo {volume} {10}},\ \bibinfo {pages} {387}
  (\bibinfo {year} {2014})}\BibitemShut {NoStop}%
\bibitem [{\citenamefont {Riley}\ \emph {et~al.}(2014)\citenamefont {Riley},
  \citenamefont {Mazzola}, \citenamefont {Dendzik}, \citenamefont {Michiardi},
  \citenamefont {Takayama}, \citenamefont {Bawden}, \citenamefont
  {Graner{\o}d}, \citenamefont {Leandersson}, \citenamefont {Balasubramanian},
  \citenamefont {Hoesch}, \citenamefont {Kim}, \citenamefont {Takagi},
  \citenamefont {Meevasana}, \citenamefont {Hofmann}, \citenamefont {Bahramy},
  \citenamefont {Wells},\ and\ \citenamefont {King}}]{Riley2014}%
  \BibitemOpen
  \bibfield  {author} {\bibinfo {author} {\bibfnamefont {J.~M.}\ \bibnamefont
  {Riley}}, \bibinfo {author} {\bibfnamefont {F.}~\bibnamefont {Mazzola}},
  \bibinfo {author} {\bibfnamefont {M.}~\bibnamefont {Dendzik}}, \bibinfo
  {author} {\bibfnamefont {M.}~\bibnamefont {Michiardi}}, \bibinfo {author}
  {\bibfnamefont {T.}~\bibnamefont {Takayama}}, \bibinfo {author}
  {\bibfnamefont {L.}~\bibnamefont {Bawden}}, \bibinfo {author} {\bibfnamefont
  {C.}~\bibnamefont {Graner{\o}d}}, \bibinfo {author} {\bibfnamefont
  {M.}~\bibnamefont {Leandersson}}, \bibinfo {author} {\bibfnamefont
  {T.}~\bibnamefont {Balasubramanian}}, \bibinfo {author} {\bibfnamefont
  {M.}~\bibnamefont {Hoesch}}, \bibinfo {author} {\bibfnamefont {T.~K.}\
  \bibnamefont {Kim}}, \bibinfo {author} {\bibfnamefont {H.}~\bibnamefont
  {Takagi}}, \bibinfo {author} {\bibfnamefont {W.}~\bibnamefont {Meevasana}},
  \bibinfo {author} {\bibfnamefont {P.}~\bibnamefont {Hofmann}}, \bibinfo
  {author} {\bibfnamefont {M.~S.}\ \bibnamefont {Bahramy}}, \bibinfo {author}
  {\bibfnamefont {J.~W.}\ \bibnamefont {Wells}}, \ and\ \bibinfo {author}
  {\bibfnamefont {P.~D.~C.}\ \bibnamefont {King}},\ }\href {\doibase
  10.1038/nphys3105} {\bibfield  {journal} {\bibinfo  {journal} {Nature
  Physics}\ }\textbf {\bibinfo {volume} {10}},\ \bibinfo {pages} {835}
  (\bibinfo {year} {2014})}\BibitemShut {NoStop}%
\bibitem [{\citenamefont {Santos-Cottin}\ \emph {et~al.}(2016)\citenamefont
  {Santos-Cottin}, \citenamefont {Casula}, \citenamefont {Lantz}, \citenamefont
  {Klein}, \citenamefont {Petaccia}, \citenamefont {Le~F{\`e}vre},
  \citenamefont {Bertran}, \citenamefont {Papalazarou}, \citenamefont {Marsi},\
  and\ \citenamefont {Gauzzi}}]{Santos-Cottin2016}%
  \BibitemOpen
  \bibfield  {author} {\bibinfo {author} {\bibfnamefont {D.}~\bibnamefont
  {Santos-Cottin}}, \bibinfo {author} {\bibfnamefont {M.}~\bibnamefont
  {Casula}}, \bibinfo {author} {\bibfnamefont {G.}~\bibnamefont {Lantz}},
  \bibinfo {author} {\bibfnamefont {Y.}~\bibnamefont {Klein}}, \bibinfo
  {author} {\bibfnamefont {L.}~\bibnamefont {Petaccia}}, \bibinfo {author}
  {\bibfnamefont {P.}~\bibnamefont {Le~F{\`e}vre}}, \bibinfo {author}
  {\bibfnamefont {F.}~\bibnamefont {Bertran}}, \bibinfo {author} {\bibfnamefont
  {E.}~\bibnamefont {Papalazarou}}, \bibinfo {author} {\bibfnamefont
  {M.}~\bibnamefont {Marsi}}, \ and\ \bibinfo {author} {\bibfnamefont
  {A.}~\bibnamefont {Gauzzi}},\ }\href {\doibase 10.1038/ncomms11258}
  {\bibfield  {journal} {\bibinfo  {journal} {Nature Communications}\ }\textbf
  {\bibinfo {volume} {7}},\ \bibinfo {pages} {11258} (\bibinfo {year}
  {2016})}\BibitemShut {NoStop}%
\bibitem [{\citenamefont {Gehlmann}\ \emph {et~al.}(2016)\citenamefont
  {Gehlmann}, \citenamefont {Aguilera}, \citenamefont {Bihlmayer},
  \citenamefont {M{\l}y{\'{n}}czak}, \citenamefont {Eschbach}, \citenamefont
  {D{\"o}ring}, \citenamefont {Gospodari{\v{c}}}, \citenamefont {Cramm},
  \citenamefont {Kardyna{\l}}, \citenamefont {Plucinski}, \citenamefont
  {Bl{\"u}gel},\ and\ \citenamefont {Schneider}}]{Gehlmann2016}%
  \BibitemOpen
  \bibfield  {author} {\bibinfo {author} {\bibfnamefont {M.}~\bibnamefont
  {Gehlmann}}, \bibinfo {author} {\bibfnamefont {I.}~\bibnamefont {Aguilera}},
  \bibinfo {author} {\bibfnamefont {G.}~\bibnamefont {Bihlmayer}}, \bibinfo
  {author} {\bibfnamefont {E.}~\bibnamefont {M{\l}y{\'{n}}czak}}, \bibinfo
  {author} {\bibfnamefont {M.}~\bibnamefont {Eschbach}}, \bibinfo {author}
  {\bibfnamefont {S.}~\bibnamefont {D{\"o}ring}}, \bibinfo {author}
  {\bibfnamefont {P.}~\bibnamefont {Gospodari{\v{c}}}}, \bibinfo {author}
  {\bibfnamefont {S.}~\bibnamefont {Cramm}}, \bibinfo {author} {\bibfnamefont
  {B.}~\bibnamefont {Kardyna{\l}}}, \bibinfo {author} {\bibfnamefont
  {L.}~\bibnamefont {Plucinski}}, \bibinfo {author} {\bibfnamefont
  {S.}~\bibnamefont {Bl{\"u}gel}}, \ and\ \bibinfo {author} {\bibfnamefont
  {C.~M.}\ \bibnamefont {Schneider}},\ }\href {\doibase 10.1038/srep26197}
  {\bibfield  {journal} {\bibinfo  {journal} {Scientific Reports}\ }\textbf
  {\bibinfo {volume} {6}},\ \bibinfo {pages} {26197} (\bibinfo {year}
  {2016})}\BibitemShut {NoStop}%
\bibitem [{\citenamefont {Wu}\ \emph {et~al.}(2017)\citenamefont {Wu},
  \citenamefont {Sumida}, \citenamefont {Miyamoto}, \citenamefont {Taguchi},
  \citenamefont {Yoshikawa}, \citenamefont {Kimura}, \citenamefont {Ueda},
  \citenamefont {Arita}, \citenamefont {Nagao}, \citenamefont {Watauchi},
  \citenamefont {Tanaka},\ and\ \citenamefont {Okuda}}]{Wu2017}%
  \BibitemOpen
  \bibfield  {author} {\bibinfo {author} {\bibfnamefont {S.-L.}\ \bibnamefont
  {Wu}}, \bibinfo {author} {\bibfnamefont {K.}~\bibnamefont {Sumida}}, \bibinfo
  {author} {\bibfnamefont {K.}~\bibnamefont {Miyamoto}}, \bibinfo {author}
  {\bibfnamefont {K.}~\bibnamefont {Taguchi}}, \bibinfo {author} {\bibfnamefont
  {T.}~\bibnamefont {Yoshikawa}}, \bibinfo {author} {\bibfnamefont
  {A.}~\bibnamefont {Kimura}}, \bibinfo {author} {\bibfnamefont
  {Y.}~\bibnamefont {Ueda}}, \bibinfo {author} {\bibfnamefont {M.}~\bibnamefont
  {Arita}}, \bibinfo {author} {\bibfnamefont {M.}~\bibnamefont {Nagao}},
  \bibinfo {author} {\bibfnamefont {S.}~\bibnamefont {Watauchi}}, \bibinfo
  {author} {\bibfnamefont {I.}~\bibnamefont {Tanaka}}, \ and\ \bibinfo {author}
  {\bibfnamefont {T.}~\bibnamefont {Okuda}},\ }\href {\doibase
  10.1038/s41467-017-02058-2} {\bibfield  {journal} {\bibinfo  {journal}
  {Nature Communications}\ }\textbf {\bibinfo {volume} {8}},\ \bibinfo {pages}
  {1919} (\bibinfo {year} {2017})}\BibitemShut {NoStop}%
\bibitem [{\citenamefont {Jones}\ \emph {et~al.}(2014)\citenamefont {Jones},
  \citenamefont {Yu}, \citenamefont {Ross}, \citenamefont {Klement},
  \citenamefont {Ghimire}, \citenamefont {Yan}, \citenamefont {Mandrus},
  \citenamefont {Yao},\ and\ \citenamefont {Xu}}]{Jones2014}%
  \BibitemOpen
  \bibfield  {author} {\bibinfo {author} {\bibfnamefont {A.~M.}\ \bibnamefont
  {Jones}}, \bibinfo {author} {\bibfnamefont {H.}~\bibnamefont {Yu}}, \bibinfo
  {author} {\bibfnamefont {J.~S.}\ \bibnamefont {Ross}}, \bibinfo {author}
  {\bibfnamefont {P.}~\bibnamefont {Klement}}, \bibinfo {author} {\bibfnamefont
  {N.~J.}\ \bibnamefont {Ghimire}}, \bibinfo {author} {\bibfnamefont
  {J.}~\bibnamefont {Yan}}, \bibinfo {author} {\bibfnamefont {D.~G.}\
  \bibnamefont {Mandrus}}, \bibinfo {author} {\bibfnamefont {W.}~\bibnamefont
  {Yao}}, \ and\ \bibinfo {author} {\bibfnamefont {X.}~\bibnamefont {Xu}},\
  }\href {\doibase 10.1038/nphys2848} {\bibfield  {journal} {\bibinfo
  {journal} {Nature Physics}\ }\textbf {\bibinfo {volume} {10}},\ \bibinfo
  {pages} {130} (\bibinfo {year} {2014})}\BibitemShut {NoStop}%
\bibitem [{\citenamefont {Sumita}\ and\ \citenamefont
  {Yanase}(2016)}]{Sumita_multipoleSC2016}%
  \BibitemOpen
  \bibfield  {author} {\bibinfo {author} {\bibfnamefont {S.}~\bibnamefont
  {Sumita}}\ and\ \bibinfo {author} {\bibfnamefont {Y.}~\bibnamefont
  {Yanase}},\ }\href {\doibase 10.1103/PhysRevB.93.224507} {\bibfield
  {journal} {\bibinfo  {journal} {Phys. Rev. B}\ }\textbf {\bibinfo {volume}
  {93}},\ \bibinfo {pages} {224507} (\bibinfo {year} {2016})}\BibitemShut
  {NoStop}%
\bibitem [{\citenamefont {Nakamura}\ and\ \citenamefont
  {Yanase}(2017)}]{Nakamura_bilayerTMD2017}%
  \BibitemOpen
  \bibfield  {author} {\bibinfo {author} {\bibfnamefont {Y.}~\bibnamefont
  {Nakamura}}\ and\ \bibinfo {author} {\bibfnamefont {Y.}~\bibnamefont
  {Yanase}},\ }\href {\doibase 10.1103/PhysRevB.96.054501} {\bibfield
  {journal} {\bibinfo  {journal} {Phys. Rev. B}\ }\textbf {\bibinfo {volume}
  {96}},\ \bibinfo {pages} {054501} (\bibinfo {year} {2017})}\BibitemShut
  {NoStop}%
\bibitem [{\citenamefont {Cavanagh}\ \emph {et~al.}(2021)\citenamefont
  {Cavanagh}, \citenamefont {Shishidou}, \citenamefont {Weinert}, \citenamefont
  {Brydon},\ and\ \citenamefont {Agterberg}}]{cavanagh2021nonsymmorphic}%
  \BibitemOpen
  \bibfield  {author} {\bibinfo {author} {\bibfnamefont {D.~C.}\ \bibnamefont
  {Cavanagh}}, \bibinfo {author} {\bibfnamefont {T.}~\bibnamefont {Shishidou}},
  \bibinfo {author} {\bibfnamefont {M.}~\bibnamefont {Weinert}}, \bibinfo
  {author} {\bibfnamefont {P.~M.~R.}\ \bibnamefont {Brydon}}, \ and\ \bibinfo
  {author} {\bibfnamefont {D.~F.}\ \bibnamefont {Agterberg}},\ }\href@noop {}
  {\  (\bibinfo {year} {2021})},\ \Eprint {http://arxiv.org/abs/2106.02698}
  {arXiv:2106.02698 [cond-mat.supr-con]} \BibitemShut {NoStop}%
\bibitem [{\citenamefont {Niu}\ \emph {et~al.}(2017)\citenamefont {Niu},
  \citenamefont {Yu}, \citenamefont {Yip}, \citenamefont {Lim}, \citenamefont
  {Kotegawa}, \citenamefont {Matsuoka}, \citenamefont {Sugawara}, \citenamefont
  {Tou}, \citenamefont {Yanase},\ and\ \citenamefont {Goh}}]{Niu_CrAs2017}%
  \BibitemOpen
  \bibfield  {author} {\bibinfo {author} {\bibfnamefont {Q.}~\bibnamefont
  {Niu}}, \bibinfo {author} {\bibfnamefont {W.~C.}\ \bibnamefont {Yu}},
  \bibinfo {author} {\bibfnamefont {K.~Y.}\ \bibnamefont {Yip}}, \bibinfo
  {author} {\bibfnamefont {Z.~L.}\ \bibnamefont {Lim}}, \bibinfo {author}
  {\bibfnamefont {H.}~\bibnamefont {Kotegawa}}, \bibinfo {author}
  {\bibfnamefont {E.}~\bibnamefont {Matsuoka}}, \bibinfo {author}
  {\bibfnamefont {H.}~\bibnamefont {Sugawara}}, \bibinfo {author}
  {\bibfnamefont {H.}~\bibnamefont {Tou}}, \bibinfo {author} {\bibfnamefont
  {Y.}~\bibnamefont {Yanase}}, \ and\ \bibinfo {author} {\bibfnamefont {S.~K.}\
  \bibnamefont {Goh}},\ }\href {\doibase 10.1038/ncomms15358} {\bibfield
  {journal} {\bibinfo  {journal} {Nature Communications}\ }\textbf {\bibinfo
  {volume} {8}},\ \bibinfo {pages} {15358} (\bibinfo {year}
  {2017})}\BibitemShut {NoStop}%
\bibitem [{\citenamefont {Ishizuka}\ and\ \citenamefont
  {Yanase}(2018)}]{Ishizuka_Sr2IrO42018}%
  \BibitemOpen
  \bibfield  {author} {\bibinfo {author} {\bibfnamefont {J.}~\bibnamefont
  {Ishizuka}}\ and\ \bibinfo {author} {\bibfnamefont {Y.}~\bibnamefont
  {Yanase}},\ }\href {\doibase 10.1103/PhysRevB.98.224510} {\bibfield
  {journal} {\bibinfo  {journal} {Phys. Rev. B}\ }\textbf {\bibinfo {volume}
  {98}},\ \bibinfo {pages} {224510} (\bibinfo {year} {2018})}\BibitemShut
  {NoStop}%
\bibitem [{\citenamefont {Bradley}\ and\ \citenamefont
  {Cracknell}(1972)}]{Bradley_book}%
  \BibitemOpen
  \bibfield  {author} {\bibinfo {author} {\bibfnamefont {C.~J.}\ \bibnamefont
  {Bradley}}\ and\ \bibinfo {author} {\bibfnamefont {A.~P.}\ \bibnamefont
  {Cracknell}},\ }\href@noop {} {\emph {\bibinfo {title} {The Mathematical
  Theory of Symmetry in Solids}}}\ (\bibinfo  {publisher} {Oxford University
  Press},\ \bibinfo {year} {1972})\BibitemShut {NoStop}%
\bibitem [{\citenamefont {Akashi}\ \emph {et~al.}(2017)\citenamefont {Akashi},
  \citenamefont {Iida}, \citenamefont {Yamamoto},\ and\ \citenamefont
  {Yoshizawa}}]{Akashi2017}%
  \BibitemOpen
  \bibfield  {author} {\bibinfo {author} {\bibfnamefont {R.}~\bibnamefont
  {Akashi}}, \bibinfo {author} {\bibfnamefont {Y.}~\bibnamefont {Iida}},
  \bibinfo {author} {\bibfnamefont {K.}~\bibnamefont {Yamamoto}}, \ and\
  \bibinfo {author} {\bibfnamefont {K.}~\bibnamefont {Yoshizawa}},\ }\href
  {\doibase 10.1103/PhysRevB.95.245401} {\bibfield  {journal} {\bibinfo
  {journal} {Phys. Rev. B}\ }\textbf {\bibinfo {volume} {95}},\ \bibinfo
  {pages} {245401} (\bibinfo {year} {2017})}\BibitemShut {NoStop}%
\bibitem [{\citenamefont {Norman}(1995)}]{Norman1995}%
  \BibitemOpen
  \bibfield  {author} {\bibinfo {author} {\bibfnamefont {M.~R.}\ \bibnamefont
  {Norman}},\ }\href {\doibase 10.1103/PhysRevB.52.15093} {\bibfield  {journal}
  {\bibinfo  {journal} {Phys. Rev. B}\ }\textbf {\bibinfo {volume} {52}},\
  \bibinfo {pages} {15093} (\bibinfo {year} {1995})}\BibitemShut {NoStop}%
\bibitem [{\citenamefont {Micklitz}\ and\ \citenamefont
  {Norman}(2009)}]{Micklitz-Norman2009}%
  \BibitemOpen
  \bibfield  {author} {\bibinfo {author} {\bibfnamefont {T.}~\bibnamefont
  {Micklitz}}\ and\ \bibinfo {author} {\bibfnamefont {M.~R.}\ \bibnamefont
  {Norman}},\ }\href {\doibase 10.1103/PhysRevB.80.100506} {\bibfield
  {journal} {\bibinfo  {journal} {Phys. Rev. B}\ }\textbf {\bibinfo {volume}
  {80}},\ \bibinfo {pages} {100506} (\bibinfo {year} {2009})}\BibitemShut
  {NoStop}%
\bibitem [{\citenamefont {Kobayashi}\ \emph {et~al.}(2016)\citenamefont
  {Kobayashi}, \citenamefont {Yanase},\ and\ \citenamefont
  {Sato}}]{Kobayashi2016}%
  \BibitemOpen
  \bibfield  {author} {\bibinfo {author} {\bibfnamefont {S.}~\bibnamefont
  {Kobayashi}}, \bibinfo {author} {\bibfnamefont {Y.}~\bibnamefont {Yanase}}, \
  and\ \bibinfo {author} {\bibfnamefont {M.}~\bibnamefont {Sato}},\ }\href
  {\doibase 10.1103/PhysRevB.94.134512} {\bibfield  {journal} {\bibinfo
  {journal} {Phys. Rev. B}\ }\textbf {\bibinfo {volume} {94}},\ \bibinfo
  {pages} {134512} (\bibinfo {year} {2016})}\BibitemShut {NoStop}%
\bibitem [{\citenamefont {Sumita}\ \emph {et~al.}(2017)\citenamefont {Sumita},
  \citenamefont {Nomoto},\ and\ \citenamefont {Yanase}}]{Sumita_Sr2IrO42017}%
  \BibitemOpen
  \bibfield  {author} {\bibinfo {author} {\bibfnamefont {S.}~\bibnamefont
  {Sumita}}, \bibinfo {author} {\bibfnamefont {T.}~\bibnamefont {Nomoto}}, \
  and\ \bibinfo {author} {\bibfnamefont {Y.}~\bibnamefont {Yanase}},\ }\href
  {\doibase 10.1103/PhysRevLett.119.027001} {\bibfield  {journal} {\bibinfo
  {journal} {Phys. Rev. Lett.}\ }\textbf {\bibinfo {volume} {119}},\ \bibinfo
  {pages} {027001} (\bibinfo {year} {2017})}\BibitemShut {NoStop}%
\bibitem [{\citenamefont {Sumita}\ and\ \citenamefont
  {Yanase}(2018)}]{Sumita-Yanase2018}%
  \BibitemOpen
  \bibfield  {author} {\bibinfo {author} {\bibfnamefont {S.}~\bibnamefont
  {Sumita}}\ and\ \bibinfo {author} {\bibfnamefont {Y.}~\bibnamefont
  {Yanase}},\ }\href {\doibase 10.1103/PhysRevB.97.134512} {\bibfield
  {journal} {\bibinfo  {journal} {Phys. Rev. B}\ }\textbf {\bibinfo {volume}
  {97}},\ \bibinfo {pages} {134512} (\bibinfo {year} {2018})}\BibitemShut
  {NoStop}%
\bibitem [{\citenamefont {Nomoto}\ and\ \citenamefont
  {Ikeda}(2017)}]{Nomoto2017}%
  \BibitemOpen
  \bibfield  {author} {\bibinfo {author} {\bibfnamefont {T.}~\bibnamefont
  {Nomoto}}\ and\ \bibinfo {author} {\bibfnamefont {H.}~\bibnamefont {Ikeda}},\
  }\href {\doibase 10.7566/JPSJ.86.023703} {\bibfield  {journal} {\bibinfo
  {journal} {Journal of the Physical Society of Japan}\ }\textbf {\bibinfo
  {volume} {86}},\ \bibinfo {pages} {023703} (\bibinfo {year} {2017})},\
  \Eprint {http://arxiv.org/abs/https://doi.org/10.7566/JPSJ.86.023703}
  {https://doi.org/10.7566/JPSJ.86.023703} \BibitemShut {NoStop}%
\bibitem [{\citenamefont {Sumita}\ \emph {et~al.}(2019)\citenamefont {Sumita},
  \citenamefont {Nomoto}, \citenamefont {Shiozaki},\ and\ \citenamefont
  {Yanase}}]{Sumita2019}%
  \BibitemOpen
  \bibfield  {author} {\bibinfo {author} {\bibfnamefont {S.}~\bibnamefont
  {Sumita}}, \bibinfo {author} {\bibfnamefont {T.}~\bibnamefont {Nomoto}},
  \bibinfo {author} {\bibfnamefont {K.}~\bibnamefont {Shiozaki}}, \ and\
  \bibinfo {author} {\bibfnamefont {Y.}~\bibnamefont {Yanase}},\ }\href
  {\doibase 10.1103/PhysRevB.99.134513} {\bibfield  {journal} {\bibinfo
  {journal} {Phys. Rev. B}\ }\textbf {\bibinfo {volume} {99}},\ \bibinfo
  {pages} {134513} (\bibinfo {year} {2019})}\BibitemShut {NoStop}%
\bibitem [{\citenamefont {Kanasugi}\ and\ \citenamefont
  {Yanase}(2020)}]{Kanasugi_bilayerTMD}%
  \BibitemOpen
  \bibfield  {author} {\bibinfo {author} {\bibfnamefont {S.}~\bibnamefont
  {Kanasugi}}\ and\ \bibinfo {author} {\bibfnamefont {Y.}~\bibnamefont
  {Yanase}},\ }\href {\doibase 10.1103/PhysRevB.102.094507} {\bibfield
  {journal} {\bibinfo  {journal} {Phys. Rev. B}\ }\textbf {\bibinfo {volume}
  {102}},\ \bibinfo {pages} {094507} (\bibinfo {year} {2020})}\BibitemShut
  {NoStop}%
\bibitem [{\citenamefont {Yoshida}\ \emph {et~al.}(2014)\citenamefont
  {Yoshida}, \citenamefont {Sigrist},\ and\ \citenamefont
  {Yanase}}]{Yoshida_parity_mixing2014}%
  \BibitemOpen
  \bibfield  {author} {\bibinfo {author} {\bibfnamefont {T.}~\bibnamefont
  {Yoshida}}, \bibinfo {author} {\bibfnamefont {M.}~\bibnamefont {Sigrist}}, \
  and\ \bibinfo {author} {\bibfnamefont {Y.}~\bibnamefont {Yanase}},\ }\href
  {\doibase 10.7566/JPSJ.83.013703} {\bibfield  {journal} {\bibinfo  {journal}
  {Journal of the Physical Society of Japan}\ }\textbf {\bibinfo {volume}
  {83}},\ \bibinfo {pages} {013703} (\bibinfo {year} {2014})},\ \Eprint
  {http://arxiv.org/abs/https://doi.org/10.7566/JPSJ.83.013703}
  {https://doi.org/10.7566/JPSJ.83.013703} \BibitemShut {NoStop}%
\bibitem [{\citenamefont {Yoshida}(2014)}]{Yoshida_thesis}%
  \BibitemOpen
  \bibfield  {author} {\bibinfo {author} {\bibfnamefont {T.}~\bibnamefont
  {Yoshida}},\ }\href@noop {} {Ph.D. thesis},\ \bibinfo  {school} {Niigata
  University} (\bibinfo {year} {2014})\BibitemShut {NoStop}%
\bibitem [{\citenamefont {Nakosai}\ \emph {et~al.}(2012)\citenamefont
  {Nakosai}, \citenamefont {Tanaka},\ and\ \citenamefont
  {Nagaosa}}]{Nakosai2012}%
  \BibitemOpen
  \bibfield  {author} {\bibinfo {author} {\bibfnamefont {S.}~\bibnamefont
  {Nakosai}}, \bibinfo {author} {\bibfnamefont {Y.}~\bibnamefont {Tanaka}}, \
  and\ \bibinfo {author} {\bibfnamefont {N.}~\bibnamefont {Nagaosa}},\ }\href
  {\doibase 10.1103/PhysRevLett.108.147003} {\bibfield  {journal} {\bibinfo
  {journal} {Phys. Rev. Lett.}\ }\textbf {\bibinfo {volume} {108}},\ \bibinfo
  {pages} {147003} (\bibinfo {year} {2012})}\BibitemShut {NoStop}%
\bibitem [{\citenamefont {M\"ockli}\ \emph {et~al.}(2018)\citenamefont
  {M\"ockli}, \citenamefont {Yanase},\ and\ \citenamefont
  {Sigrist}}]{Moecli_PDW2018}%
  \BibitemOpen
  \bibfield  {author} {\bibinfo {author} {\bibfnamefont {D.}~\bibnamefont
  {M\"ockli}}, \bibinfo {author} {\bibfnamefont {Y.}~\bibnamefont {Yanase}}, \
  and\ \bibinfo {author} {\bibfnamefont {M.}~\bibnamefont {Sigrist}},\ }\href
  {\doibase 10.1103/PhysRevB.97.144508} {\bibfield  {journal} {\bibinfo
  {journal} {Phys. Rev. B}\ }\textbf {\bibinfo {volume} {97}},\ \bibinfo
  {pages} {144508} (\bibinfo {year} {2018})}\BibitemShut {NoStop}%
\bibitem [{\citenamefont {Schertenleib}\ \emph {et~al.}(2021)\citenamefont
  {Schertenleib}, \citenamefont {Fischer},\ and\ \citenamefont
  {Sigrist}}]{Schertenleib2021}%
  \BibitemOpen
  \bibfield  {author} {\bibinfo {author} {\bibfnamefont {E.~G.}\ \bibnamefont
  {Schertenleib}}, \bibinfo {author} {\bibfnamefont {M.~H.}\ \bibnamefont
  {Fischer}}, \ and\ \bibinfo {author} {\bibfnamefont {M.}~\bibnamefont
  {Sigrist}},\ }\href {\doibase 10.1103/PhysRevResearch.3.023179} {\bibfield
  {journal} {\bibinfo  {journal} {Phys. Rev. Research}\ }\textbf {\bibinfo
  {volume} {3}},\ \bibinfo {pages} {023179} (\bibinfo {year}
  {2021})}\BibitemShut {NoStop}%
\bibitem [{\citenamefont {M\"ockli}\ and\ \citenamefont
  {Ramires}(2021{\natexlab{a}})}]{moeckli2021}%
  \BibitemOpen
  \bibfield  {author} {\bibinfo {author} {\bibfnamefont {D.}~\bibnamefont
  {M\"ockli}}\ and\ \bibinfo {author} {\bibfnamefont {A.}~\bibnamefont
  {Ramires}},\ }\href {\doibase 10.1103/PhysRevResearch.3.023204} {\bibfield
  {journal} {\bibinfo  {journal} {Phys. Rev. Research}\ }\textbf {\bibinfo
  {volume} {3}},\ \bibinfo {pages} {023204} (\bibinfo {year}
  {2021}{\natexlab{a}})}\BibitemShut {NoStop}%
\bibitem [{\citenamefont {Skurativska}\ \emph {et~al.}(2021)\citenamefont
  {Skurativska}, \citenamefont {Sigrist},\ and\ \citenamefont
  {Fischer}}]{skurativska2021spin}%
  \BibitemOpen
  \bibfield  {author} {\bibinfo {author} {\bibfnamefont {A.}~\bibnamefont
  {Skurativska}}, \bibinfo {author} {\bibfnamefont {M.}~\bibnamefont
  {Sigrist}}, \ and\ \bibinfo {author} {\bibfnamefont {M.~H.}\ \bibnamefont
  {Fischer}},\ }\href {\doibase 10.1103/PhysRevResearch.3.033133} {\bibfield
  {journal} {\bibinfo  {journal} {Phys. Rev. Research}\ }\textbf {\bibinfo
  {volume} {3}},\ \bibinfo {pages} {033133} (\bibinfo {year}
  {2021})}\BibitemShut {NoStop}%
\bibitem [{\citenamefont {M\"ockli}\ and\ \citenamefont
  {Ramires}(2021{\natexlab{b}})}]{Moeckli_disorder}%
  \BibitemOpen
  \bibfield  {author} {\bibinfo {author} {\bibfnamefont {D.}~\bibnamefont
  {M\"ockli}}\ and\ \bibinfo {author} {\bibfnamefont {A.}~\bibnamefont
  {Ramires}},\ }\href {\doibase 10.1103/PhysRevB.104.134517} {\bibfield
  {journal} {\bibinfo  {journal} {Phys. Rev. B}\ }\textbf {\bibinfo {volume}
  {104}},\ \bibinfo {pages} {134517} (\bibinfo {year}
  {2021}{\natexlab{b}})}\BibitemShut {NoStop}%
\bibitem [{\citenamefont {Watanabe}\ \emph {et~al.}(2015)\citenamefont
  {Watanabe}, \citenamefont {Yoshida},\ and\ \citenamefont
  {Yanase}}]{Watanabe_PDW2015}%
  \BibitemOpen
  \bibfield  {author} {\bibinfo {author} {\bibfnamefont {T.}~\bibnamefont
  {Watanabe}}, \bibinfo {author} {\bibfnamefont {T.}~\bibnamefont {Yoshida}}, \
  and\ \bibinfo {author} {\bibfnamefont {Y.}~\bibnamefont {Yanase}},\ }\href
  {\doibase 10.1103/PhysRevB.92.174502} {\bibfield  {journal} {\bibinfo
  {journal} {Phys. Rev. B}\ }\textbf {\bibinfo {volume} {92}},\ \bibinfo
  {pages} {174502} (\bibinfo {year} {2015})}\BibitemShut {NoStop}%
\bibitem [{\citenamefont {Saito}\ \emph {et~al.}(2016)\citenamefont {Saito},
  \citenamefont {Nakamura}, \citenamefont {Bahramy}, \citenamefont {Kohama},
  \citenamefont {Ye}, \citenamefont {Kasahara}, \citenamefont {Nakagawa},
  \citenamefont {Onga}, \citenamefont {Tokunaga}, \citenamefont {Nojima},
  \citenamefont {Yanase},\ and\ \citenamefont {Iwasa}}]{Saito2016}%
  \BibitemOpen
  \bibfield  {author} {\bibinfo {author} {\bibfnamefont {Y.}~\bibnamefont
  {Saito}}, \bibinfo {author} {\bibfnamefont {Y.}~\bibnamefont {Nakamura}},
  \bibinfo {author} {\bibfnamefont {M.~S.}\ \bibnamefont {Bahramy}}, \bibinfo
  {author} {\bibfnamefont {Y.}~\bibnamefont {Kohama}}, \bibinfo {author}
  {\bibfnamefont {J.}~\bibnamefont {Ye}}, \bibinfo {author} {\bibfnamefont
  {Y.}~\bibnamefont {Kasahara}}, \bibinfo {author} {\bibfnamefont
  {Y.}~\bibnamefont {Nakagawa}}, \bibinfo {author} {\bibfnamefont
  {M.}~\bibnamefont {Onga}}, \bibinfo {author} {\bibfnamefont {M.}~\bibnamefont
  {Tokunaga}}, \bibinfo {author} {\bibfnamefont {T.}~\bibnamefont {Nojima}},
  \bibinfo {author} {\bibfnamefont {Y.}~\bibnamefont {Yanase}}, \ and\ \bibinfo
  {author} {\bibfnamefont {Y.}~\bibnamefont {Iwasa}},\ }\href {\doibase
  10.1038/nphys3580} {\bibfield  {journal} {\bibinfo  {journal} {Nature
  Physics}\ }\textbf {\bibinfo {volume} {12}},\ \bibinfo {pages} {144}
  (\bibinfo {year} {2016})}\BibitemShut {NoStop}%
\bibitem [{\citenamefont {Lu}\ \emph {et~al.}(2015)\citenamefont {Lu},
  \citenamefont {Zheliuk}, \citenamefont {Leermakers}, \citenamefont {Yuan},
  \citenamefont {Zeitler}, \citenamefont {Law},\ and\ \citenamefont
  {Ye}}]{Lu_Ising_SC}%
  \BibitemOpen
  \bibfield  {author} {\bibinfo {author} {\bibfnamefont {J.~M.}\ \bibnamefont
  {Lu}}, \bibinfo {author} {\bibfnamefont {O.}~\bibnamefont {Zheliuk}},
  \bibinfo {author} {\bibfnamefont {I.}~\bibnamefont {Leermakers}}, \bibinfo
  {author} {\bibfnamefont {N.~F.~Q.}\ \bibnamefont {Yuan}}, \bibinfo {author}
  {\bibfnamefont {U.}~\bibnamefont {Zeitler}}, \bibinfo {author} {\bibfnamefont
  {K.~T.}\ \bibnamefont {Law}}, \ and\ \bibinfo {author} {\bibfnamefont
  {J.~T.}\ \bibnamefont {Ye}},\ }\href {\doibase 10.1126/science.aab2277}
  {\bibfield  {journal} {\bibinfo  {journal} {Science}\ }\textbf {\bibinfo
  {volume} {350}},\ \bibinfo {pages} {1353} (\bibinfo {year}
  {2015})}\BibitemShut {NoStop}%
\bibitem [{\citenamefont {Xi}\ \emph {et~al.}(2016)\citenamefont {Xi},
  \citenamefont {Wang}, \citenamefont {Zhao}, \citenamefont {Park},
  \citenamefont {Law}, \citenamefont {Berger}, \citenamefont {Forr{\'o}},
  \citenamefont {Shan},\ and\ \citenamefont {Mak}}]{Xi_Ising_SC}%
  \BibitemOpen
  \bibfield  {author} {\bibinfo {author} {\bibfnamefont {X.}~\bibnamefont
  {Xi}}, \bibinfo {author} {\bibfnamefont {Z.}~\bibnamefont {Wang}}, \bibinfo
  {author} {\bibfnamefont {W.}~\bibnamefont {Zhao}}, \bibinfo {author}
  {\bibfnamefont {J.-H.}\ \bibnamefont {Park}}, \bibinfo {author}
  {\bibfnamefont {K.~T.}\ \bibnamefont {Law}}, \bibinfo {author} {\bibfnamefont
  {H.}~\bibnamefont {Berger}}, \bibinfo {author} {\bibfnamefont
  {L.}~\bibnamefont {Forr{\'o}}}, \bibinfo {author} {\bibfnamefont
  {J.}~\bibnamefont {Shan}}, \ and\ \bibinfo {author} {\bibfnamefont {K.~F.}\
  \bibnamefont {Mak}},\ }\href {\doibase 10.1038/nphys3538} {\bibfield
  {journal} {\bibinfo  {journal} {Nature Physics}\ }\textbf {\bibinfo {volume}
  {12}},\ \bibinfo {pages} {139} (\bibinfo {year} {2016})}\BibitemShut
  {NoStop}%
\bibitem [{\citenamefont {Aoki}\ \emph {et~al.}(2021)\citenamefont {Aoki},
  \citenamefont {Brison}, \citenamefont {Flouquet}, \citenamefont {Ishida},
  \citenamefont {Knebel}, \citenamefont {Tokunaga},\ and\ \citenamefont
  {Yanase}}]{UTe2_review}%
  \BibitemOpen
  \bibfield  {author} {\bibinfo {author} {\bibfnamefont {D.}~\bibnamefont
  {Aoki}}, \bibinfo {author} {\bibfnamefont {J.~P.}\ \bibnamefont {Brison}},
  \bibinfo {author} {\bibfnamefont {J.}~\bibnamefont {Flouquet}}, \bibinfo
  {author} {\bibfnamefont {K.}~\bibnamefont {Ishida}}, \bibinfo {author}
  {\bibfnamefont {G.}~\bibnamefont {Knebel}}, \bibinfo {author} {\bibfnamefont
  {Y.}~\bibnamefont {Tokunaga}}, \ and\ \bibinfo {author} {\bibfnamefont
  {Y.}~\bibnamefont {Yanase}},\ }\href@noop {} {\  (\bibinfo {year} {2021})},\
  \Eprint {http://arxiv.org/abs/2110.10451} {arXiv:2110.10451
  [cond-mat.str-el]} \BibitemShut {NoStop}%
\bibitem [{\citenamefont {Yoshida}\ \emph {et~al.}(2015)\citenamefont
  {Yoshida}, \citenamefont {Sigrist},\ and\ \citenamefont
  {Yanase}}]{TomoYoshida_SupLatt_PRL15}%
  \BibitemOpen
  \bibfield  {author} {\bibinfo {author} {\bibfnamefont {T.}~\bibnamefont
  {Yoshida}}, \bibinfo {author} {\bibfnamefont {M.}~\bibnamefont {Sigrist}}, \
  and\ \bibinfo {author} {\bibfnamefont {Y.}~\bibnamefont {Yanase}},\ }\href
  {\doibase 10.1103/PhysRevLett.115.027001} {\bibfield  {journal} {\bibinfo
  {journal} {Phys. Rev. Lett.}\ }\textbf {\bibinfo {volume} {115}},\ \bibinfo
  {pages} {027001} (\bibinfo {year} {2015})}\BibitemShut {NoStop}%
\bibitem [{\citenamefont {Yoshida}\ \emph {et~al.}(2017)\citenamefont
  {Yoshida}, \citenamefont {Daido}, \citenamefont {Yanase},\ and\ \citenamefont
  {Kawakami}}]{Yoshida_ZxZtoZxZ8superlattice_PRL17}%
  \BibitemOpen
  \bibfield  {author} {\bibinfo {author} {\bibfnamefont {T.}~\bibnamefont
  {Yoshida}}, \bibinfo {author} {\bibfnamefont {A.}~\bibnamefont {Daido}},
  \bibinfo {author} {\bibfnamefont {Y.}~\bibnamefont {Yanase}}, \ and\ \bibinfo
  {author} {\bibfnamefont {N.}~\bibnamefont {Kawakami}},\ }\href {\doibase
  10.1103/PhysRevLett.118.147001} {\bibfield  {journal} {\bibinfo  {journal}
  {Phys. Rev. Lett.}\ }\textbf {\bibinfo {volume} {118}},\ \bibinfo {pages}
  {147001} (\bibinfo {year} {2017})}\BibitemShut {NoStop}%
\bibitem [{\citenamefont {Chiu}\ \emph {et~al.}(2013)\citenamefont {Chiu},
  \citenamefont {Yao},\ and\ \citenamefont {Ryu}}]{Chiu2013}%
  \BibitemOpen
  \bibfield  {author} {\bibinfo {author} {\bibfnamefont {C.-K.}\ \bibnamefont
  {Chiu}}, \bibinfo {author} {\bibfnamefont {H.}~\bibnamefont {Yao}}, \ and\
  \bibinfo {author} {\bibfnamefont {S.}~\bibnamefont {Ryu}},\ }\href {\doibase
  10.1103/PhysRevB.88.075142} {\bibfield  {journal} {\bibinfo  {journal} {Phys.
  Rev. B}\ }\textbf {\bibinfo {volume} {88}},\ \bibinfo {pages} {075142}
  (\bibinfo {year} {2013})}\BibitemShut {NoStop}%
\bibitem [{\citenamefont {Morimoto}\ and\ \citenamefont
  {Furusaki}(2013)}]{Morimoto2013}%
  \BibitemOpen
  \bibfield  {author} {\bibinfo {author} {\bibfnamefont {T.}~\bibnamefont
  {Morimoto}}\ and\ \bibinfo {author} {\bibfnamefont {A.}~\bibnamefont
  {Furusaki}},\ }\href {\doibase 10.1103/PhysRevB.88.125129} {\bibfield
  {journal} {\bibinfo  {journal} {Phys. Rev. B}\ }\textbf {\bibinfo {volume}
  {88}},\ \bibinfo {pages} {125129} (\bibinfo {year} {2013})}\BibitemShut
  {NoStop}%
\bibitem [{\citenamefont {Shiozaki}\ and\ \citenamefont
  {Sato}(2014)}]{Shiozaki2013}%
  \BibitemOpen
  \bibfield  {author} {\bibinfo {author} {\bibfnamefont {K.}~\bibnamefont
  {Shiozaki}}\ and\ \bibinfo {author} {\bibfnamefont {M.}~\bibnamefont
  {Sato}},\ }\href {\doibase 10.1103/PhysRevB.90.165114} {\bibfield  {journal}
  {\bibinfo  {journal} {Phys. Rev. B}\ }\textbf {\bibinfo {volume} {90}},\
  \bibinfo {pages} {165114} (\bibinfo {year} {2014})}\BibitemShut {NoStop}%
\bibitem [{\citenamefont {Shiozaki}\ \emph {et~al.}(2016)\citenamefont
  {Shiozaki}, \citenamefont {Sato},\ and\ \citenamefont {Gomi}}]{Shiozaki2016}%
  \BibitemOpen
  \bibfield  {author} {\bibinfo {author} {\bibfnamefont {K.}~\bibnamefont
  {Shiozaki}}, \bibinfo {author} {\bibfnamefont {M.}~\bibnamefont {Sato}}, \
  and\ \bibinfo {author} {\bibfnamefont {K.}~\bibnamefont {Gomi}},\ }\href
  {\doibase 10.1103/PhysRevB.93.195413} {\bibfield  {journal} {\bibinfo
  {journal} {Phys. Rev. B}\ }\textbf {\bibinfo {volume} {93}},\ \bibinfo
  {pages} {195413} (\bibinfo {year} {2016})}\BibitemShut {NoStop}%
\bibitem [{\citenamefont {Shiozaki}\ \emph {et~al.}(2017)\citenamefont
  {Shiozaki}, \citenamefont {Sato},\ and\ \citenamefont {Gomi}}]{Shiozaki2017}%
  \BibitemOpen
  \bibfield  {author} {\bibinfo {author} {\bibfnamefont {K.}~\bibnamefont
  {Shiozaki}}, \bibinfo {author} {\bibfnamefont {M.}~\bibnamefont {Sato}}, \
  and\ \bibinfo {author} {\bibfnamefont {K.}~\bibnamefont {Gomi}},\ }\href
  {\doibase 10.1103/PhysRevB.95.235425} {\bibfield  {journal} {\bibinfo
  {journal} {Phys. Rev. B}\ }\textbf {\bibinfo {volume} {95}},\ \bibinfo
  {pages} {235425} (\bibinfo {year} {2017})}\BibitemShut {NoStop}%
\bibitem [{\citenamefont {Teo}\ \emph {et~al.}(2008)\citenamefont {Teo},
  \citenamefont {Fu},\ and\ \citenamefont {Kane}}]{Teo_MCN_PRB08}%
  \BibitemOpen
  \bibfield  {author} {\bibinfo {author} {\bibfnamefont {J.~C.~Y.}\
  \bibnamefont {Teo}}, \bibinfo {author} {\bibfnamefont {L.}~\bibnamefont
  {Fu}}, \ and\ \bibinfo {author} {\bibfnamefont {C.~L.}\ \bibnamefont
  {Kane}},\ }\href {\doibase 10.1103/PhysRevB.78.045426} {\bibfield  {journal}
  {\bibinfo  {journal} {Phys. Rev. B}\ }\textbf {\bibinfo {volume} {78}},\
  \bibinfo {pages} {045426} (\bibinfo {year} {2008})}\BibitemShut {NoStop}%
\bibitem [{\citenamefont {Nogaki}\ \emph {et~al.}(2021)\citenamefont {Nogaki},
  \citenamefont {Daido}, \citenamefont {Ishizuka},\ and\ \citenamefont
  {Yanase}}]{nogaki2021topological}%
  \BibitemOpen
  \bibfield  {author} {\bibinfo {author} {\bibfnamefont {K.}~\bibnamefont
  {Nogaki}}, \bibinfo {author} {\bibfnamefont {A.}~\bibnamefont {Daido}},
  \bibinfo {author} {\bibfnamefont {J.}~\bibnamefont {Ishizuka}}, \ and\
  \bibinfo {author} {\bibfnamefont {Y.}~\bibnamefont {Yanase}},\ }\href
  {\doibase 10.1103/PhysRevResearch.3.L032071} {\bibfield  {journal} {\bibinfo
  {journal} {Phys. Rev. Research}\ }\textbf {\bibinfo {volume} {3}},\ \bibinfo
  {pages} {L032071} (\bibinfo {year} {2021})}\BibitemShut {NoStop}%
\bibitem [{\citenamefont {Fidkowski}\ and\ \citenamefont
  {Kitaev}(2010)}]{Z_to_Zn_Fidkowski_10}%
  \BibitemOpen
  \bibfield  {author} {\bibinfo {author} {\bibfnamefont {L.}~\bibnamefont
  {Fidkowski}}\ and\ \bibinfo {author} {\bibfnamefont {A.}~\bibnamefont
  {Kitaev}},\ }\href {\doibase 10.1103/PhysRevB.81.134509} {\bibfield
  {journal} {\bibinfo  {journal} {Phys. Rev. B}\ }\textbf {\bibinfo {volume}
  {81}},\ \bibinfo {pages} {134509} (\bibinfo {year} {2010})}\BibitemShut
  {NoStop}%
\bibitem [{\citenamefont {Turner}\ \emph {et~al.}(2011)\citenamefont {Turner},
  \citenamefont {Pollmann},\ and\ \citenamefont {Berg}}]{Turner11}%
  \BibitemOpen
  \bibfield  {author} {\bibinfo {author} {\bibfnamefont {A.~M.}\ \bibnamefont
  {Turner}}, \bibinfo {author} {\bibfnamefont {F.}~\bibnamefont {Pollmann}}, \
  and\ \bibinfo {author} {\bibfnamefont {E.}~\bibnamefont {Berg}},\ }\href
  {\doibase 10.1103/PhysRevB.83.075102} {\bibfield  {journal} {\bibinfo
  {journal} {Phys. Rev. B}\ }\textbf {\bibinfo {volume} {83}},\ \bibinfo
  {pages} {075102} (\bibinfo {year} {2011})}\BibitemShut {NoStop}%
\bibitem [{\citenamefont {Fidkowski}\ and\ \citenamefont
  {Kitaev}(2011)}]{Fidkowski_1Dclassificatin_11}%
  \BibitemOpen
  \bibfield  {author} {\bibinfo {author} {\bibfnamefont {L.}~\bibnamefont
  {Fidkowski}}\ and\ \bibinfo {author} {\bibfnamefont {A.}~\bibnamefont
  {Kitaev}},\ }\href {\doibase 10.1103/PhysRevB.83.075103} {\bibfield
  {journal} {\bibinfo  {journal} {Phys. Rev. B}\ }\textbf {\bibinfo {volume}
  {83}},\ \bibinfo {pages} {075103} (\bibinfo {year} {2011})}\BibitemShut
  {NoStop}%
\bibitem [{\citenamefont {Yao}\ and\ \citenamefont
  {Ryu}(2013)}]{YaoRyu_Z_to_Z8_2013}%
  \BibitemOpen
  \bibfield  {author} {\bibinfo {author} {\bibfnamefont {H.}~\bibnamefont
  {Yao}}\ and\ \bibinfo {author} {\bibfnamefont {S.}~\bibnamefont {Ryu}},\
  }\href {\doibase 10.1103/PhysRevB.88.064507} {\bibfield  {journal} {\bibinfo
  {journal} {Phys. Rev. B}\ }\textbf {\bibinfo {volume} {88}},\ \bibinfo
  {pages} {064507} (\bibinfo {year} {2013})}\BibitemShut {NoStop}%
\bibitem [{\citenamefont {Ryu}\ and\ \citenamefont
  {Zhang}(2012)}]{Ryu_Z_to_Z8_2013}%
  \BibitemOpen
  \bibfield  {author} {\bibinfo {author} {\bibfnamefont {S.}~\bibnamefont
  {Ryu}}\ and\ \bibinfo {author} {\bibfnamefont {S.-C.}\ \bibnamefont
  {Zhang}},\ }\href {\doibase 10.1103/PhysRevB.85.245132} {\bibfield  {journal}
  {\bibinfo  {journal} {Phys. Rev. B}\ }\textbf {\bibinfo {volume} {85}},\
  \bibinfo {pages} {245132} (\bibinfo {year} {2012})}\BibitemShut {NoStop}%
\bibitem [{\citenamefont {Qi}(2013)}]{Qi_Z_to_Z8_2013}%
  \BibitemOpen
  \bibfield  {author} {\bibinfo {author} {\bibfnamefont {X.-L.}\ \bibnamefont
  {Qi}},\ }\href {\doibase 10.1088/1367-2630/15/6/065002} {\bibfield  {journal}
  {\bibinfo  {journal} {New J. Phys.}\ }\textbf {\bibinfo {volume} {15}},\
  \bibinfo {pages} {065002} (\bibinfo {year} {2013})}\BibitemShut {NoStop}%
\bibitem [{\citenamefont {Gu}\ and\ \citenamefont
  {Wen}(2014)}]{gu_supercohomology}%
  \BibitemOpen
  \bibfield  {author} {\bibinfo {author} {\bibfnamefont {Z.-C.}\ \bibnamefont
  {Gu}}\ and\ \bibinfo {author} {\bibfnamefont {X.-G.}\ \bibnamefont {Wen}},\
  }\href {\doibase 10.1103/PhysRevB.90.115141} {\bibfield  {journal} {\bibinfo
  {journal} {Phys. Rev. B}\ }\textbf {\bibinfo {volume} {90}},\ \bibinfo
  {pages} {115141} (\bibinfo {year} {2014})}\BibitemShut {NoStop}%
\bibitem [{\citenamefont {Kapustin}\ \emph {et~al.}(2015)\citenamefont
  {Kapustin}, \citenamefont {Thorngren}, \citenamefont {Turzillo},\ and\
  \citenamefont {Wang}}]{kapustin_fermionic_cobordisms2014}%
  \BibitemOpen
  \bibfield  {author} {\bibinfo {author} {\bibfnamefont {A.}~\bibnamefont
  {Kapustin}}, \bibinfo {author} {\bibfnamefont {R.}~\bibnamefont {Thorngren}},
  \bibinfo {author} {\bibfnamefont {A.}~\bibnamefont {Turzillo}}, \ and\
  \bibinfo {author} {\bibfnamefont {Z.}~\bibnamefont {Wang}},\ }\href {\doibase
  10.1007/JHEP12(2015)052} {\bibfield  {journal} {\bibinfo  {journal} {J. High
  Energy Phys.}\ }\textbf {\bibinfo {volume} {2015}},\ \bibinfo {pages} {52}
  (\bibinfo {year} {2015})}\BibitemShut {NoStop}%
\bibitem [{\citenamefont {Lu}\ and\ \citenamefont
  {Vishwanath}(2012)}]{Lu_CS_2011}%
  \BibitemOpen
  \bibfield  {author} {\bibinfo {author} {\bibfnamefont {Y.-M.}\ \bibnamefont
  {Lu}}\ and\ \bibinfo {author} {\bibfnamefont {A.}~\bibnamefont
  {Vishwanath}},\ }\href {\doibase 10.1103/PhysRevB.86.125119} {\bibfield
  {journal} {\bibinfo  {journal} {Phys. Rev. B}\ }\textbf {\bibinfo {volume}
  {86}},\ \bibinfo {pages} {125119} (\bibinfo {year} {2012})}\BibitemShut
  {NoStop}%
\bibitem [{\citenamefont {Levin}\ and\ \citenamefont
  {Stern}(2012)}]{Levin_CS_2012}%
  \BibitemOpen
  \bibfield  {author} {\bibinfo {author} {\bibfnamefont {M.}~\bibnamefont
  {Levin}}\ and\ \bibinfo {author} {\bibfnamefont {A.}~\bibnamefont {Stern}},\
  }\href {\doibase 10.1103/PhysRevB.86.115131} {\bibfield  {journal} {\bibinfo
  {journal} {Phys. Rev. B}\ }\textbf {\bibinfo {volume} {86}},\ \bibinfo
  {pages} {115131} (\bibinfo {year} {2012})}\BibitemShut {NoStop}%
\bibitem [{\citenamefont {Hsieh}\ \emph {et~al.}(2014)\citenamefont {Hsieh},
  \citenamefont {Morimoto},\ and\ \citenamefont {Ryu}}]{Hsieh_CS_CPT_PRB14}%
  \BibitemOpen
  \bibfield  {author} {\bibinfo {author} {\bibfnamefont {C.-T.}\ \bibnamefont
  {Hsieh}}, \bibinfo {author} {\bibfnamefont {T.}~\bibnamefont {Morimoto}}, \
  and\ \bibinfo {author} {\bibfnamefont {S.}~\bibnamefont {Ryu}},\ }\href
  {\doibase 10.1103/PhysRevB.90.245111} {\bibfield  {journal} {\bibinfo
  {journal} {Phys. Rev. B}\ }\textbf {\bibinfo {volume} {90}},\ \bibinfo
  {pages} {245111} (\bibinfo {year} {2014})}\BibitemShut {NoStop}%
\bibitem [{\citenamefont {Isobe}\ and\ \citenamefont
  {Fu}(2015)}]{Isobe_ZtoZ82015}%
  \BibitemOpen
  \bibfield  {author} {\bibinfo {author} {\bibfnamefont {H.}~\bibnamefont
  {Isobe}}\ and\ \bibinfo {author} {\bibfnamefont {L.}~\bibnamefont {Fu}},\
  }\href@noop {} {\bibfield  {journal} {\bibinfo  {journal} {arXiv preprint
  arXiv:1502.06962}\ } (\bibinfo {year} {2015})}\BibitemShut {NoStop}%
\bibitem [{\citenamefont {Yoshida}\ and\ \citenamefont
  {Furusaki}(2015)}]{Yoshida_ZtoZ8PRB15}%
  \BibitemOpen
  \bibfield  {author} {\bibinfo {author} {\bibfnamefont {T.}~\bibnamefont
  {Yoshida}}\ and\ \bibinfo {author} {\bibfnamefont {A.}~\bibnamefont
  {Furusaki}},\ }\href {\doibase 10.1103/PhysRevB.92.085114} {\bibfield
  {journal} {\bibinfo  {journal} {Phys. Rev. B}\ }\textbf {\bibinfo {volume}
  {92}},\ \bibinfo {pages} {085114} (\bibinfo {year} {2015})}\BibitemShut
  {NoStop}%
\bibitem [{\citenamefont {Fidkowski}\ \emph {et~al.}(2013)\citenamefont
  {Fidkowski}, \citenamefont {Chen},\ and\ \citenamefont
  {Vishwanath}}]{Fidkowski_Z162013}%
  \BibitemOpen
  \bibfield  {author} {\bibinfo {author} {\bibfnamefont {L.}~\bibnamefont
  {Fidkowski}}, \bibinfo {author} {\bibfnamefont {X.}~\bibnamefont {Chen}}, \
  and\ \bibinfo {author} {\bibfnamefont {A.}~\bibnamefont {Vishwanath}},\
  }\href {\doibase 10.1103/PhysRevX.3.041016} {\bibfield  {journal} {\bibinfo
  {journal} {Phys. Rev. X}\ }\textbf {\bibinfo {volume} {3}},\ \bibinfo {pages}
  {041016} (\bibinfo {year} {2013})}\BibitemShut {NoStop}%
\bibitem [{\citenamefont {Wang}\ \emph {et~al.}(2014)\citenamefont {Wang},
  \citenamefont {Potter},\ and\ \citenamefont
  {Senthil}}]{Wang_Potter_Senthil2014}%
  \BibitemOpen
  \bibfield  {author} {\bibinfo {author} {\bibfnamefont {C.}~\bibnamefont
  {Wang}}, \bibinfo {author} {\bibfnamefont {A.~C.}\ \bibnamefont {Potter}}, \
  and\ \bibinfo {author} {\bibfnamefont {T.}~\bibnamefont {Senthil}},\ }\href
  {\doibase 10.1126/science.1243326} {\bibfield  {journal} {\bibinfo  {journal}
  {Science}\ }\textbf {\bibinfo {volume} {343}},\ \bibinfo {pages} {629}
  (\bibinfo {year} {2014})}\BibitemShut {NoStop}%
\bibitem [{\citenamefont {Metlitski}\ \emph {et~al.}(2014)\citenamefont
  {Metlitski}, \citenamefont {Fidkowski}, \citenamefont {Chen},\ and\
  \citenamefont {Vishwanath}}]{Metlitski_3Dinteraction2014}%
  \BibitemOpen
  \bibfield  {author} {\bibinfo {author} {\bibfnamefont {M.~A.}\ \bibnamefont
  {Metlitski}}, \bibinfo {author} {\bibfnamefont {L.}~\bibnamefont
  {Fidkowski}}, \bibinfo {author} {\bibfnamefont {X.}~\bibnamefont {Chen}}, \
  and\ \bibinfo {author} {\bibfnamefont {A.}~\bibnamefont {Vishwanath}},\
  }\href@noop {} {\bibfield  {journal} {\bibinfo  {journal} {arXiv:1406.3032}\
  } (\bibinfo {year} {2014})}\BibitemShut {NoStop}%
\bibitem [{\citenamefont {Wang}\ and\ \citenamefont
  {Senthil}(2014)}]{Wang_Senthil2014}%
  \BibitemOpen
  \bibfield  {author} {\bibinfo {author} {\bibfnamefont {C.}~\bibnamefont
  {Wang}}\ and\ \bibinfo {author} {\bibfnamefont {T.}~\bibnamefont {Senthil}},\
  }\href {\doibase 10.1103/PhysRevB.89.195124} {\bibfield  {journal} {\bibinfo
  {journal} {Phys. Rev. B}\ }\textbf {\bibinfo {volume} {89}},\ \bibinfo
  {pages} {195124} (\bibinfo {year} {2014})}\BibitemShut {NoStop}%
\bibitem [{\citenamefont {You}\ and\ \citenamefont {Xu}(2014)}]{You_Cenke2014}%
  \BibitemOpen
  \bibfield  {author} {\bibinfo {author} {\bibfnamefont {Y.-Z.}\ \bibnamefont
  {You}}\ and\ \bibinfo {author} {\bibfnamefont {C.}~\bibnamefont {Xu}},\
  }\href {\doibase 10.1103/PhysRevB.90.245120} {\bibfield  {journal} {\bibinfo
  {journal} {Phys. Rev. B}\ }\textbf {\bibinfo {volume} {90}},\ \bibinfo
  {pages} {245120} (\bibinfo {year} {2014})}\BibitemShut {NoStop}%
\bibitem [{\citenamefont {She}\ and\ \citenamefont
  {Balatsky}(2012)}]{superlattice_proximity_12}%
  \BibitemOpen
  \bibfield  {author} {\bibinfo {author} {\bibfnamefont {J.-H.}\ \bibnamefont
  {She}}\ and\ \bibinfo {author} {\bibfnamefont {A.~V.}\ \bibnamefont
  {Balatsky}},\ }\href {\doibase 10.1103/PhysRevLett.109.077002} {\bibfield
  {journal} {\bibinfo  {journal} {Phys. Rev. Lett.}\ }\textbf {\bibinfo
  {volume} {109}},\ \bibinfo {pages} {077002} (\bibinfo {year}
  {2012})}\BibitemShut {NoStop}%
\bibitem [{\citenamefont {Yamanaka}\ \emph {et~al.}(2015)\citenamefont
  {Yamanaka}, \citenamefont {Shimozawa}, \citenamefont {Endo}, \citenamefont
  {Mizukami}, \citenamefont {Shishido}, \citenamefont {Terashima},
  \citenamefont {Shibauchi}, \citenamefont {Matsuda},\ and\ \citenamefont
  {Ishida}}]{Yamashita_NMR_confinement_PRB15}%
  \BibitemOpen
  \bibfield  {author} {\bibinfo {author} {\bibfnamefont {T.}~\bibnamefont
  {Yamanaka}}, \bibinfo {author} {\bibfnamefont {M.}~\bibnamefont {Shimozawa}},
  \bibinfo {author} {\bibfnamefont {R.}~\bibnamefont {Endo}}, \bibinfo {author}
  {\bibfnamefont {Y.}~\bibnamefont {Mizukami}}, \bibinfo {author}
  {\bibfnamefont {H.}~\bibnamefont {Shishido}}, \bibinfo {author}
  {\bibfnamefont {T.}~\bibnamefont {Terashima}}, \bibinfo {author}
  {\bibfnamefont {T.}~\bibnamefont {Shibauchi}}, \bibinfo {author}
  {\bibfnamefont {Y.}~\bibnamefont {Matsuda}}, \ and\ \bibinfo {author}
  {\bibfnamefont {K.}~\bibnamefont {Ishida}},\ }\href {\doibase
  10.1103/PhysRevB.92.241105} {\bibfield  {journal} {\bibinfo  {journal} {Phys.
  Rev. B}\ }\textbf {\bibinfo {volume} {92}},\ \bibinfo {pages} {241105}
  (\bibinfo {year} {2015})}\BibitemShut {NoStop}%
\bibitem [{\citenamefont {Matsuda}\ \emph
  {et~al.}(2006{\natexlab{b}})\citenamefont {Matsuda}, \citenamefont {Izawa},\
  and\ \citenamefont {Vekhter}}]{Matsuda_AngResTherm_JPCM06}%
  \BibitemOpen
  \bibfield  {author} {\bibinfo {author} {\bibfnamefont {Y.}~\bibnamefont
  {Matsuda}}, \bibinfo {author} {\bibfnamefont {K.}~\bibnamefont {Izawa}}, \
  and\ \bibinfo {author} {\bibfnamefont {I.}~\bibnamefont {Vekhter}},\ }\href
  {http://stacks.iop.org/0953-8984/18/i=44/a=R01} {\bibfield  {journal}
  {\bibinfo  {journal} {Journal of Physics: Condensed Matter}\ }\textbf
  {\bibinfo {volume} {18}},\ \bibinfo {pages} {R705} (\bibinfo {year}
  {2006}{\natexlab{b}})}\BibitemShut {NoStop}%
\bibitem [{\citenamefont {Tayama}\ \emph {et~al.}(2002)\citenamefont {Tayama},
  \citenamefont {Harita}, \citenamefont {Sakakibara}, \citenamefont {Haga},
  \citenamefont {Shishido}, \citenamefont {Settai},\ and\ \citenamefont
  {Onuki}}]{Tayama_CeCoIn5PauliLimPRB02}%
  \BibitemOpen
  \bibfield  {author} {\bibinfo {author} {\bibfnamefont {T.}~\bibnamefont
  {Tayama}}, \bibinfo {author} {\bibfnamefont {A.}~\bibnamefont {Harita}},
  \bibinfo {author} {\bibfnamefont {T.}~\bibnamefont {Sakakibara}}, \bibinfo
  {author} {\bibfnamefont {Y.}~\bibnamefont {Haga}}, \bibinfo {author}
  {\bibfnamefont {H.}~\bibnamefont {Shishido}}, \bibinfo {author}
  {\bibfnamefont {R.}~\bibnamefont {Settai}}, \ and\ \bibinfo {author}
  {\bibfnamefont {Y.}~\bibnamefont {Onuki}},\ }\href {\doibase
  10.1103/PhysRevB.65.180504} {\bibfield  {journal} {\bibinfo  {journal} {Phys.
  Rev. B}\ }\textbf {\bibinfo {volume} {65}},\ \bibinfo {pages} {180504}
  (\bibinfo {year} {2002})}\BibitemShut {NoStop}%
\bibitem [{Pre()}]{Prefac_R_op_footnote}%
  \BibitemOpen
  \href@noop {} {}\bibinfo {note} {{ We note that prefactor of the reflection
  operator $\mathcal{R}$ is not relevant to the topological classification
  }}\BibitemShut {NoStop}%
\bibitem [{\citenamefont {Haldane}(1995)}]{Haldane_nullvector_PRL95}%
  \BibitemOpen
  \bibfield  {author} {\bibinfo {author} {\bibfnamefont {F.}~\bibnamefont
  {Haldane}},\ }\href@noop {} {\bibfield  {journal} {\bibinfo  {journal} {Phys.
  Rev. Lett.}\ }\textbf {\bibinfo {volume} {74}},\ \bibinfo {pages} {2090}
  (\bibinfo {year} {1995})}\BibitemShut {NoStop}%
\bibitem [{\citenamefont {Yoshida}\ and\ \citenamefont
  {Kawakami}(2016)}]{Yoshida_TMI2D_PRB2016}%
  \BibitemOpen
  \bibfield  {author} {\bibinfo {author} {\bibfnamefont {T.}~\bibnamefont
  {Yoshida}}\ and\ \bibinfo {author} {\bibfnamefont {N.}~\bibnamefont
  {Kawakami}},\ }\href {\doibase 10.1103/PhysRevB.94.085149} {\bibfield
  {journal} {\bibinfo  {journal} {Phys. Rev. B}\ }\textbf {\bibinfo {volume}
  {94}},\ \bibinfo {pages} {085149} (\bibinfo {year} {2016})}\BibitemShut
  {NoStop}%
\bibitem [{\citenamefont {Zhang}\ \emph
  {et~al.}(2016{\natexlab{b}})\citenamefont {Zhang}, \citenamefont {Xu},\ and\
  \citenamefont {Liu}}]{Zhang_TMI2D_PRB2016}%
  \BibitemOpen
  \bibfield  {author} {\bibinfo {author} {\bibfnamefont {R.-X.}\ \bibnamefont
  {Zhang}}, \bibinfo {author} {\bibfnamefont {C.}~\bibnamefont {Xu}}, \ and\
  \bibinfo {author} {\bibfnamefont {C.-X.}\ \bibnamefont {Liu}},\ }\href
  {\doibase 10.1103/PhysRevB.94.235128} {\bibfield  {journal} {\bibinfo
  {journal} {Phys. Rev. B}\ }\textbf {\bibinfo {volume} {94}},\ \bibinfo
  {pages} {235128} (\bibinfo {year} {2016}{\natexlab{b}})}\BibitemShut
  {NoStop}%
\bibitem [{\citenamefont {Haze}\ \emph {et~al.}(2018)\citenamefont {Haze},
  \citenamefont {Torii}, \citenamefont {Peters}, \citenamefont {Kasahara},
  \citenamefont {Kasahara}, \citenamefont {Shibauchi}, \citenamefont
  {Terashima},\ and\ \citenamefont {Matsuda}}]{Haze2018}%
  \BibitemOpen
  \bibfield  {author} {\bibinfo {author} {\bibfnamefont {M.}~\bibnamefont
  {Haze}}, \bibinfo {author} {\bibfnamefont {Y.}~\bibnamefont {Torii}},
  \bibinfo {author} {\bibfnamefont {R.}~\bibnamefont {Peters}}, \bibinfo
  {author} {\bibfnamefont {S.}~\bibnamefont {Kasahara}}, \bibinfo {author}
  {\bibfnamefont {Y.}~\bibnamefont {Kasahara}}, \bibinfo {author}
  {\bibfnamefont {T.}~\bibnamefont {Shibauchi}}, \bibinfo {author}
  {\bibfnamefont {T.}~\bibnamefont {Terashima}}, \ and\ \bibinfo {author}
  {\bibfnamefont {Y.}~\bibnamefont {Matsuda}},\ }\href {\doibase
  10.7566/JPSJ.87.034702} {\bibfield  {journal} {\bibinfo  {journal} {Journal
  of the Physical Society of Japan}\ }\textbf {\bibinfo {volume} {87}},\
  \bibinfo {pages} {034702} (\bibinfo {year} {2018})},\ \Eprint
  {http://arxiv.org/abs/https://doi.org/10.7566/JPSJ.87.034702}
  {https://doi.org/10.7566/JPSJ.87.034702} \BibitemShut {NoStop}%
\bibitem [{\citenamefont {Yoshida}\ \emph {et~al.}(2018)\citenamefont
  {Yoshida}, \citenamefont {Danshita}, \citenamefont {Peters},\ and\
  \citenamefont {Kawakami}}]{Yoshida_ZtoZ4ColdAtom_PRL18}%
  \BibitemOpen
  \bibfield  {author} {\bibinfo {author} {\bibfnamefont {T.}~\bibnamefont
  {Yoshida}}, \bibinfo {author} {\bibfnamefont {I.}~\bibnamefont {Danshita}},
  \bibinfo {author} {\bibfnamefont {R.}~\bibnamefont {Peters}}, \ and\ \bibinfo
  {author} {\bibfnamefont {N.}~\bibnamefont {Kawakami}},\ }\href {\doibase
  10.1103/PhysRevLett.121.025301} {\bibfield  {journal} {\bibinfo  {journal}
  {Phys. Rev. Lett.}\ }\textbf {\bibinfo {volume} {121}},\ \bibinfo {pages}
  {025301} (\bibinfo {year} {2018})}\BibitemShut {NoStop}%
\bibitem [{\citenamefont {Can}\ \emph {et~al.}(2021{\natexlab{a}})\citenamefont
  {Can}, \citenamefont {Tummuru}, \citenamefont {Day}, \citenamefont {Elfimov},
  \citenamefont {Damascelli},\ and\ \citenamefont {Franz}}]{Can2021}%
  \BibitemOpen
  \bibfield  {author} {\bibinfo {author} {\bibfnamefont {O.}~\bibnamefont
  {Can}}, \bibinfo {author} {\bibfnamefont {T.}~\bibnamefont {Tummuru}},
  \bibinfo {author} {\bibfnamefont {R.~P.}\ \bibnamefont {Day}}, \bibinfo
  {author} {\bibfnamefont {I.}~\bibnamefont {Elfimov}}, \bibinfo {author}
  {\bibfnamefont {A.}~\bibnamefont {Damascelli}}, \ and\ \bibinfo {author}
  {\bibfnamefont {M.}~\bibnamefont {Franz}},\ }\href {\doibase
  10.1038/s41567-020-01142-7} {\bibfield  {journal} {\bibinfo  {journal}
  {Nature Physics}\ } (\bibinfo {year} {2021}{\natexlab{a}}),\
  10.1038/s41567-020-01142-7}\BibitemShut {NoStop}%
\bibitem [{\citenamefont {Can}\ \emph {et~al.}(2021{\natexlab{b}})\citenamefont
  {Can}, \citenamefont {Zhang}, \citenamefont {Kallin},\ and\ \citenamefont
  {Franz}}]{can2021probing}%
  \BibitemOpen
  \bibfield  {author} {\bibinfo {author} {\bibfnamefont {O.}~\bibnamefont
  {Can}}, \bibinfo {author} {\bibfnamefont {X.-X.}\ \bibnamefont {Zhang}},
  \bibinfo {author} {\bibfnamefont {C.}~\bibnamefont {Kallin}}, \ and\ \bibinfo
  {author} {\bibfnamefont {M.}~\bibnamefont {Franz}},\ }\href {\doibase
  10.1103/PhysRevLett.127.157001} {\bibfield  {journal} {\bibinfo  {journal}
  {Phys. Rev. Lett.}\ }\textbf {\bibinfo {volume} {127}},\ \bibinfo {pages}
  {157001} (\bibinfo {year} {2021}{\natexlab{b}})}\BibitemShut {NoStop}%
\bibitem [{\citenamefont {Barzykin}\ and\ \citenamefont
  {Gor'kov}(2002)}]{Barzykin_Gorkov2002}%
  \BibitemOpen
  \bibfield  {author} {\bibinfo {author} {\bibfnamefont {V.}~\bibnamefont
  {Barzykin}}\ and\ \bibinfo {author} {\bibfnamefont {L.~P.}\ \bibnamefont
  {Gor'kov}},\ }\href {\doibase 10.1103/PhysRevLett.89.227002} {\bibfield
  {journal} {\bibinfo  {journal} {Phys. Rev. Lett.}\ }\textbf {\bibinfo
  {volume} {89}},\ \bibinfo {pages} {227002} (\bibinfo {year}
  {2002})}\BibitemShut {NoStop}%
\bibitem [{\citenamefont {Dimitrova}\ and\ \citenamefont
  {Feigel'man}(2003)}]{Dimitrova_Feigelman2003}%
  \BibitemOpen
  \bibfield  {author} {\bibinfo {author} {\bibfnamefont {O.~V.}\ \bibnamefont
  {Dimitrova}}\ and\ \bibinfo {author} {\bibfnamefont {M.~V.}\ \bibnamefont
  {Feigel'man}},\ }\href {\doibase 10.1134/1.1644308} {\bibfield  {journal}
  {\bibinfo  {journal} {Journal of Experimental and Theoretical Physics
  Letters}\ }\textbf {\bibinfo {volume} {78}},\ \bibinfo {pages} {637}
  (\bibinfo {year} {2003})}\BibitemShut {NoStop}%
\bibitem [{\citenamefont {Agterberg}\ and\ \citenamefont
  {Kaur}(2007)}]{Agterberg2007}%
  \BibitemOpen
  \bibfield  {author} {\bibinfo {author} {\bibfnamefont {D.~F.}\ \bibnamefont
  {Agterberg}}\ and\ \bibinfo {author} {\bibfnamefont {R.~P.}\ \bibnamefont
  {Kaur}},\ }\href {\doibase 10.1103/PhysRevB.75.064511} {\bibfield  {journal}
  {\bibinfo  {journal} {Phys. Rev. B}\ }\textbf {\bibinfo {volume} {75}},\
  \bibinfo {pages} {064511} (\bibinfo {year} {2007})}\BibitemShut {NoStop}%
\bibitem [{\citenamefont {Agterberg}(2012)}]{Agterberg2012}%
  \BibitemOpen
  \bibfield  {author} {\bibinfo {author} {\bibfnamefont {D.~F.}\ \bibnamefont
  {Agterberg}},\ }\enquote {\bibinfo {title} {Magnetoelectric effects, helical
  phases, and fflo phases},}\ in\ \href {\doibase 10.1007/978-3-642-24624-1_5}
  {\emph {\bibinfo {booktitle} {Non-Centrosymmetric Superconductors:
  Introduction and Overview}}},\ \bibinfo {editor} {edited by\ \bibinfo
  {editor} {\bibfnamefont {E.}~\bibnamefont {Bauer}}\ and\ \bibinfo {editor}
  {\bibfnamefont {M.}~\bibnamefont {Sigrist}}}\ (\bibinfo  {publisher}
  {Springer Berlin Heidelberg},\ \bibinfo {address} {Berlin, Heidelberg},\
  \bibinfo {year} {2012})\ pp.\ \bibinfo {pages} {155--170}\BibitemShut
  {NoStop}%
\end{thebibliography}%
\end{document}